\documentclass{aa}  

\pdfoutput=1

\usepackage{xcolor}
\usepackage{ulem}
\usepackage{pgffor}
\usepackage{xcolor}
\usepackage{rotating}
\usepackage{tikz}
\usepackage{morefloats}
\usepackage{float}
\usepackage{tablefootnote}
\usepackage{graphics}
\usepackage{epsfig}
\usepackage{pdfpages}
\usepackage{balance}
\usepackage[utf8]{inputenc}

\usepackage{graphicx}
\usepackage{caption}
\usepackage{subcaption}
\usepackage{txfonts}
\usepackage{csvsimple}
\usepackage{longtable}

\begin{document}

   \title{Chemical diversity of gas in distant galaxies}

   \subtitle{The metal and dust enrichment and variations within absorbing galaxies}

   \author{T. Ramburuth-Hurt \inst{1}
          \and
          A. De Cia \inst{1}%\fnmsep\thanks{Just to show the usage of the elements in the author field}
          \and
          J.-K. Krogager \inst{1, 2} 
          \and
          C. Ledoux \inst{3}
          \and
          P. Petitjean \inst{4}
          \and
          C. Péroux \inst{5,6}
          \and
          M. Dessauges-Zavadsky \inst{1}
          \and
          J. Fynbo \inst{7,8}
          \and
          M. Wendt \inst{9}
          \and
          N. F. Bouch{\'e} \inst{2}
          \and
          C. Konstantopoulou \inst{1}
          \and
          I. Jermann \inst{1}
          }

   \institute{Department of Astronomy, University of Geneva, Chemin Pegasi 51, Versoix, Switzerland %1
   \and
   Univ. Lyon, Univ. Lyon1, Ens de Lyon, CNRS, Centre de Recherche Astrophysique de Lyon UMR5574, 69230 Saint-Genis-Laval, France   %2
   \and
   European Southern Observatory, Alonso de C{\'o}rdova 3107, Vitacura, Casilla 19001, Santiago, Chile %3
   \and
   Institut d’Astrophysique de Paris, Sorbonne Universités and CNRS, 98bis boulevard Arago, 75014 Paris, France %4
   \and
   European Southern Observatory, Karl-Schwarzschild-Str. 2, D-85748, Garching, Germany %5
   \and
   Aix Marseille Université, CNRS, LAM (Laboratoire d’Astrophysique de Marseille) UMR 7326, 13388, Marseille, France %6
   \and
   Cosmic Dawn Center (DAWN), Copenhagen, Denmark %7
   \and
   Niels Bohr Institute, University of Copenhagen, Jagtvej 128, DK-2200 Copenhagen N, Denmark %8
   \and
   Institut für Physik und Astronomie, Universität Potsdam, Karl-Liebknecht-Str. 24/25, 14476 Golm, Germany %9
   } 
   
              %\email{tanita.ramburuth-hurt@unige.ch}
         %\and
             %University of Geneva \\
             %\email{annalisa.decia@unige.ch}
             %\thanks{The university of heaven temporarily does not accept e-mails}

   \date{Received xxx; accepted yyy}

% \abstract{}{}{}{}{} 
% 5 {} token are mandatory
\abstract{The chemical composition of gas in galaxies can be measured in great detail from absorption spectroscopy. By studying gas in galaxies in this way, it is possible to investigate the small and faint galaxies, which are the most numerous in the universe. In particular, the chemical distribution of gas in absorbing systems gives us insight into cycles of gas in and around galaxies. In this work we study chemical enrichment within 64 Damped Lyman-$\alpha$ Absorption (DLA) systems between redshifts $1.7 < z < 4.2$. We use high-resolution spectra from VLT/UVES to infer dust depletion from relative abundances of several metals. Specifically, we perform a component-by-component analysis within DLAs, and characterise variations in their chemical enrichment. Unlike hydrogen, the metal columns can be characterised for individual components. We use them to derive the dust depletion, which is an indicator for chemical enrichment. Our main results are as follows. Firstly, we find that some DLAs are chemically diverse within themselves (with the measure of dust depletion -- [Zn/Fe]$_{\mathrm{fit}}$ -- ranging up to 0.62 dex within a single system), suggesting that the absorbing gas within these galaxies is chemically diverse. Secondly, although we do not find a clear trend of decreasing dust depletion with redshift, we do see that the most chemically enriched systems are at lower redshifts. We also observe evidence for dust-poor components at all redshifts, which may be due to the accretion of pristine gas onto galaxies. By combining the chemical and kinematic properties of the individual gas components, we observe potential signatures of infalling gas, with low depletion at velocities below $\sim 100$ km/s, and outflows, with high depletion and velocities of $\sim$ 600 km/s. Finally, we find over-abundances of $\alpha$-elements (an enhancement of $\sim$ 0.3 dex) and under-abundances of Mn in several gas components, which is likely a signature of core-collapse supernovae nucleosythesis in the ISM. We observe these effects mostly at lower levels of chemical enrichment.}


   \maketitle
%
%-------------------------------------------------------------------

\section{Introduction} \label{sec. intro}
    
    Chemical abundances in galaxies across cosmic time play a crucial role in our understanding of galactic baryon cycles \citep{MaiolinoMannucci2019_metals}. In particular, damped Lyman-$\alpha$ absorption systems (DLAs), defined by having a neutral hydrogen ($\ion{H}{i}$) column density of $N (\ion{H}{i}) > 10^{20.3} $cm$^{-2}$ \citep[e.g.,][]{Wolfe+1986_DLAs} are excellent laboratories for studying chemical abundances in galaxies out to high-$z$. This criteria ensures shielding of the gas cloud from ionising radiation. DLAs are $\ion{H}{i}$-rich absorption systems, and are associated with galaxies with a wide range of masses \citep[$10^6 - 10^{11}$~M$_{\odot}$ e.g.][]{Christensen+2014_DLAs, Augustin+2018_DLA-masses}. This includes low-mass galaxies, which are the most numerous in the Universe \citep[e.g.][]{Grazian+2015}, but are very difficult to observe in emission \citep[e.g.][]{Lowenthal+1995_DLA-emiss-z1-25, Bouche+2001_DLA-emiss-z06, Christensen+2009_DLA-emiss-z34,  York+2012_DLA-emiss-z038, Krogager+2017_DLAsEmiss}. %, and are suspected to be the main fuel for star formation \citep{Peroux+2020_metalscycle}. 
    Not only are DLAs the largest cosmic reservoirs of $\ion{H}{i}$, but also of cosmic metals up to high $z$ \citep{Peroux+2020_metalscycle}. Therefore, these absorbing systems can give us great insight into the evolution and composition of both galaxies and the neutral gas in the Universe.
    
    The DLA population is expected to be quite diverse in principle, due to a range of different masses, star-formation histories (and therefore nucleosynthesis enrichment), and dust content \citep[e.g.][]{Prochaska+2003_age-met-rel, Noterdaeme+2008, Christensen+2014_DLAs}. Further, the metal content of DLA systems increases across cosmic time \citep{Prochaska+2002, Rafelski+2012, DeCia+2018}, due to the production of metals by different nucleosynthetic processes. There is also growing observational evidence that the interstellar medium (ISM) in galaxies is not chemically homogeneous, for example from observations of the Milky Way \citep[][]{Welty+2020, DeCia+2021_Nature} and nearby galaxies \citep[e.g.,][]{SanchezDeAlmeida+2015_dwarfgals, Wang+2021}.
    %\citet{Guber+2016, Prochaska+2003_age-met-rel} and others show that the metal content of DLA systems tend to increase in general across cosmic time, due the expulsion of metals from supernovae \citep{Peroux+2020_metalscycle}.
    
    While most studies of DLAs have focused on their global properties, variations of metal content inside individual systems have been observed in a few cases through component-by-component analysis of the absorption-line profiles \citep{Prochaska2003_uniformDLAs, Dessauges-Zavadsky+2006, Rodriguez+2006_6DLAs, Wiseman+2017_GRBflows, deUgartePostigo+2018, Noterdaeme+2017, Guber+2018}. These gas components, which have been identified using Voigt-profile fitting techniques, relate to clouds of gas that move at different velocities. While we sometimes call them ``clouds", individual components may still have relatively large velocity broadening, indicating that they are probably a mix of multiple clouds themselves. With very high resolution (e.g. 1 km/s), it is possible to isolate individual gas components such as the translucent cloud observed in the Milky Way by \citet{Welty+2020}. However, while there is general agreement on the properties of gas in simulations and observations along whole line of sight (LOS) profiles \citep{Marra+2021_simulations_whole}, on a component-by-component basis, there can be degeneracies between physical parameters such as temperature, $N(\mathrm{H})$, and metallicity \citep{Marra+2022_simulations_comp}. It is therefore important to note that, when studying DLAs on a component-by-component basis, components could indeed be a mix of sub-components with different physical parameters.

    %\textcolor{blue}{A caveat with studying DLAs is that it is not easy to tell whether the gas we are probing is indeed part of the ISM, or whether it is gas in the surrounding medium, or circumgalactic medium (CGM). }
    
    %In particular, they showed that for low-ionisation phases of the CGM there is a 52\% chance that an Si II absorption component contains a single gas cloud. Further, they show that there are multiple possible hydrogen column densities, temperatures and metallicities (with Si II as a proxy) for a gas components identified through Voigt profile fitting techniques.
    
    In a component-by-component analysis, the total metallicity cannot be characterised because the information on the $\ion{H}{i}$ is normally obtained from the Lyman-alpha line, which is a broad blend of the individual components. Despite this challenge, it is still viable to obtain information of the metal enrichment of components by studying their dust depletion. This is the phenomenon whereby metals form dust grains and are no longer observable in the gas phase \citep{Field1974, Phillips+1982, Jenkins+1986, SavageSembach1996, Jenkins2009, DeCia+2016, Roman-Duval+2021, DeCia+2021_Nature}. The depletion of different elements correlate with each other to varying degrees depending on how easily they form dust grains \citep{Jenkins2009, DeCia+2016, Roman-Duval+2021}. It has been shown that the amount of dust in a system is indicative, to some extent, of its total metallicity \citep{Ledoux+2002, Vladilo+2002, Noterdaeme+2008, DeCia+2016}. Therefore, although we cannot accurately know the total metallicity of individual gas clouds in DLAs, it is still possible to study their chemical enrichment through characterising their dust depletion. 
    % PUT INTO METHODS SECTION
    %In general, the overall strength of dust depletion in a gas cloud can be estimated from the observed relative abundances of two metals that have very different refractory properties: $[X/Y] = \log(N(X)/N(Y)) - \log(N(X)_\odot/N(Y)_\odot)$. Metals $X$ and $Y$ are chosen such that $X$ is refractory (forms dust grains readily) and $Y$ is volatile (does not form dust grains readily). 
   
    When studying chemical abundances, it is also important to take into consideration possible nucleosynthetic effects, which could impact observed metal abundances.
    %An additional phenomenon that has an effect on chemical abundances we observe over cosmic time is nucleosynthesis. 
    Here, we refer to nucleosynthetic effects as over- or under-abundance of certain metals due to their formation processes. $\alpha$-elements, e.g. S, Si, O, Mg, Ti, Ca, tend to be more abundant in the early part of a galaxy's evolution \citep{Tinsley1997_nuc, McWilliam1997_nucleo}, because they are formed mostly by core-collapse supernovae, and these are the first supernovae to explode \citep{Nomoto+2006_nucleo}. This over-abundance of $\alpha$-elements decreases with time, when metals formed mostly by Type Ia supernovae (e.g. Fe, Ni, Co, Mn) begin to contribute more to the overall metal content of the galaxy. %Observations of stellar chemical abundances of the Local Group have shown that, due to nucleosynthetic effects, older stellar populations with a lower metallicity have an over-abundance of $\alpha$-elements, or $\alpha$-element enhancement \citep{Lambert1987_nucleo, McWilliam1997_nucleo, Tolstoy2009_nucleo, Boer+2014_nucleo}. 
    The observation of $\alpha$-element enhancement is mostly limited to the Local Group from measurements of chemical abundances in stars, where older stellar populations with a lower metallicity have an over-abundance of $\alpha$-elements and an under-abundance of Mn \citep{Lambert1987_nucleo, McWilliam1997_nucleo, Tolstoy2009_nucleo, Boer+2014_nucleo}. At higher redshift, it is more challenging to observe $\alpha$-element enhancement. In some cases it is possible to observe them as signatures in the ISM of galaxies \citep{Cullen+2015} and metal-poor DLAs, where the effects of dust depletion are minimal and it is therefore more straightforward to disentangle dust depletion from nucleosynthetic effects \citep[e.g.,][]{Dessauges-Zavadsky+2002_Mn-underabundance, Dessauges-Zavadsky+2006, Cooke+2011_nucl-DLAs-metal-poor, Becker+2012_nucl,Ledoux+2002_nucl-DLAs, DeCia+2016}. When gas components have a higher metallicity, however, it is important to be able to make considerations for both dust depletion and nucleosynthesis. 
    
    Observations of the galactic gas cycles in the form of signatures of outflowing and inflowing gas from the circumgalactic medium (CGM) has been seen in the Milky Way in the form of high velocity clouds (HVCs) \citep[e.g.][]{Fox+2019_HVCs}. In extragalactic studies, correlations between metallicity tracers and azimuthal angle of galaxies have been found, indicating metal-rich outflowing gas along the minor axis \citep[e.g.][]{Bouche+2012_MgII_galflows, Peroux+2020_angulardep, Wendt+2021_gasflows}. In DLAs, however, this task is more difficult without the emission counterpart of the galaxy: we do not know their morphologies. It is not straightforward to assess whether the DLA gas is indeed part of the ISM, or whether it is gas in the CGM. Indeed, there is evidence from studies of DLA samples \textit{with} galaxy emission counterparts that DLA absorption systems are built up of gas extending to the full halo of their host galaxies \citep[e.g.][]{Christensen+2019_darkmatterinDLAs}. It is possible to use the combination of the chemical enrichment of individual gas components with their relative velocities to start investigating signatures of inflow and outflow to and from DLA galaxies.

    In this work we do a component-by-component analysis of the column densities measured from absorption-line spectra of 70 DLAs by \citet{DeCia+2016}. We analyse the metal and dust-depletion properties of individual gas components within DLAs over the redshift range $z_{abs} =  1.7 - 4.2$ and investigate their chemical diversity and the possible effects of nucleosynthesis. We combine the chemical enrichment properties with the kinematics of individual components to investigate possible evidence for inflow and outflow of gas. We denote the relative abundances of metals $X$ with respect to the metal $Y$ by $[X/Y] = \log(N(X)/N(Y)) - \log(N(X)_\odot/N(Y)_\odot)$, where $\log(N(X))$ is the column density of $X$, measured in units of cm$^{-2}$, and $N(X)_\odot$ is the solar abundance of $X$ \citep{Asplund+2021_sol-abnds}, and using the recommendations of \citep{Lodders+2009_solar-abd}). This paper is structured as follows: in Section \ref{sec. method} we describe the methodology used to study dust-depletion; in Section \ref{sec. res-disc} we present our results and discuss their implications; and finally, we present our concluding remarks in Section \ref{sec. concl}.

%--------------------------------------------------------------------
\section{Methodology}\label{sec. method}

%-------------------------------------- 

%
   We use the column density measurements of \citet{DeCia+2016}, who studied the abundances and relative abundances for a sample of 70 DLAs towards QSOs at redshifts $1.7 \leq z_{\mathrm{abs}} \leq 4.2$ using data from the Very Large Telescope (VLT) and the Ultraviolet and Visual Echelle Spectrograph (UVES). The average resolution of the spectra is $R \sim$ 40 000. All DLA systems have at least $\log{N(\mathrm{H})} \geq 20$, to ensure overall shielding from ionising radiation. \citet{DeCia+2016} studied depletion sequences for systems with $\log{N(\mathrm{H})} \geq 20$, ensuring that ionisation effects do not play a significant role.  However, the hydrogen column density is not available for individual clouds, and therefore we cannot ensure the criteria of $\log{N(\mathrm{H})} \geq 20$. Here we make the assumption that the depletion sequences are still valid at the level of individual components because we still observe strong correlations in their depletion sequences. Figure A.1. shows that indeed the depletion sequences for the individual components are very similar to the depletion sequences of \citet{DeCia+2016}, with an overall larger scatter. This scatter possibly holds information on ionisation and nucleosynthetic effects, which are important aspects of this work.
   
   In this work, we perform a component-by-component analysis of these line profiles. Individual gas components were identified by \citet{DeCia+2016} through Voigt profile fitting to the spectral lines available. To avoid misidentification of individual components as well as the effects of blending between lines, several lines were modelled simultaneously, both transitions from the same ion and from different ions, to determine individual components. This means that we are not necessarily identifying individual separated clouds, but rather groups of clouds at similar velocities. The zero velocity is chosen to be the component with the highest $\ion{Fe}{ii}$ column density because we assume this to be a component closest to the inner parts of the galaxy. The full data set is listed in Table F1 of \citet{DeCia+2016}. From the 70 DLA systems, we identified 64 DLAs with one or more components for which we could measure the level of dust depletion within an error of 0.5 dex. From this, there are 25 DLAs with two or more components which we use to study the chemical diversity within systems.
   
    \subsection{Measuring dust depletion from depletion patterns} \label{sec. measuring DD}

    Several methods have been used in the literature to estimate the amount of dust depletion in gas clouds \citep[for example][]{Jenkins2009, DeCia+2016, DeCia+2021_Nature}. The overall strength of dust depletion can be estimated from the differences in the observed column densities of metals that have different refractory properties, regardless of the H content. The analysis in this work is based on the analysis of relative abundances developed in \citet{DeCia+2016}, who found correlations between the relative abundances of metals and the relative abundances of zinc and iron, [Zn/Fe] -- also called depletion sequences. We describe the method we use in this work below.

    The depletion of an element $X$ in a system, $\delta_X$, can be expressed as
    \begin{equation} \label{eq:meth1}
        \delta_X = A2_X + B2_X \times \mathrm{[Zn/Fe]}_{\rm fit},
    \end{equation}
    
    where [Zn/Fe]$_{\rm fit}$ is the overall strength of depletion, or the depletion factor. $A2_X$ and $B2_X$ are coefficients of the depletion sequence fits in \citet{Konstantopoulou2022} and are specific for each element (see Table \ref{table: depletion sequences De Cia 2016}). In principle, $A2_X$ can be assumed to be zero because we expect to have no dust depletion when there is no depletion, [Zn/Fe]$_{\rm fit} = 0$ (i.e. positive depletion is unphysical). In this work we make use of [Zn/Fe]$_{\rm fit}$ instead of [Zn/Fe], as is used in \citet{DeCia+2016}, because this new quantity is derived from information of all the available metals and is therefore more solid. We show a comparison of the two procedures in Figure \ref{fig:ZnFe_vs_fit}.
    
    $B2_X$ can be interpreted as a \textit{refractory index}, which is closer to zero for elements that do not form dust grains easily (volatile elements), and further from zero for elements that are able to form dust grains more readily (refractory elements).  

    The total dust-corrected abundance of metal $X$ is given by
    \begin{equation} \label{eq:meth2}
       [X/\mathrm{H}]_{\mathrm{tot}} = [X/\mathrm{H}] - \delta_X, 
    \end{equation}
    where $[X/\mathrm{H}]$ are the observed gas-phase abundances. Combining these equations, we obtain a linear relation:
    \begin{align}
        y_i =& a_i + B2_X \times \rm [Zn/Fe]_{fit},
    \end{align}
    
    which is a linear relation, $y_i = a_i + bx$, where 
    \begin{align}
        y_i =& \log{N(X)} - X_{\odot} + 12 - A2_X \\
        a_i =& [\mathrm{M/H}]_{\mathrm{tot}} + \log{N(\mathrm{H})}
    \end{align}
    
    The value of $y_i$ represents the amount of each metal, normalised by its solar abundance. We use the notation $y_i$, instead of $y$, to highlight the difference from the notation of \citet{DeCia+2021_Nature}. Here $y_i$ does not represent abundances, but rather normalised metal column densities, which can be characterised in individual gas components $i$. If there is no dust depletion in a system, or any additional sources of variation, $y_i$ for each metal would be similar. If there is only dust depletion at play, we would expect the $y_i$ values to align on a straight line with a gradient [Zn/Fe]$_{\rm fit}$, which represents the overall strength of depletion. The $y$-intercept of this straight line, $a_i = \mathrm{[M/H]_{tot}} + \log{N(\mathrm{H})}$, represents an ``equivalent metal column density''. This is the total amount of metals after correcting for dust depletion. Additional deviations to the straight line could be due to other processes, such as nucleosynthesis (e.g. alpha-element enhancement). 
    
    In this work, we perform two sets of linear fits to the depletion patterns. In the first we fit a straight line to all the available metals, and in the second we include only Cr, Fe, Zn, and/or P (excluding $\alpha$-elements and Mn; which we also refer to as \textit{NAM} for ``non-$\alpha$ elements and Mn''). In both cases, we require at least three constrained data points for a linear fit to the data and we do not include the upper limits in the fitting procedure. The second approach highlights any deviations due to nucleosysnthesis and ensures that these are not biasing the estimate of the dust depletion. We obtained the depletion factors [Zn/Fe]$_{\rm fit}$ and equivalent metal columns, $\mathrm{[M/H]_{tot}} + \log{N(\mathrm{H})}$ from the linear fits to the depletion patterns for individual gas components within the DLAs in our sample. We used the Python package ODR to fit the lines, which uses orthogonal distance regression (ODR) and considers uncertainties on both $x$ and $y$.

    \subsection{Deviations from depletion patterns}
    
    Deviations from the linear fits to the depletion patterns could occur for at least two possible reasons: ionisation and/or nucleosynthetic effects. Because we do not know if the H column density is sufficient to shield each individual gas cloud from ionising radiation, we could expect ions with lower ionisation potentials to be more easily ionised and fall below the linear fits to the depletion patterns. On the other hand, if all the metals were evenly ionised, there would not be any deviations from the straight line, and the depletion factor [Zn/Fe]$_{\rm fit}$ would still be an accurate representation of the dust depletion in the system. We investigate the possibility of ionisation effects by comparing ionisation potential to deviation from the depletion pattern in Section \ref{sec. results ion}. If ionisation effects do occur in our sample, then we expect some correlation between ionisation potential and the amount of deviation from the depletion pattern.
    
    Secondly, deviations from the linear fit to the depletion pattern could also be due to nucleosynthetic effects. For example, some galaxies could have an enhancement of $\alpha$-elements due to the contribution of $\alpha$-elements from Type II supernovae. In particular, if this is the case for some for our galaxies, we may observe over-abundances from Si, S, Mg and O, and an under-abundance of Mn similarly to what is observed in the Milky Way and nearby dwarf galaxies. If nucleosynthetic effects are indeed present in our data, then including these four metals in the fitting procedure could produce an inaccurate fit to the depletion patterns. In an attempt to investigate any nucleosynthetic effects, we exclude the $\alpha$-elements and Mn from the depletion pattern fit. We describe deviations from the linear fit due to nucleosynthesis as the difference between the deviation of $X$ and the deviation of Fe: [X/Fe]$_{\mathrm{nucl}} = \mathrm{X}_{\mathrm{dev}} - \mathrm{Fe}_{\mathrm{dev}}$, where the deviation is the vertical distance from the observed point to the linear fit to the depletion pattern. In other words, [X/Fe]$_{\mathrm{nucl}}$ describes the relative abundances observed in the gas after correcting for dust depletion, and likely representing the product of stellar nucleosynthesis on the gas.

% ===========================================================================
   %\input{tables/DepSeq_DeCia+16}

    \begin{table}
        \caption{Coefficients of the depletion sequences for metals $X$ \citep{Konstantopoulou2022}. The values for P that we use here are being revised and updated in Konstantopoulou et al. (in prep.)}             % title of Table
        \label{table: depletion sequences De Cia 2016}      % is used to refer this table in the text
        \centering                          % used for centering table
        \begin{tabular}{c | c c}        % centered columns (4 columns)
            \hline\hline                 % inserts double horizontal lines
            $X$ & $A2$ & $B2$ \\    % table heading 
            \hline                        % inserts single horizontal line
            $\delta_{\mathrm{Zn}}$ & 0.00 $\pm$ 0.01    & $-0.27$ $\pm$ 0.03 \\
            $\delta_{\mathrm{O}}$  & 0.00 $\pm$ 0.00    & $-0.20$ $\pm$ 0.05 \\
            $\delta_{\mathrm{P}}$  & 0.08 $\pm$ 0.05    & -0.26 $\pm$ 0.08  \\
            $\delta_{\mathrm{S}}$  & 0.01 $\pm$ 0.02    & $-0.48$ $\pm$ 0.04  \\
            $\delta_{\mathrm{Si}}$ & $-0.04$ $\pm$ 0.02 & $-0.75$ $\pm$ 0.03 \\ 
            $\delta_{\mathrm{Mg}}$ & 0.01 $\pm$ 0.03    & $-0.66$ $\pm$ 0.04 \\ 
            $\delta_{\mathrm{Mn}}$ & 0.07 $\pm$ 0.02    & $-1.03$ $\pm$ 0.03 \\ 
            $\delta_{\mathrm{Cr}}$ & 0.12 $\pm$ 0.01    & $-1.30$ $\pm$ 0.01 \\ 
            $\delta_{\mathrm{Fe}}$ & $-0.01$ $\pm$ 0.03 & $-1.26$ $\pm$ 0.04 \\ 
             
            \hline                                   %inserts single line
        \end{tabular}
    \end{table}

% ===========================================================================
    
\section{Results and discussion}\label{sec. res-disc}
 
    \subsection{Diversity of dust depletion within DLA systems} \label{sec. diversity of dust depletion}

\begin{sidewaysfigure*}
    \centering
    \includegraphics[width=\textwidth]{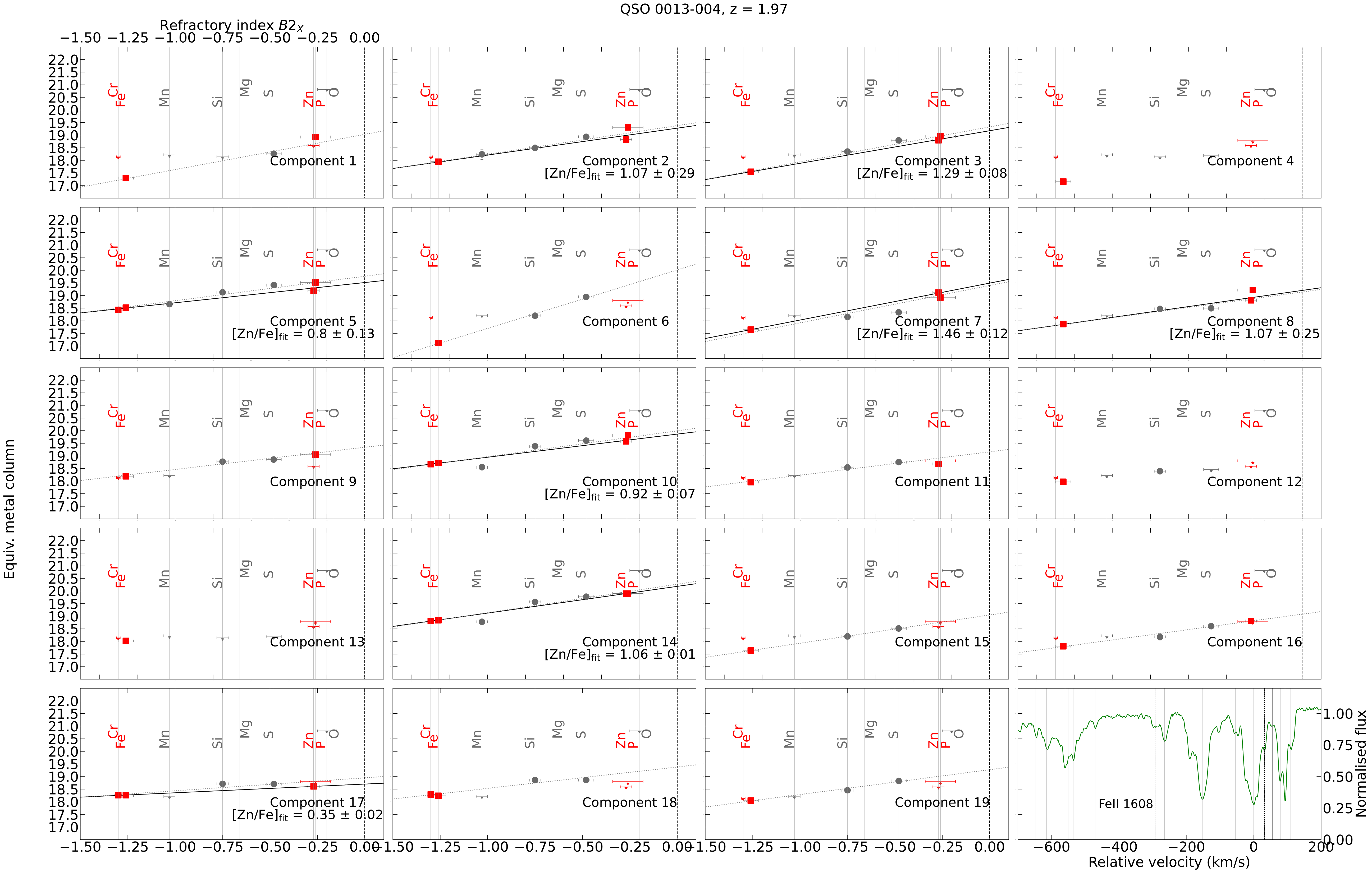}
    \caption{The depletion pattern and fitted straight lines to DLA system QSO~0013-004 with at least 19 individual gas components. Components are numbered increasingly with increasing velocities (i.e., number one being the bluest component). Here the solid black line shows the linear fit to the depletion pattern for only the non-$\alpha$ elements Cr, Fe, Zn, P (in red squares). The depletion factor [Zn/Fe]$_{\rm fit}$ is the slope of the fit. The dotted line shows the linear fit to the depletion patterns when all the available metals are included. The dashed vertical line is where $B2_X = 0$. The last panel shows the spectrum of the quasar, and indicates the positions of the individual components as determined with a Voigt-profile fitting technique. We see diversity in the depletion strengths of the components. The dashed black line shows the components in which H$_2$ is detected by \citet{Petitjean+2002_Q0013-004}.}
    \label{fig:depl-pattern_sys3}
\end{sidewaysfigure*}

\begin{figure}
    \centering
    \includegraphics[width=0.425\textwidth]{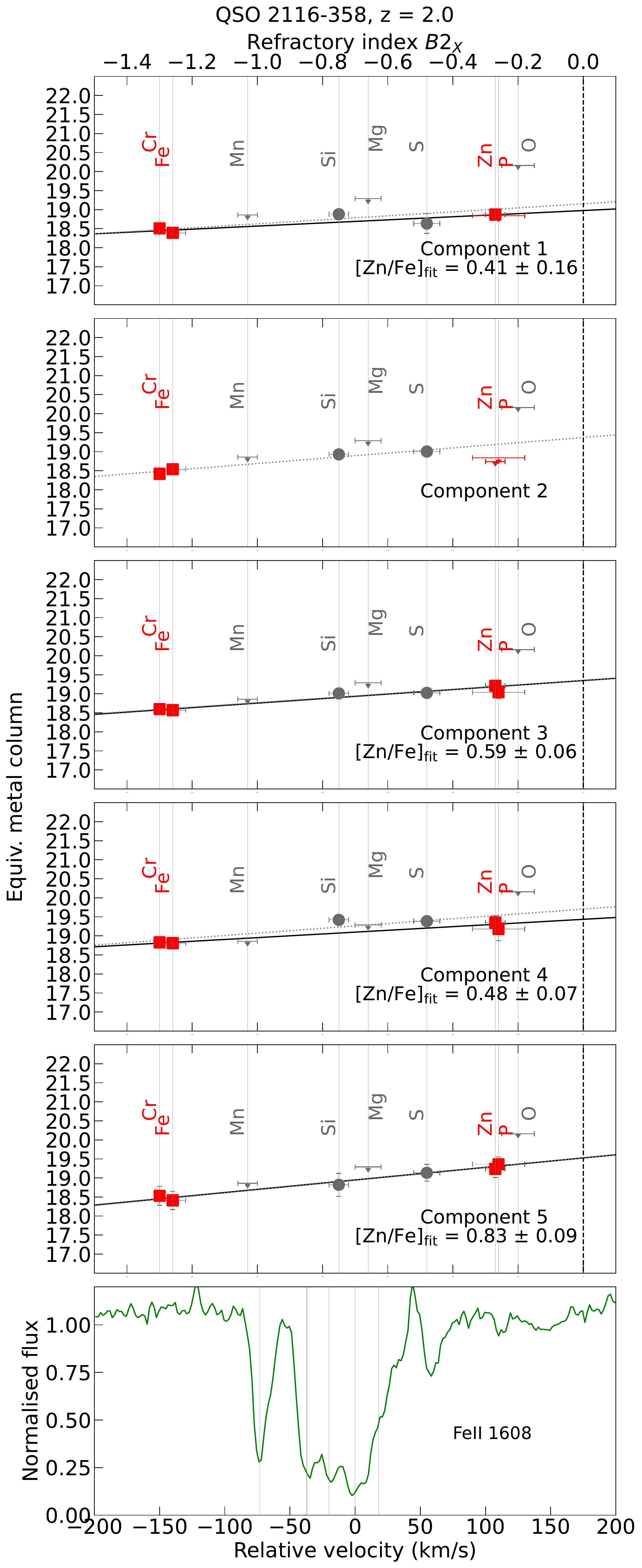}
    \caption{Same as Figure \ref{fig:depl-pattern_sys3}, but for QSO 2116-358. The label for relative velocity only refers to the bottom panel. }
    \label{fig:depl-pattern_sys55}
\end{figure}

\begin{figure}
    \centering
    \includegraphics[width=0.42\textwidth]{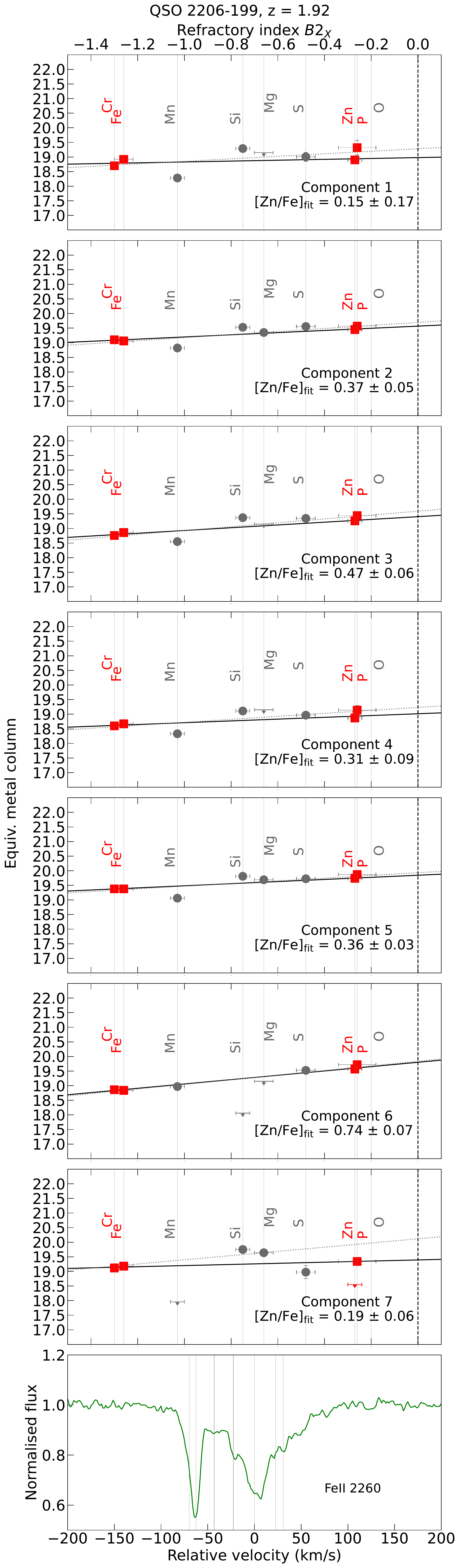}
    \caption{Same as Figure \ref{fig:depl-pattern_sys3}, but for QSO 2206-199. The label for relative velocity only refers to the bottom panel. }
    \label{fig:depl-pattern_sys61}
\end{figure}

\begin{figure}
    \centering
    \includegraphics[width=0.425\textwidth]{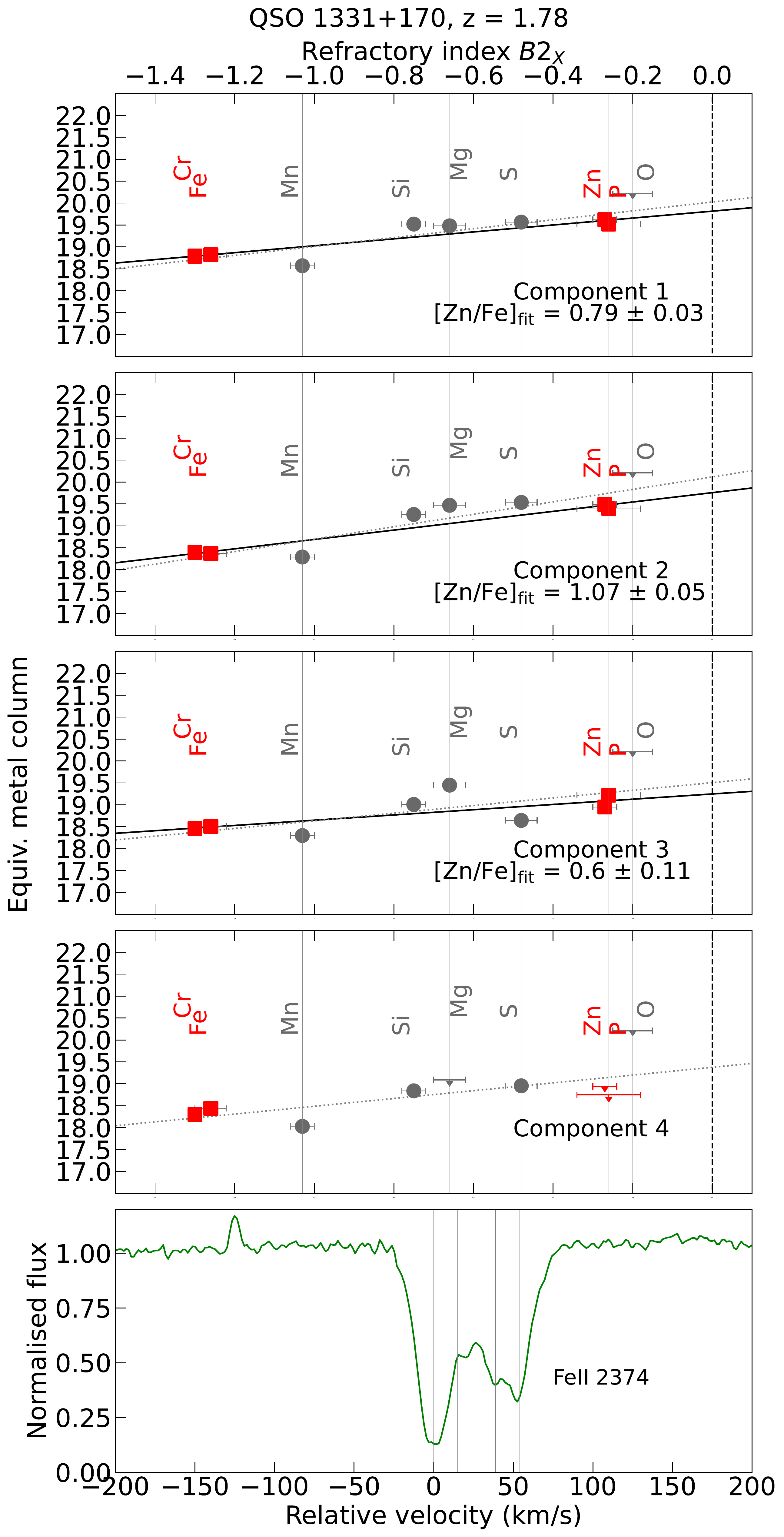}
    \caption{Same as Figure \ref{fig:depl-pattern_sys3}, but for QSO 1331+170. The label for relative velocity only refers to the bottom panel. }
    \label{fig:depl-pattern_sys41}
\end{figure}

    In Figures \ref{fig:depl-pattern_sys3} to \ref{fig:depl-pattern_sys41} we present the depletion patterns and their linear fits for obtaining the overall strength of dust depletion [Zn/Fe]$_{\mathrm {fit}}$, and equivalent metal columns $\mathrm{[M/H]_{tot}} + \log{N(\mathrm{H})}$ for individual components in systems towards QSO~0013-004, QSO~2116-358, QSO~2206-199 and QSO~1331+170 respectively. Figures \ref{fig:depl-pattern_sys3} and \ref{fig:depl-pattern_sys55} are selected to highlight chemical diversity in the level of depletion within single DLA systems, while Figures \ref{fig:depl-pattern_sys61} and \ref{fig:depl-pattern_sys41} highlight the deviations due to nucleosynthesis. The depletion patterns for the remaining individual systems are included in the Appendix. 

    Figures \ref{fig:dust-dep-by-DLA_all} and \ref{fig:dust-dep-by-DLA_NA} show the diversity in dust depletion across our DLA population, when we perform the linear fits to the depletion patterns with and without the $\alpha$-elements and Mn respectively. In 6 cases, fewer than three metals are constrained, so we do not perform a linear fit to the depletion patterns to ensure robustness of this analysis. We have not included these cases in Table \ref{tab:data}, and this leaves us with 64 DLA systems of the total of 70.
    
    The system towards QSO~0013-004 has the largest range in depletion, where the least depleted component has [Zn/Fe]$_{\mathrm {fit}} = 0.35 \pm 0.004$ dex, and the most depleted has [Zn/Fe]$_{\mathrm {fit}} = 1.43 \pm 0.19$ dex. In Table \ref{tab:data} we present the values for [Zn/Fe]$_{\mathrm {fit}}$ and $\mathrm{[M/H]_{tot}} + \log{N(\mathrm{H})}$ for each DLA system, both when fitting to all available metals and when we exclude the $\alpha$-elements and Mn.

    We include a statistical measure of diversity within each system by performing a \textit{z-test}. For each component within the system, we calculate the weighted difference between the depletion and that of the reference component:
    
    \begin{align}
        \sigma_{\mathrm{z-test}} = \dfrac{|\mathrm{[Zn/Fe]}_{\mathrm{fit}, i} - \mathrm{[Zn/Fe]}_{\mathrm{fit, ref}}|}{\sqrt{\sigma_{\mathrm{[Zn/Fe]}_{\mathrm{fit}, i}}^2 + \sigma_{\mathrm{[Zn/Fe]}_{\mathrm{fit, ref}}}^2}}.
    \end{align}
    
    We choose the reference to be the component with the minimum uncertainty on its depletion, so as to optimise the sensitivity of the z-test\footnote{The test is sensitive to the chosen reference point. We find that the use of the maximum uncertainty produces fewer (6) components with $\sigma_{\mathrm{z-test}} \geq 3$. One could also choose the component with the minimum (or maximum) depletion as a reference in order to maximise the difference in depletion between the minimum/maximum. However, in case the component with minimum/maximum depletion has a very large uncertainty, this choice would not capture diversity between all the other components.}. We plot $\sigma_{\mathrm{z-test}}$ for each component in each system in Figure \ref{fig:z-test}. There are 10 systems with $\sigma_{\mathrm{z-test}} \geq 3\sigma$, implying statistically significant diversity in dust depletion. Namely, these are systems towards QSO~0013-004 ($z = 1.97$), QSO~0058-292 ($z = 2.67$), QSO~0405-443 ($z = 1.91$), QSO~0528-250 ($z =2.81 $), QSO~1157+014 ($z = 1.94$), QSO~1223+178 ($z = 2.47$), QSO~1331+170 ($z = 1.78$), QSO~1444+014 ($z = 2.09$), QSO~2206-199 ($z = 1.92$), and QSO~2243-605 ($z = 2.33$). We observe diversity in dust depletion within these systems regardless of whether Si, S, Mg, O and Mn are included in the fit to the depletion patterns. The remaining 15 DLA systems seem to be chemically homogeneous within the limit of $3\sigma$. In Section \ref{sec. kinematics} we discuss further the depletion distribution of 4 of the systems that have statistically significant variations of depletion strength alongside their kinematics.
    
    The depletion factor is broadly correlated with the metallicity of DLAs \cite[e.g.][]{Noterdaeme+2008, DeCia+2016}, and in general the amount of dust in the ISM can be considered as a tracer of metal enrichment. The relation between the dust depletion and metallicity is theoretically expected in galaxies where dust is mostly built through grain growth in the ISM \citep{Triani+2020_dustcosmictime}. Thus, variations that we observe in the depletion factor in individual absorbing clouds indicate variations in the metal enrichment within distant galaxies. These results add further evidence (and at higher redshifts) that gas in galaxies can be, in general, chemically inhomogeneous, as observed in the recent years in the Milky Way and other nearby galaxies \citep[e.g.][]{DeCia+2021_Nature, SanchezDeAlmeida+2015_dwarfgals}. This inhomogeneity could have several causes. One could be local variations in star formation within the disk of the galaxy itself. There is evidence for metallicity gradients in galaxies, out to $z \sim 3$ \citep[e.g.][]{Cresci+2010_vel-grads-z3, Patricio+2019_vel-grad-z1, Kreckel+2019, Arellano+2020}. Another could be the infall and ouflow of gas to and from the galaxy \citep[e.g.][]{Fox+2017, Tumlinson+2017}. \citet{Wendt+2021_gasflows} showed the first chemical characterisation of the circumgalactic medium indicating dust and metal-rich outflows and dust and metal-poor infalling gas using a similar technique to the one we use in this work. Diversity of depletion could also be due to local variations in metal production, and dust production and destruction within the galaxy itself \citep{Dessauges-Zavadsky+2017_galclumps, Triani+2020_dustcosmictime, Slavin+2020}. 
    
% ==================================================
    %\input{figures/depl-patterns_sept22.tex}

% ==================================================
    
    %\input{figures/ZnFe-by-DLA_all}

    \begin{figure*}
     \centering
     \begin{subfigure}[b]{\linewidth}
         \centering
         \includegraphics[width=\textwidth]{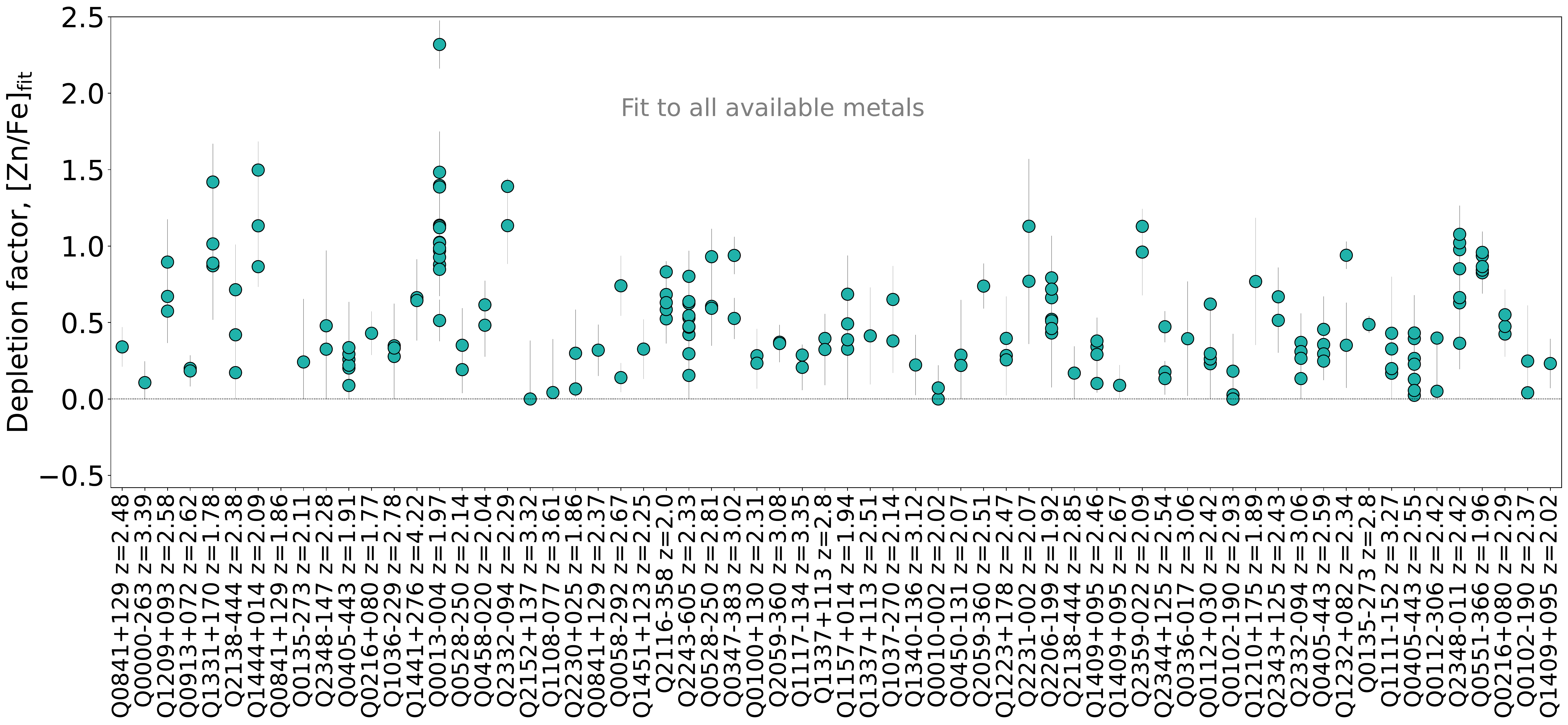}
         \caption{Depletion strengths calculated with a linear fit to all available metals. }
         \label{fig:dust-dep-by-DLA_all}
     \end{subfigure}
     \hfill
     \begin{subfigure}[b]{\textwidth}
         \centering
         \includegraphics[width=\textwidth]{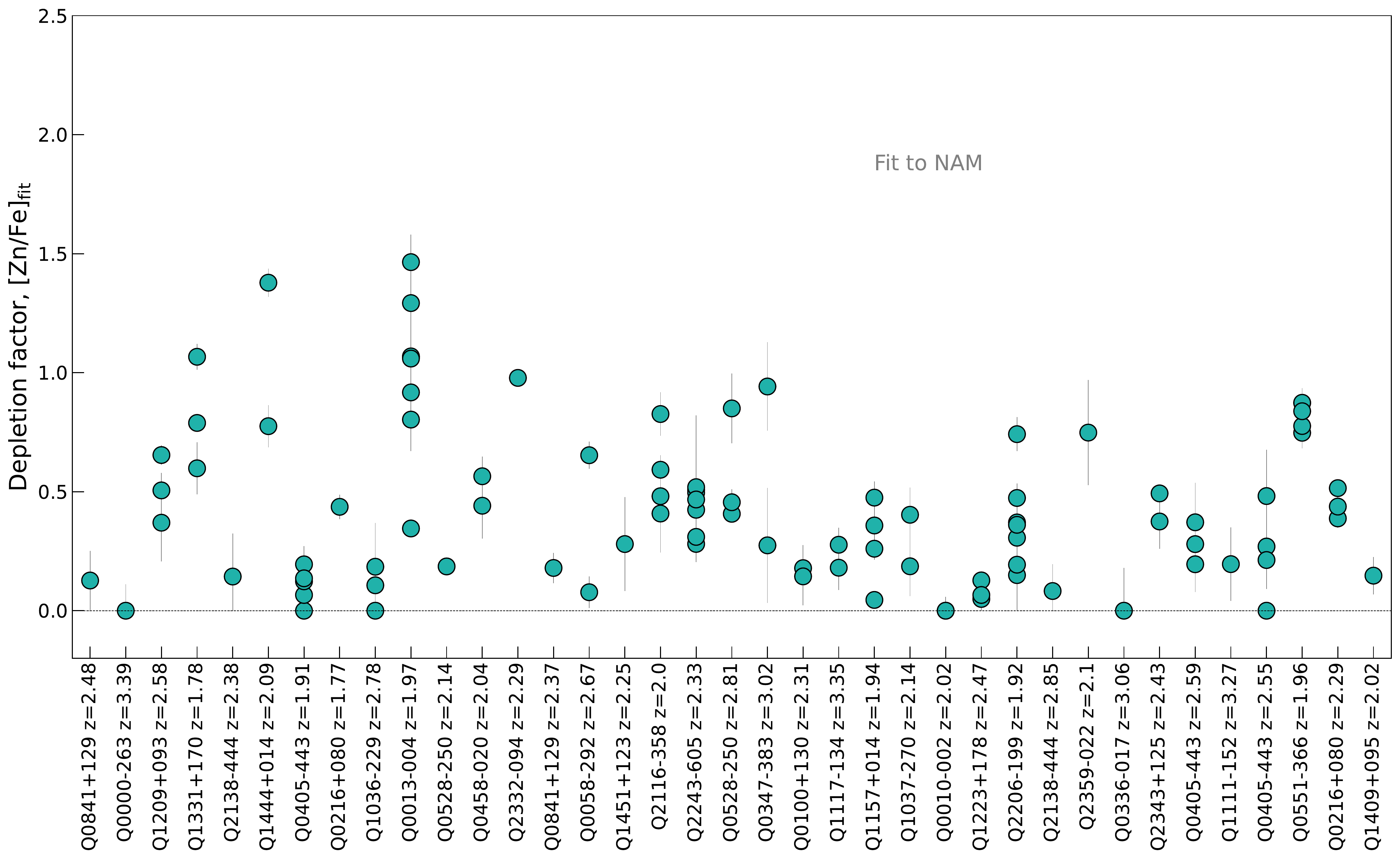}
         \caption{Depletion strengths calculated with a linear fit to the depletion patterns, excluding $\alpha$-elements and Mn (NAM). }
         \label{fig:dust-dep-by-DLA_NA}
     \end{subfigure}
     \caption{The range of dust depletion strength [Zn/Fe]$_{\rm fit}$ within  each DLA system. In the top panel, the depletion factor was calculated by fitting to the depletion patterns for all available metals. In the bottom panel, the depletion factor was calculated by fitting to the non-$\alpha$ elements and Mn (NAM).}
    \end{figure*}

% ==================================================

    %\input{figures/z-test}

        \begin{figure}
        \centering
        \includegraphics[width=0.5\textwidth]{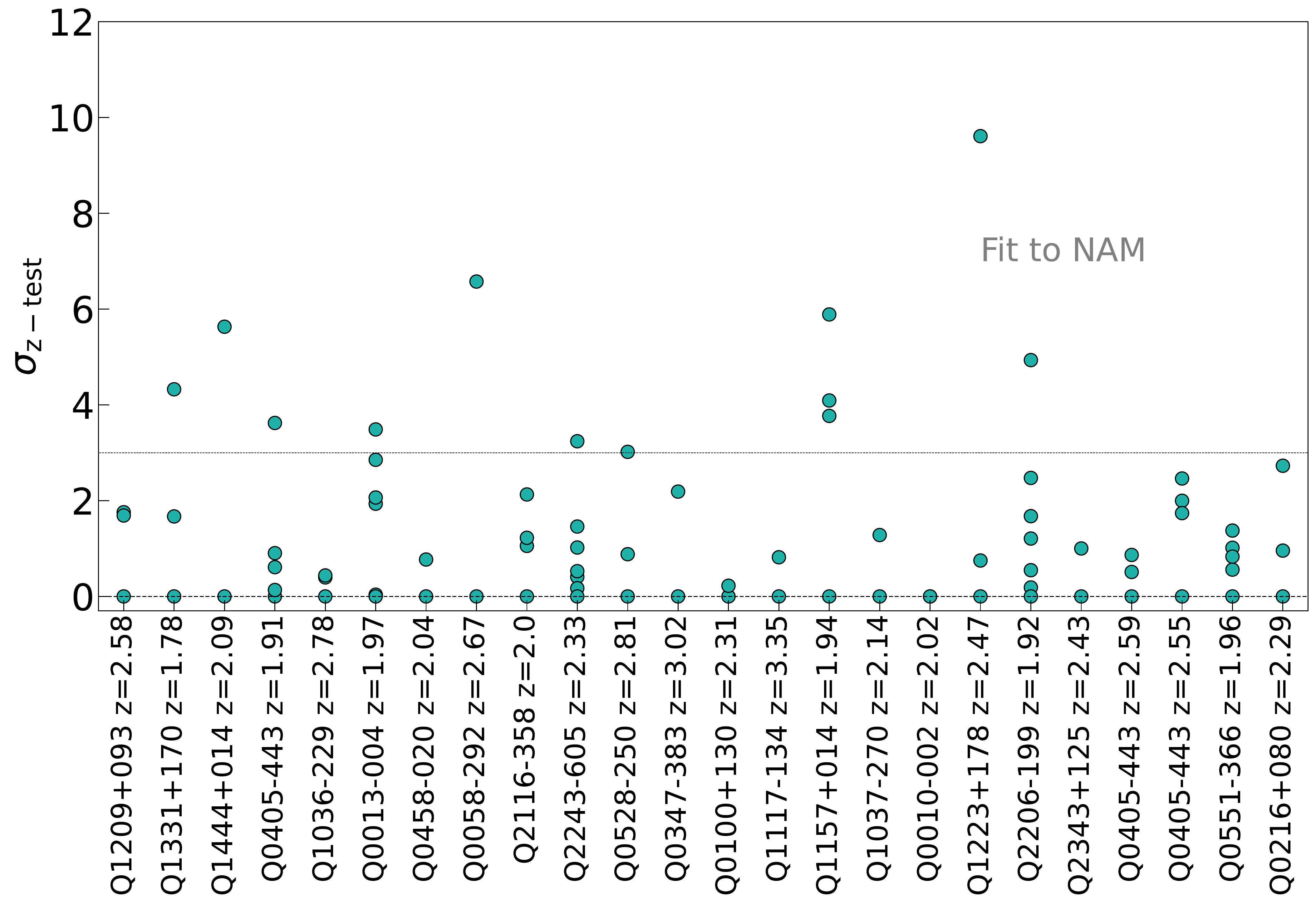}
        \caption{Statistical significance of diversity in [Zn/Fe]$_{\mathrm{fit}}$ within DLAs using the z-test: $\sigma_{\mathrm{z-test}} = \big{|} \mathrm{[Zn/Fe]}_{\mathrm{fit}, i} - \mathrm{[Zn/Fe]}_{\mathrm{fit, ref}} \big{|} \Big{/} \sqrt{\sigma_{\mathrm{[Zn/Fe]}_{\mathrm{fit}, i}}^2 + \sigma_{\mathrm{[Zn/Fe]}_{\mathrm{fit, ref}}}^2}$, where we choose the reference component $\mathrm{[Zn/Fe]}_{\mathrm{fit, ref}}$ to be the one with the minimum uncertainty on its depletion. Here, the linear fits to the depletion patterns exclude the $\alpha$-elements and Mn (NAM). QSO~0013-004 has an additional point at sigma $\sim 55$, which is not shown here for clarity.}
        \label{fig:z-test}
    \end{figure}

% ==================================================

    \subsection{Including or excluding $\alpha$-elements and Mn}
    
\begin{figure}
        \centering
        \includegraphics[width=\hsize]{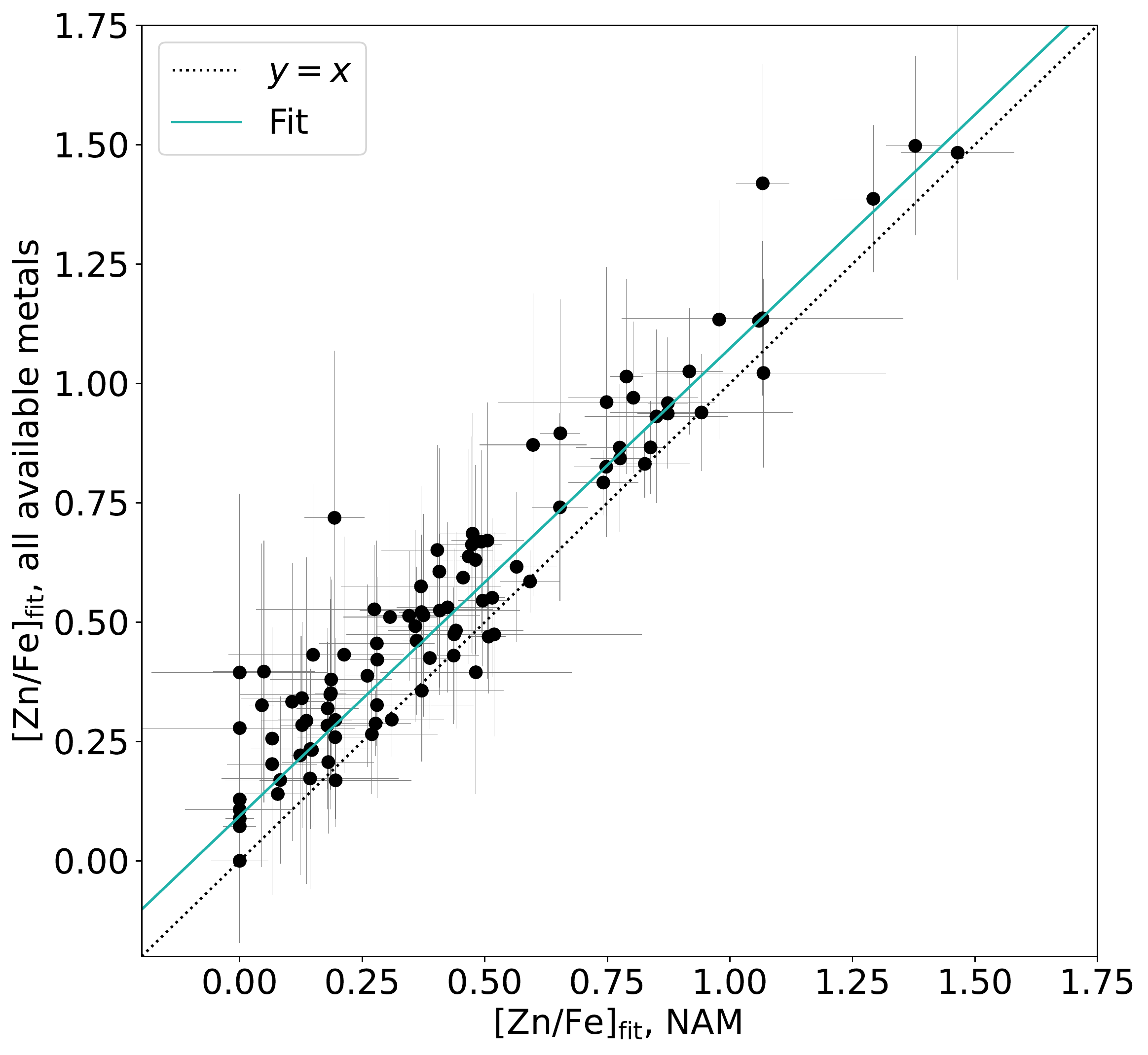}
        \caption{[Zn/Fe]$_{\mathrm{fit}}$ values calculated from fitting a straight line to the depletion patterns. The $x$-axis has [Zn/Fe]$_{\mathrm{fit}}$ where we exclude the $\alpha$-elements and Mn (NAM) in the fitting process. The $y$-axis has [Zn/Fe]$_{\mathrm{fit}}$ derived using all the available metals in the fit. Fitting a straight line to these two values for [Zn/Fe]$_{\mathrm{fit}}$ gives $ y = 0.98 (\pm 0.03) x + 0.09 (\pm 0.01)$. This is different from the $y=x$ line, which implies that including or excluding the $\alpha$-elements and Mn does notably affect the dust depletion factor that we calculate.}
        \label{fig:comparing_all-NA}
    \end{figure}

   At first we included all of the available metals in a linear fit to the depletion patterns. From these fits we often observed an over-abundance of the $\alpha$-elements, and an under-abundance of Mn, for example in Figures \ref{fig:depl-pattern_sys61} and \ref{fig:depl-pattern_sys41}. Therefore, nucleosynthetic effects seemed to add a level of complexity to a simple linear fit to the data. The linear fits to the depletion patterns describe only the effect of dust depletion. Excluding the $\alpha$-elements and Mn from the linear fit provides a way to characterise dust depletion independently from nucleosynthetic effects by removing deviations that are not due to dust depletion. It also allows us to estimate the potential nucleosynthetic effects after having taken dust depletion into account. In general, excluding the $\alpha$-elements and Mn flattens the slope of the linear fit to the depletion patterns because the $\alpha$ elements considered here are mainly volatile. This causes an off-set towards lower column densities. In some cases, slightly negative [Zn/Fe]$_{\mathrm{fit}}$ result from the linear fit to the depletion patterns. This cannot be physically caused by depletion, but rather due to the observational uncertainties. Thus, we limit the possible [Zn/Fe]$_{\mathrm{fit}}$ values to be non-negative. Figure \ref{fig:comparing_all-NA} shows the comparison between the depletion factors derived with the two methods. The two deviate significantly from a one-to-one line ($ y = 0.98 (\pm 0.03) x + 0.09 (\pm 0.01)$), with [Zn/Fe]$_{\mathrm{fit}}$, \textit{NAM} being lower in general. This implies that the systems are on average affected by nucleosynthetic differences in $\alpha$-element enrichment. However, there is some scatter and some systems are indeed consistent with no $\alpha$-element enhancement. There are several points that fit the $y=x$ line in Figure \ref{fig:comparing_all-NA} well. We found that these are components with either little deviation of the $\alpha$-elements from the linear fit to the depletion patterns, which we interpret as not having evidence for nucleosynthesis effects, or components with few constrained data points.
   
   Figures \ref{fig:dust-dep-by-DLA_all} and \ref{fig:dust-dep-by-DLA_NA} show, respectively, the depletion factors for all of the gas components calculated using all the available metals, and using only Zn, Fe, Cr, and/or P. 

% ==================================================
    %\input{figures/compare_NA-all}

% ==================================================

        \subsection{Depletion properties across redshift}
    
\begin{figure*}
        \begin{subfigure}[b]{\linewidth}
        \centering
        \includegraphics[width=\hsize]{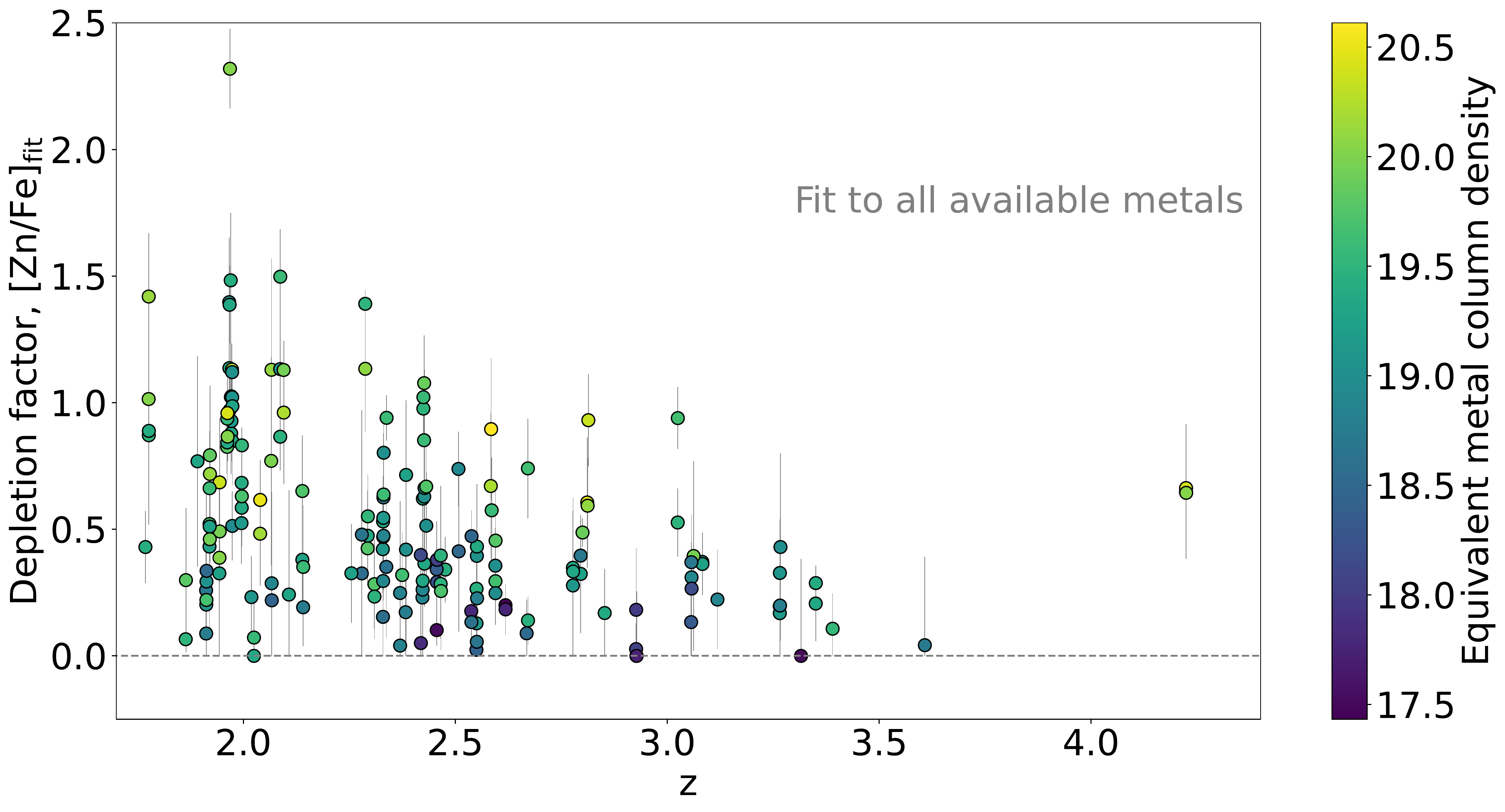}
        \caption{Depletion factors in individual components as a function of redshift. Here, the depletion factors were calculated by fitting a straight line to all the available metals in the depletion patterns. }
        \label{fig:cbar_all}
        \end{subfigure}
        \hfill
        \begin{subfigure}[b]{\linewidth}
         \centering
        \includegraphics[width=\hsize]{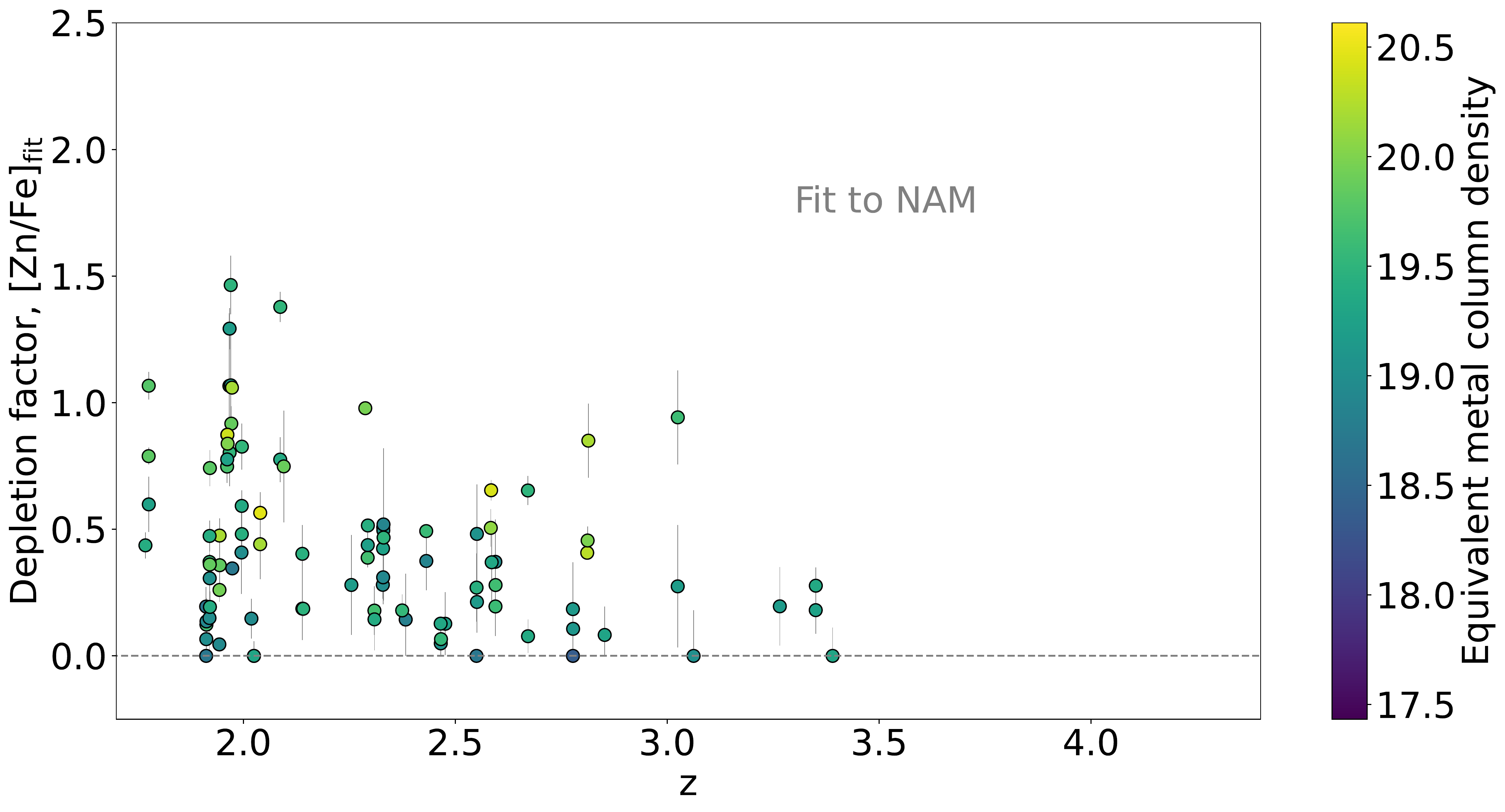}
        \caption{Depletion factors as a function of redshift, calculated from linear fits to the depletion patterns excluding the $\alpha$-elements and Mn (NAM).}
        \label{fig:cbar_NA}
         \end{subfigure}
         \caption{Depletion factors as a function of redshift. The colour bar represents the equivalent metal column densities $([\mathrm{M/H]}_{\mathrm{TOT}}+\log{N(\mathrm{H})})$.}
    \end{figure*}

    Figures \ref{fig:cbar_all} and \ref{fig:cbar_NA} show the depletion factors for each individual component as a function of redshift using both fitting methods. The colour bar shows the equivalent metal column densities. A particularly interesting result that we see in both figures is that there are components with little depletion at all redshifts. This could be the contribution from infalling pristine (metal-poor) gas either from gas inflows or within the galaxy itself. The colour bar highlights that low equivalent metal column densities are observed most frequently with low depletion factors, indicating either a low metallicity or an overall low amount of gas. Although there is no clear trend of increasing chemical enrichment with decreasing redshift, we do observe the most depleted components more frequently at lower redshifts. We also see that the upper bound limit of depletion is in general decreasing with increasing $z$, which is an indication of evolution of dust depletion with redshift. These results are consistent with observations of decreasing metallicity with increasing redshift \citep[][]{Rafelski+2012, DeCia+2018, Peroux+2020_metalscycle}. Overall, there are two effects that result in the differences between Figures \ref{fig:cbar_all} and \ref{fig:cbar_NA}. First, Figure \ref{fig:cbar_all} includes more components than \ref{fig:cbar_NA} because in some components it was not possible to perform a NAM fit due to a lack of data. Secondly, omitting alpha elements from the linear fits to the depletion patterns, which are over-abundant and mainly volatile in this data set, causes an offset towards lower metal columns. Fitting without the alpha elements and Mn causes the slope to decrease, but there is no clear trend for the $y$-intercept (i.e. the equivalent metal column density). In 55 out of 96 cases (57\%) the $y$-intercept decreases, while in 41 of 96 cases (43\%) it increases.

% ======================================================
    %\input{figures/cbar_all}

% ======================================================

    \subsection{Ionisation effects} \label{sec. results ion}
    
\begin{figure} 
    \centering
    \includegraphics[width=0.5\textwidth]{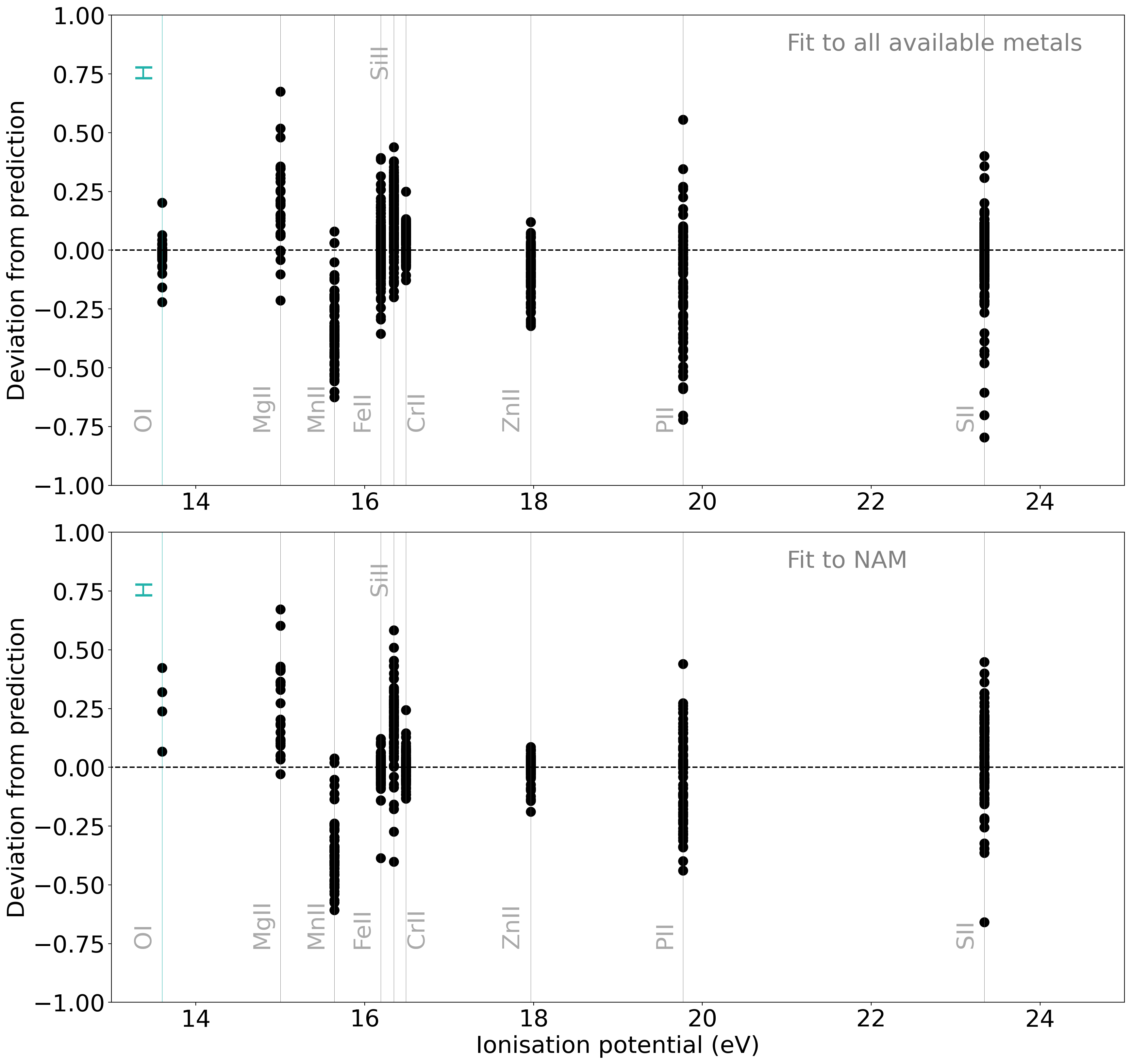}
    \caption{Ionisation potential against the residuals, including and excluding the $\alpha$-elements and Mn (NAM) respectively. We do not see a clear correlation between deviation from the fit to the depletion patterns and the ionisation potential. Although this is not concrete evidence that ionisation is completely negligible, it is enough to rule out any major ionisation effects.}
    \label{fig:ionisation}
\end{figure}

    Metal abundances in DLAs are not strongly affected by ionisation \citep[e.g.][]{Vladilo+2001_ionisation}. However, the individual components studied in this work could be clouds or groups of clouds at lower columns densities than DLAs, where ionisation may play a non-negligible role. To identify any potential effects of ionisation, we plotted the ionisation potential of each metal against their residual from the abundance pattern. If the gas clouds are indeed affected by ionising radiation, we expect a correlation in these plots: metals with a low ionisation potential should deviate more from the linear fits to the depletion pattern than the metals that require more ionisation energy. Figure \ref{fig:ionisation} shows that this is not the case. Although this is not final evidence that the gas clouds are not significantly affected by ionisation, the lack of correlation is enough to rule-out any major ionisation effects.

% ============================================================
    %\input{figures/ionisation}

% ============================================================

    \subsection{$\alpha$-element enhancement and Mn underabundance}
    
\begin{figure}
        \centering
        \includegraphics[width=0.46\textwidth]{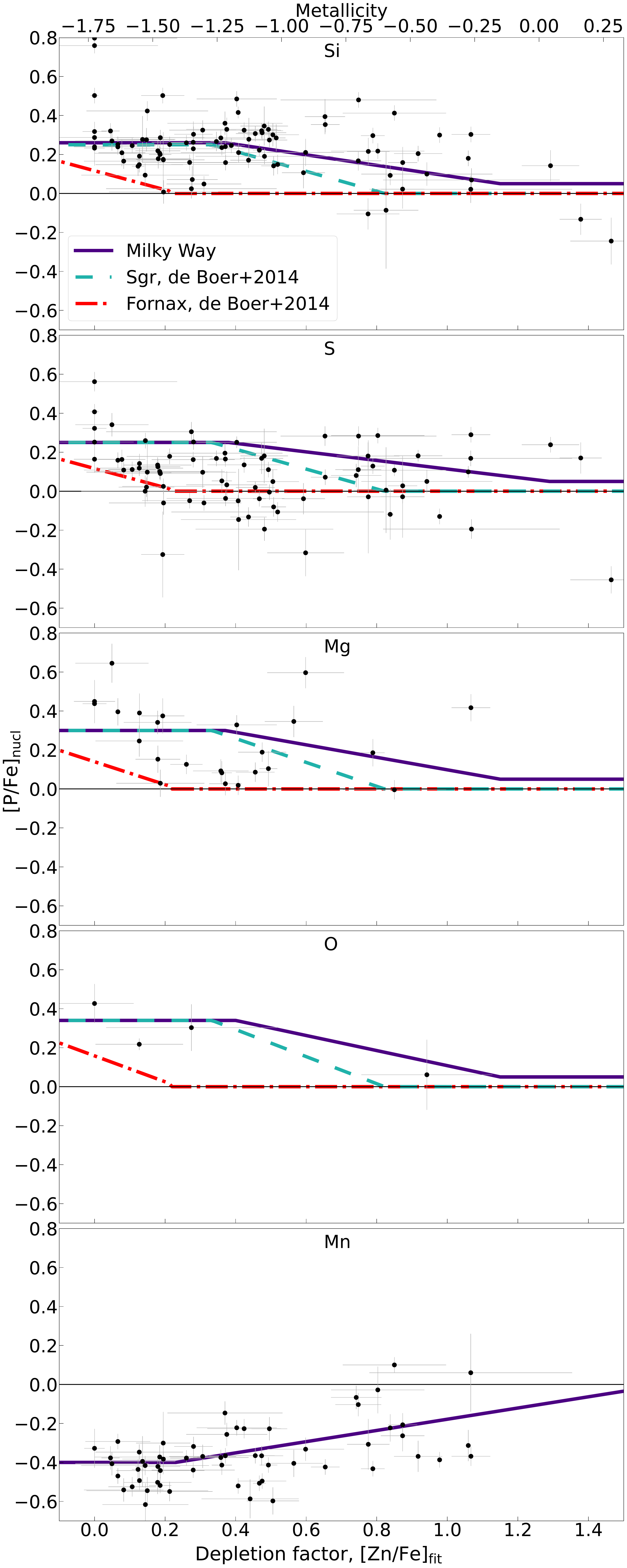}
        \caption{[X/Fe]$_{\mathrm{nucl}}$ plotted against the depletion factor. Here, we excluded the $\alpha$-elements and Mn from the depletion pattern fitting to reduce potential nucleosynthesis bias. The assumed nucleosynthetic under- and over-abundance in units of the depletion factor [Zn/Fe]$_{\mathrm{fit}}$ from Figure 7 of \cite{DeCia+2016} is shown in the solid purple line. We also include nucleosynthesis over-abundances for local dwarf galaxies, Fornax, and Sgr from \citet{Boer+2014_nucleo}. Here we only include points for which the absolute value of the error on [X/Fe]$_{\mathrm{nucl}}$ is less than 0.5.}
        \label{fig:nucleosynthesis}
    \end{figure}

    Evidence for nucleosythesis effects are apparent from the observation of systematic over-abundances of S, Si, Mg and O and an under-abundance of Mn in several systems, such as in Figures \ref{fig:depl-pattern_sys61} and \ref{fig:depl-pattern_sys41}. The deviations from the linear fit to the depletion patterns are even more pronounced after excluding the $\alpha$-elements and Mn from the depletion pattern fit, because it separates the possible nucleosynthetic effects from dust depletion. In Figure \ref{fig:nucleosynthesis} we present [X/Fe]$_{\mathrm{nucl}}$ as a function of [Zn/Fe]$_{\mathrm{fit}}$ for individual components for each of S, Si, O, Mn and Mg. Assuming a linear conversion between [Zn/Fe]$_{\mathrm{fit}}$ and metallicity \citep{DeCia+2016}, we derive the [X/Fe]$_{\mathrm{nucl}}$ distribution as a function of metallicity. We can then compare this directly with stellar metallicity measurements. We observe an over-abundance of the $\alpha$-elements and under-abundance of Mn at low levels of dust depletion, which begins to level-out at higher levels of depletion, and thus higher levels of chemical enrichment. This result is similar to studies of stellar abundances in the Milky Way and nearby dwarf galaxies (e.g. \citet{Lambert1987_nucleo, McWilliam1997_nucleo, Tolstoy2009_nucleo, Boer+2014_nucleo}.

    The enhancement of alpha elements is somewhat more pronounced for Mg, and possibly O, which have a lower ionisation potential (Figure \ref{fig:ionisation}). This effect, if real, is not fully explained. Ionisation of the dominant species in the warm neutral medium would cause the contrary effect, because ions with a lower ionisation potential would be more easily ionised.
    
    The position of the knee of the distribution of the $\alpha$-element enhancement with the metallicity (ie. the $\alpha$-element knee) depends on the star-formation history (SFH) of a galaxy and its efficiency in producing metals \citep[e.g.][]{Tinsley1997_nuc, McWilliam1997_nucleo}. From measurements of nearby dwarfs, it has been observationally shown that the $\alpha$-element knee varies with stellar mass \citep{Boer+2014_nucleo}. The amplitude of the plateau depends on the initial mass function (IMF) because if there are more massive stars then there are more core-collapse supernovae, and therefore more $\alpha$-elements. In Figure \ref{fig:nucleosynthesis}, we include stellar nucleosynthesis curves observed for the Galaxy and two satellite galaxies, Fornax and Sagittarius from \citet{Boer+2014_nucleo}.
    
    There is some debate about whether we can assume that Zn and Fe follow each other nucleosynthetically because of the complex origin of Zn \citep[e.g.][]{Skulaottir+2017_ZnNucl}. The stellar nucleosynthesis of Zn and Fe follow each other, to a first approximation, within the metallicity range $-2 \leq$ [Fe/H] $\leq 0$, which is most relevant for DLAs. Thus, to first approximation [Zn/Fe] is a solid tracer of dust content. More details are discussed in \citet{DeCia2018}.
    
    There is a large scatter in Figure \ref{fig:nucleosynthesis}, and our data points do not closely follow the lines for the Milky Way and local dwarf galaxies. However, it should be noted that these assumed curves are based on stellar observations, also with broad scatter. Moreover, DLAs (and their sub-components) represent a heterogeneous mix of galaxies with different masses and SFHs. Nevertheless, $\alpha$-element enhancements and Mn under-abundance in DLAs are evident in Figure \ref{fig:nucleosynthesis}.

    $\alpha$-element enhancements have been conclusively observed before for systems with depletion levels close to zero \citep[e.g.][]{Dessauges-Zavadsky+2006, Cooke+2011_nucl-DLAs-metal-poor, Becker+2012_nucl, Ledoux+2002_nucl-DLAs, DeCia+2016} or systems with [Zn/Fe] $\sim 0.5$ \citep{Konstantopoulou2022}. For systems with [Zn/Fe] $> 0.5$  it has been difficult to disentangle dust depletion from nucleosynthetic effects. Here we have been able to perform this disentanglement and we show that some DLA sub-components with higher metal enrichment (depletion, and possibly metallicity) have $\alpha$-element enhancements.
    
    We show that their extent is not dissimilar to what we observe in local galaxies. So, although there is indeed scatter in our plot, this seems to be a common feature in $\alpha$-element enhancement observed in stellar spectra \citep{Boer+2014_nucleo, Tolstoy2009_nucleo}. Nevertheless, our points do cluster around the lines, and, despite the uncertainties, the observations of the $\alpha$-element enhancement in DLAs seem to be consistent with the plateau observed in the Milky Way and local galaxies. In general, the $\alpha$-element plateau is influenced by the IMF, \citep[e.g.][]{Tinsley1997_nuc}. The location of the $\alpha$-element knee cannot be further constrained with the current sample, especially since it represents a mixed-bag of galaxies and their sub-components. Nevertheless, our observations are the first hints on the distribution of the $\alpha$-element enhancement with metallicity in distant galaxies.

    It could be argued that these nucleosynthetic effects would exist by construction of the method because our data is a subset of the total column densities from \citet{DeCia+2016}, which were used to derive the coefficients $A2_X$ and $B2_X$ and made an assumption on the $\alpha$-element enhancement. These coefficients are in turn used to derive the $y_i$ in this work. However, we do not believe this is a concern for several reasons. Firstly, the individual components can be considered as independent from the data set used by \citet{DeCia+2016}, who consider only integrated column densities. Secondly, although \citet{Konstantopoulou2022} found small values for $A2_X$, we argue that in principle, we can assume a value of 0 for $A2_X$ because a positive depletion is non-physical. This is indeed consistent with the uncertainties on $A2_X$, with the exception of Cr. This would remove the dependence of $y_i$ on $A2_X$. 

% =================================================
    
    %\input{figures/nucleosynthesis}

% =================================================

    \subsection{Kinematics of individual gas components} \label{sec. kinematics}
    
    We search for potential evidence for infalling and outflowing gas by comparing relative velocities of individual components and their level of depletion. In Figures \ref{fig:vel-depl3}, \ref{fig:vel-depl17}, \ref{fig:vel-depl24}, and \ref{fig:vel-depl36} we show the [Zn/Fe]$_{\mathrm{fit}}$ distribution with velocity of each component in the left panel of the figure, and their z-test in the right panel for systems towards QSO~0013-004, QSO~0405-443, QSO~0528-250 and QSO~1157+014 respectively. These four systems show statistically-significant chemical diversity (i.e. with $\sigma_{\mathrm{z-test}} \geq 3$, as discussed in Section \ref{sec. diversity of dust depletion}), as well as show some diversity in the kinematic properties. Similar plots for the other systems are included in the Appendix. For most systems, velocities of components are between $-100$ and 100 km/s. An exception is the system towards QSO~0013-004, which is shown in Figure 11, where we see components with both high blue-shifted velocities and high depletion. We discuss this system further in Section \ref{sec. QSO0013-004}.

    The system towards QSO~0528-250 (Figure \ref{fig:vel-depl24}) also has a component with velocity $\sim 200$ km/s, but with high depletion, which could be related to outflows. This system has a detected galaxy counterpart \citep[][]{MollerWarren1993_Q0528-250}, which allows for more detailed studies. In an in-depth study of this system, \citet{Balachez+2020_sys0528_gasorigin} find that this system is likely at a physical distance of 150 -- 200 kpc away from the quasar. This makes it more likely that the metal-rich DLA sub-component is associated with a companion galaxy, rather than with a metal-rich outflow. In this work, we do find a statistically significant variation of depletion strength, with $\sigma_{\mathrm{z-test}} \sim 3$ for component 3. This results supports those of \citet{Balachez+2020_sys0528_gasorigin}.
    
    Our observations could also probe the accretion of metal-poor gas, for example the system towards QSO~2243-605, in Figure \ref{fig:vel-depl65}. \citet{Bouche+2013_in-outflows} found evidence for infalling gas in UVES spectra of this QSO-galaxy pair, where the infalling components are blue-shifted $\sim -150$ to $-100$ km/s with respect to the galaxy counter-part observed in emission. In their work, they measure a component with lower-depletion and lower-``metallicity'' (using \ion{Si}{II} as a proxy for the \ion{H}{I} column density at the level of individual components) compared to the strongest component (component 6 compared to component 9 in Figure \ref{fig:vel-depl65}), which they attribute to infalling gas onto the galaxy. In our work, we do not see large diversity in the depletion between these two components. We do, however, see statistically significant diversity for the outer-most component (component 1).
    
    In Figure \ref{fig:depl-distribution} we show the distribution of the absolute values of the velocities of the individual components. We make a distinction between zero depletion and low- and high-depletion components at a [Zn/Fe]$_{\mathrm{fit}}$ value of 0.25. In these figures we see different distributions for the low- vs high-depletion groups. Interestingly, the low-depletion components have velocities below 100 km/s, while the high-depletion components show a wider distribution of velocities up to $\sim 600$ km/s. It is possible that some of the components with high-depletion are associated with metal-rich outflowing gas that is being ejected from the galaxy due to star formation. On the other hand, some of the components with low dust depletion could be consistent with infalling gas, with typically lower infalling velocities. The dust-free components are not exclusively the components with the highest column density, which are likely innermost parts of the galaxy (at 0 km/s), and the distribution is largely flat. This is consistent with evidence for pockets of dust-free gas within galaxies. We note that these conclusions should be drawn loosely because we do not have a statistically significant representation for low- high- and zero-depletion components.
    
    Further studies can be done on the systems towards QSO~0458-020 and QSO~0528-250 because these have confirmed galaxy counterparts \citep[][]{Moller+Christensen2020_DLAgalaxies}. However, without knowing the morphologies for most of the host galaxies in our DLA sample, it is not straightforward for us to interpret the DLA kinematics, and the values to be expected are highly dependent on the geometry of the system. 

    \subsubsection{Proximate DLAs}

    There are four DLA systems in our sample for which the redshift difference between the quasar and the DLA corresponds to a velocity difference smaller than 5000 km/s, i.e. $ v_{diff} = c|z_{em} - z_{abs}|/(1 + z_{abs}) \leq 5000$ km/s. These are the systems towards QSO~0528-250 ($v_{diff} = 3225$ km/s), QSO~0841+129 ($v_{diff} = 2070$ km/s), QSO~1157+014 ($v_{diff} = 4684$ km/s), and QSO~2059-360 ($v_{diff} = 514$ km/s), and are called proximate DLAs. 

    There are several possible scenarios we can use to interpret the depletion-kinematics relationship. One is if we assume that the absorbing galaxy is the QSO host itself. In this case, the interpretation of the geometry of the absorbing gas should be easier because we are necessarily on the closest side of the absorbing galaxy. We could naively expect inflowing gas clouds to be red-shifted, and outflows blue-shifted. We also expect outflowing gas to be more metal rich, and inflowing gas to be more metal-poor. We check for a trend among these proximate DLAs to see if they differ from the sample. For QSO~0528-250 (Figure \ref{fig:vel-depl24}) and QSO~1157+014 (Figure \ref{fig:vel-depl36}) we see instead components with higher depletion being red-shifted compared to low-depletion components, which is counter-intuitive in this scenario. We are not able to perform this analysis for the remaining two proximate DLAs because there are too few absorbing components to make such a comparison. In comparing the kinematics-depletion trend for QSO~0528-250 and QSO~1157+014 to the overall sample, we see a similar trend in the systems towards QSO~0216+080 (Figure \ref{fig:vel-depl-0216+080}), QSO~0347-383 (Figure \ref{fig:vel-depl-0347-383}) and QSO~2116-358 (Figure \ref{fig:vel-depl-2116-258}), although with less significant variations in the dust depletion (z$_{\mathrm{test}} < 3\sigma$). We see this trend in 2/2 of PDLAs, and in 3/12 of standard DLAs (for systems with 3 or more components). This interpretation is based on a very small sample, and the system towards QSO~0528-250 is a special case, as we explain below.

    Since a velocity limit of 5000 km/s does not ensure that the absorbing galaxy is indeed the QSO host, another scenario is that the DLA may be a nearby companion galaxy to the QSO. QSO~0528-250 is shown to be part of a small galaxy group, and the absorber is likely to not to be associated with the QSO, which makes this scenario more likely for this system \citep{MollerWarren1993_Q0528-250, ProchaskaWolfe1997b_Q0528-250, SrianandPetitjean1998_Q0528-250, Ledoux+1998_Q0528-250, Balachez+2020_sys0528_gasorigin}. Further, $z_{abs} > z_{em}$ for this DLA, which makes it particularly interesting, although the QSO redshift may have been underestimated. In this interpretation, it could be that the red-shifted, highly depleted components are metal-rich outflows from the far side of the companion galaxy. 

    However, relative velocities of gas clouds are not robust indicators of their physical locations. For example, gas clouds on either side of a galaxy can both produce redshifted velocities due to the motion of the galaxy with respect to the QSO. In theory, it is possible to have a combination of the two scenarios described in the above two paragraphs that could produce red-shifted highly-depleted components. In this case, part of the absorption could be located within the QSO and part of it could be in a different galaxy with a peculiar velocity towards the QSO. In velocity space, the absorption lines of these two parts could coincidentally coincide. It is therefore difficult to make any conclusive remarks about the relationship between kinematics, as interpreted in velocity space, and depletion of individual gas components.
    
    \subsubsection{QSO~0013-004} \label{sec. QSO0013-004}
    
\begin{figure}
    \centering
    \includegraphics[width=0.5\textwidth]{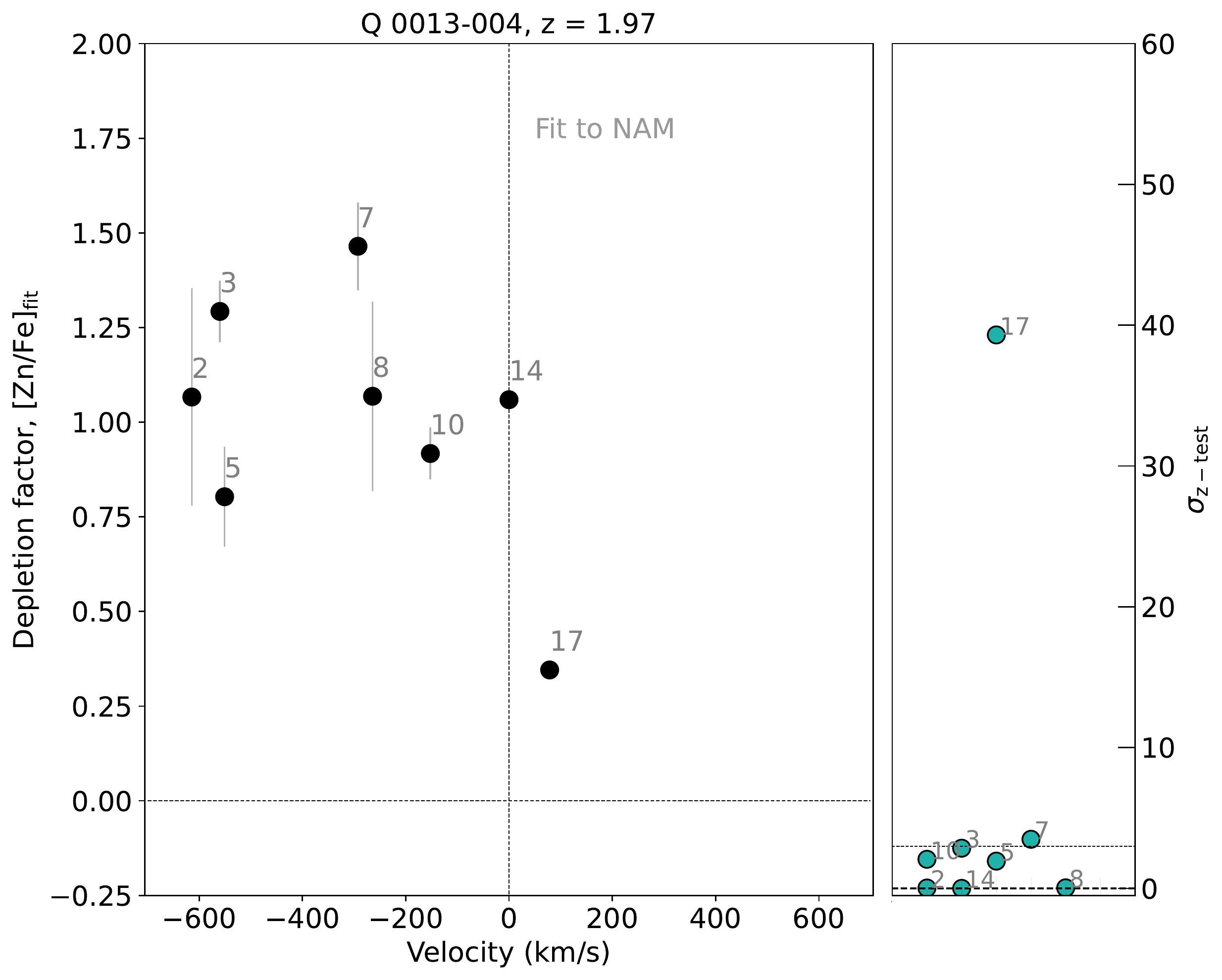}
    \caption{The velocity of individual components within DLA system QSO 0013-004 against their depletion strengths. The depletion factor is derived from a fit to the non-$\alpha$ elements and Mn (NAM) in the depletion patterns. The component with the highest Fe column density was chosen as the zero-velocity point, because this is likely the component closest to the inner parts of the galaxy. This system has the highest number of components in our sample, and the highest velocities. Each component is numbered.}
    \label{fig:vel-depl3}
\end{figure}

\begin{figure}
    \centering
    \includegraphics[width=0.5\textwidth]{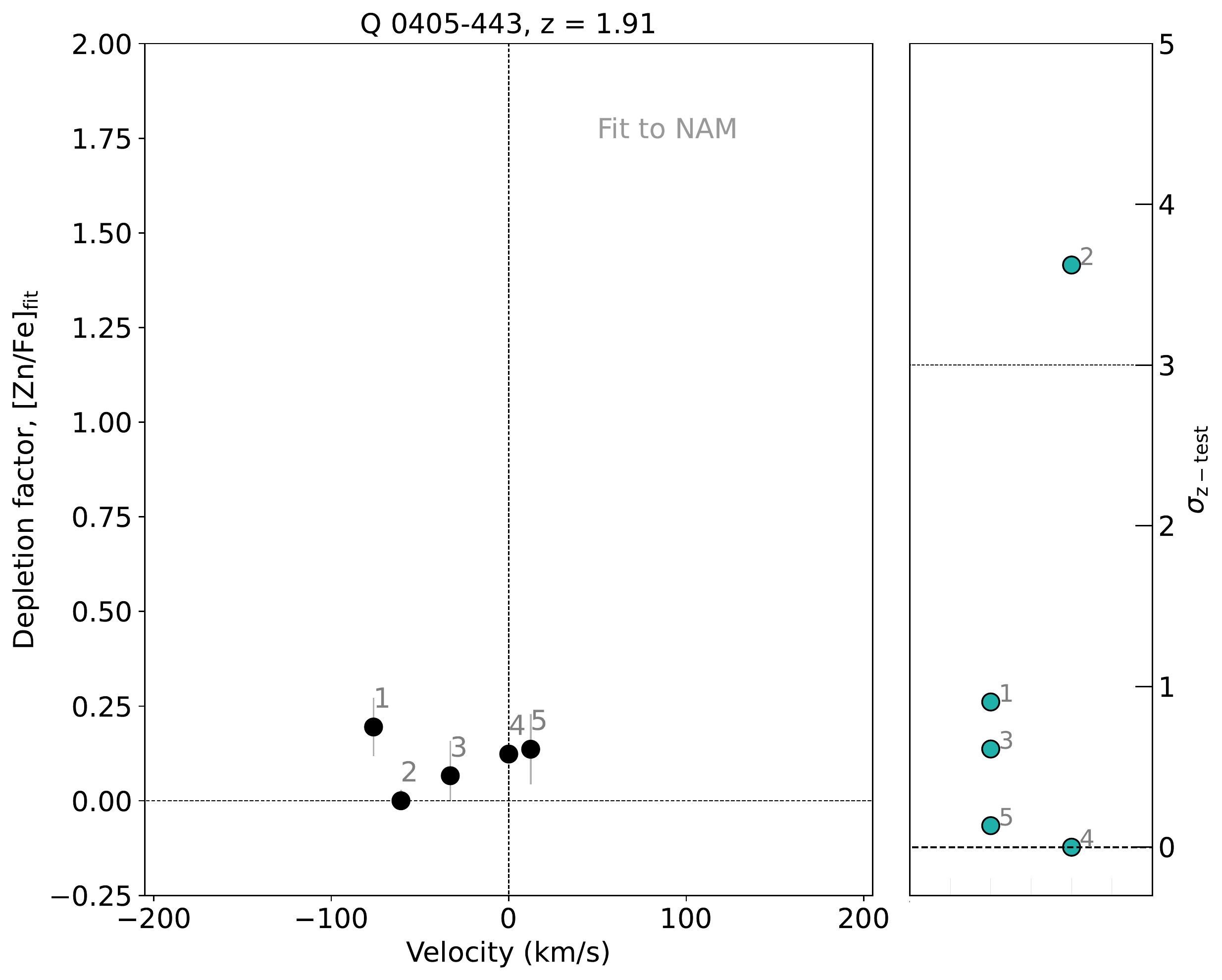}
    \caption{Same as Figure \ref{fig:vel-depl3} but for DLA system QSO~0405-443.}
    \label{fig:vel-depl17}
\end{figure}

\begin{figure}
    \centering
    \includegraphics[width=0.5\textwidth]{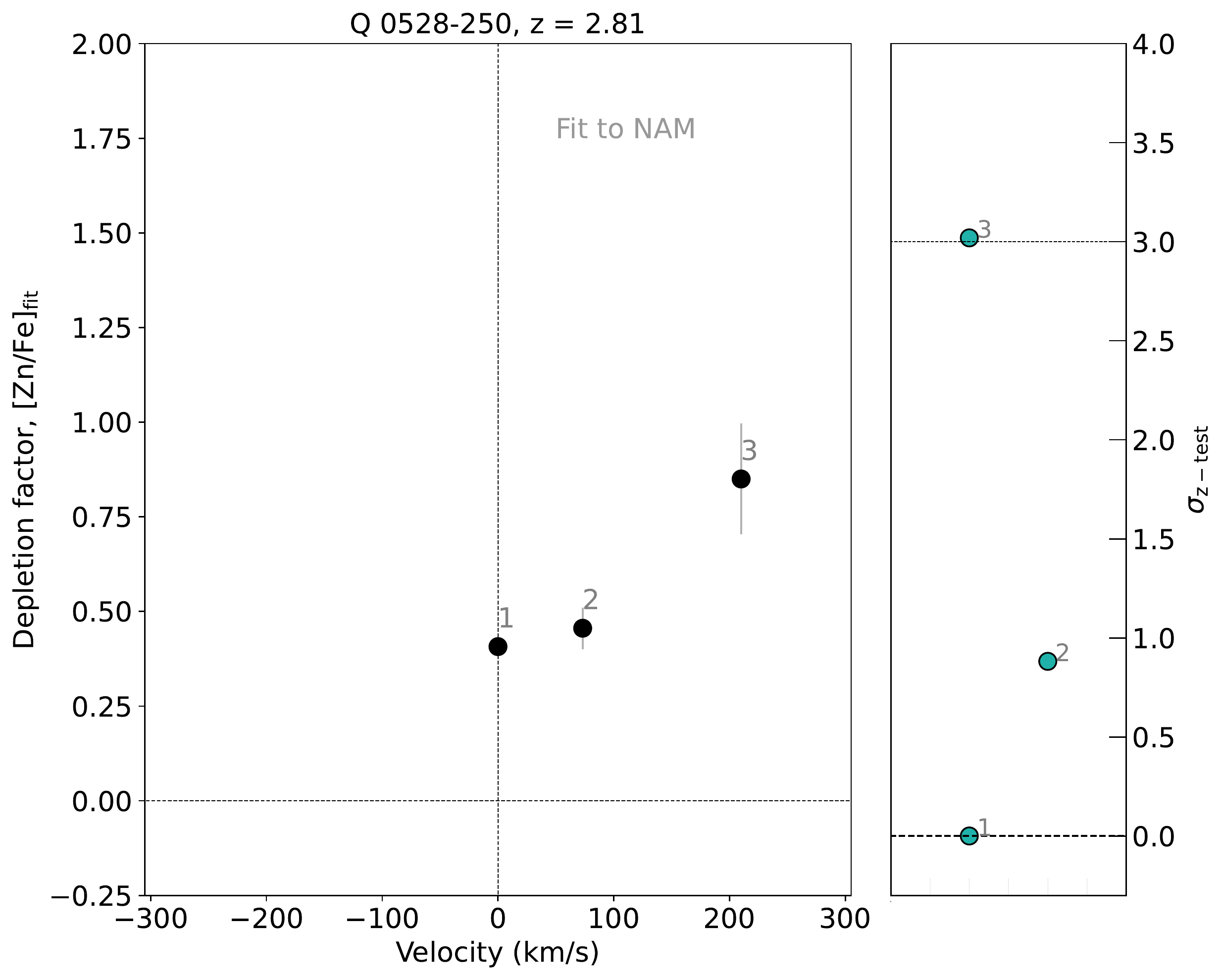}
    \caption{Same as Figure \ref{fig:vel-depl3} but for DLA system towards QSO~0528-250. This system is a proximate DLA, where the redshift difference between the quasar and the DLA corresponds to a velocity difference smaller than 5000 km/s, i.e. $c|zem - zabs|/(1 + zabs)$. Here we see a relatively high-velocity red-shifted component with high depletion (component 3).}
    \label{fig:vel-depl24}
\end{figure}

\begin{figure}
    \centering
    \includegraphics[width=0.5\textwidth]{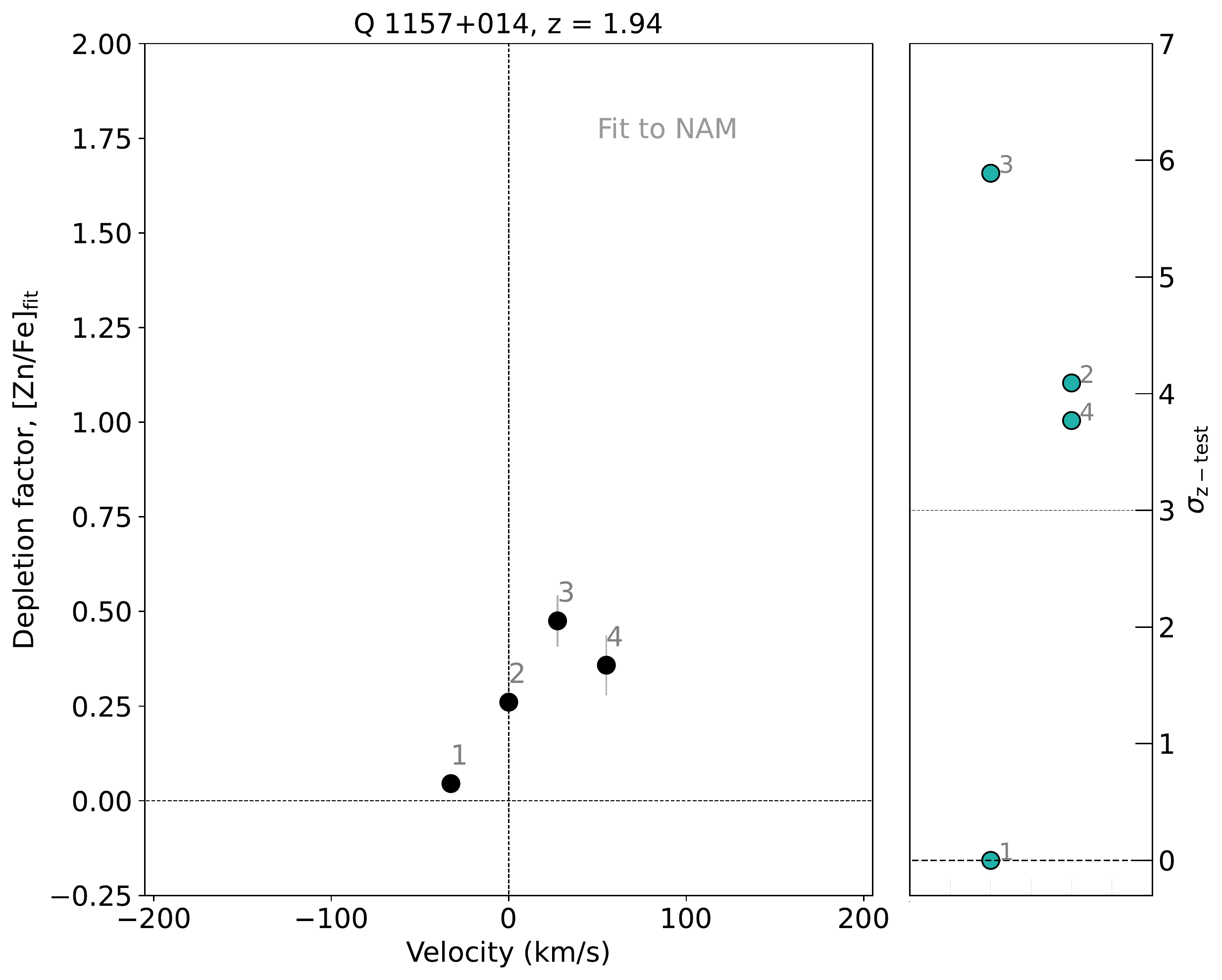}
    \caption{Same as Figure \ref{fig:vel-depl3} but for DLA system QSO~1157+014. This system is a proximate DLA.}
    \label{fig:vel-depl36}
\end{figure}

\begin{figure}
    \centering
    \includegraphics[width=0.5\textwidth]{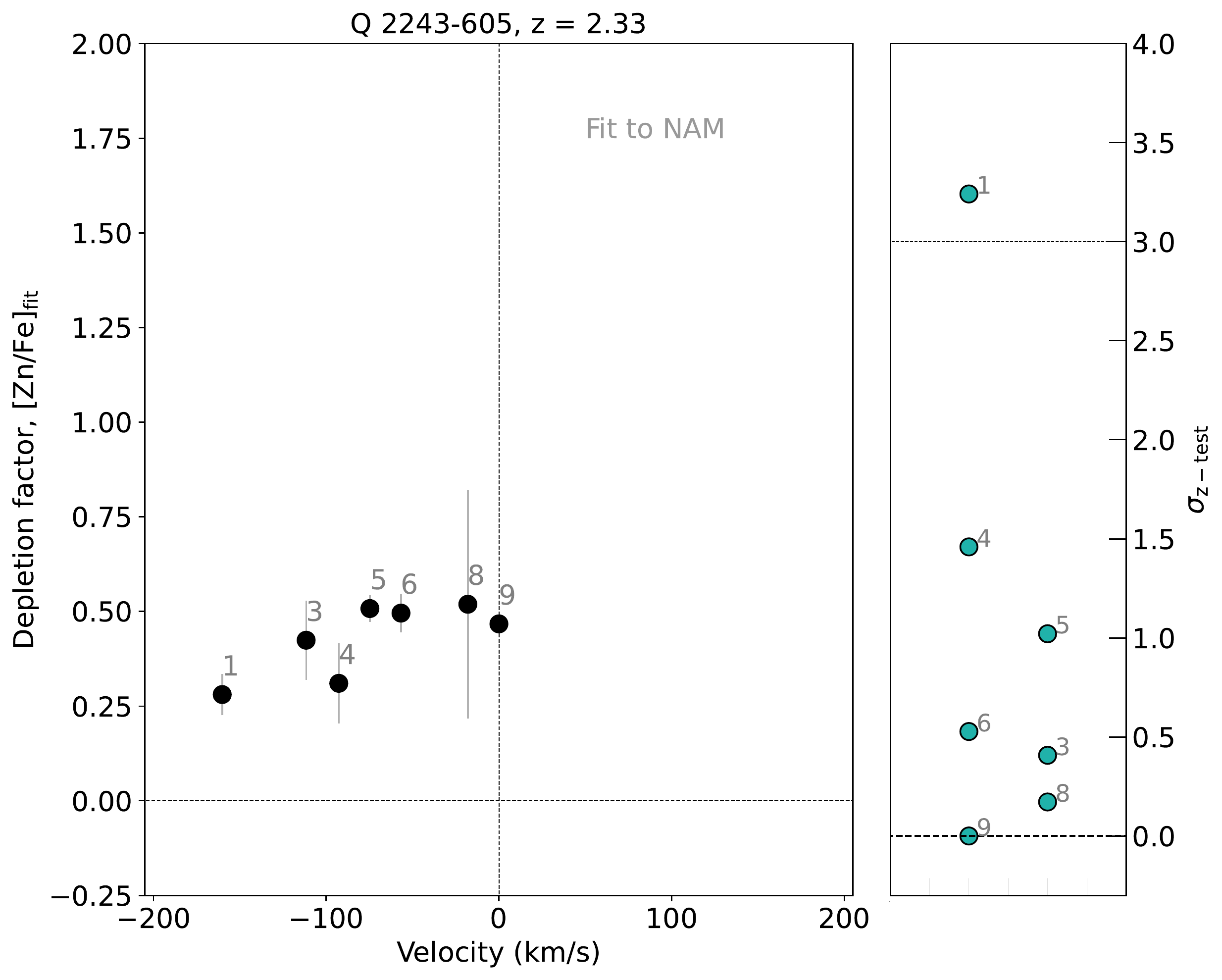}
    \caption{Same as Figure \ref{fig:vel-depl3} but for DLA system QSO 2243-605. This system shows a general trend of the depletion decreasing with blue-shifted velocities.}
    \label{fig:vel-depl65}
\end{figure}

\begin{figure}
        \centering
        \includegraphics[width=\hsize]{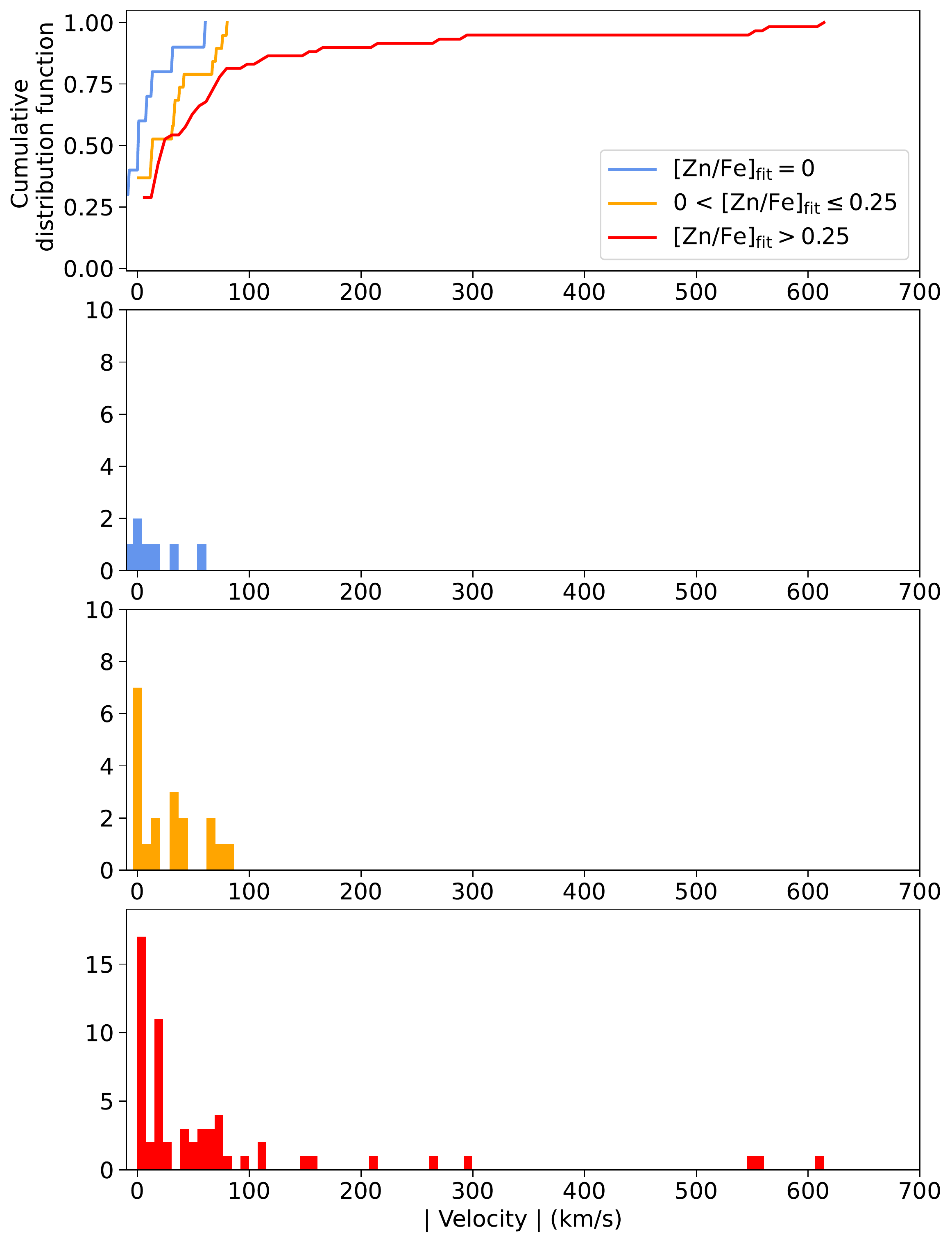}
        \caption{Absolute value velocity distribution for all components. Here we make a distinction between zero depletion, and low- and higher-depletion components at a [Zn/Fe]$_{\mathrm{fit}}$ value of 0.25. The top panel shows the cumulative distributions for all three distribution. The following three panels show histograms for zero, low- and high-depletion components respectively. We see a similar distribution for the proximate DLAs.}
        \label{fig:depl-distribution}
    \end{figure}

\begin{figure}
        \centering
        \includegraphics[width=0.46\textwidth]{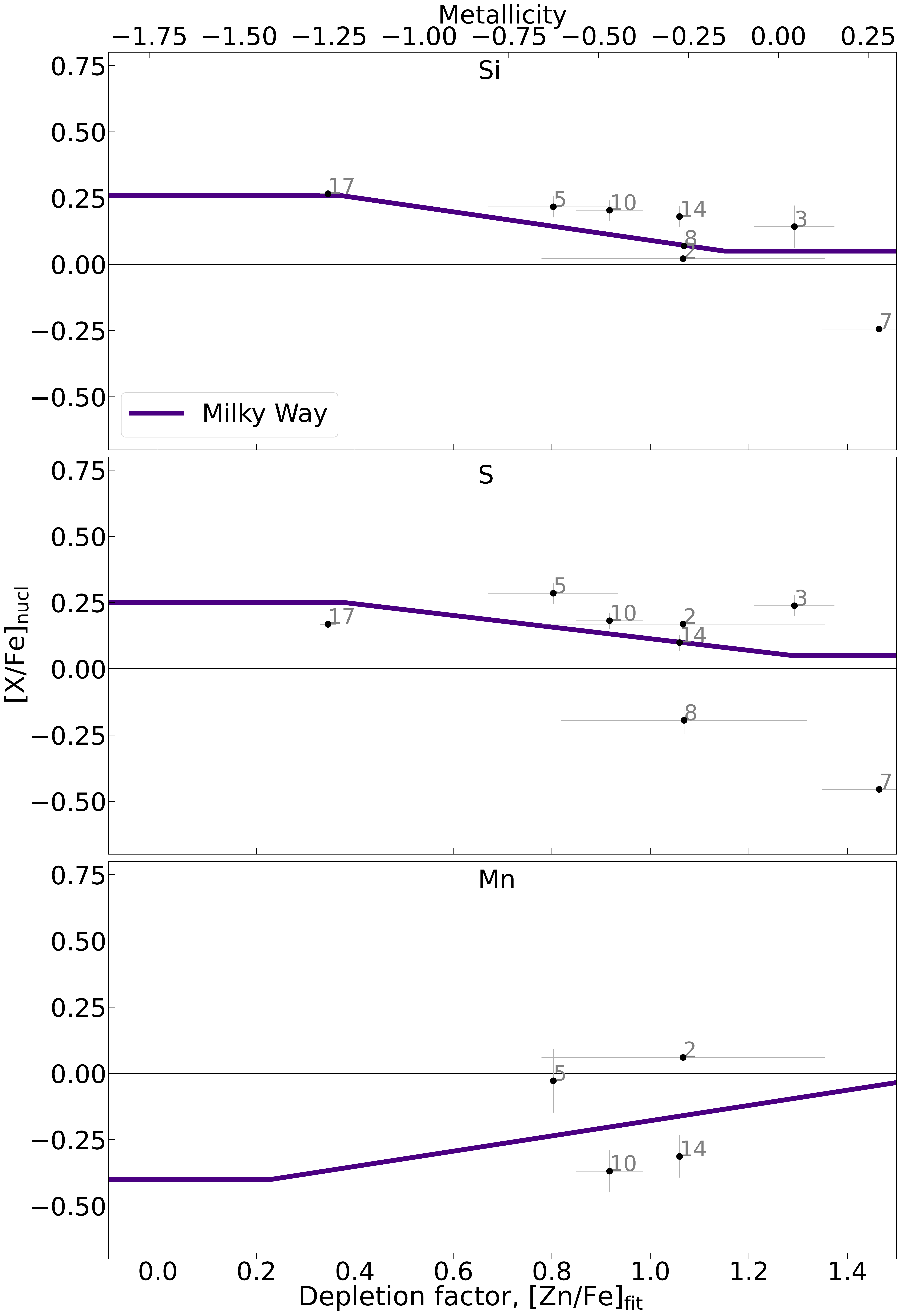}
        \caption{The same as Figure \ref{fig:nucleosynthesis}, but only for the DLA system towards QSO 0013-004.}
        \label{fig:nucleosynthesis_Q0013}
    \end{figure}

    This system has many components (see Figure \ref{fig:depl-pattern_sys3}), some with high velocities (up to $-600$ km/s) and high depletion strengths (mostly between 0.75 and 1.5 dex). It has also been studied in the literature, for example by \citet{Petitjean+2002_Q0013-004, Ledoux+2002_nucl-DLAs, Rodriguez+2006_6DLAs, Noterdaeme+2021_sys0013}. Given its relatively high metallicity for a DLA, this system is likely to be associated with a relatively massive galaxy according to the mass-metallicity relation in DLAs \citep{Christensen+2014_DLAs}, possibly of the order $M_{*} \sim $ $10^{10} M_{\odot}$. Furthermore, the system is an outlier from the velocity-metallicity relation of \citet{Ledoux+2006_massmet}: it seems to have a higher velocity width of their low-ionisation absorption-line profiles, $\Delta V = 720$ km/s, for its metallicity, [M/H] $= -0.4 \pm 0.1$ \citep{DeCia+2016}. In a detailed study, \citet{Petitjean+2002_Q0013-004} found that at least 4 of the components have H$_2$ absorption, which they attribute to intense star formation activity in the regions around the absorbing system (components 2, 6, 14 and 17 in this paper). In their paper, they postulate that the system is composed of 2 main sub-systems centred at 0 and $-480$ km/s. From this work, we calculate that the first sub-component would have an average [Zn/Fe]$_{\mathrm{fit}}$ of 1.03 $\pm 0.21$, and the second an average of 0.96 $\pm 0.21$, which are indeed very similar. This is the only system that contributes to the very high velocities ($\sim 600$ km/s) in the top and bottom panels of Figure \ref{fig:depl-distribution}. However, if this is indeed two sub-systems, the system centred at $480$ km/s would fall right on top of the other system centred at 0 km/s in the cumulative distribution if we were to split the systems. This could be motivation for QSO~0013-004 being a set of two sub-systems, as postulated by \citet{Petitjean+2002_Q0013-004}, instead of for out-flowing gas. This system shows $\alpha$-element enhancement and Mn under-abundance in several components, as shown in Fig. \ref{fig:nucleosynthesis_Q0013}. This may also indicate some recent star formation in the system. Intriguingly, the distribution of the $\alpha$-element enhancement for this system follows closely the curve for the Galaxy. Furthermore, we compare the velocities, depletion, and the $\alpha$-element enhancement by looking at Figures \ref{fig:depl-pattern_sys3}, \ref{fig:vel-depl3} and \ref{fig:nucleosynthesis_Q0013} simultaneously. For example, component 17 has the lowest chemical enrichment, but a large $\alpha$-element enhancement. This may suggest that this gas component could be associated with a more metal-poor part of the galaxy or a satellite galaxy, and could possibly have recent star-formation. Our observation suggests that this gas is not well-mixed or homogeneously distributed.

% ====================================================================

    %\input{figures/vel-depl}

% ====================================================================

    %\input{figures/vel_distribution}

% ====================================================================
    
    %\input{figures/nucleosynthesis_Q0013}

% ====================================================================
    
    \subsection{Dust-to-metal ratio}
    
\begin{figure*}
    \centering
    \includegraphics[width=\hsize]{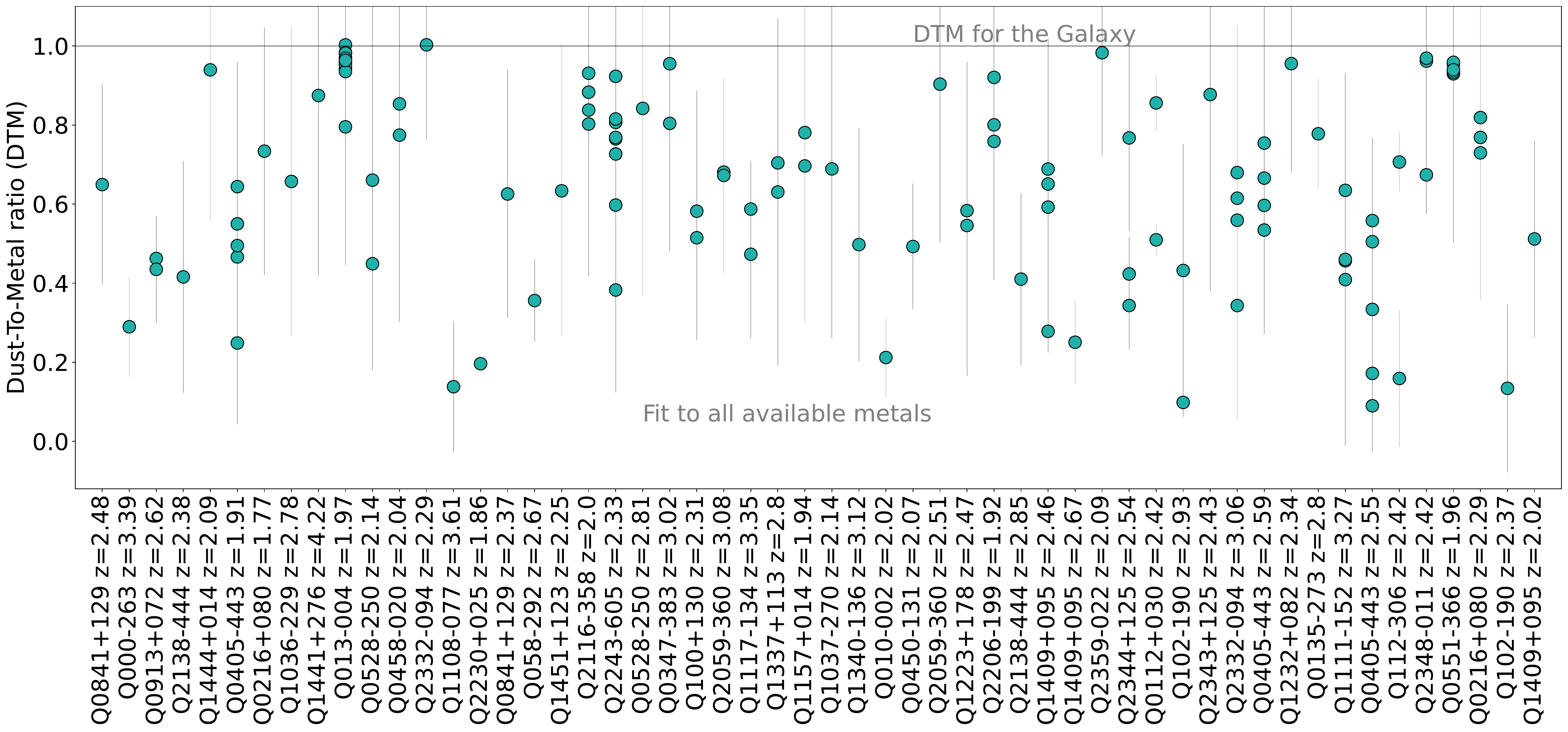}
    \caption{Dust-to-metal ratios (DTM) for each individual component in each DLA. Here, the DTM ratio was calculated by fitting to all the available metals in the depletion patterns. Notably, we see many individual components with a DTM similar to that of the Milky Way.}
    \label{fig:DTM_all}
    \end{figure*}

\begin{figure*}
        \centering
        \includegraphics[width=\hsize]{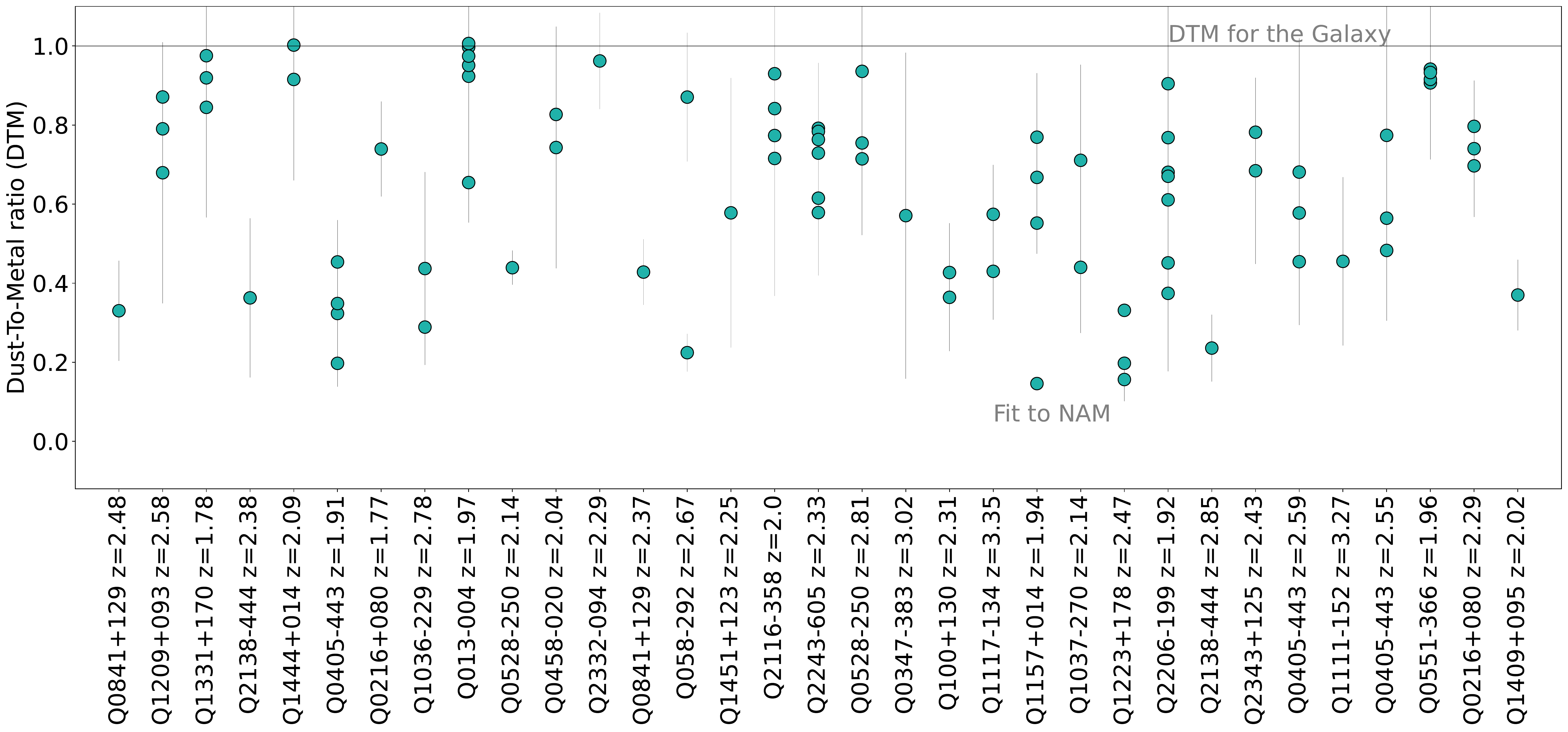}
        \caption{Similar to Figure \ref{fig:DTM_all} calculated from a fit excluding the non-$\alpha$-elements and Mn.}
        \label{fig:DTM_NA}
    \end{figure*}

    The dust-to-metal ratio can be defined as the fraction of a metal that is in the dust phase, and can be written in terms of number of atoms as follows:
    
    \begin{align}
        dtm_{\mathrm{Fe}} = \frac{N(\mathrm{Fe})_{\mathrm{dust}}}{N(\mathrm{Fe})_{\mathrm{tot}}} =& 1 - 10^{\delta_{\mathrm{Fe}}} \\
        DTM_{\mathrm{Fe}} =& \frac{dtm_{\mathrm{Fe}}}{dtm_{\mathrm{Gal}}},
    \end{align}
    
    where DTM is the selective dust-to-metal ratio, based on Fe, normalised by the value for the Galaxy $dtm_{\mathrm{Gal}}$ = 0.98 \citep{DeCia+2016}, and $\delta_X$ is the depletion of element $X$, as defined in Equation \ref{eq:meth1}. We adopt $\delta_{Fe} = A2_{Fe} + B2_{Fe} \times [\mathrm{Zn/Fe}]_{\mathrm{fit}} = -0.01 + (-1.26) \times [\mathrm{Zn/Fe}]_{\mathrm{fit}}$ \citep{DeCia+2016}. These values of DTM are not absolute mass ratios, so they are not directly comparable to those of \citet{Konstantopoulou2022}. The DTM is plotted for each individual component in the DLA systems in Figures \ref{fig:DTM_all} and \ref{fig:DTM_NA}\footnote{For some systems, there are a different number of components in the \textit{NAM} fit compared to the fit with all available metals. This is because we neglect points where the observational uncertainties are too large, i.e. for which the absolute value of the error on the DTM calculation is greater than 0.5. This is the case for the system towards towards QSO~1444+014, for example.} Notably, we see DTM ratios that are similar to the Galactic value in several DLA systems. The diversity of depletion strengths within DLAs is also reflected in the range of values we obtain for the DTM ratio although the uncertainty is large. 

% ==========================================================
    %\input{figures/DTM-by-DLA_NA}

 % ==========================================================
    
\section{Summary and conclusions} \label{sec. concl}

   We perform a component-by-component analysis of the relative abundances of different metals in the ISM/CGM of a sample of 64 DLAs in the redshift range 1.7 -- 4.2. Below we summarise the main results from this work.

   \begin{enumerate}
  
     \item Our results show that chemical enrichment, as traced by dust depletion ([Zn/Fe]$_{\mathrm{fit}}$), is diverse within DLA systems. We see 10 systems with statistically significant diversity ($\geq 3\sigma_{\mathrm{z-test}}$). This indicates that these distant galaxies are chemically diverse. This diversity is also reflected in the DTM ratios that we derive from the depletion.
    
    \item There are components with very low depletion at all redshifts, including DLAs at $z \sim 2$, albeit with large uncertainties. This suggests the presence of dust-poor gas, likely with low metallicity, potentially falling onto and contributing to the build up of DLA galaxies. 
    
    \item We performed a check on the effect of ionisation as a potential cause for deviations from the linear fit to the depletion patterns, and can exclude strong ionisation effects.

    \item Our measurements reveal an over-abundance of $\alpha$-elements (Si, S, O and Mg), and an under-abundance of Mn in some gas clouds in our DLA sample, to a level very similar to the Galaxy and local dwarf galaxies. We interpret this as an effect of SNe nucleosynthesis. These are the first observations of the distribution of $\alpha$-elements with chemical enrichment in the gas for distant galaxies.   
    
    \item We compare the chemical properties of the individual gas components with their kinematic information. We see different distributions for the zero, low-, and high-depletion components. This could be a result of some of the high-depletion components being associated with metal-rich outflows. On the other hand, some of the low-depletion components are consistent with metal-poor infalling gas.
    
    \item In the system towards QSO~0013-004 we observe components with high depletion (up to [Zn/Fe]$_{\mathrm{fit}} \sim 1.25$) and high velocities (up to $\sim -600$ km/s). This system could be a massive DLA galaxy, and the high-depletion, high-velocity components may suggest evidence for metal-enriched outflows. On the other hand, this could be a galaxy system of two sub-systems centred at 0 km/s and $\sim -480$ km/s with similar chemical enrichment. The system towards QSO~2243-605 has low-depletion, high-velocity components, which could be evidence for infalling metal-poor gas onto the galaxy. However, without knowing the morphology of this galaxy, it is not a straight-forward conclusion to make. 
    
   \end{enumerate}

\begin{acknowledgements}
T.R.-H., A.D.C., J.-K.K. , and C.K. acknowledge support by the Swiss National Science Foundation under grant 185692. This work is based on observations carried out at the European Organisation for Astronomical Research in the Southern Hemisphere under ESO programmes 065.P-0038, 065.O-0063, 066.A-0624, 067.A-0078, and 068.A-0600. This research has made use of NASA’s Astrophysics Data System. N.B. acknowledges
support from the grant 3DGasFlows (ANR-17-CE31-0017) from the Agence Nationale de Recherche (ANR).
\end{acknowledgements}

\bibliographystyle{aa}
\bibliography{main}

\begin{appendix} \label{sec. appendix}

\section{Depletion sequences}
\begin{figure*}
        \centering
        \includegraphics[width=0.9\textwidth]{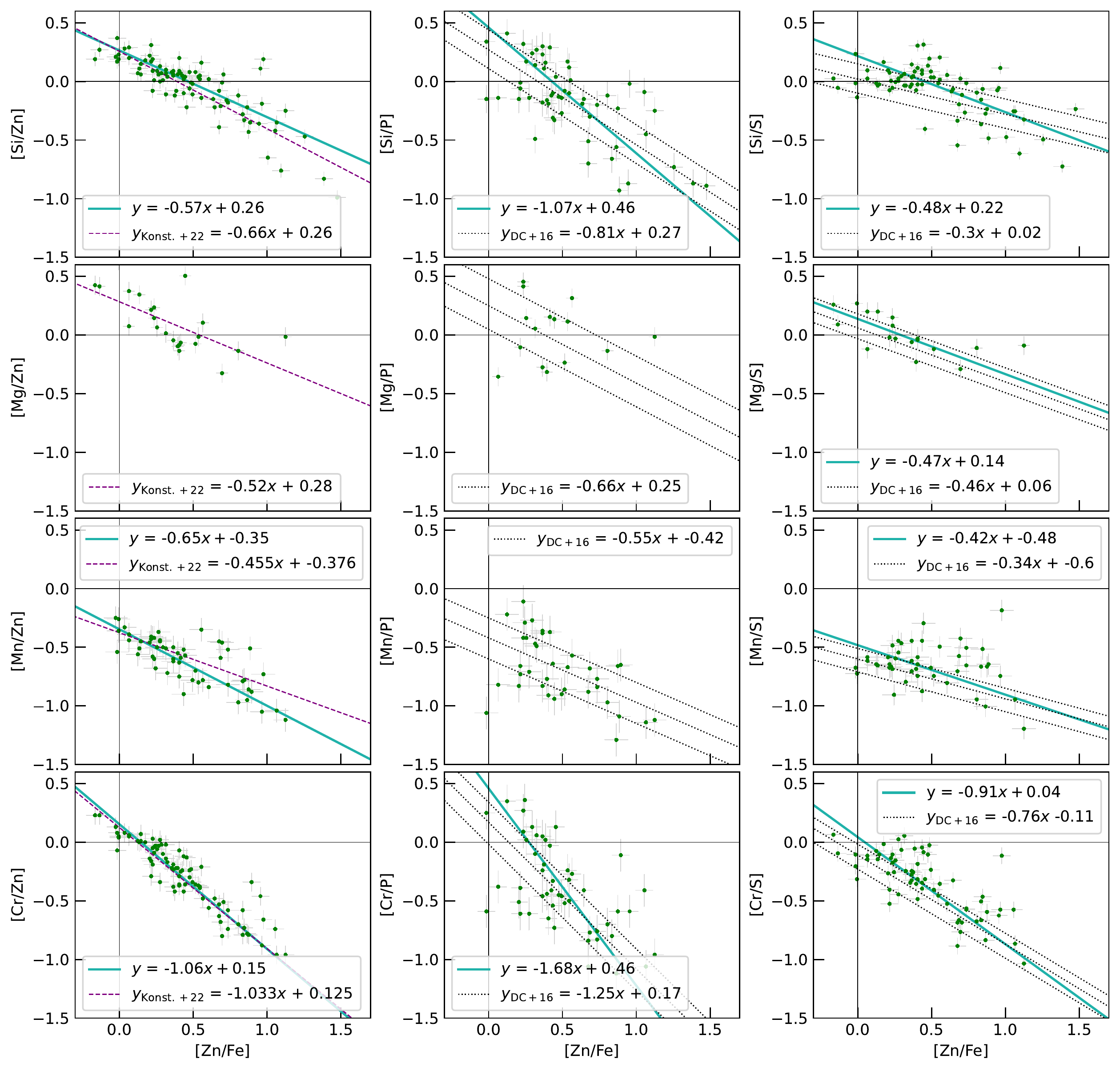}
        \caption{Depletion sequences (solid cyan line) fit to individual components (green points), with a comparison to those derived by \citet{DeCia+2016} and \citet{Konstantopoulou2022} (black dotted and purple dashed lines respectively). We do not include linear fits for [Mg/P], [Mg/Zn] and [Mn/P] because the linear correlation coefficient is too small.}
        \label{fig:depl-sequences}
    \end{figure*}

Figure \ref{fig:depl-sequences} shows the relation between the dust tracer [Zn/Fe] and the relative abundances of metals with respect to a volatile metal (Zn, S, or P). We do this for individual DLA components, while the original sequences were discovered and characterised for full DLAs \citep{DeCia+2016, Konstantopoulou2022}. The correlations are evident also for individual gas components, which justifies our use of the slopes of these correlations for our sample as well. 

% =========================================================
%\input{figures/depl-seq.tex}

% =========================================================

\section{[Zn/Fe] vs [Zn/Fe]$_{\mathrm{fit}}$}
\begin{figure}[H]
        \centering
        \includegraphics[width=\hsize]{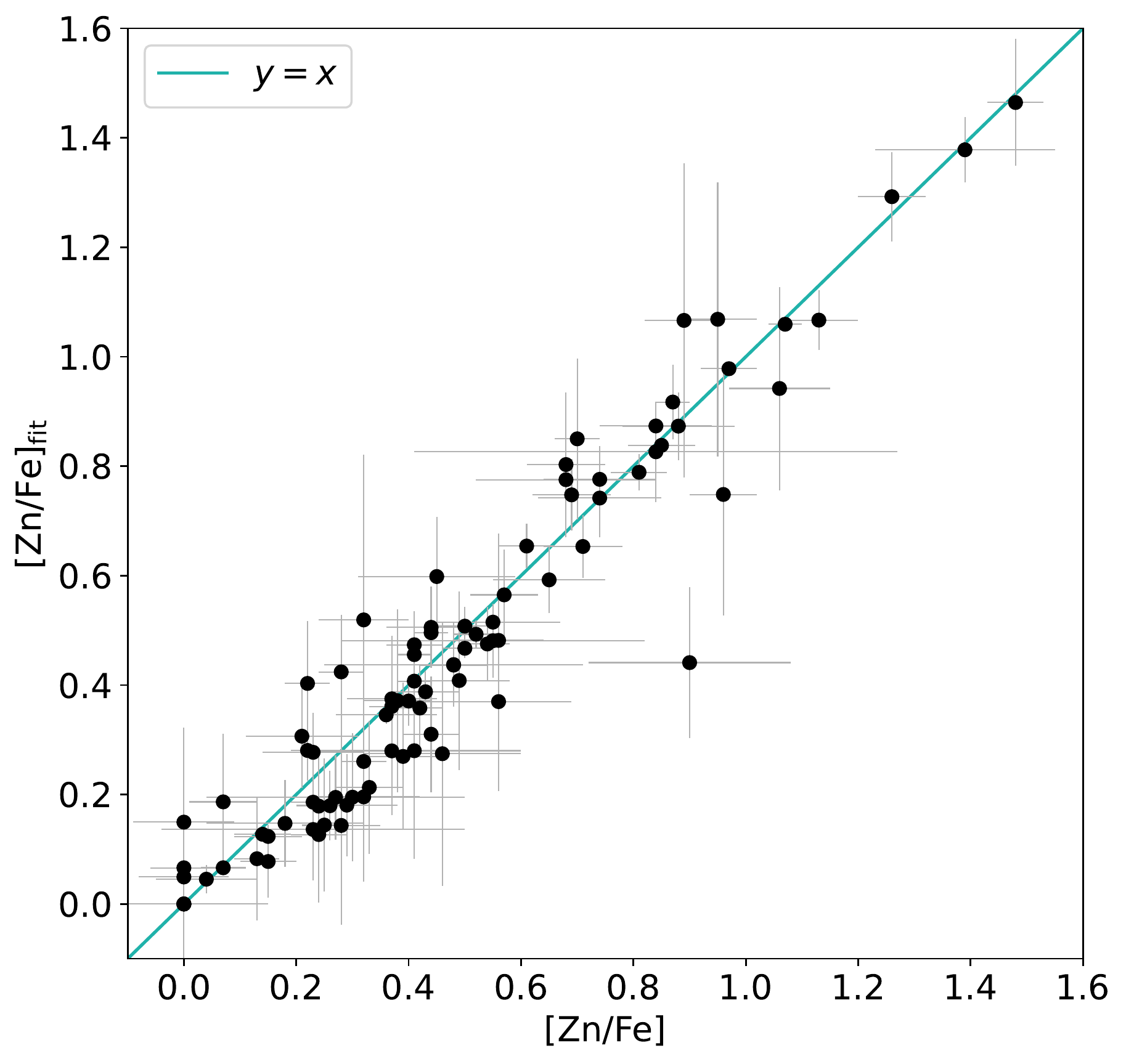}
        \caption{Comparison between two determinations of the depletion factor [Zn/Fe] and [Zn/Fe]$_{\mathrm{fit}}$}
        \label{fig:ZnFe_vs_fit}
    \end{figure}

Figure \ref{fig:ZnFe_vs_fit} shows a comparison between two determinations of the depletion factor. We gain a better understanding of the level of depletion when we consider more than two metals. 

% =========================================================

%\input{figures/Zn_vs_FIT.tex}

% =========================================================

\newpage

\section{Depletion patterns}

\begin{figure}[H]
    \centering
    \includegraphics[width=0.425\textwidth]{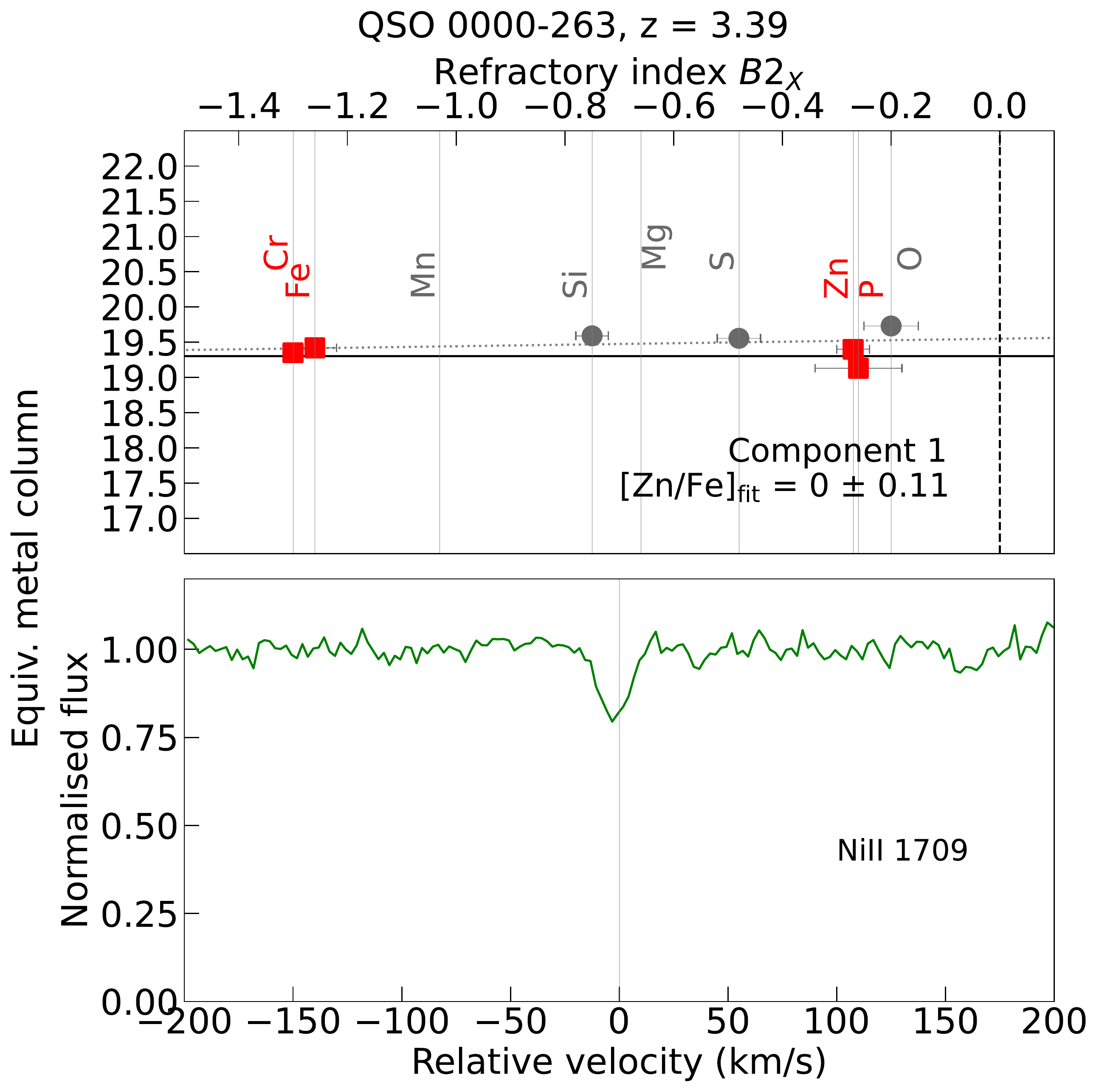}
    \caption{Depletion patterns and respective spectrum for QSO~0000-263} 
    \label{fig:depl-patt1}
    \end{figure}

Figures \ref{fig:depl-patt1} to \ref{fig: depl_patt_last} show the depletion patterns of the individual components for the 70 DLAs in our sample, except those already shown in Figs. \ref{fig:depl-pattern_sys3} to \ref{fig:depl-pattern_sys41}. A selected line profile to exemplify the line decomposition is shown in the bottom panel for each system. We fit linear relations to the data, with the exception of those components with two or less constrained metal column densities, for a total of 64 systems. From the total of 70 systems, we have not included those for which all of the individual components have only one or two constrained measurements of ionic column densities.

% =====================================================
%\input{figures/depl-patterns_app_sept22.tex}

\clearpage

\balance

%\maxdeadcycles=400

\begin{figure}[H]
    \centering
    \includegraphics[width=0.425\textwidth]{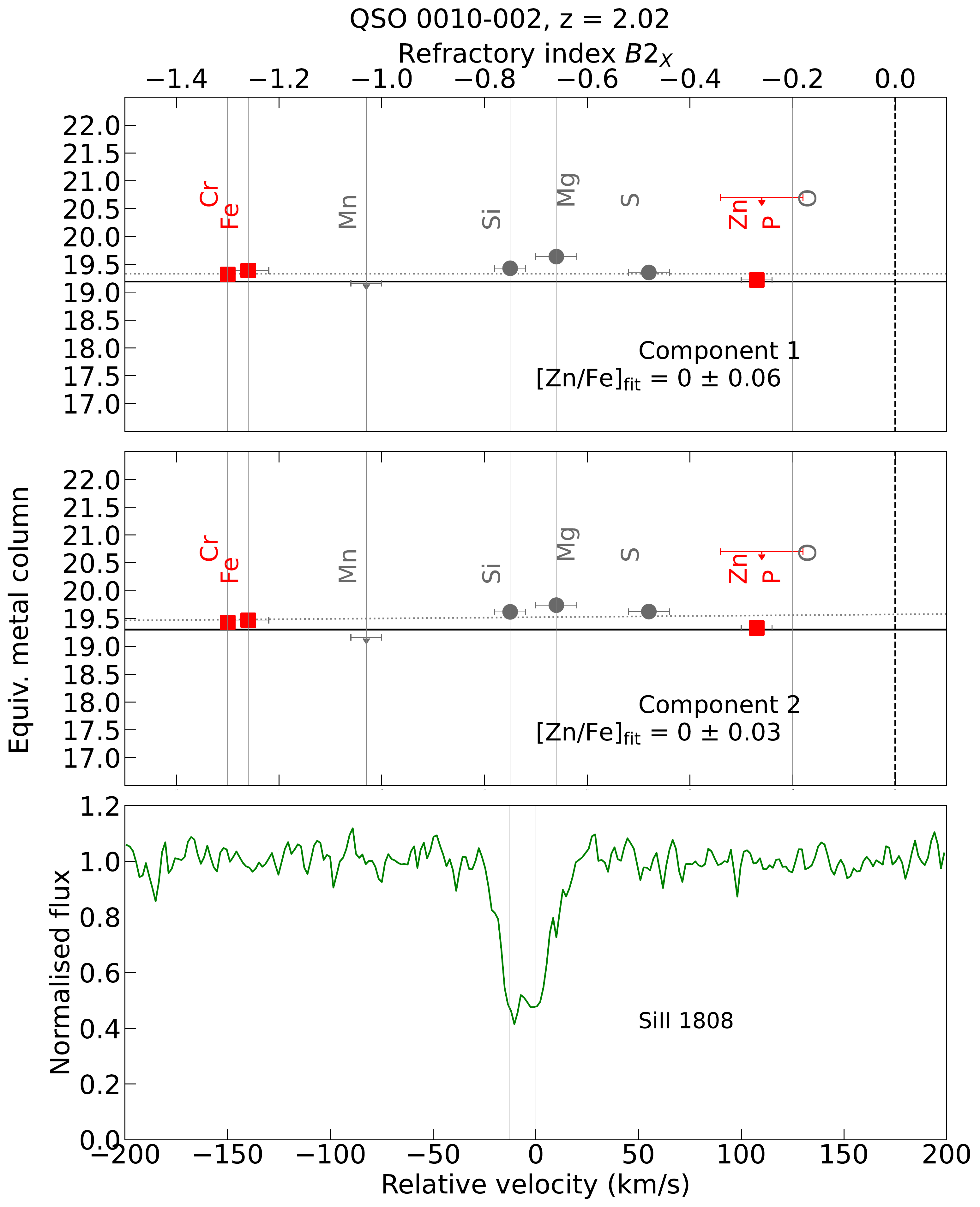}
    \caption{Depletion patterns and respective spectrum for Q~0010-002} 
    \end{figure}

\begin{figure}[H]
    \centering
    \includegraphics[width=0.425\textwidth]{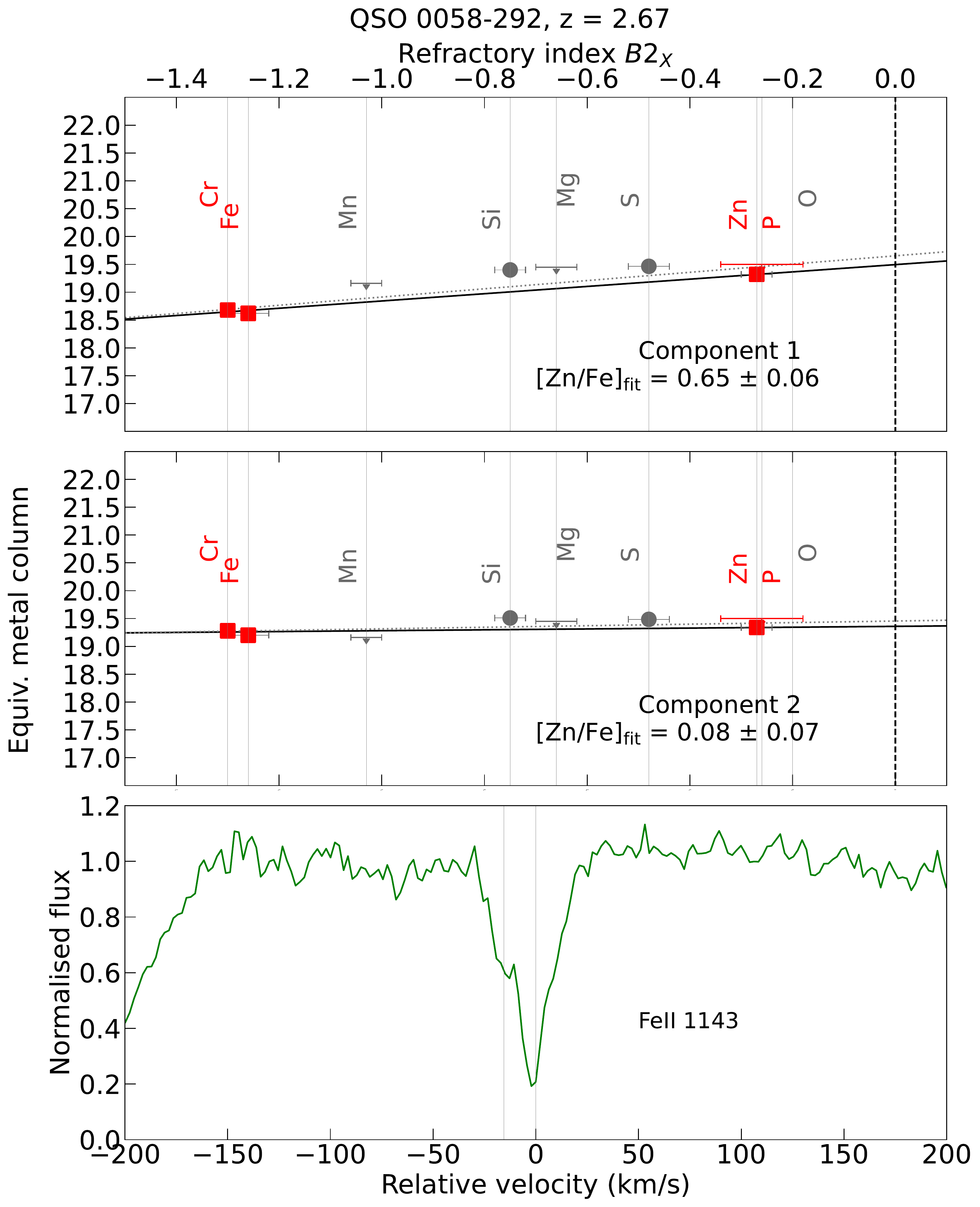}
    \caption{Depletion patterns and respective spectrum for QSO~0058-292} 
    \end{figure}

\begin{figure}[H]
    \centering
    \includegraphics[width=0.425\textwidth]{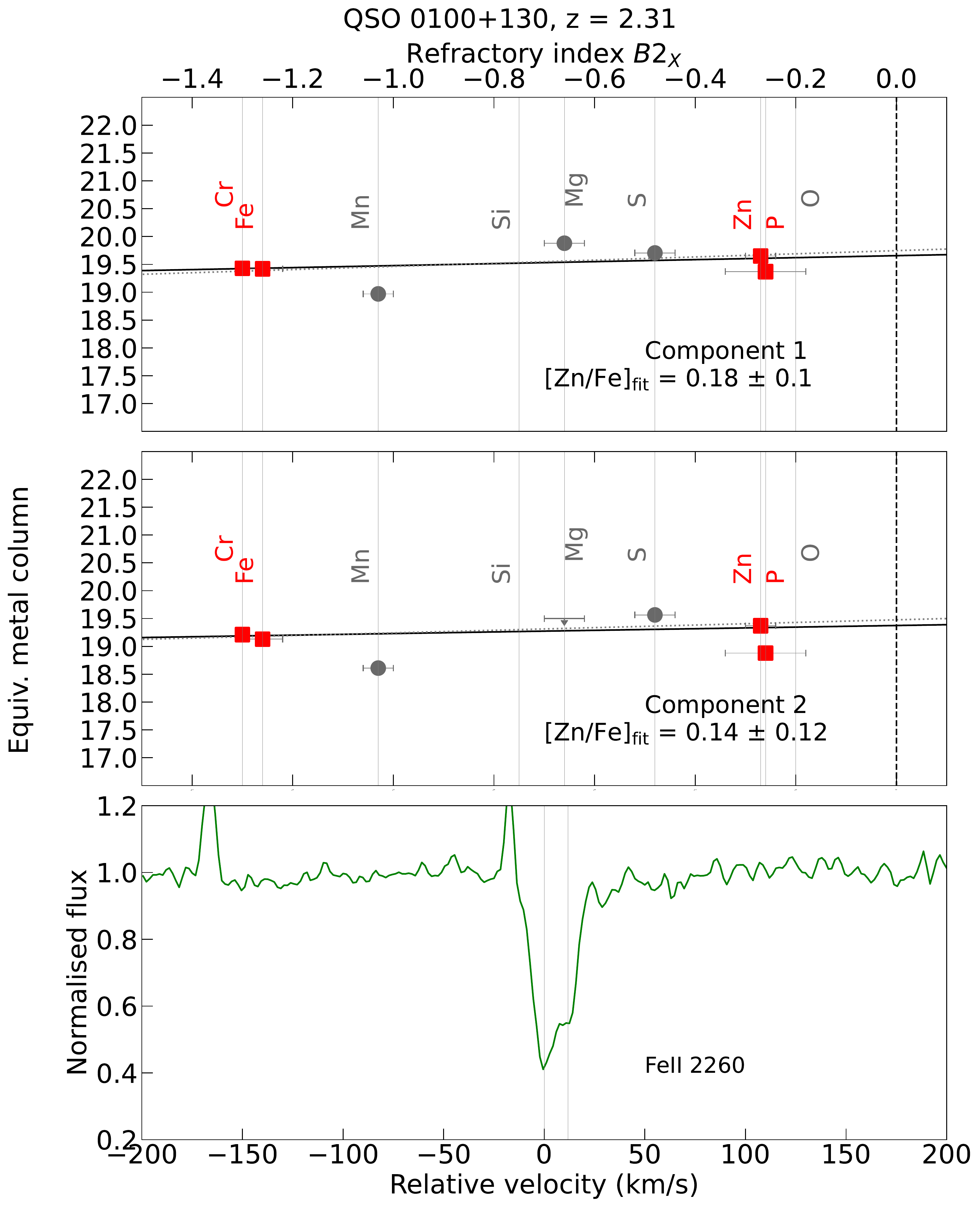}
    \caption{Depletion patterns and respective spectrum for QSO~0100+130} \end{figure}

\begin{figure}[H]
    \centering
    \includegraphics[width=0.425\textwidth]{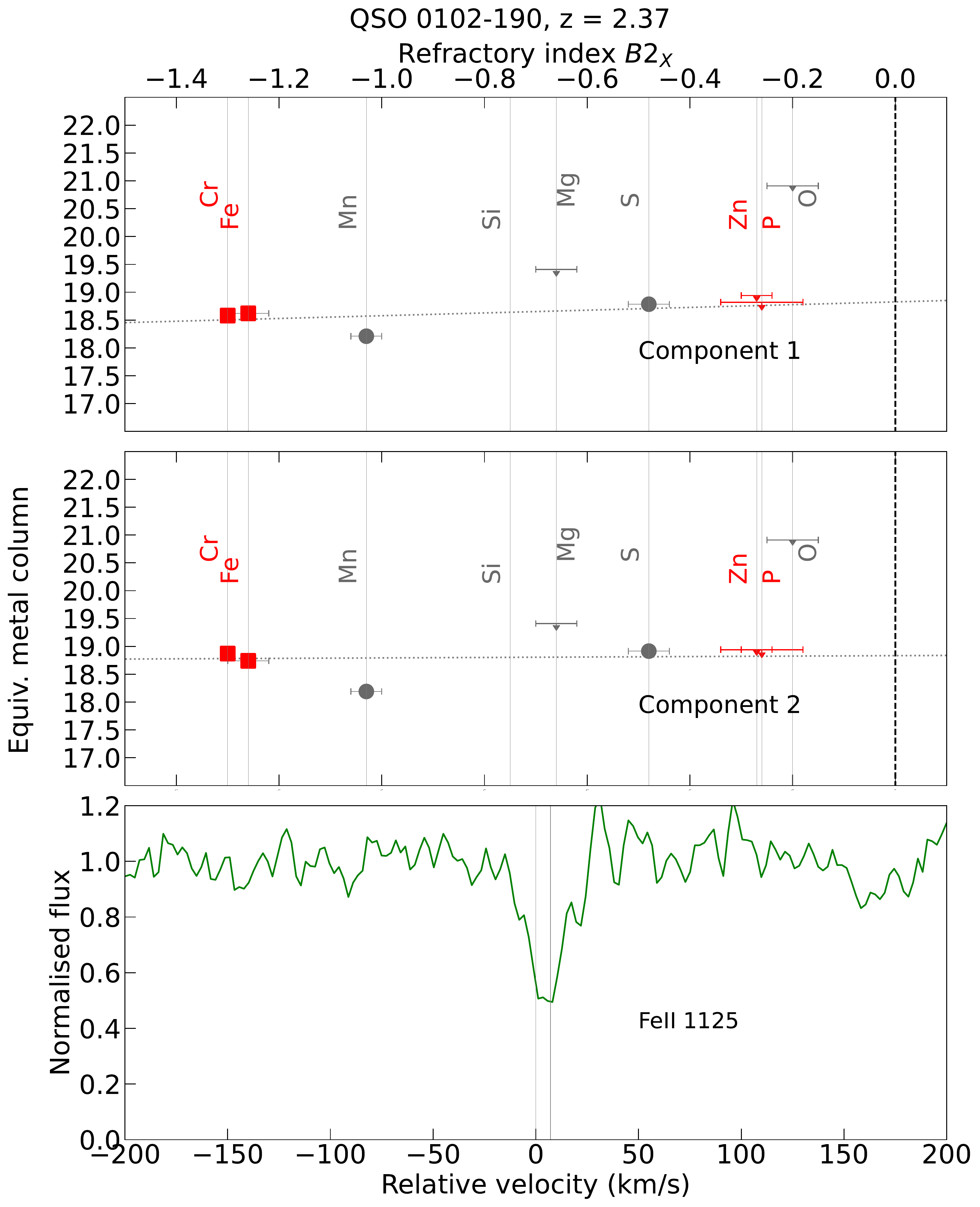}
    \caption{Depletion patterns and respective spectrum for QSO~0102-190}  \end{figure}

\begin{figure}[H]
    \centering
    \includegraphics[width=0.425\textwidth]{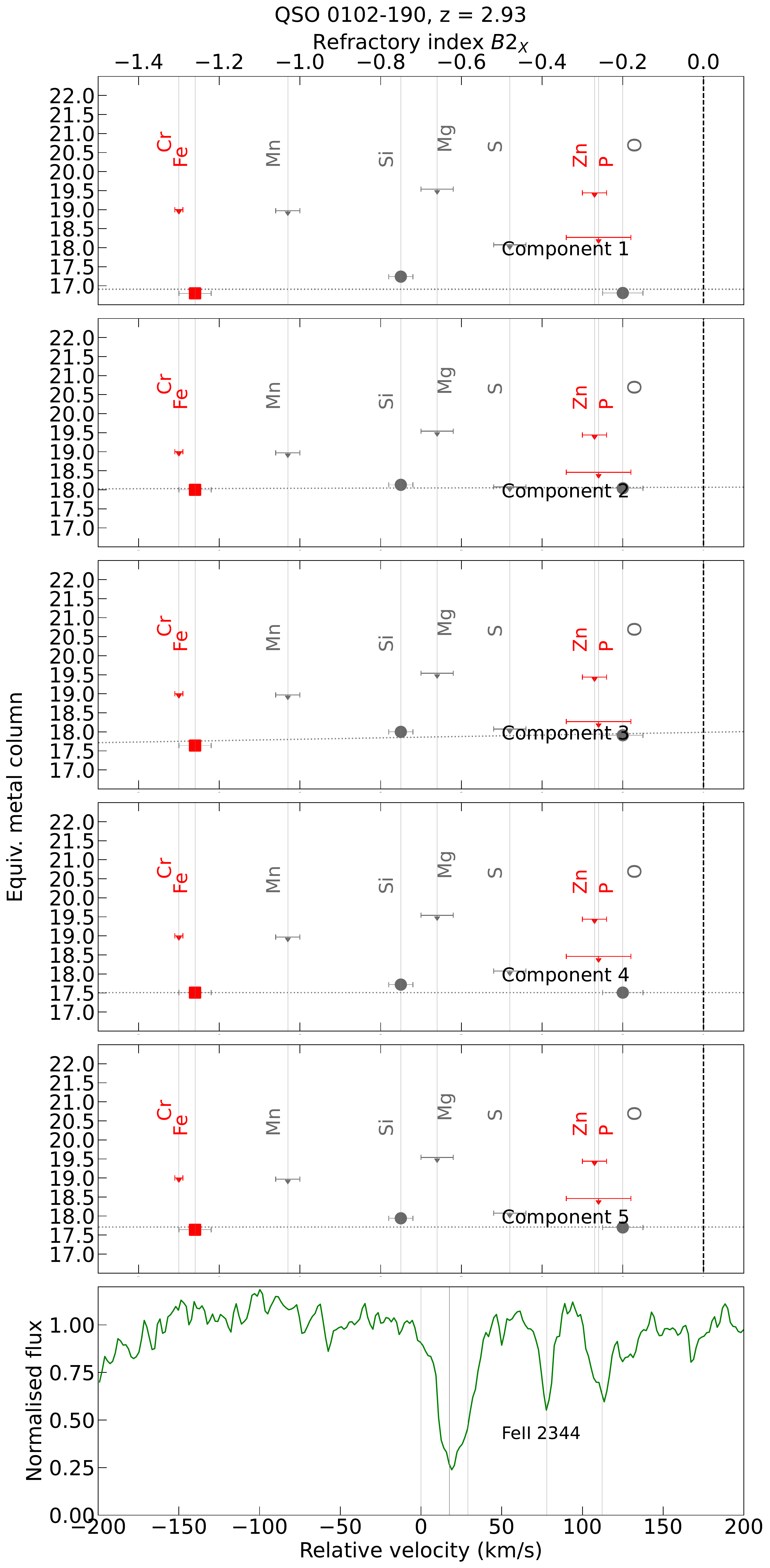}
    \caption{Depletion patterns and respective spectrum for QSO~0102-190} \end{figure}

\clearpage

\begin{figure}[H]
    \centering
    \includegraphics[width=0.425\textwidth]{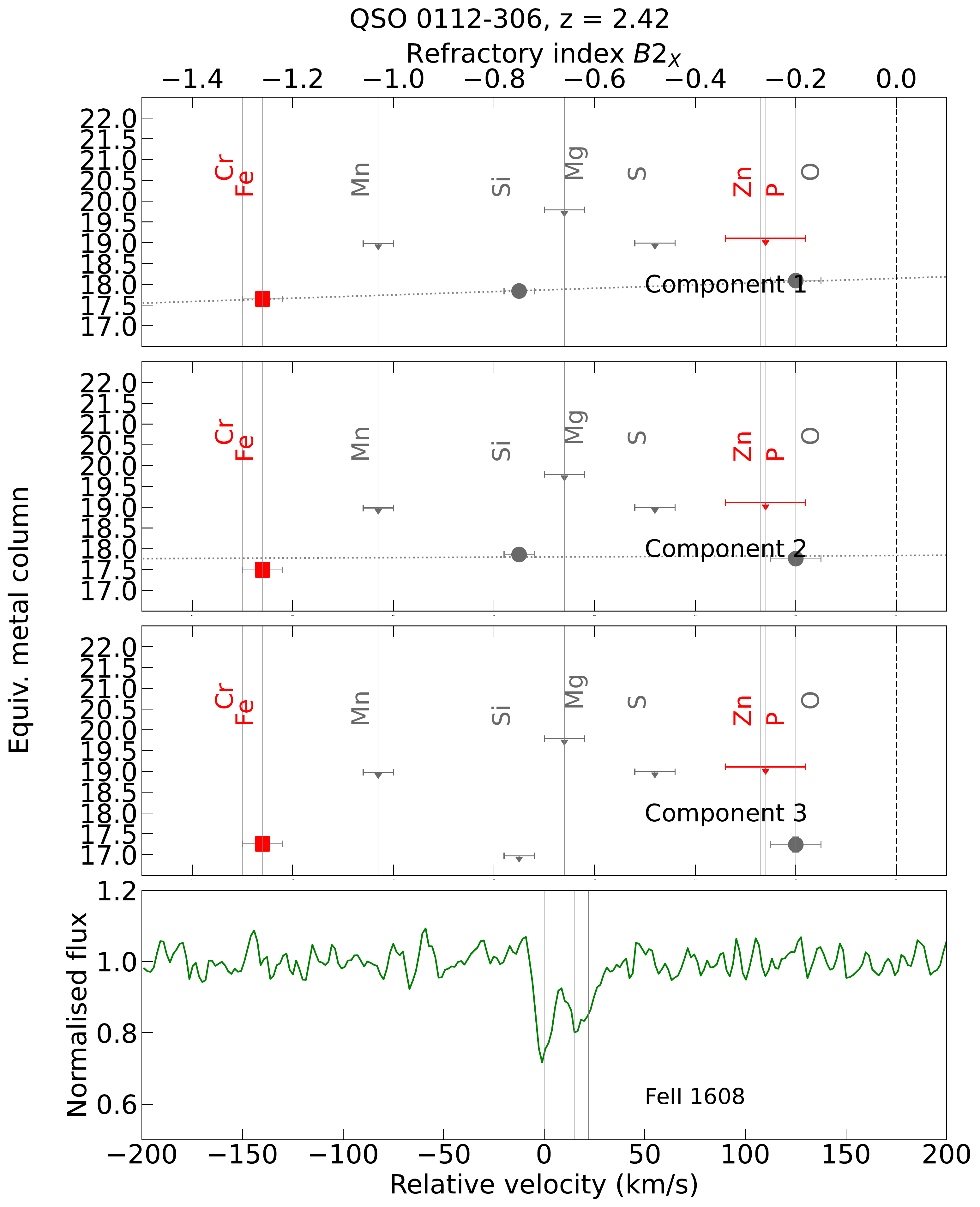}
    \caption{Depletion patterns and respective spectrum for QSO~0112-306}  \end{figure}

\begin{figure*}[h!]
    \centering
    \includegraphics[width=0.8\textwidth]{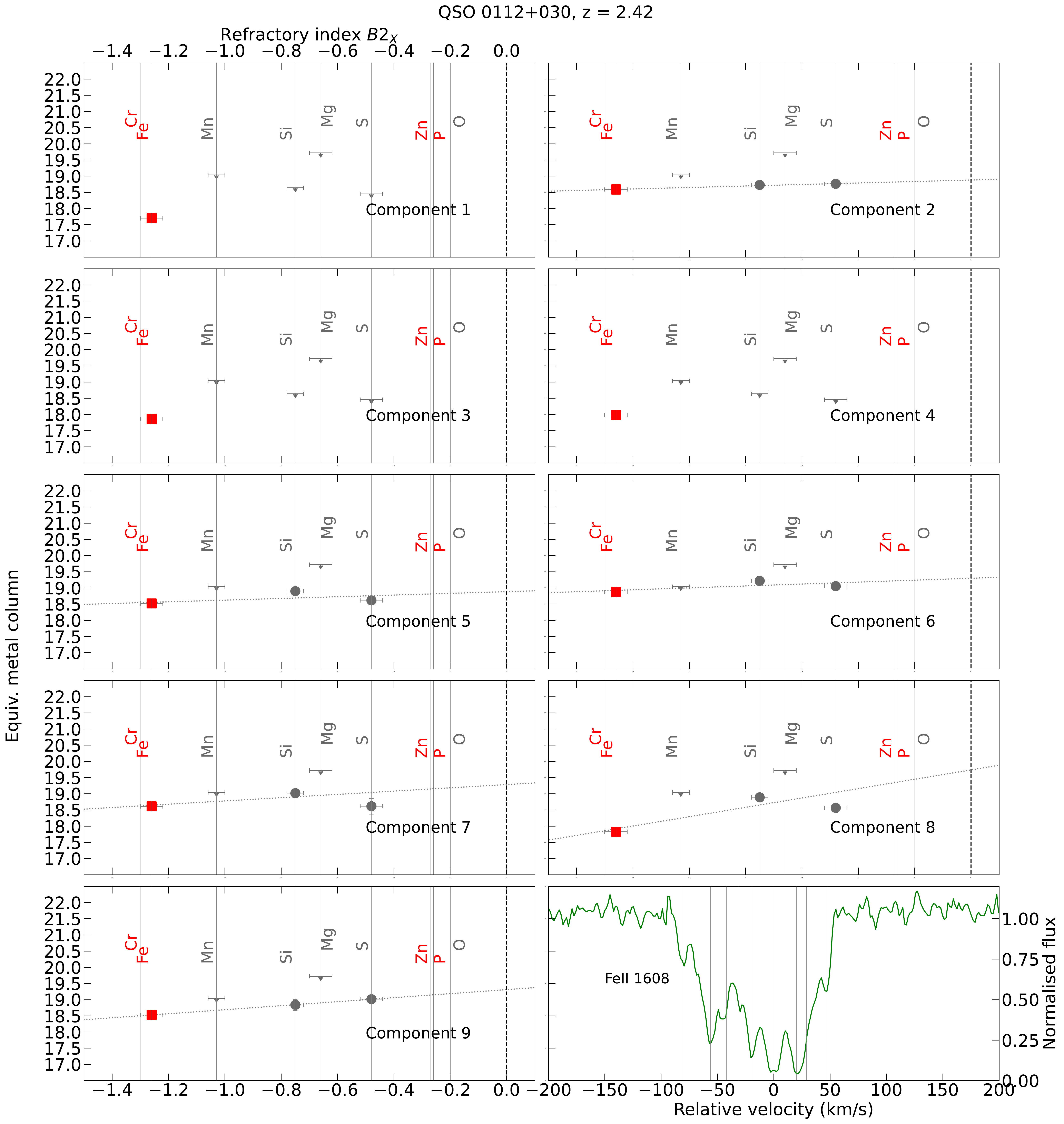}
    \caption{Depletion patterns and respective spectrum for QSO~0112+030} 
    \end{figure*}

\clearpage

\begin{figure}[H]
    \centering
    \includegraphics[width=0.425\textwidth]{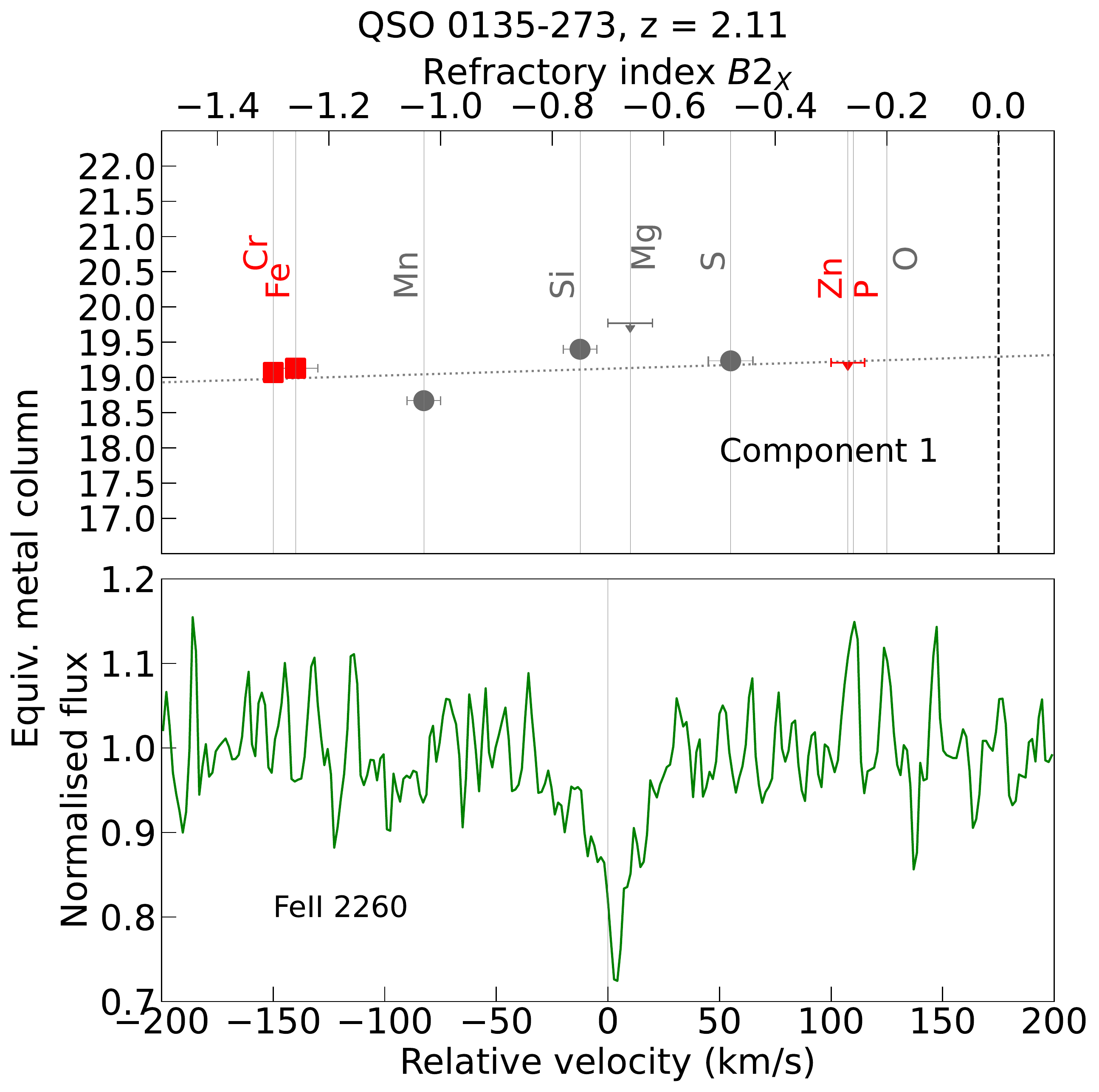}
    \caption{Depletion patterns and respective spectrum for QSO~0135-273} \end{figure}

\begin{figure}[H]
    \centering
    \includegraphics[width=0.425\textwidth]{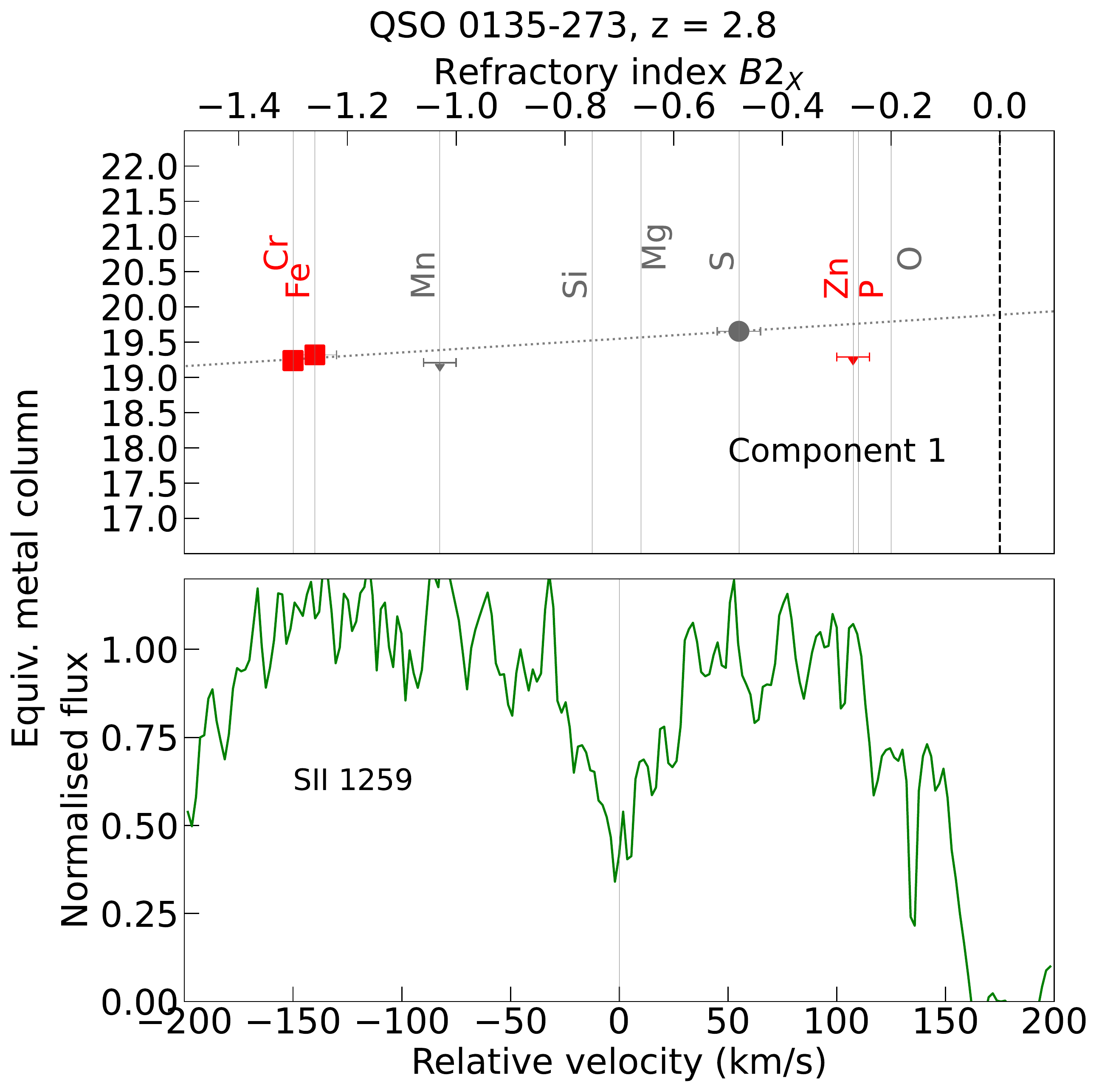}

\caption{Depletion patterns and respective spectrum for QSO~0135-273} \end{figure}

\begin{figure}[H]
    \centering
    \includegraphics[width=0.42\textwidth]{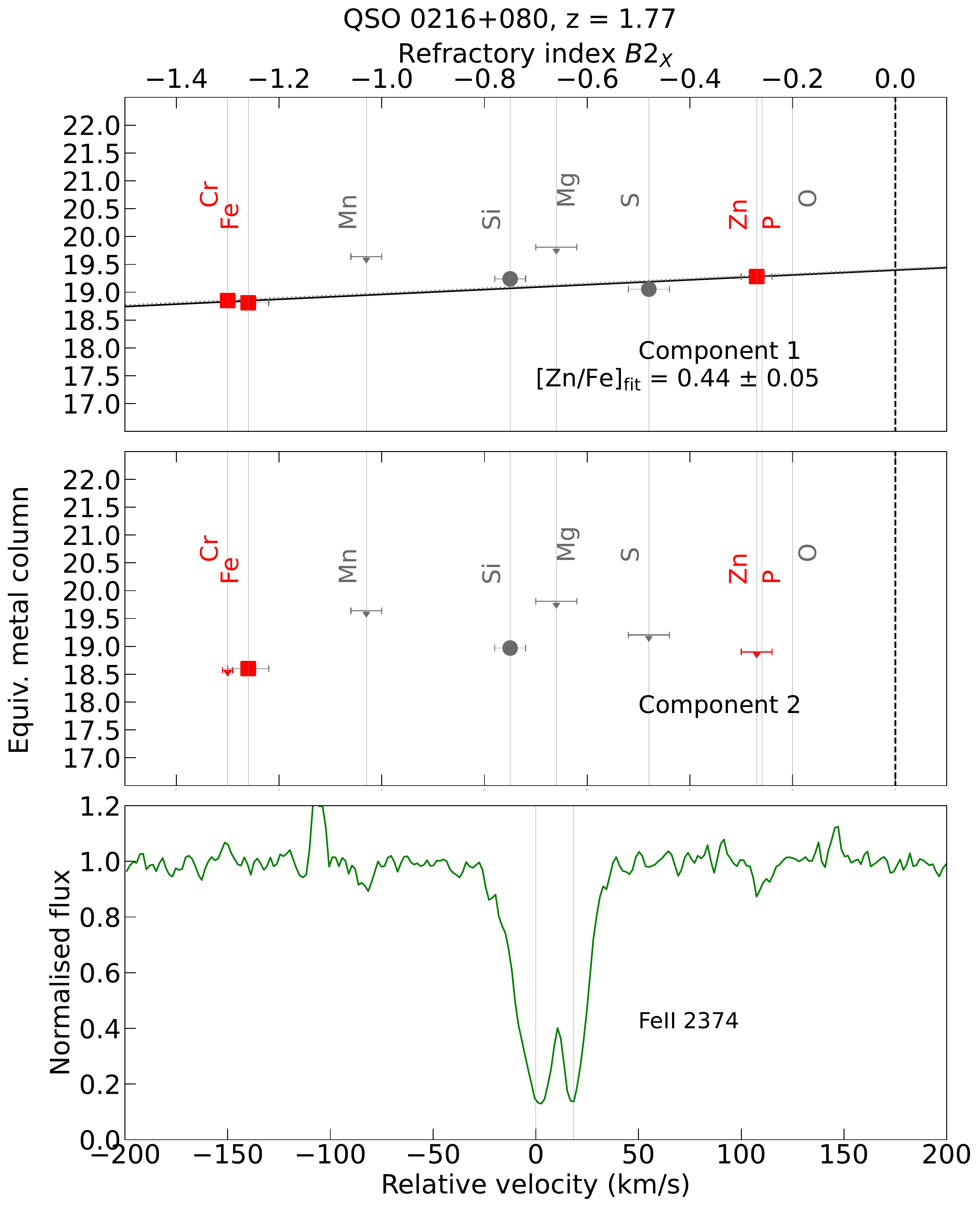}
    \caption{Depletion patterns and respective spectrum for QSO~0216+080} \end{figure}
\clearpage

\begin{figure}[H]
    \centering
    \includegraphics[width=0.4\textwidth]{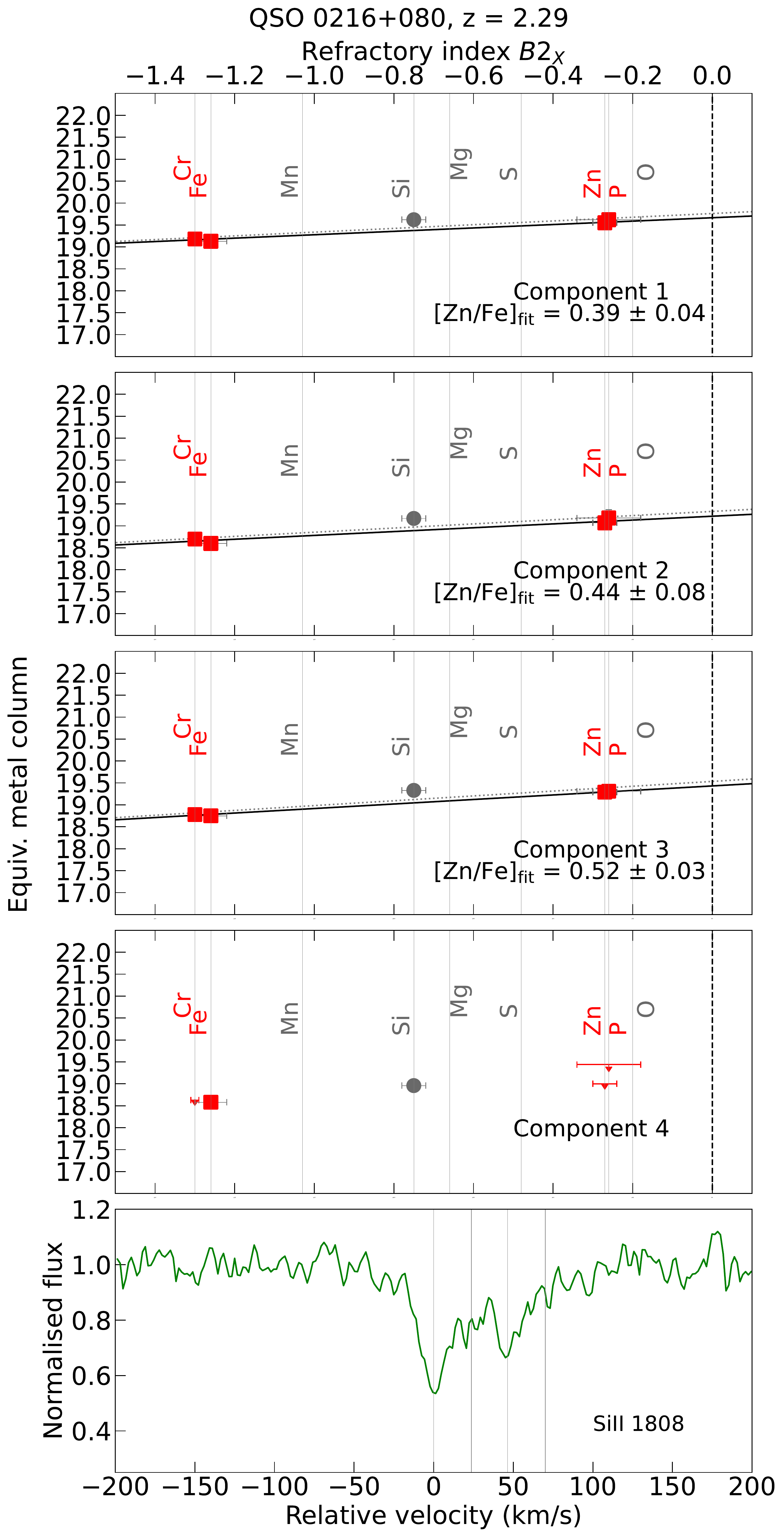}
    \caption{Depletion patterns and respective spectrum for QSO~0216+080} \end{figure}

\begin{figure}[H]
    \centering
    \includegraphics[width=0.425\textwidth]{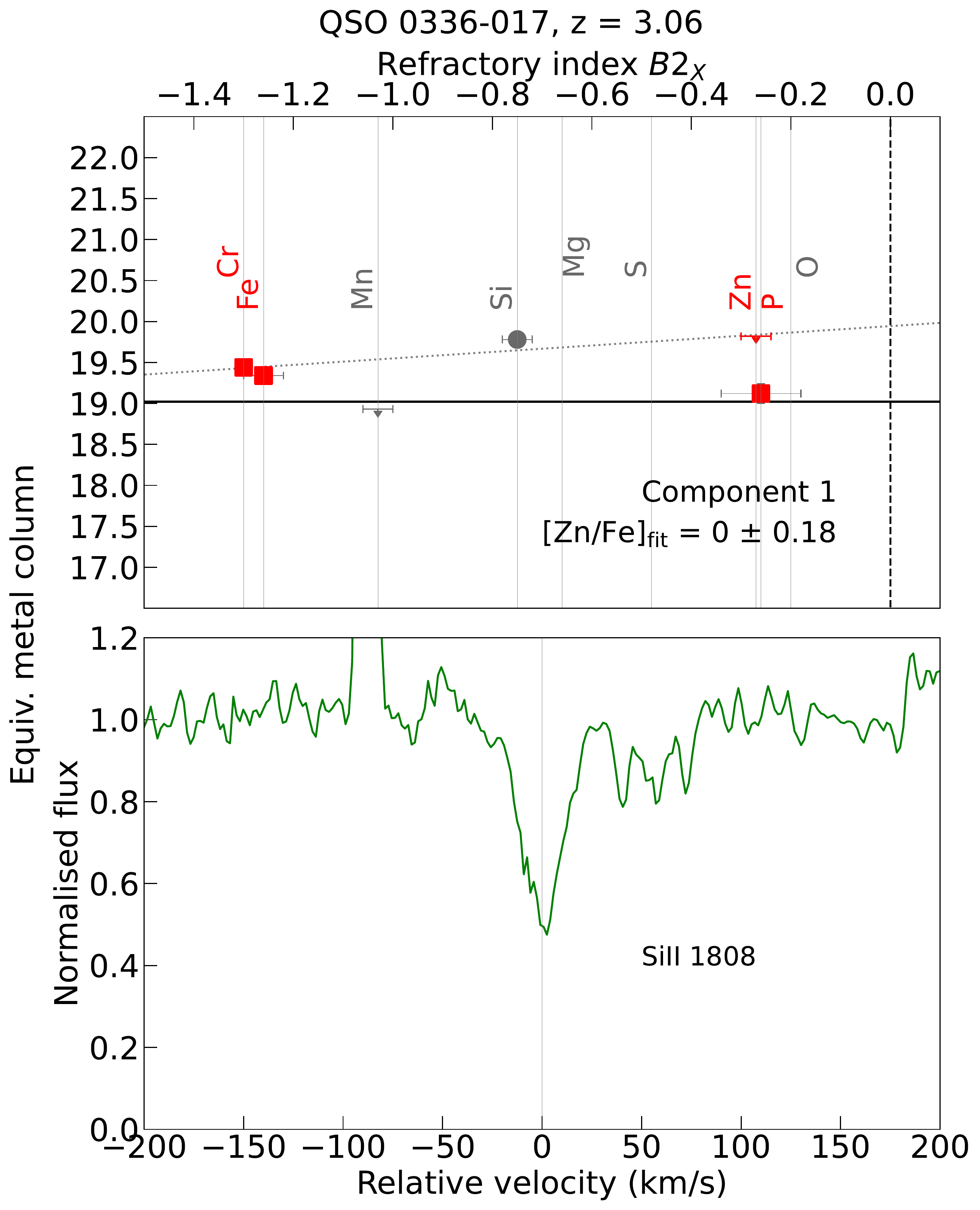}
    \caption{Depletion patterns and respective spectrum for QSO~0336-017} \end{figure}

\begin{figure}[H]
    \centering
    \includegraphics[width=0.425\textwidth]{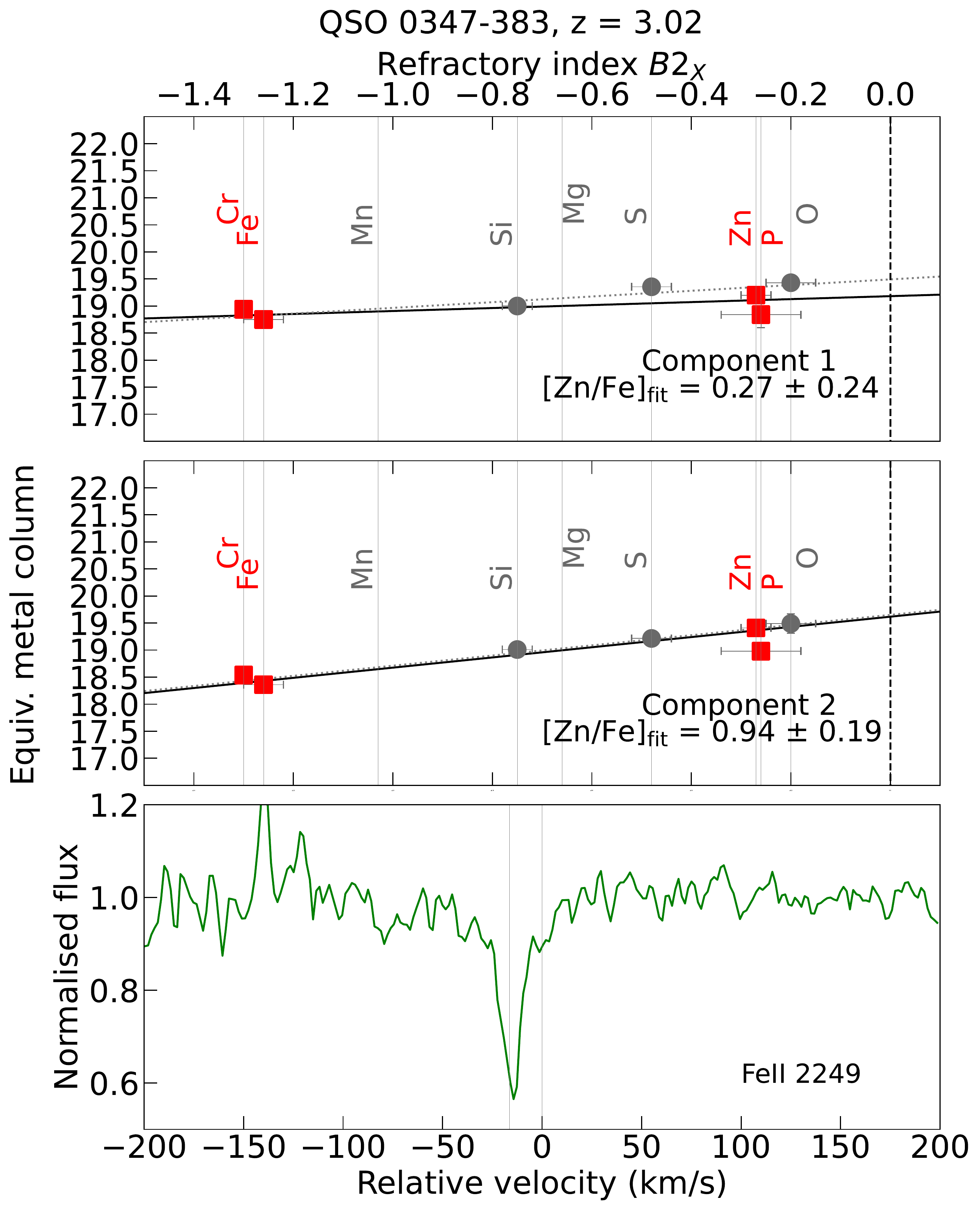}
    \caption{Depletion patterns and respective spectrum for QSO~0347-383} \end{figure}

\begin{figure}[H]
    \centering
    \includegraphics[width=0.4\textwidth]{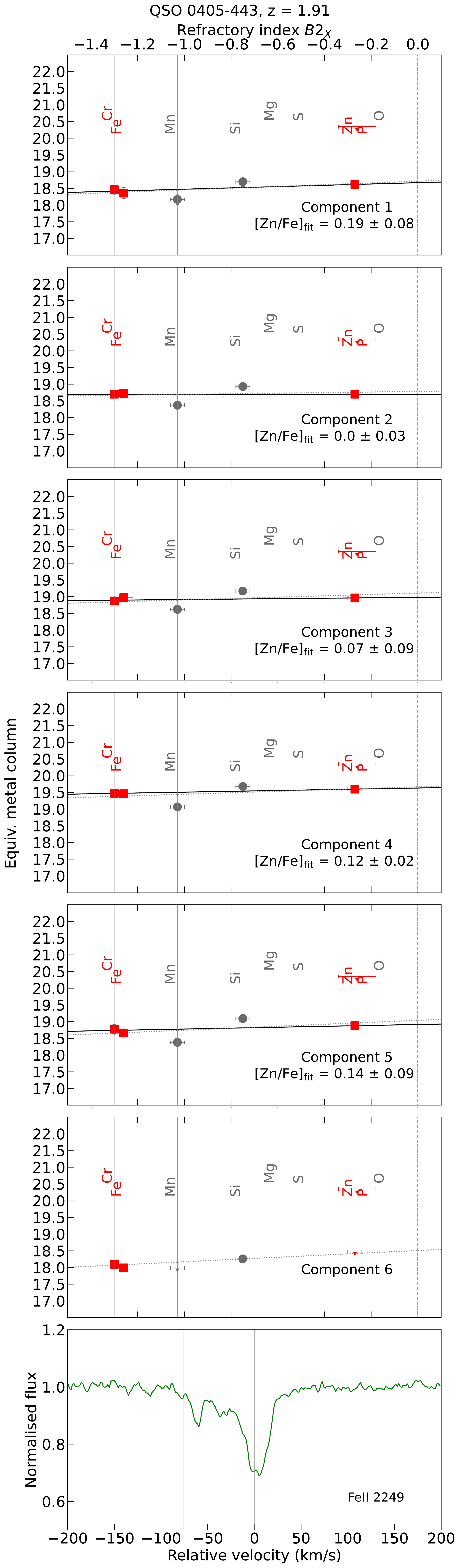}
    \caption{Depletion patterns and respective spectrum for QSO~0405-443} \end{figure}

\begin{figure}[H]
    \centering
    \includegraphics[width=0.38\textwidth]{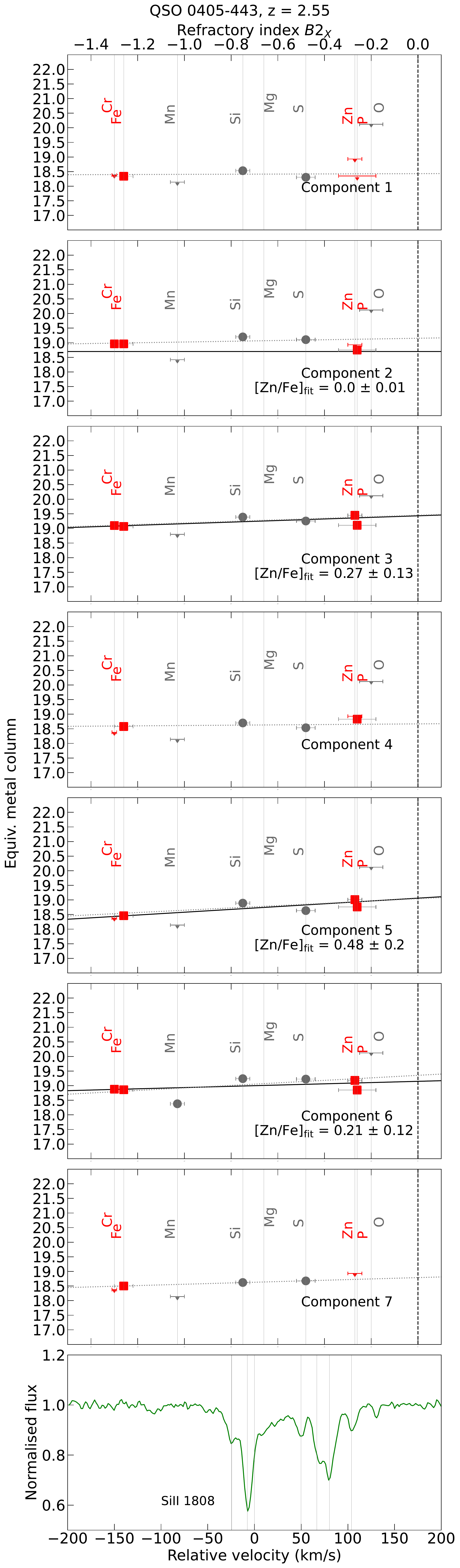}
    \caption{Depletion patterns and respective spectrum for QSO~0405-443} \end{figure}

\begin{figure}[H]
    \centering
    \includegraphics[width=0.425\textwidth]{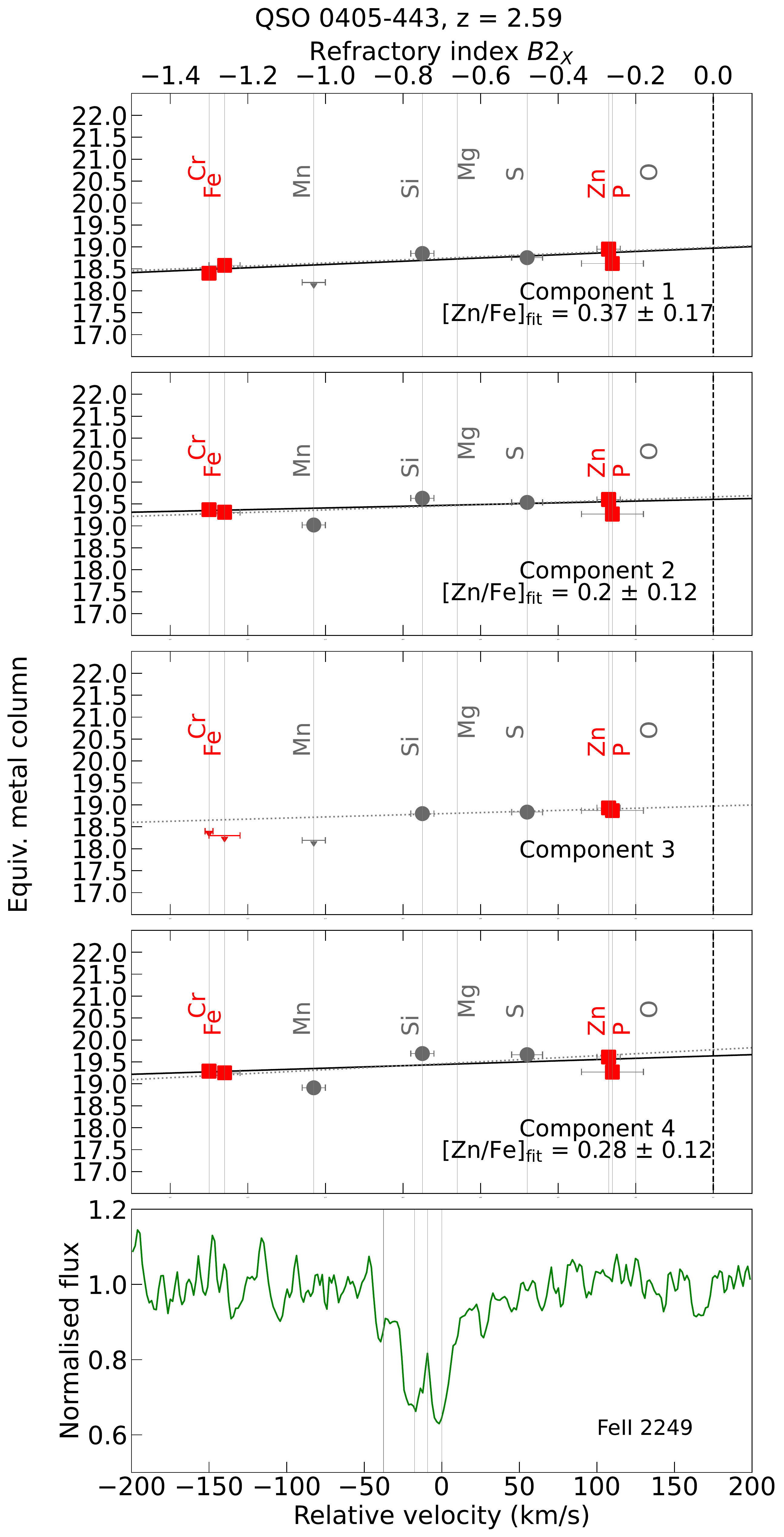}
    \caption{Depletion patterns and respective spectrum for QSO~0405-443} \end{figure}

\begin{figure}[H]
    \centering
    \includegraphics[width=0.4\textwidth]{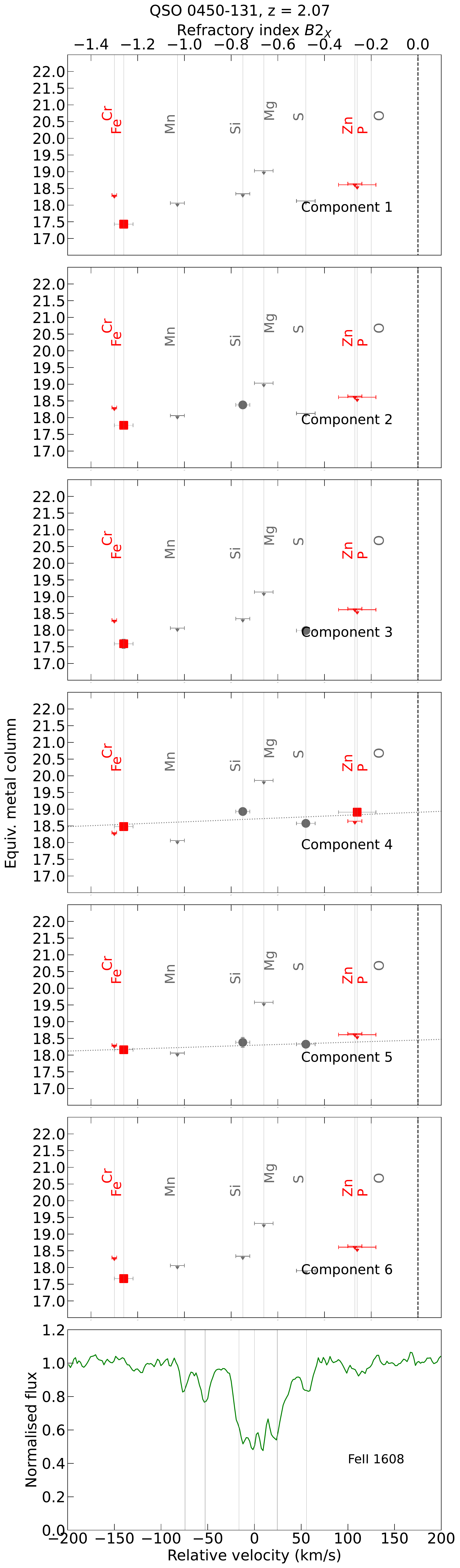}
    \caption{Depletion patterns and respective spectrum for QSO~0450-131} \end{figure}

\begin{figure}[H]
    \centering
    \includegraphics[width=0.425\textwidth]{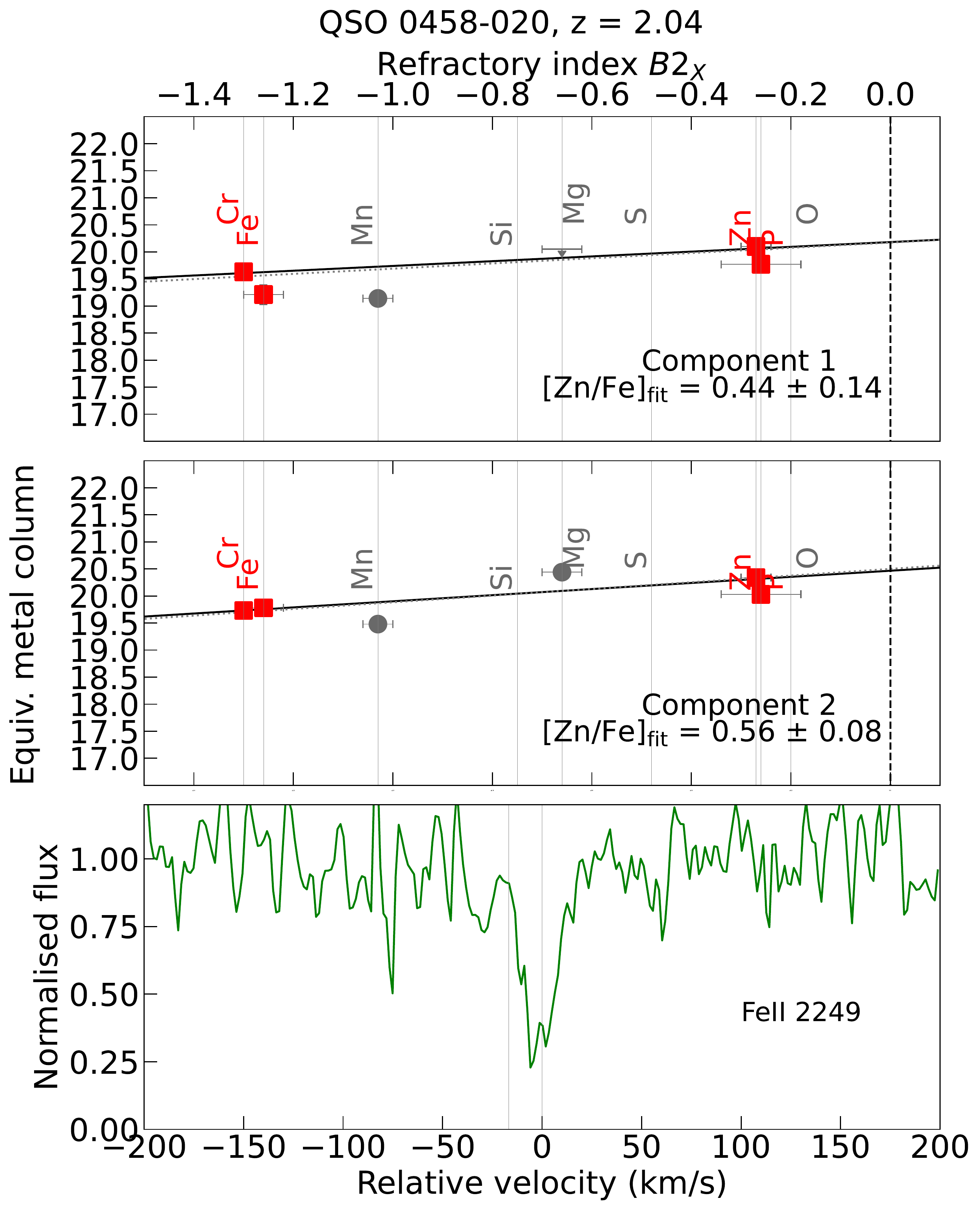}
    \label{fig:vel-depl-q0458-020}
\caption{Depletion patterns and respective spectrum for QSO~0458-020} \end{figure}

\begin{figure}[H]
    \centering
    \label{fig:vel-depl-q0528-250}
    \includegraphics[width=0.425\textwidth]{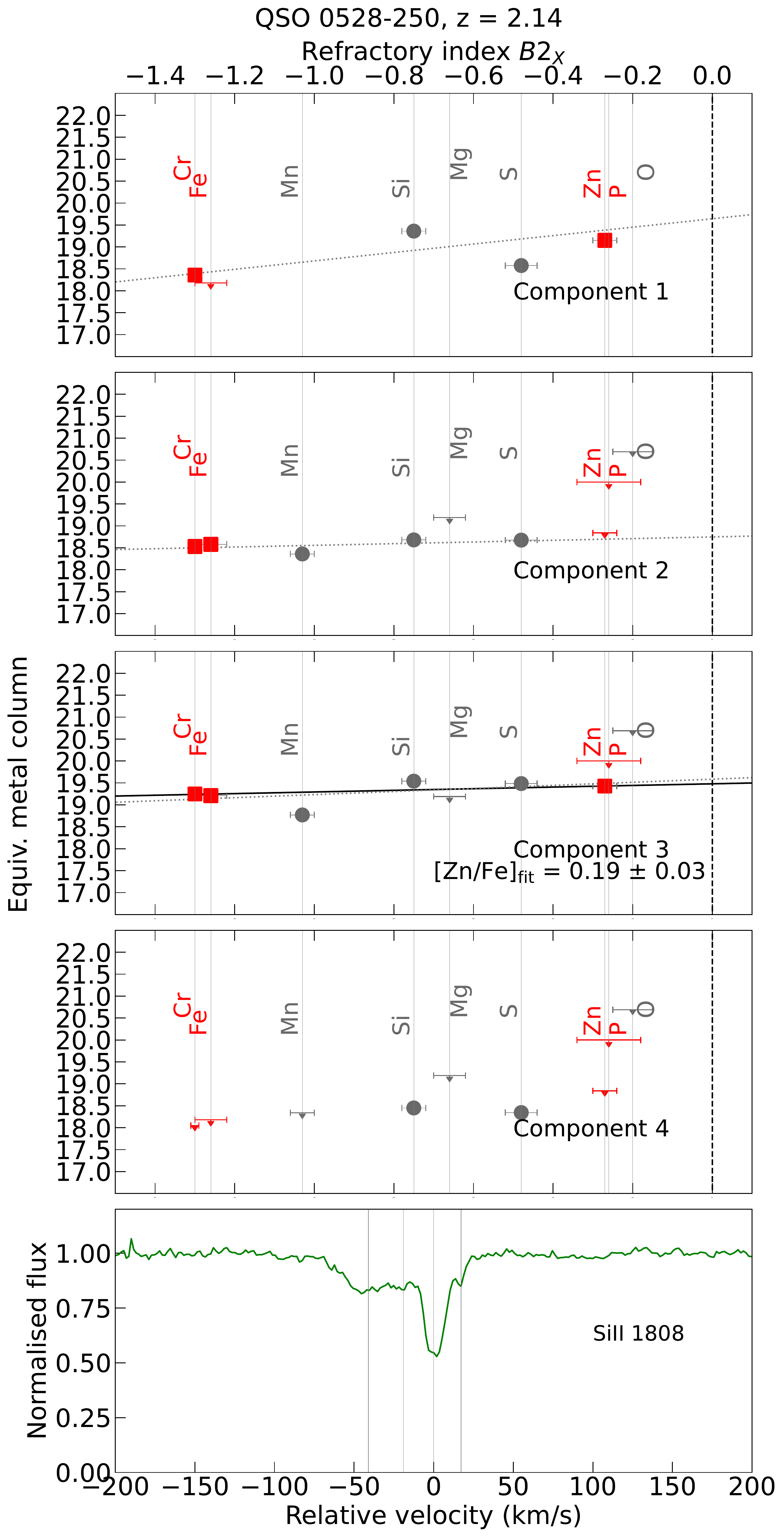}

\caption{Depletion patterns and respective spectrum for QSO~0528-250} \end{figure}

\begin{figure}[H]
    \centering
    \includegraphics[width=0.425\textwidth]{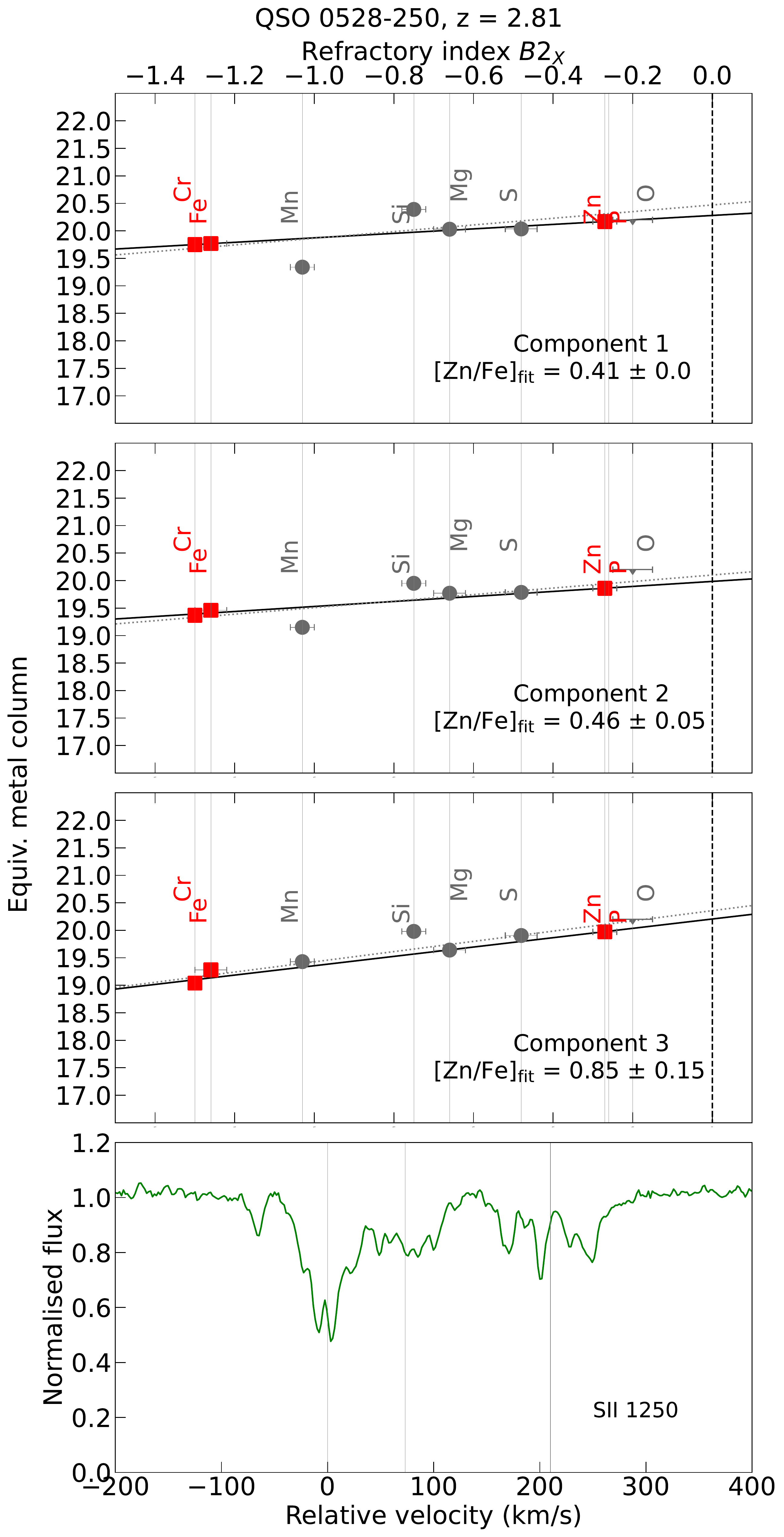}

\caption{Depletion patterns and respective spectrum for QSO~0528-250} \end{figure}

\begin{figure}[H]
    \centering
    \includegraphics[width=0.425\textwidth]{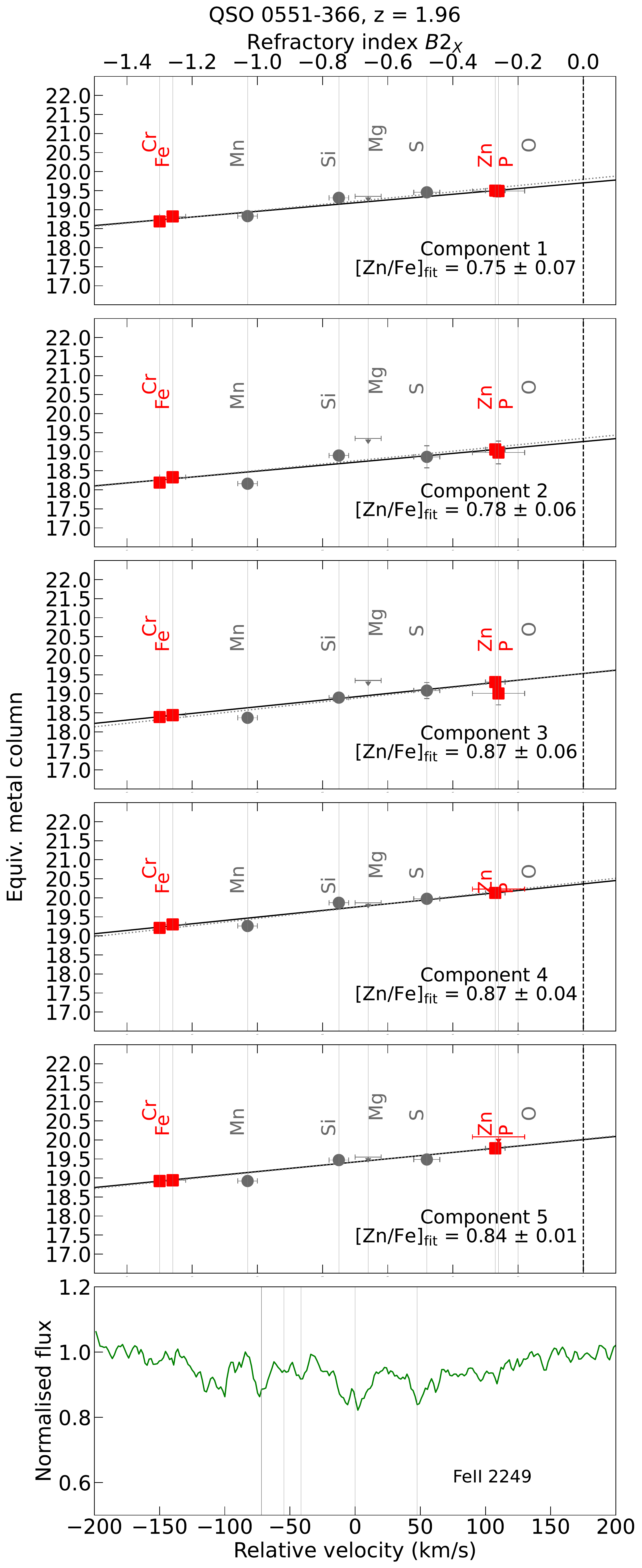}

\caption{Depletion patterns and respective spectrum for QSO~0551-366} \end{figure}

\begin{figure}[H]
    \centering
    \includegraphics[width=0.425\textwidth]{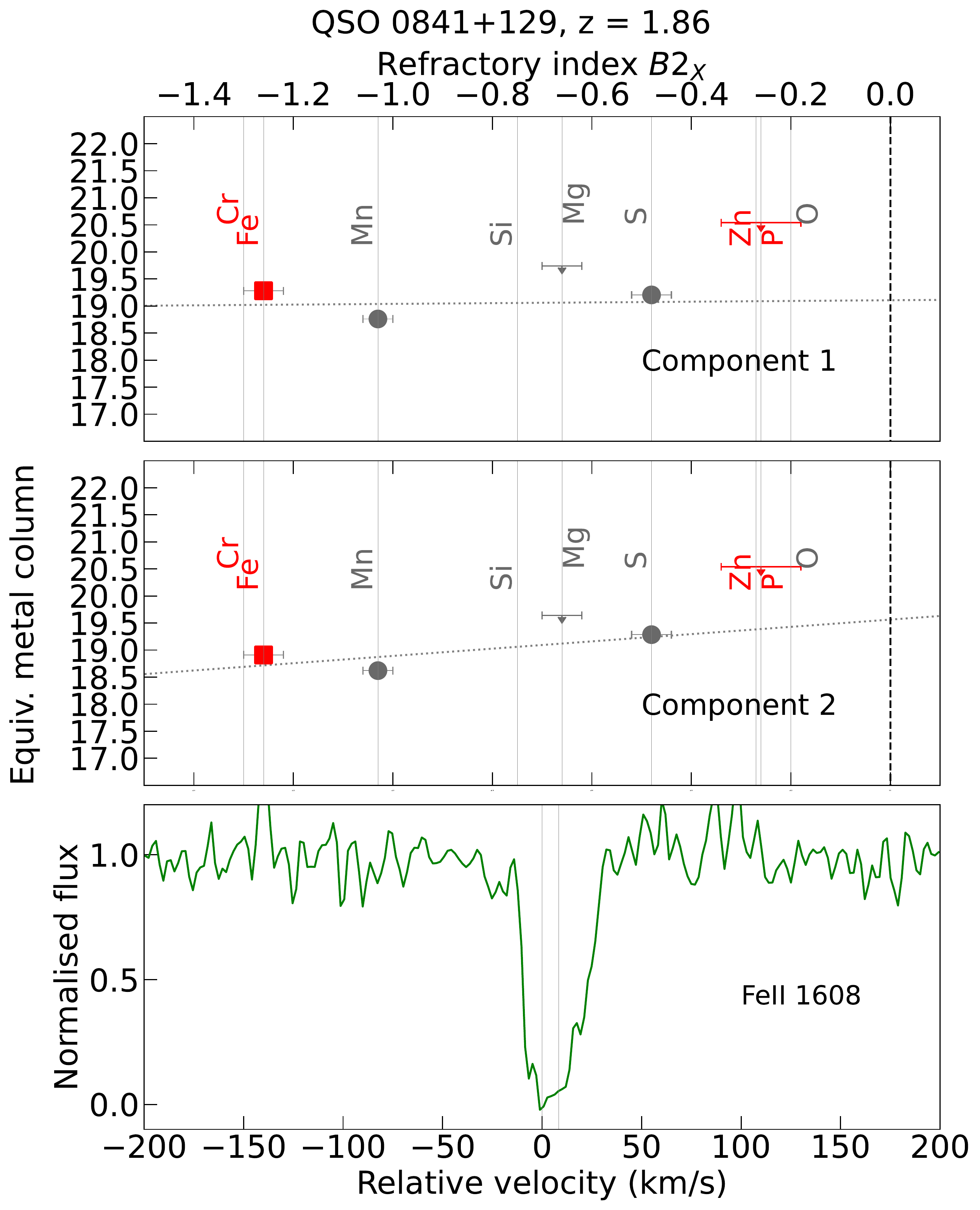}

\caption{Depletion patterns and respective spectrum for QSO~0841+129} \end{figure}

\begin{figure}[H]
    \centering
    \includegraphics[width=0.425\textwidth]{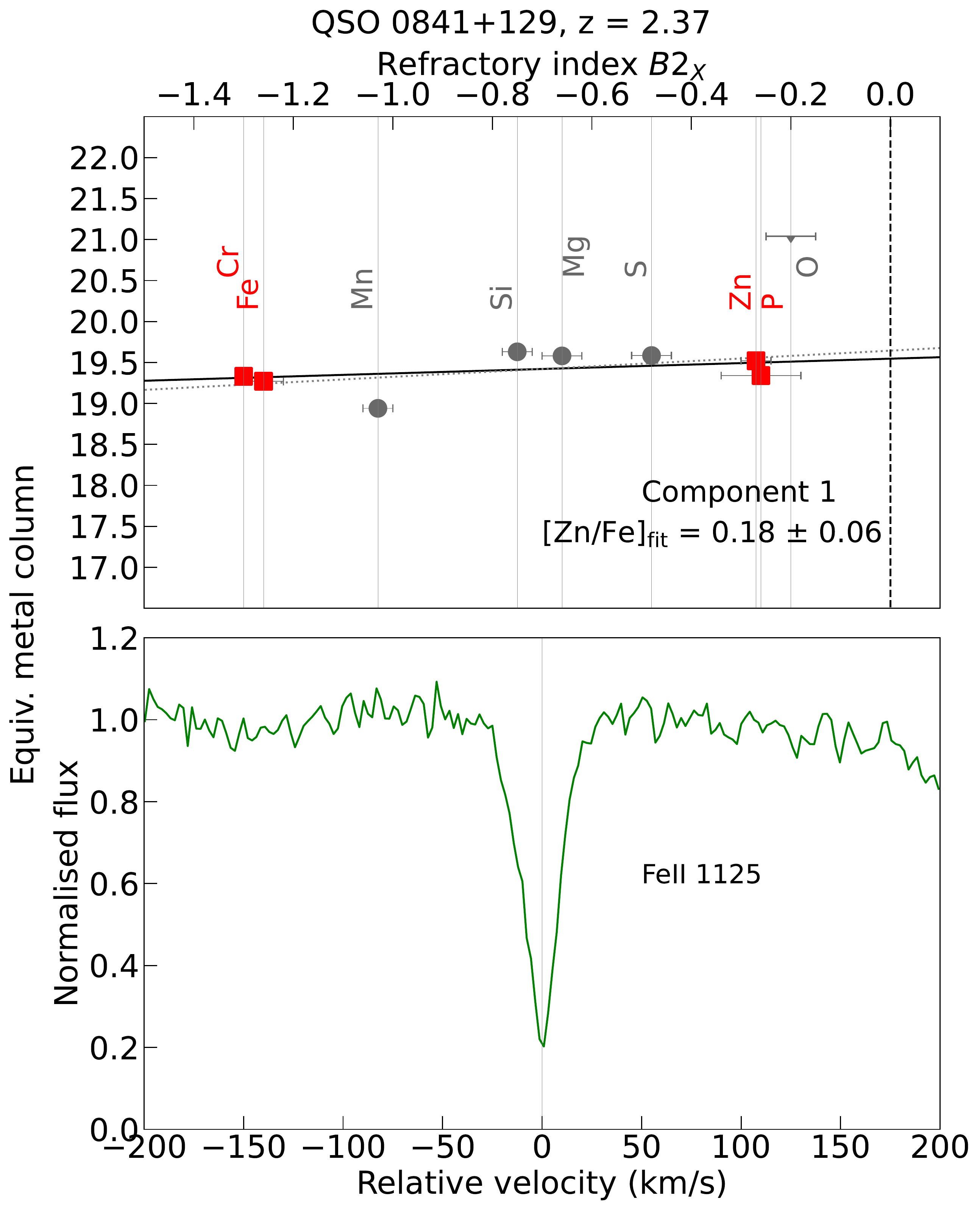}

\caption{Depletion patterns and respective spectrum for QSO~0841+129} \end{figure}

\begin{figure}[H]
    \centering
    \includegraphics[width=0.42\textwidth]{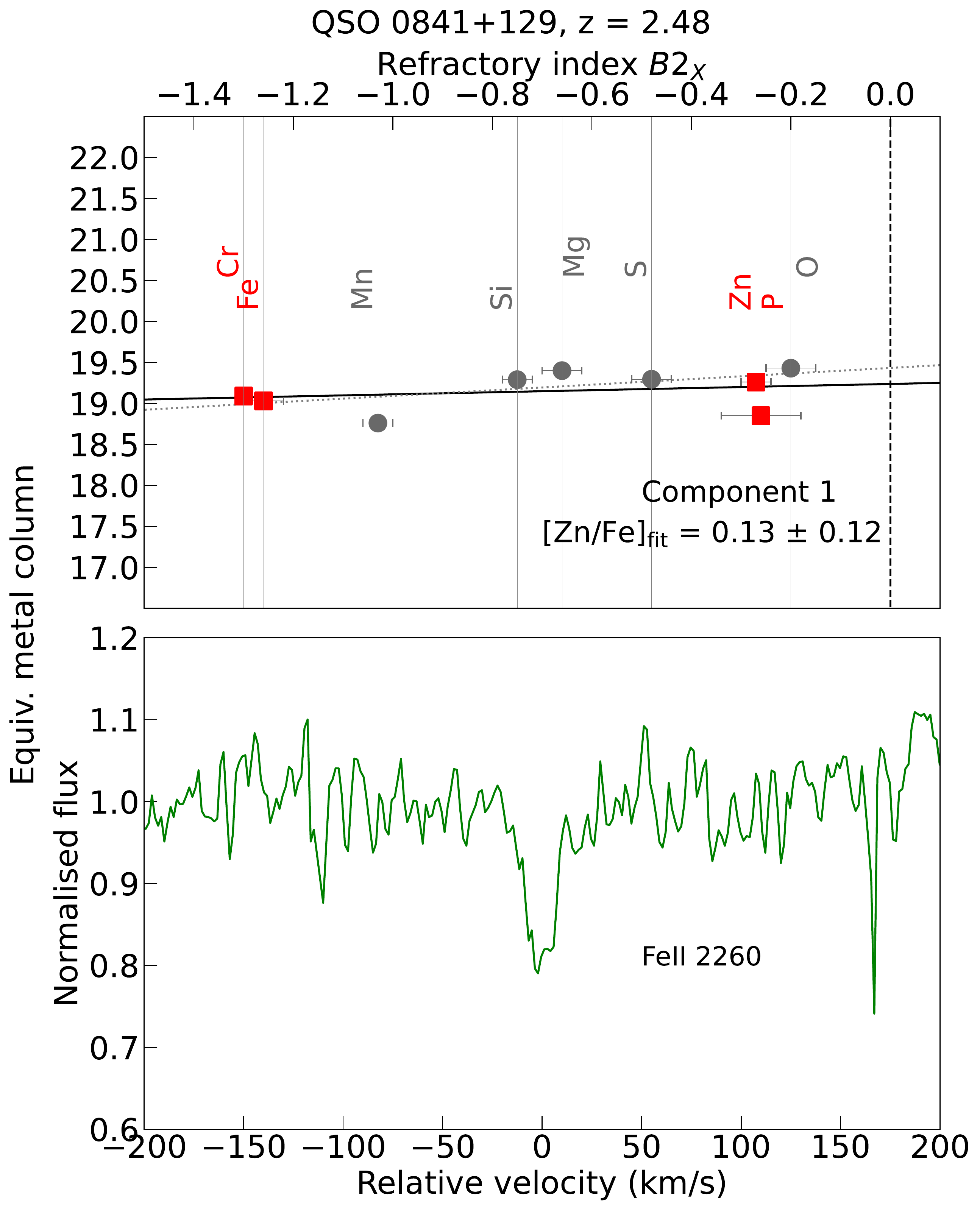}
    \caption{Depletion patterns and respective spectrum for QSO~0841+129} \end{figure}

\begin{figure}[H]
    \centering
    \includegraphics[width=0.4\textwidth]{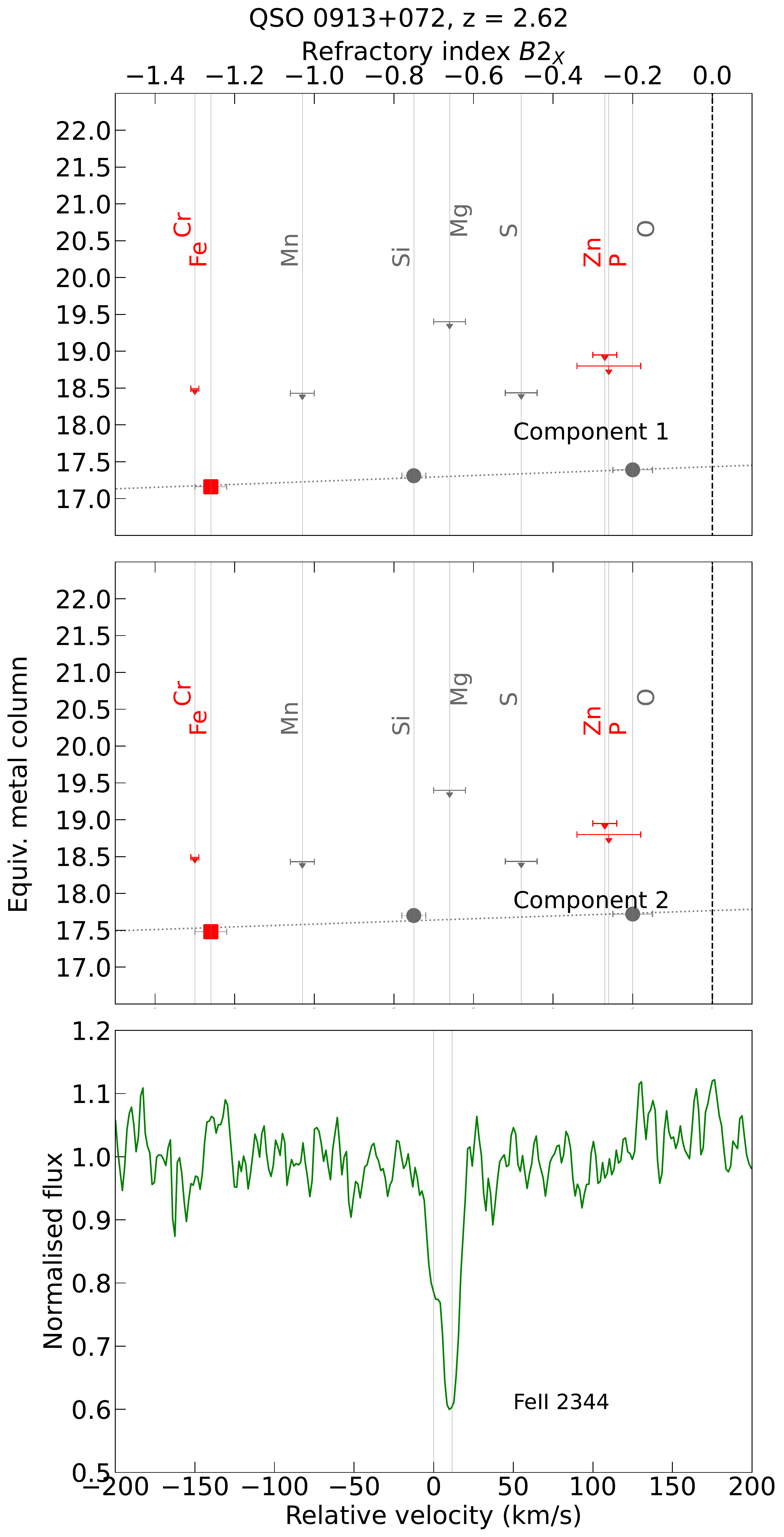}
    \caption{Depletion patterns and respective spectrum for QSO~0913+072} \end{figure}

\clearpage

\begin{figure}[H]
    \centering
    \includegraphics[width=0.425\textwidth]{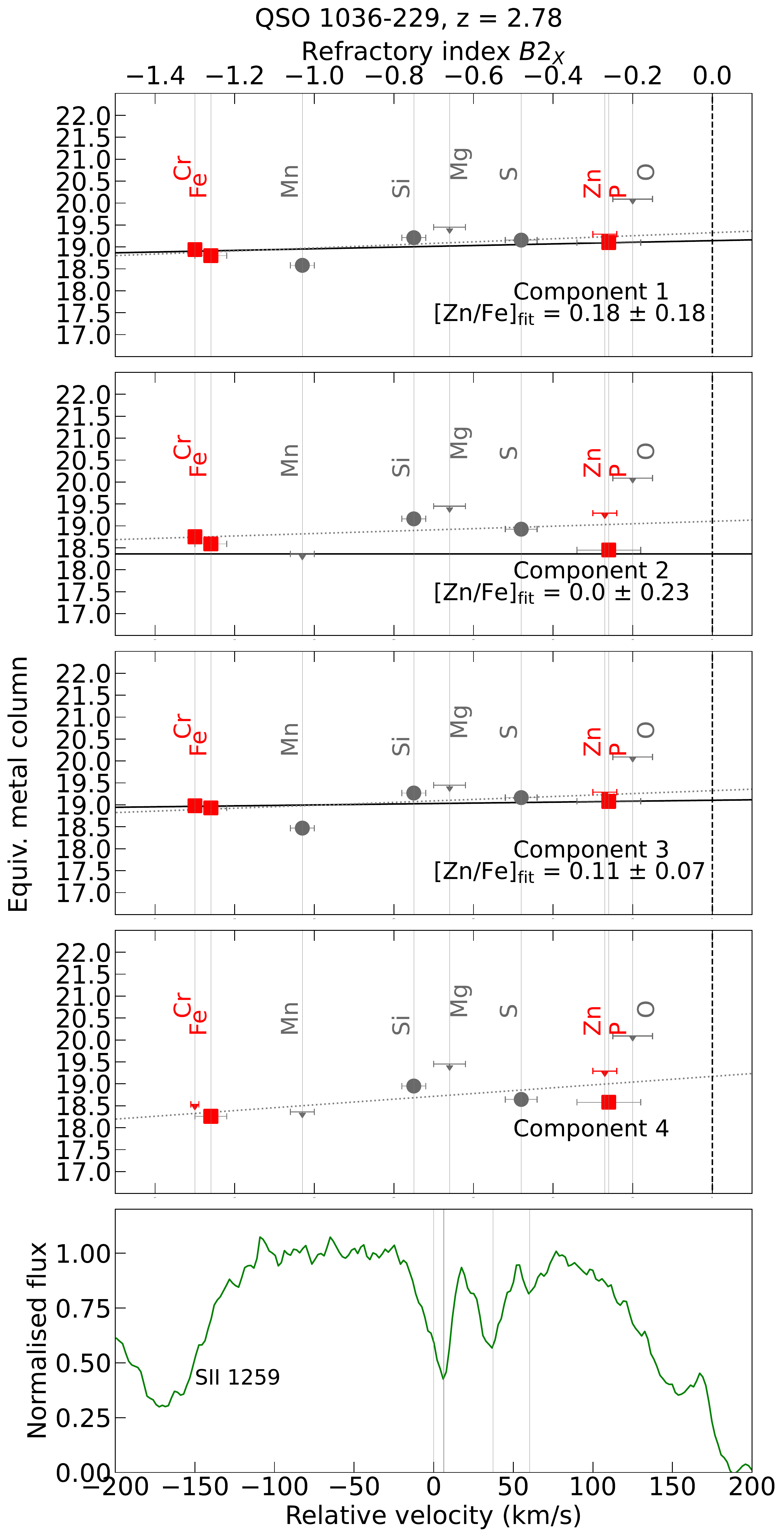}

\caption{Depletion patterns and respective spectrum for QSO~1036-229} \end{figure}

\begin{figure}[H]
    \centering
    \includegraphics[width=0.425\textwidth]{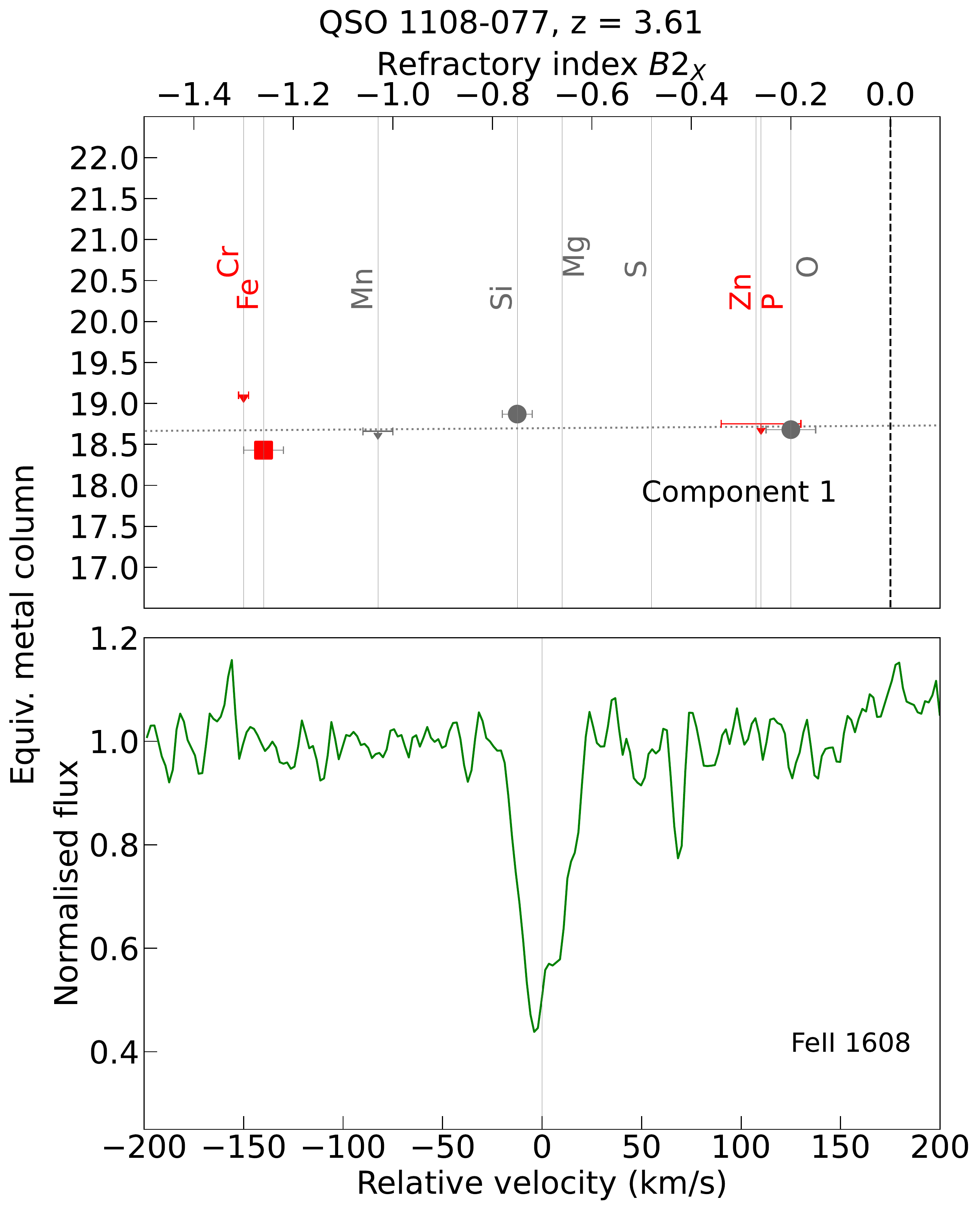}

\caption{Depletion patterns and respective spectrum for QSO~1108-077} \end{figure}

\begin{figure}[H]
    \centering
    \includegraphics[width=0.38\textwidth]{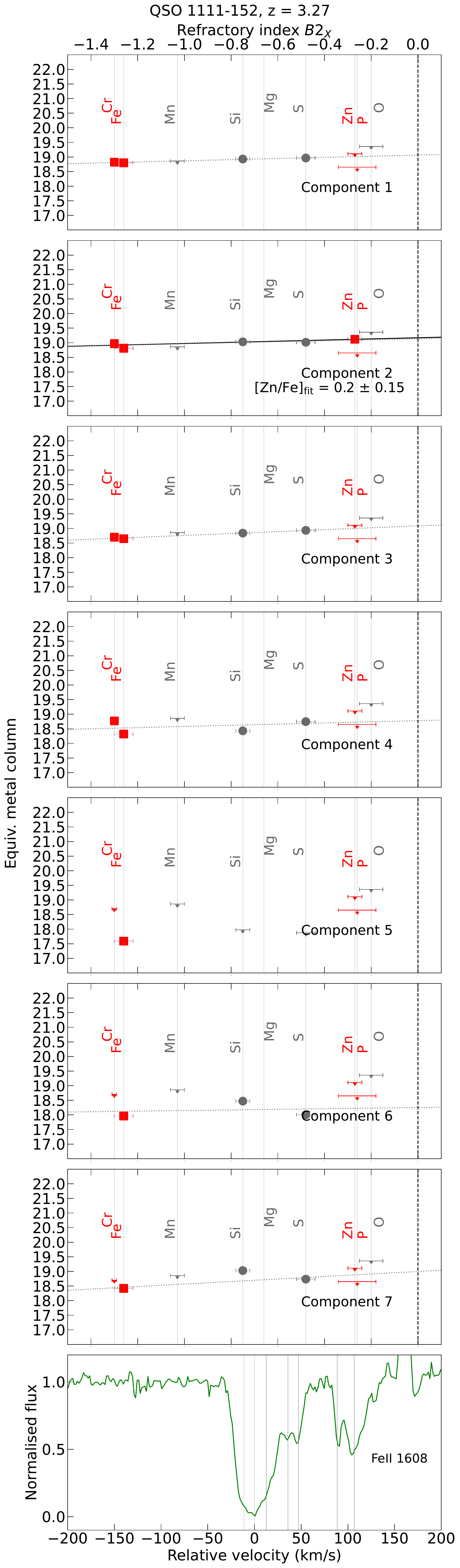}
    \caption{Depletion patterns and respective spectrum for QSO~1111-152} \end{figure}

\begin{figure}[H]
    \centering
    \includegraphics[width=0.425\textwidth]{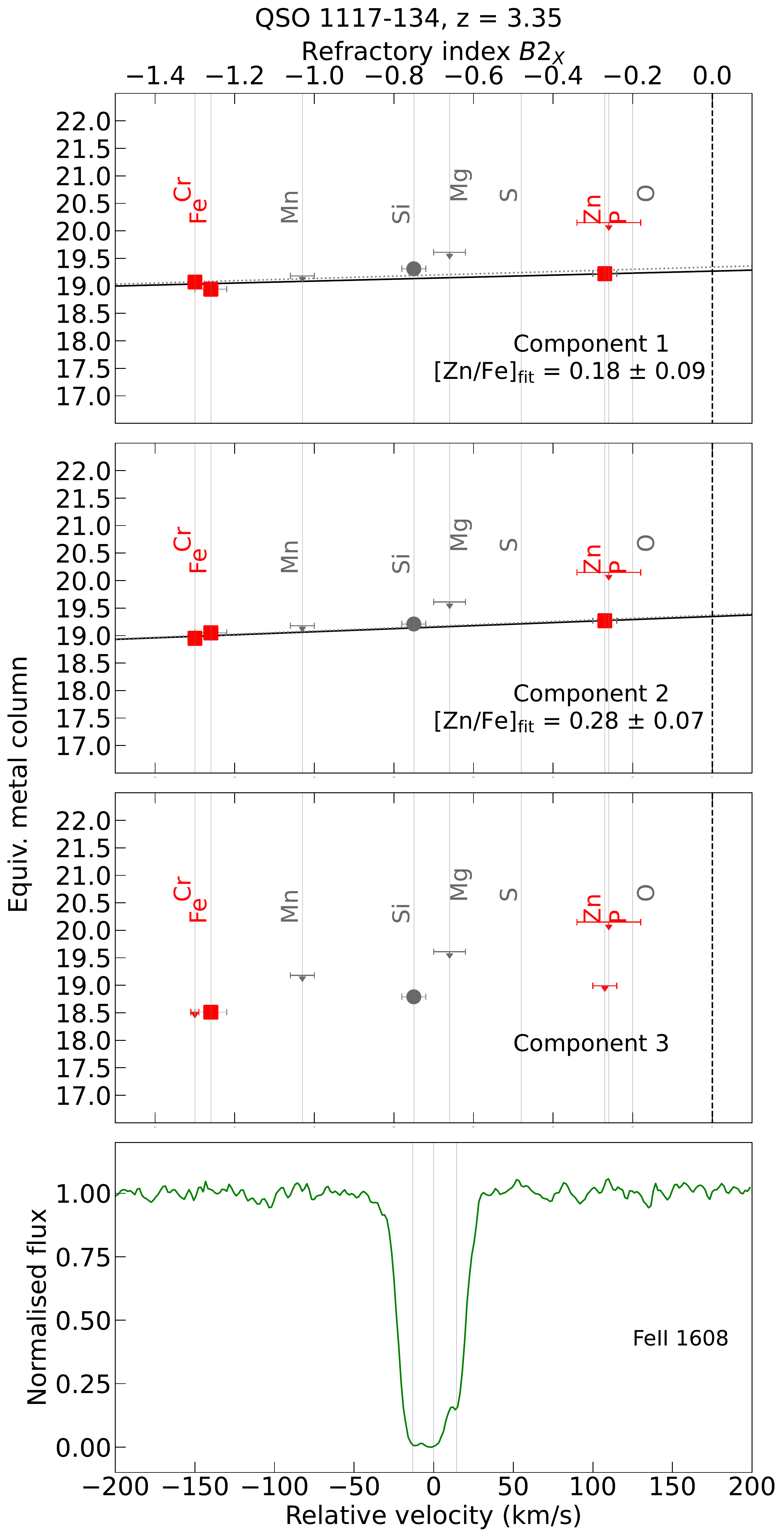}

\caption{Depletion patterns and respective spectrum for QSO~1117-134} \end{figure}

\begin{figure}[H]
    \centering
    \includegraphics[width=0.425\textwidth]{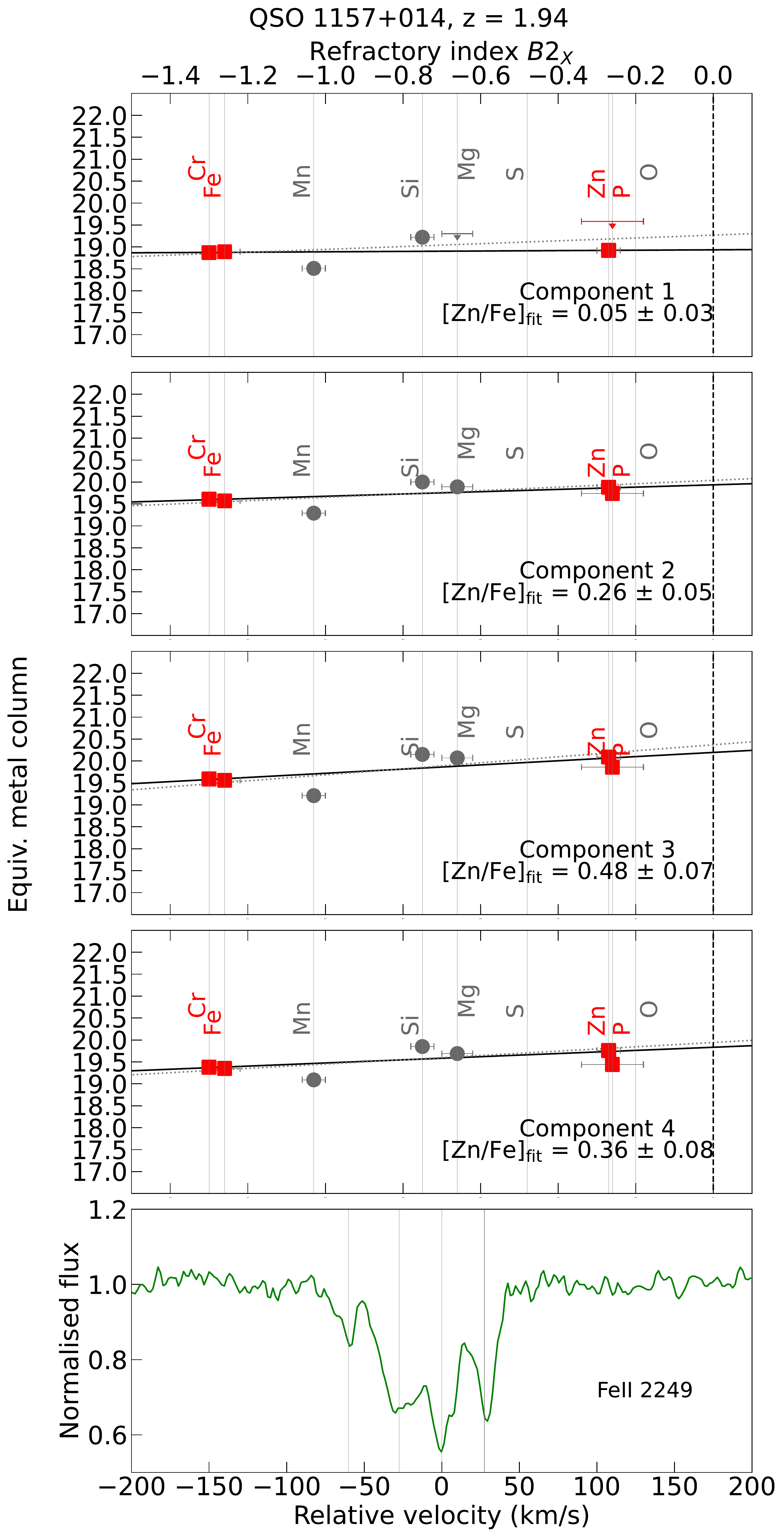}

\caption{Depletion patterns and respective spectrum for QSO~1157+014} \end{figure}

\begin{figure}[H]
    \centering
    \includegraphics[width=0.425\textwidth]{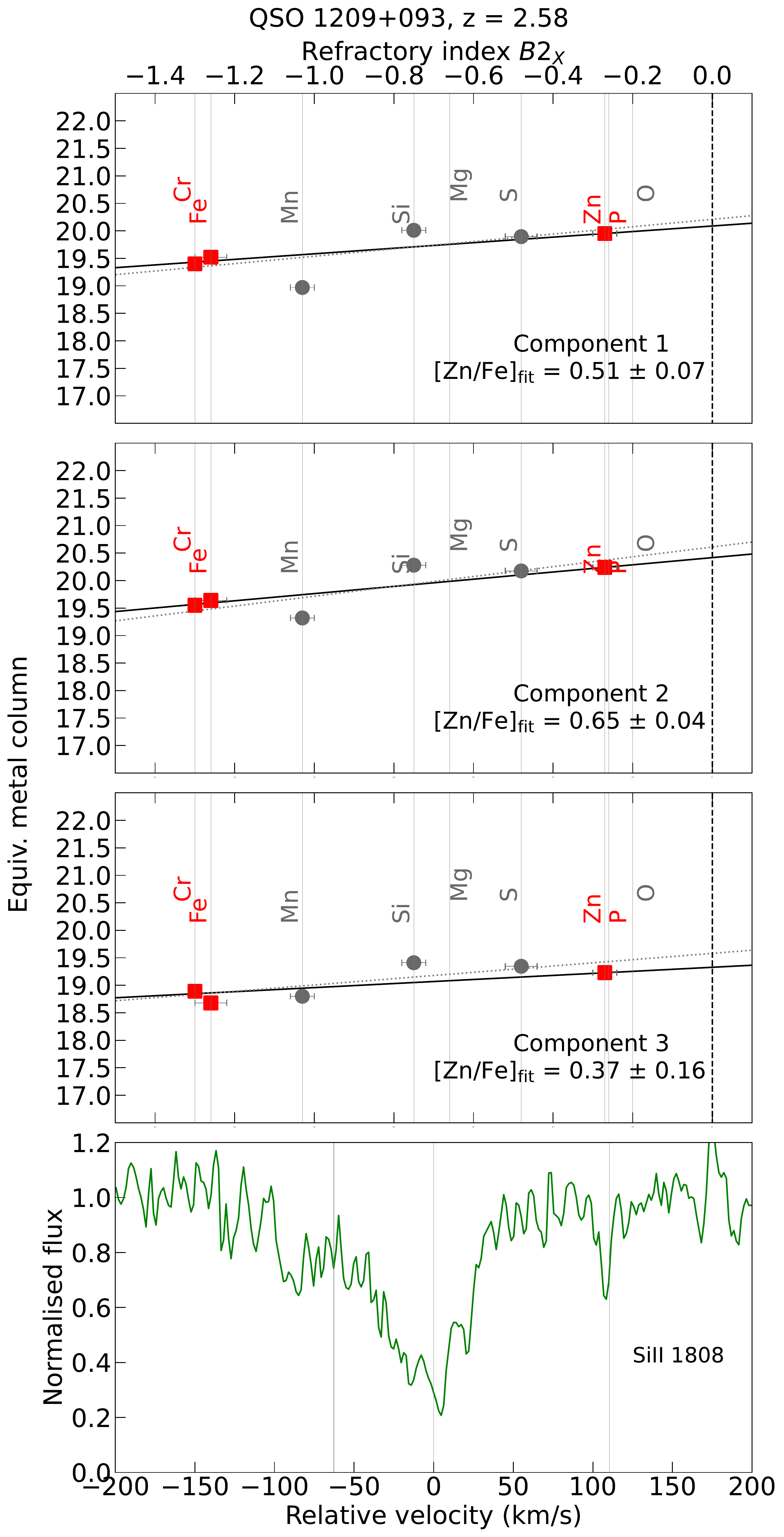}

\caption{Depletion patterns and respective spectrum for QSO~1209+093} \end{figure}

\begin{figure}[H]
    \centering
    \includegraphics[width=0.425\textwidth]{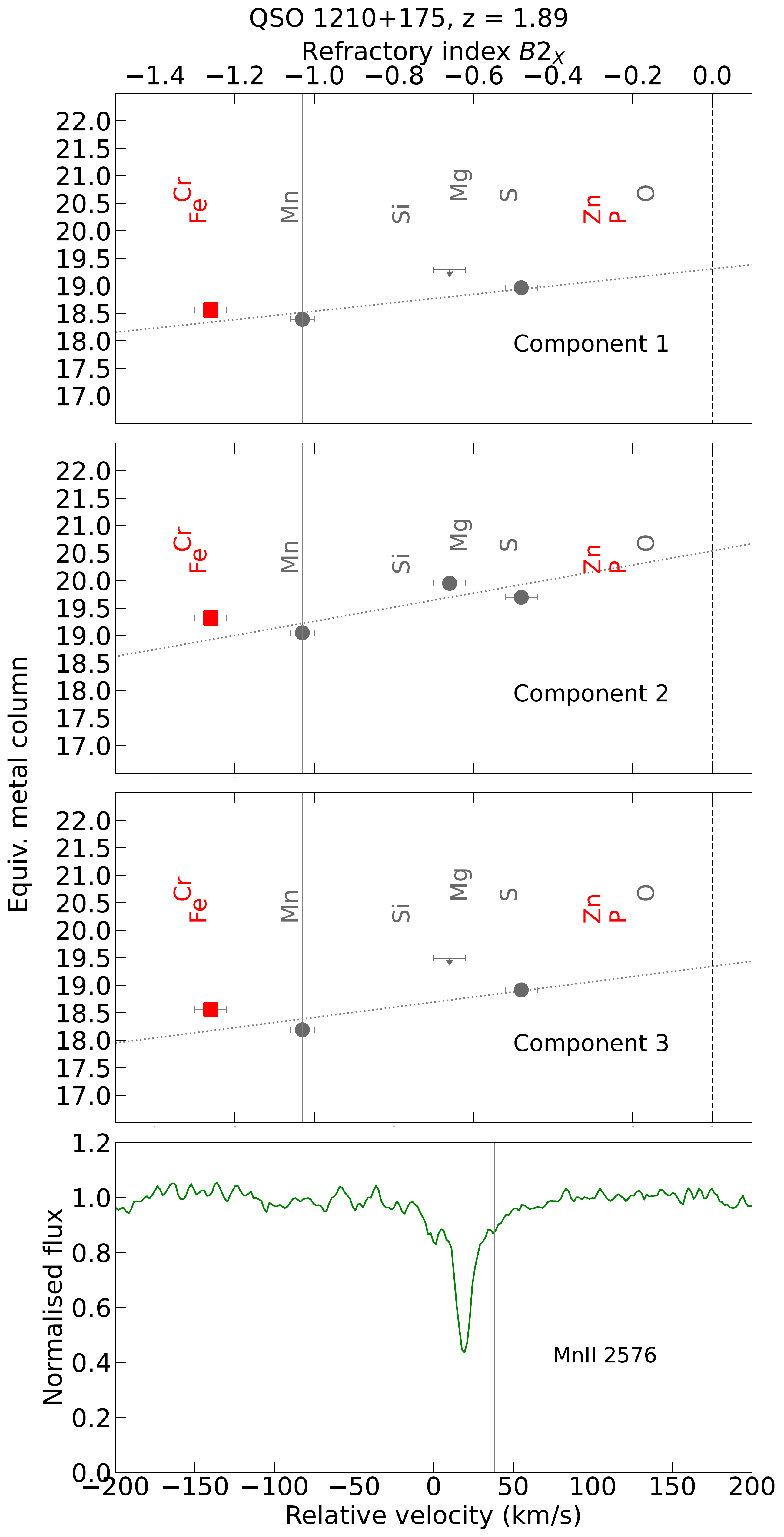}

\caption{Depletion patterns and respective spectrum for QSO~1210+175} \end{figure}

\begin{figure}[H]
    \centering
    \includegraphics[width=0.425\textwidth]{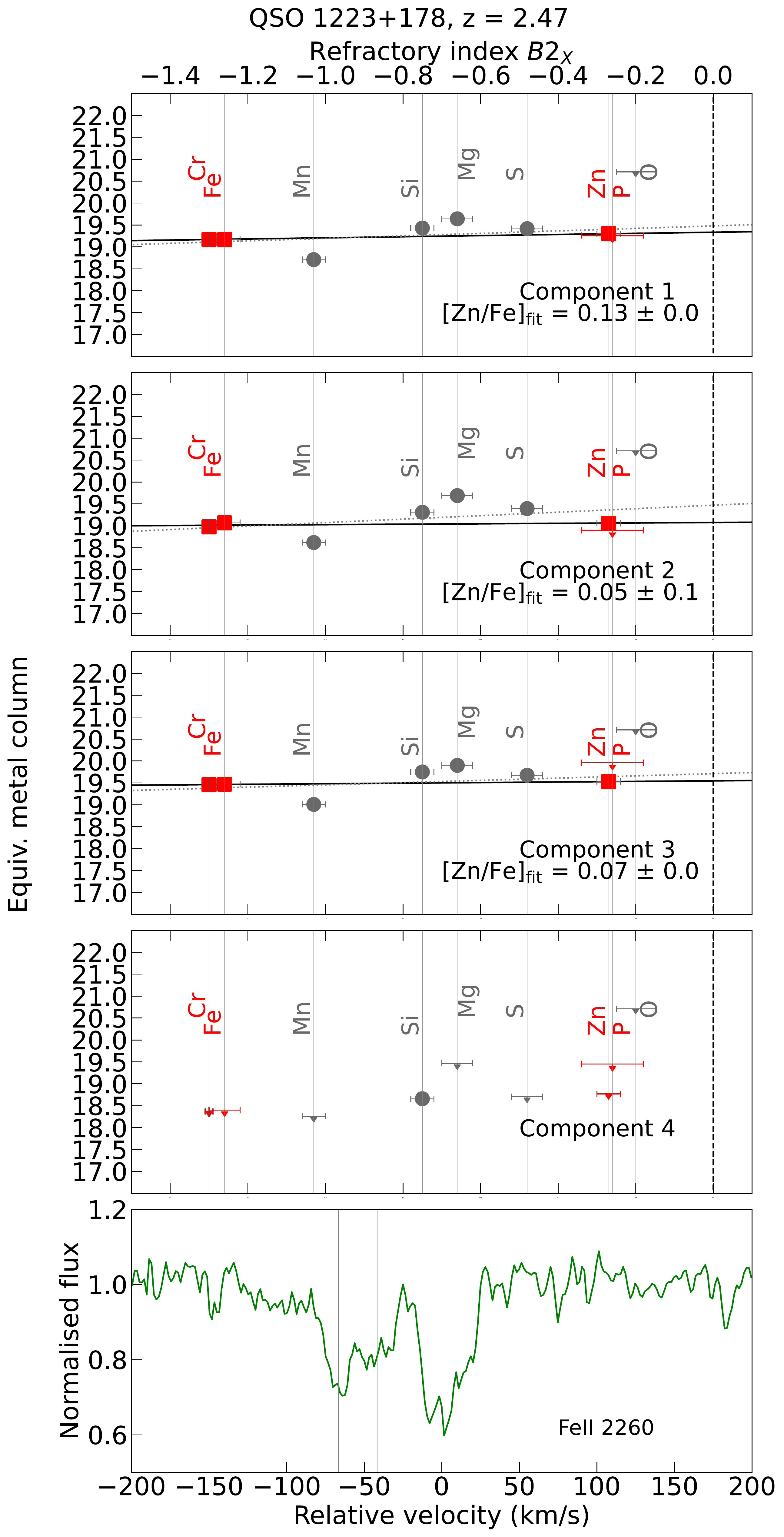}

\caption{Depletion patterns and respective spectrum for QSO~1223+178} \end{figure}

\begin{figure}[H]
    \centering
    \includegraphics[width=0.425\textwidth]{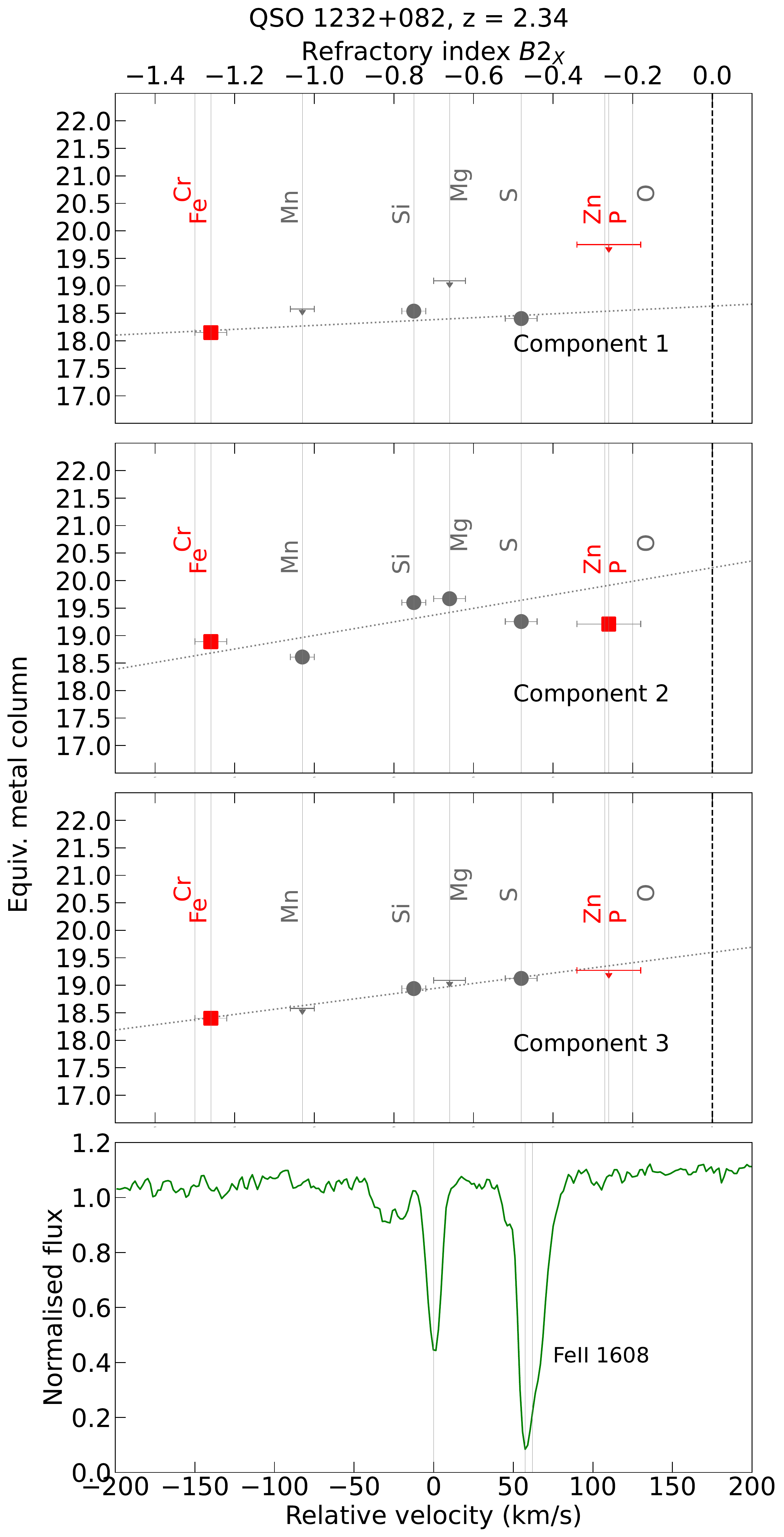}

\caption{Depletion patterns and respective spectrum for QSO~1232+082} \end{figure}

\begin{figure}[H]
    \centering
    \includegraphics[width=0.4\textwidth]{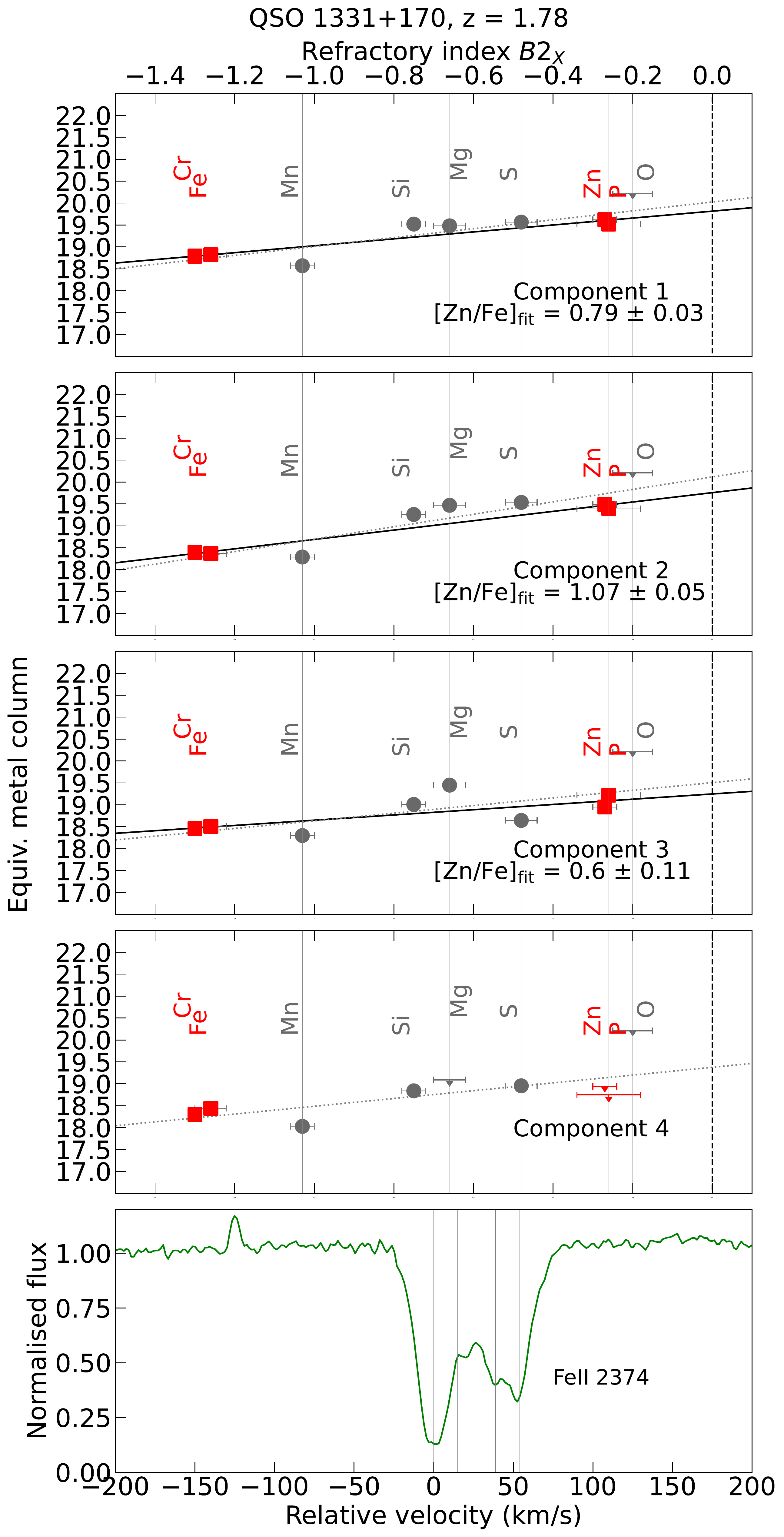}
    \caption{Depletion patterns and respective spectrum for QSO~1331+170} \end{figure}

\clearpage

\begin{figure}[H]
    \centering
    \includegraphics[width=0.4\textwidth]{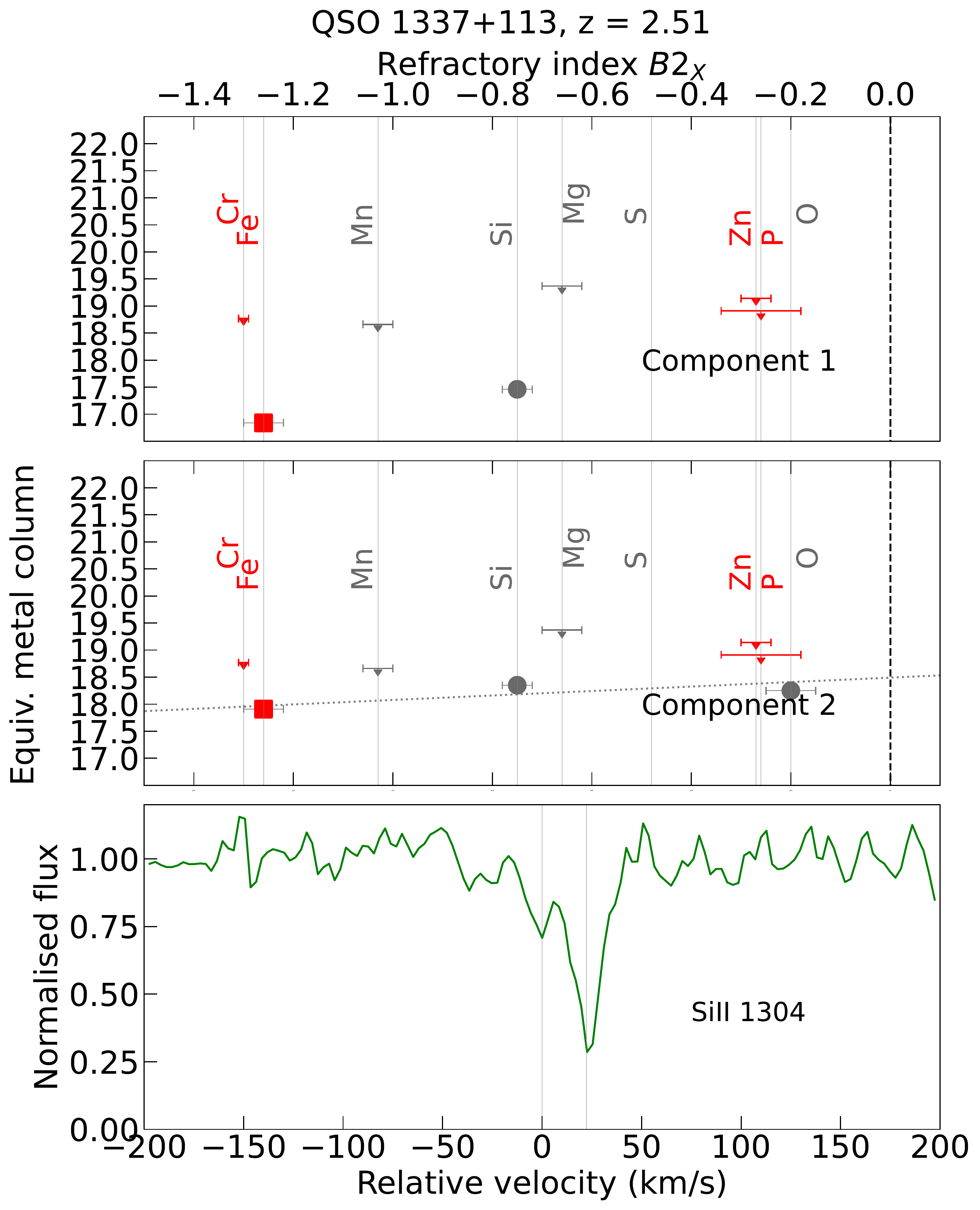}
    \caption{Depletion patterns and respective spectrum for QSO~1337+113} \end{figure}

\begin{figure}[H]
    \centering
    \includegraphics[width=0.425\textwidth]{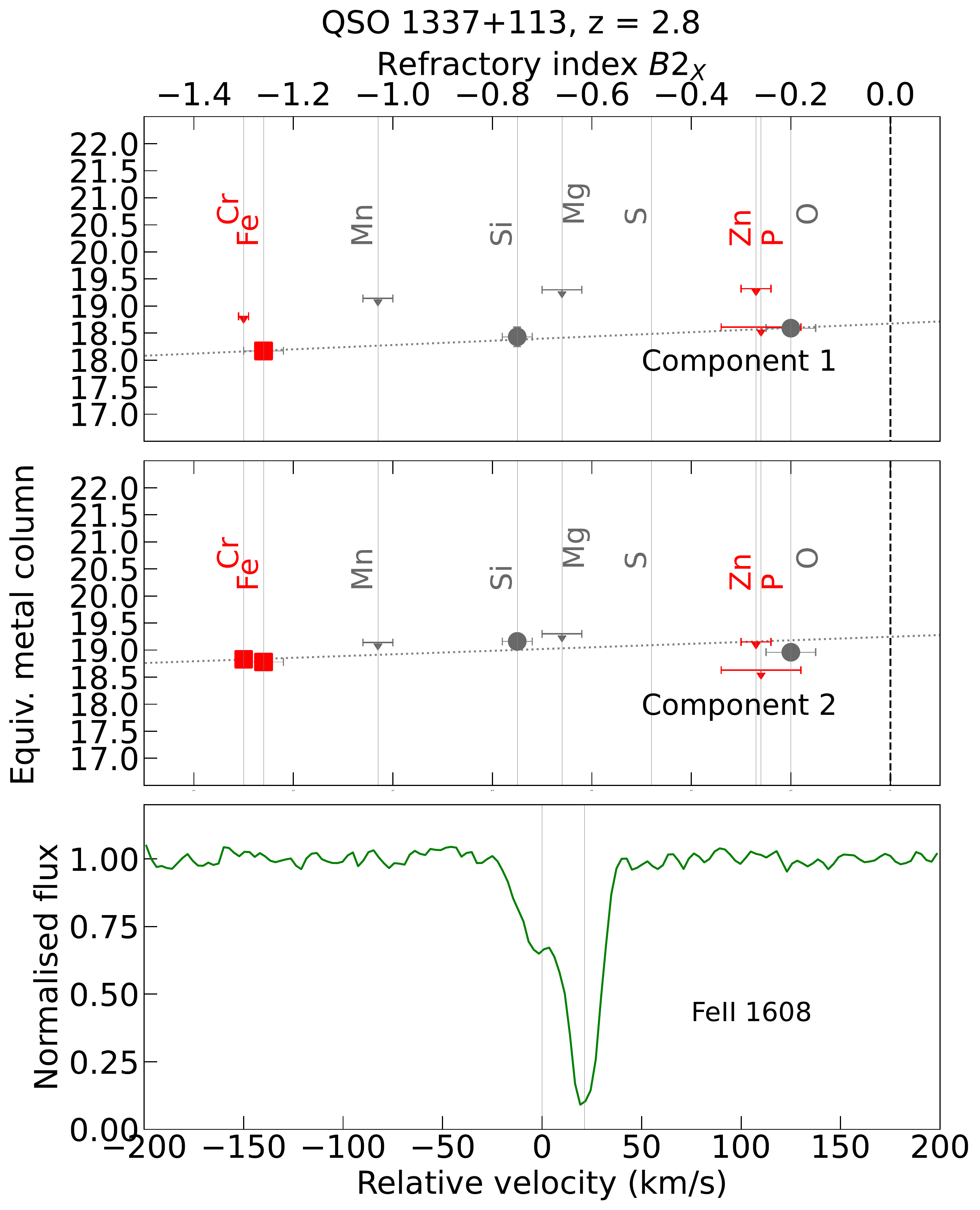}
    \caption{Depletion patterns and respective spectrum for QSO~1337+113} \end{figure}

\begin{figure}[H]
    \centering
    \includegraphics[width=0.425\textwidth]{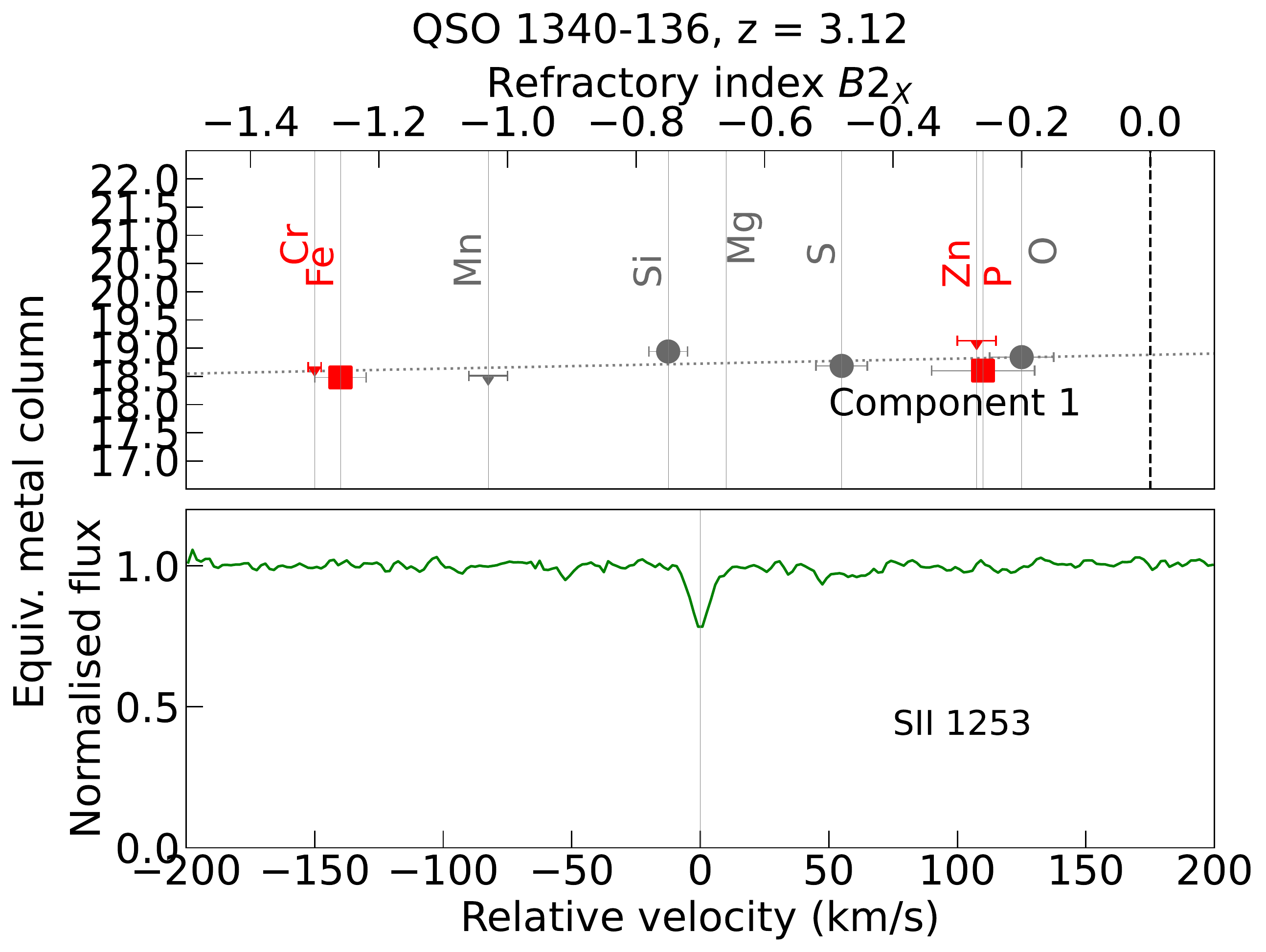}

\caption{Depletion patterns and respective spectrum for QSO~1340-136} \end{figure}

\begin{figure}[H]
    \centering
    \includegraphics[width=0.425\textwidth]{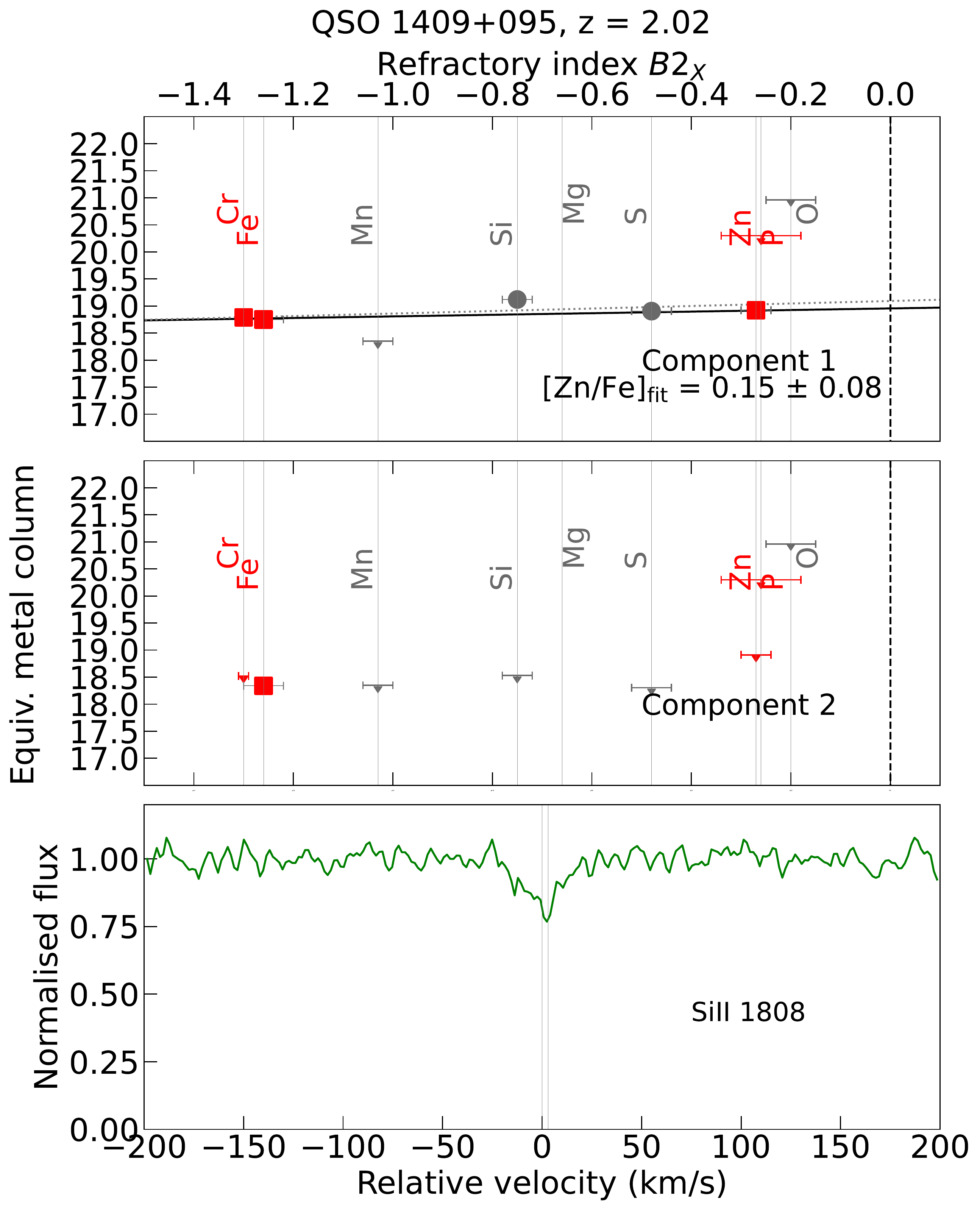}

\caption{Depletion patterns and respective spectrum for QSO~1409+095} \end{figure}

\begin{figure}[H]
    \centering
    \includegraphics[width=0.425\textwidth]{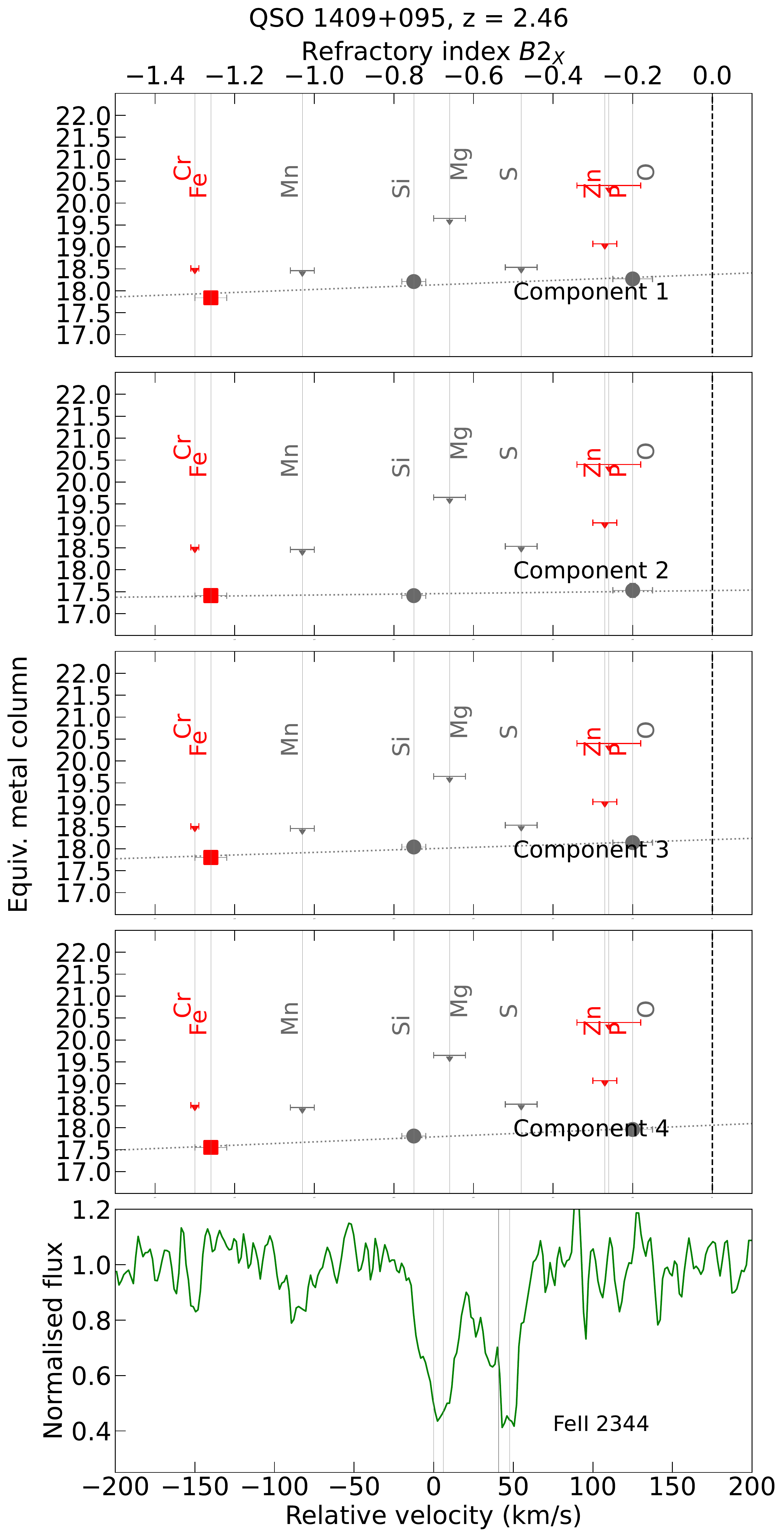}
    \caption{Depletion patterns and respective spectrum for QSO~1409+095} \end{figure}

\begin{figure}[H]
    \centering
    \includegraphics[width=0.425\textwidth]{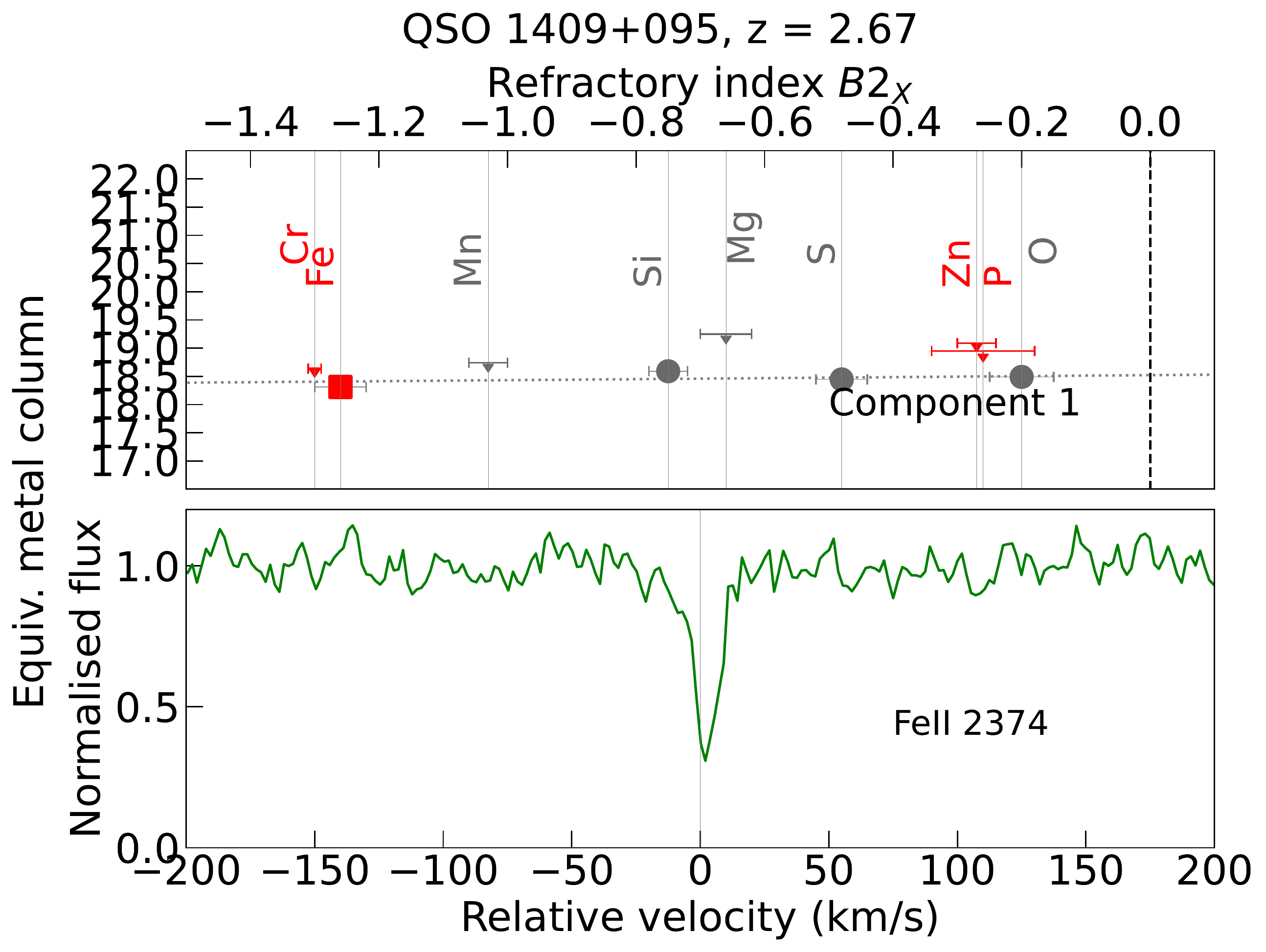}
\caption{Depletion patterns and respective spectrum for QSO~1409+095} \end{figure}

\begin{figure}[H]
    \centering
    \includegraphics[width=0.425\textwidth]{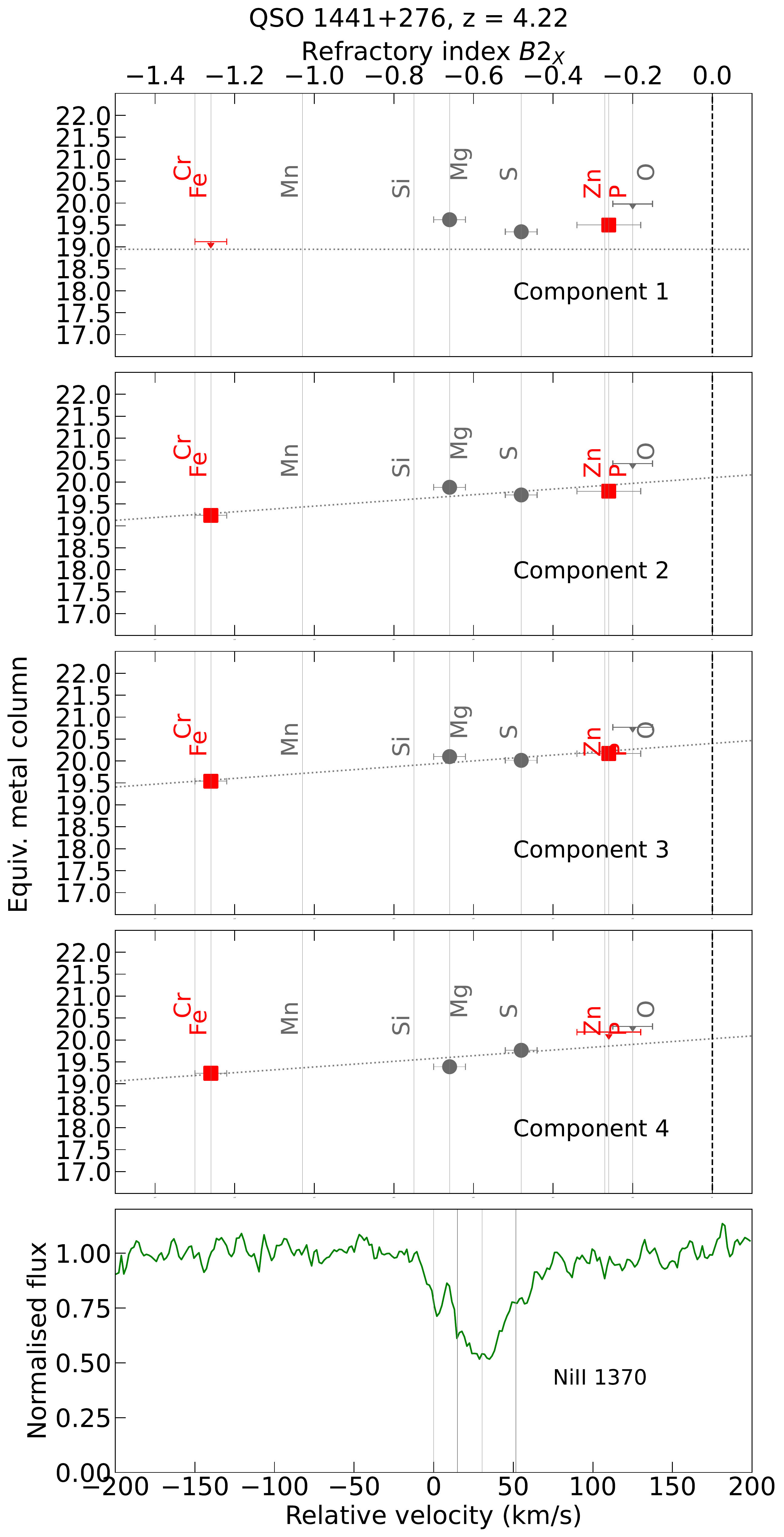}

\caption{Depletion patterns and respective spectrum for QSO~1441+276} \end{figure}

\begin{figure}[H]
    \centering
    \includegraphics[width=0.425\textwidth]{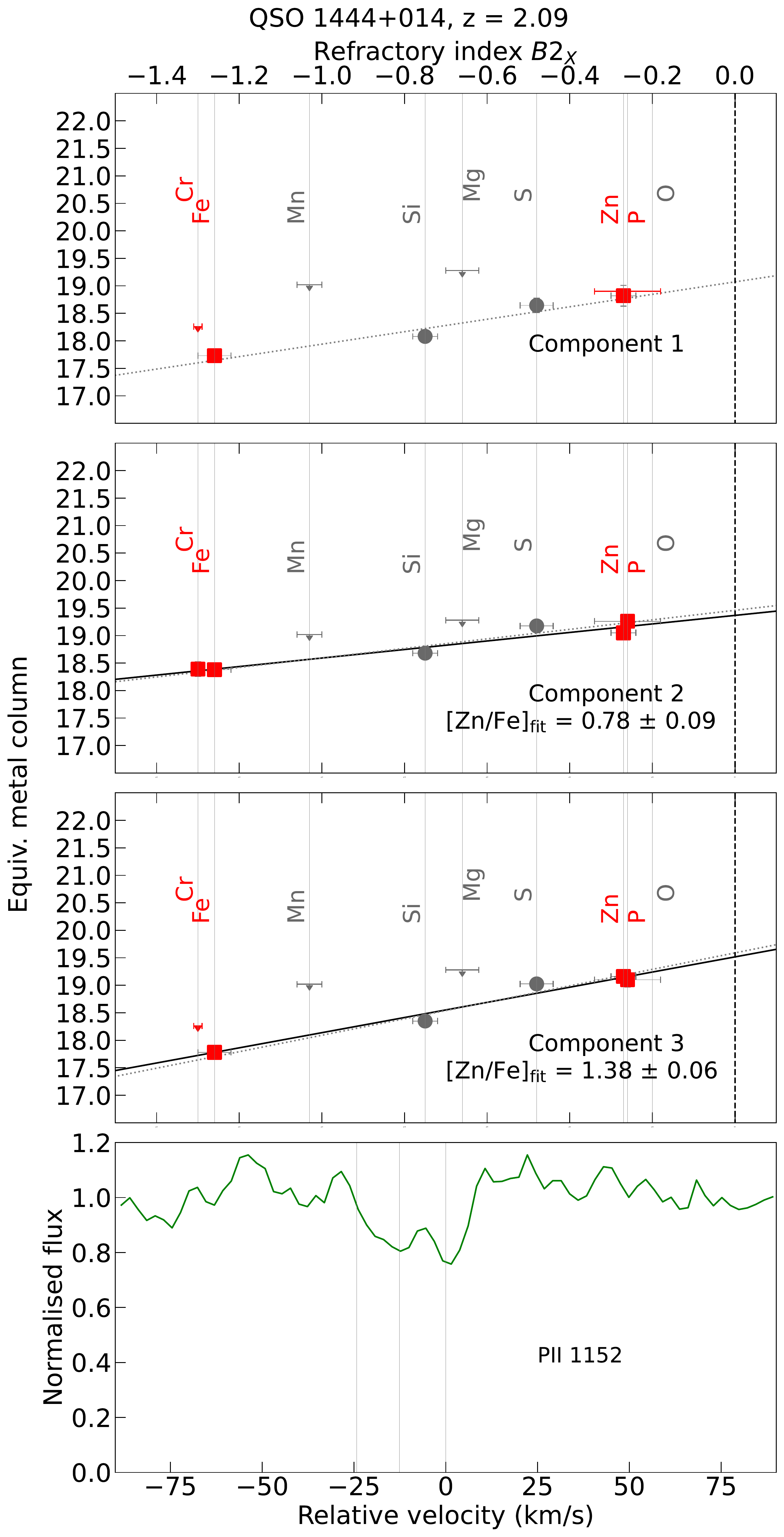}

\caption{Depletion patterns and respective spectrum for QSO~1444+014} \end{figure}

\begin{figure}[H]
    \centering
    \includegraphics[width=0.4\textwidth]{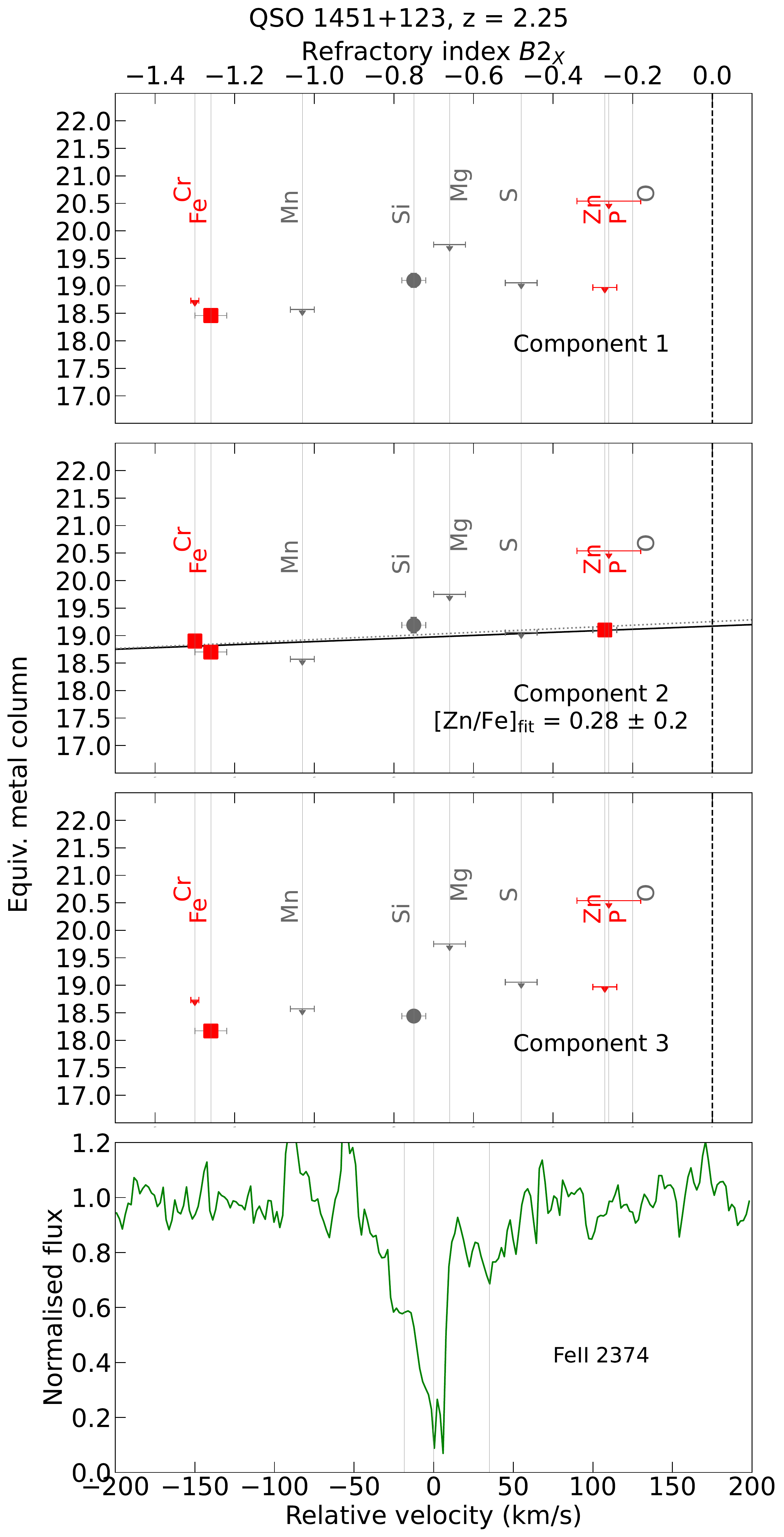}
    \caption{Depletion patterns and respective spectrum for QSO~1451+123} \end{figure}

\begin{figure}[H]
    \centering
    \includegraphics[width=0.42\textwidth]{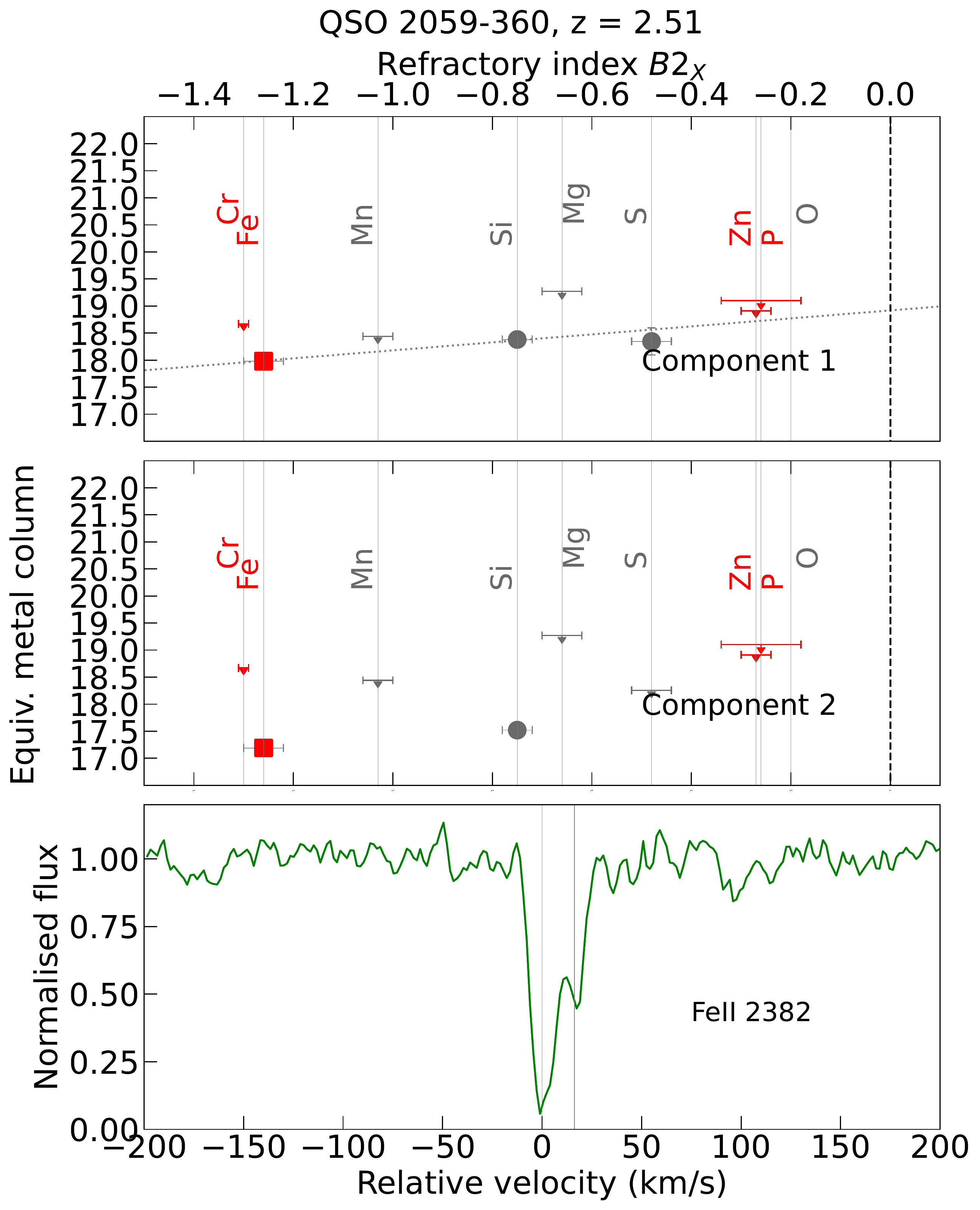}

\caption{Depletion patterns and respective spectrum for QSO~2059-360} \end{figure}

\begin{figure}[H]
    \centering
    \includegraphics[width=0.425\textwidth]{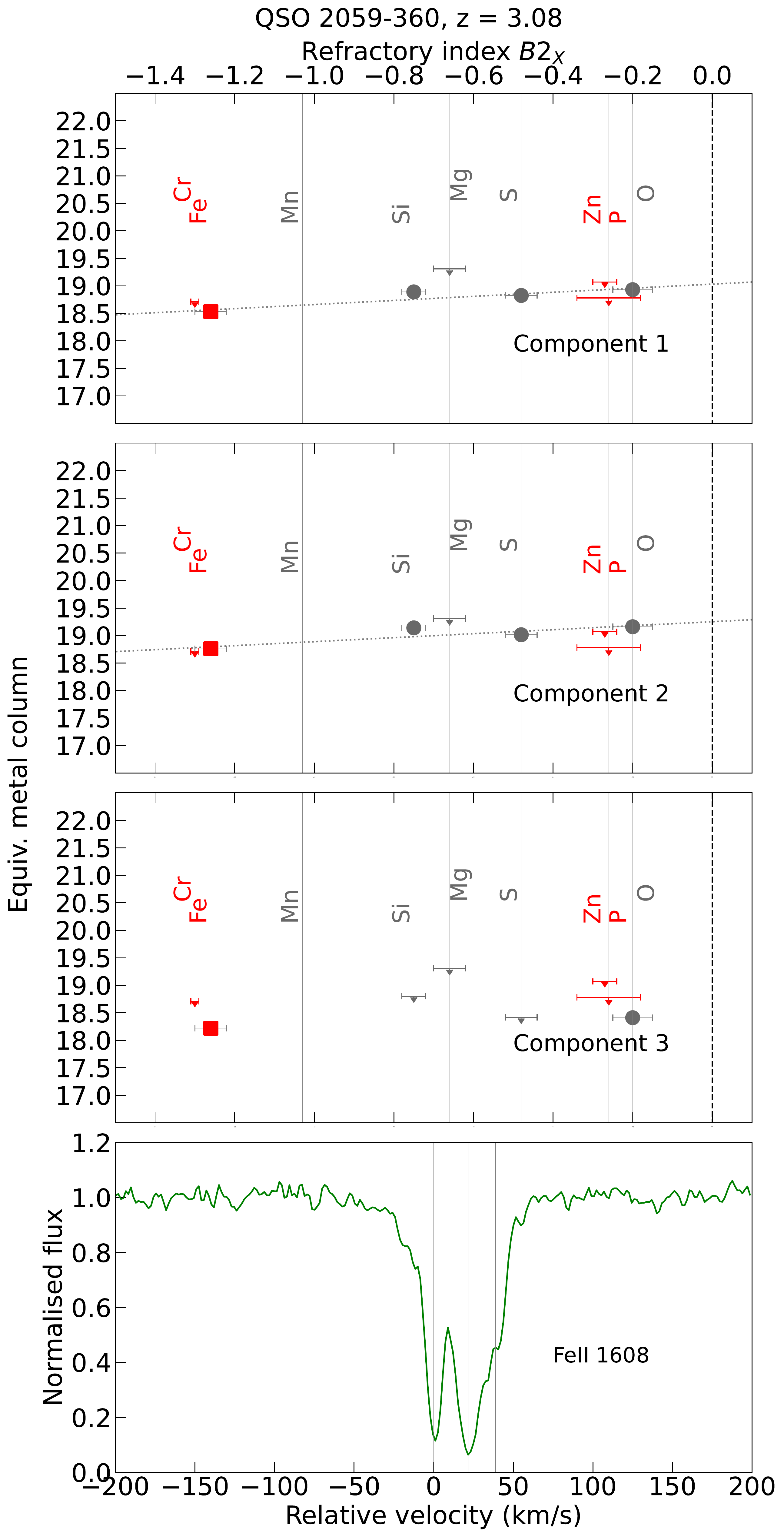}
\caption{Depletion patterns and respective spectrum for QSO~2059-360} \end{figure}

\begin{figure*}[h!]
    \centering
    \includegraphics[width=0.8\textwidth]{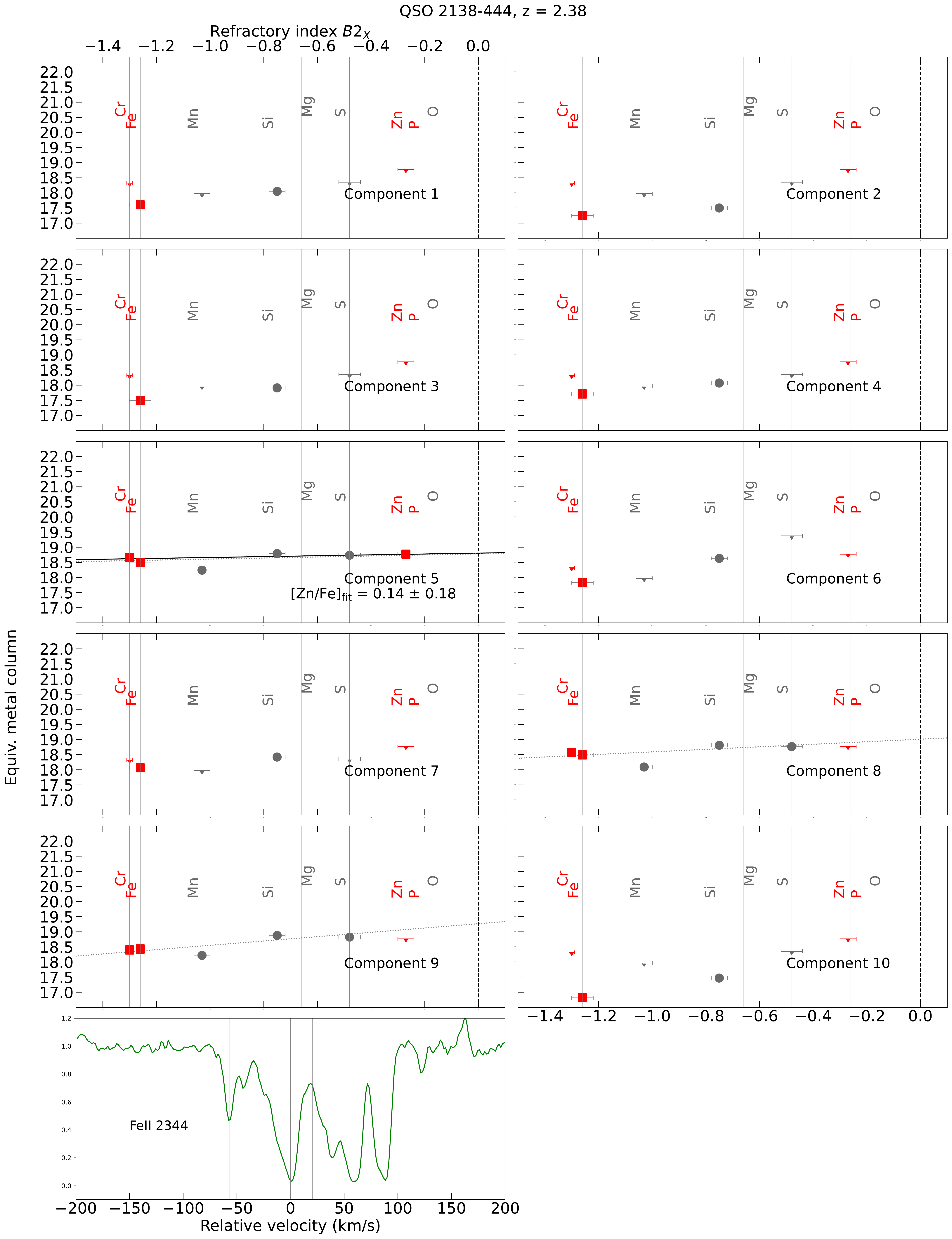}
    \caption{Depletion patterns and respective spectrum for QSO~2138-444}
\end{figure*}

\clearpage

\begin{figure}[H]
    \centering
    \includegraphics[width=0.4\textwidth]{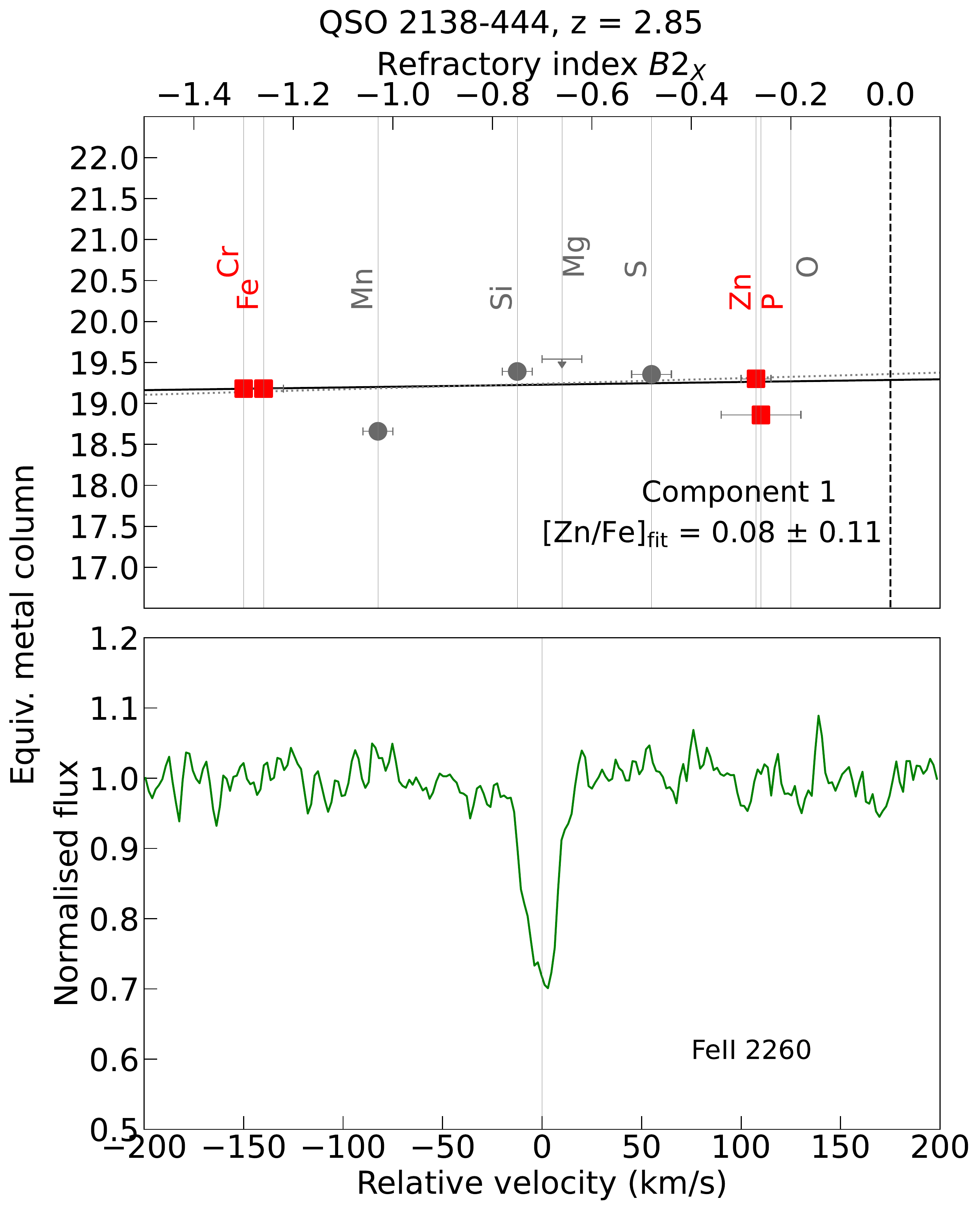}
\caption{Depletion patterns and respective spectrum for QSO~2138-444} \end{figure}

\begin{figure}[H]
    \centering
    \includegraphics[width=0.42\textwidth]{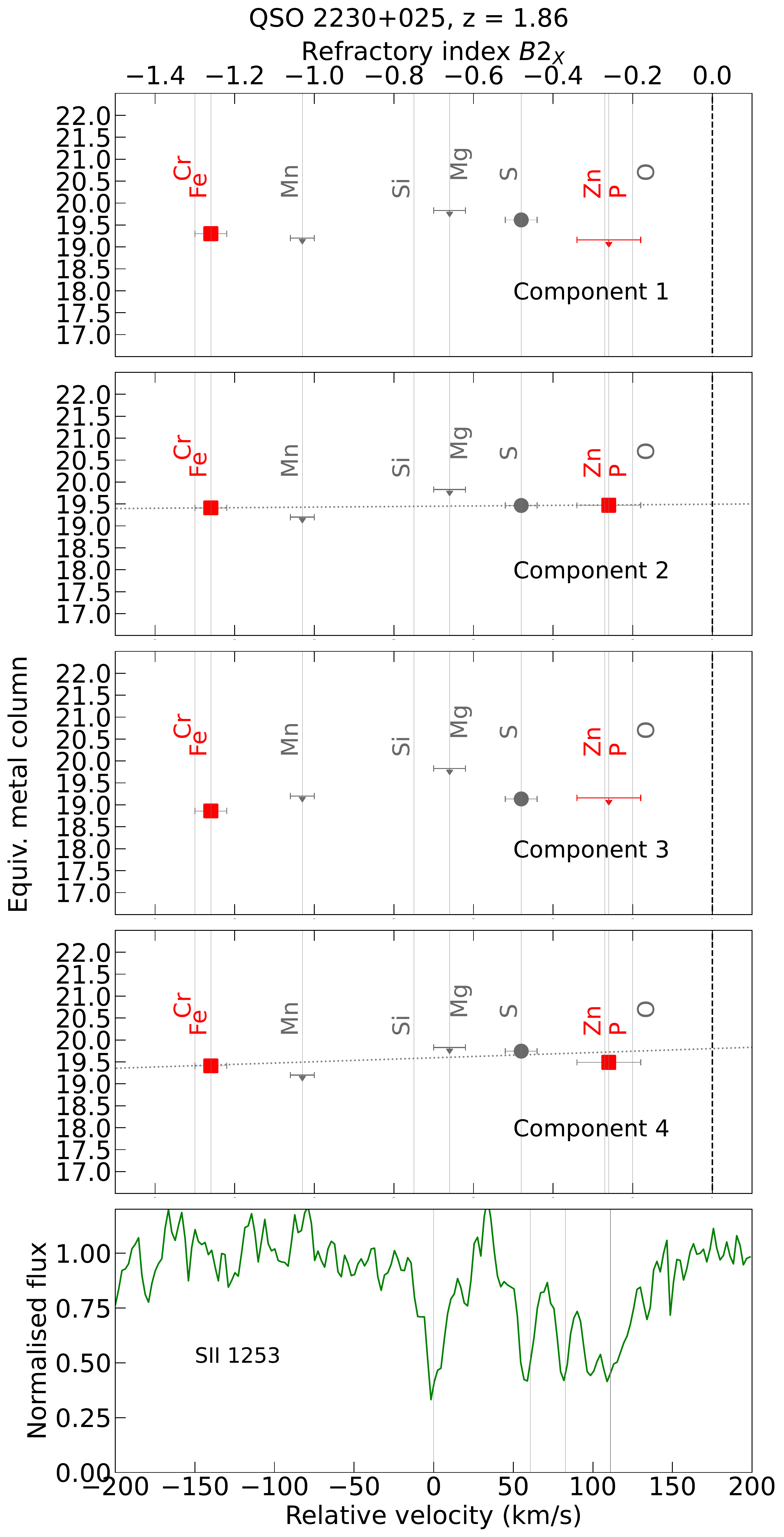}
\caption{Depletion patterns and respective spectrum for QSO~2230+025} \end{figure}

\begin{figure}[H]
    \centering    
    \includegraphics[width=0.425\textwidth]{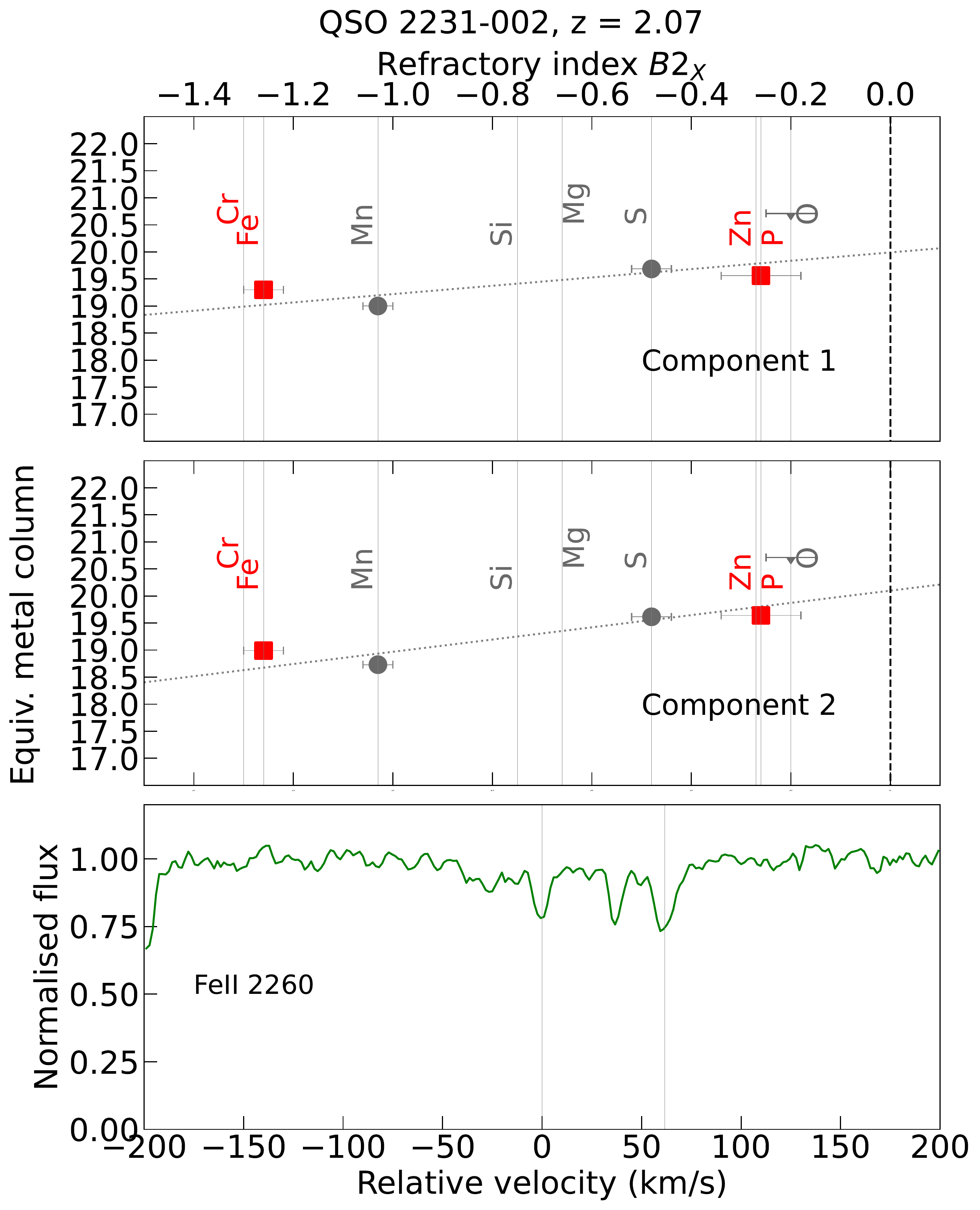}
\caption{Depletion patterns and respective spectrum for QSO~2231-002} \end{figure}

\begin{figure*}[h!]
    \centering
    \includegraphics[width=0.8\textwidth]{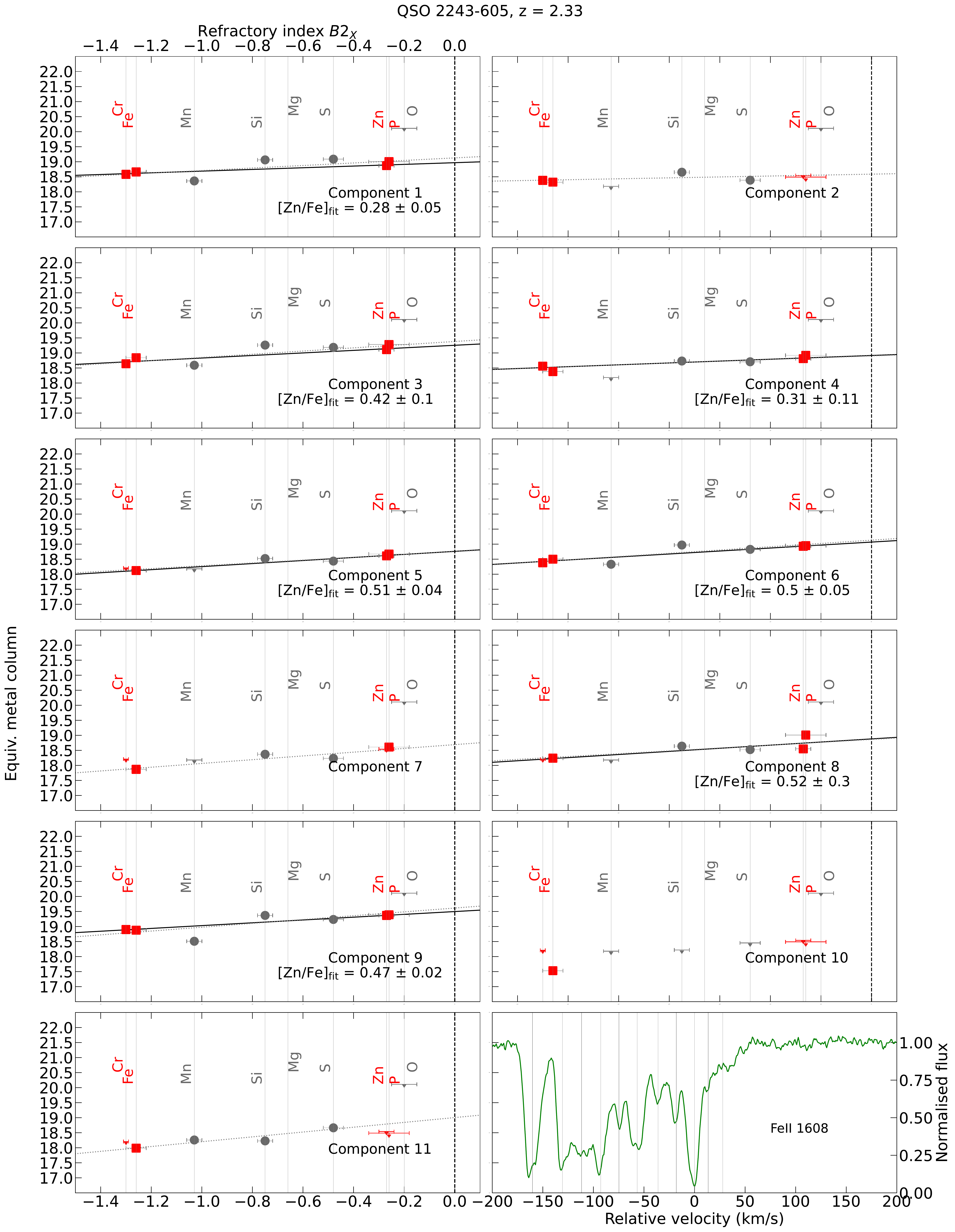}
    \caption{Depletion patterns and respective spectrum for QSO~2243-605} 
\end{figure*}

\clearpage

\begin{figure}[H]
    \centering
    \includegraphics[width=0.425\textwidth]{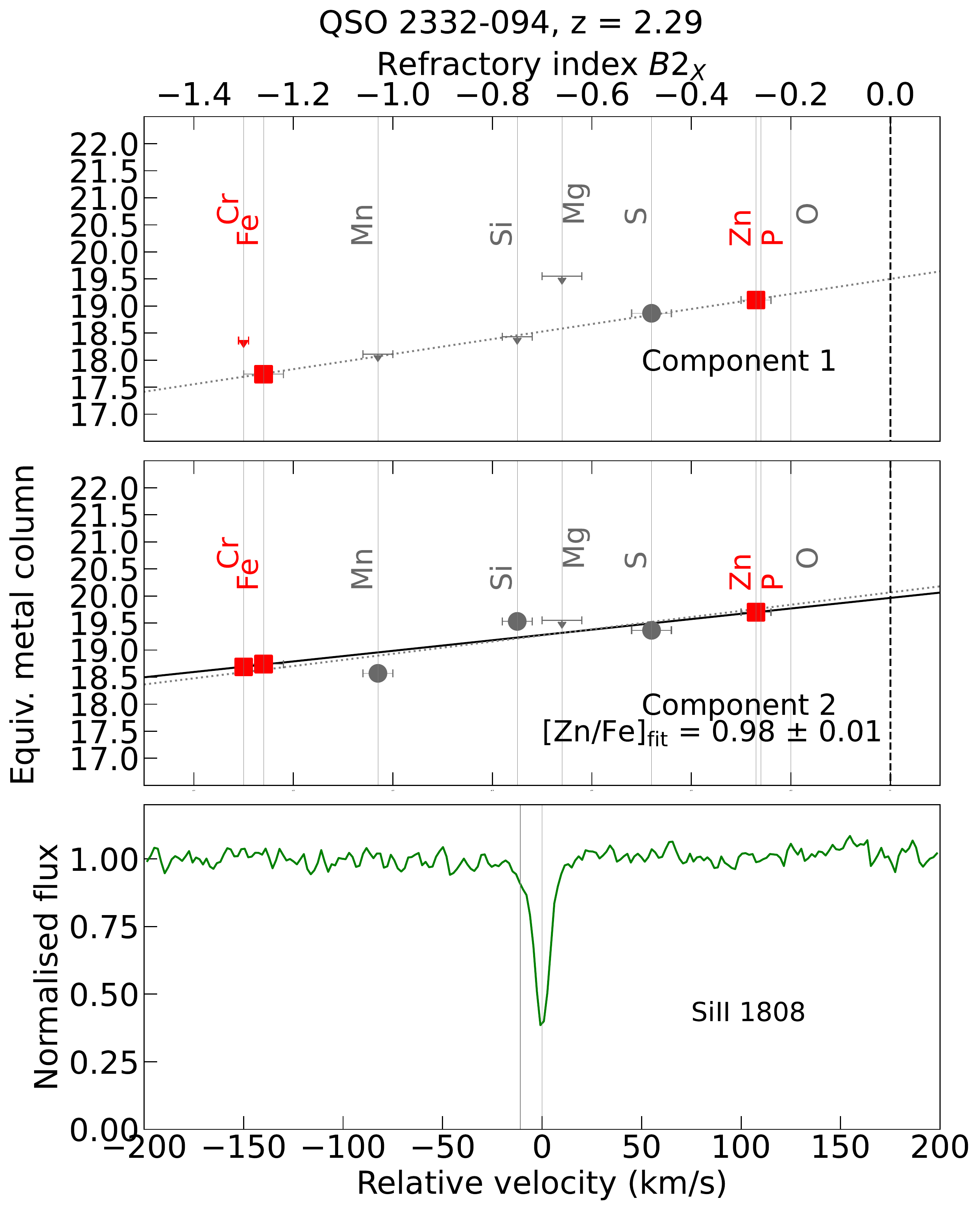}
\caption{Depletion patterns and respective spectrum for QSO~2332-094} \end{figure}

\begin{figure}[H]
    \centering
    \includegraphics[width=0.385\textwidth]{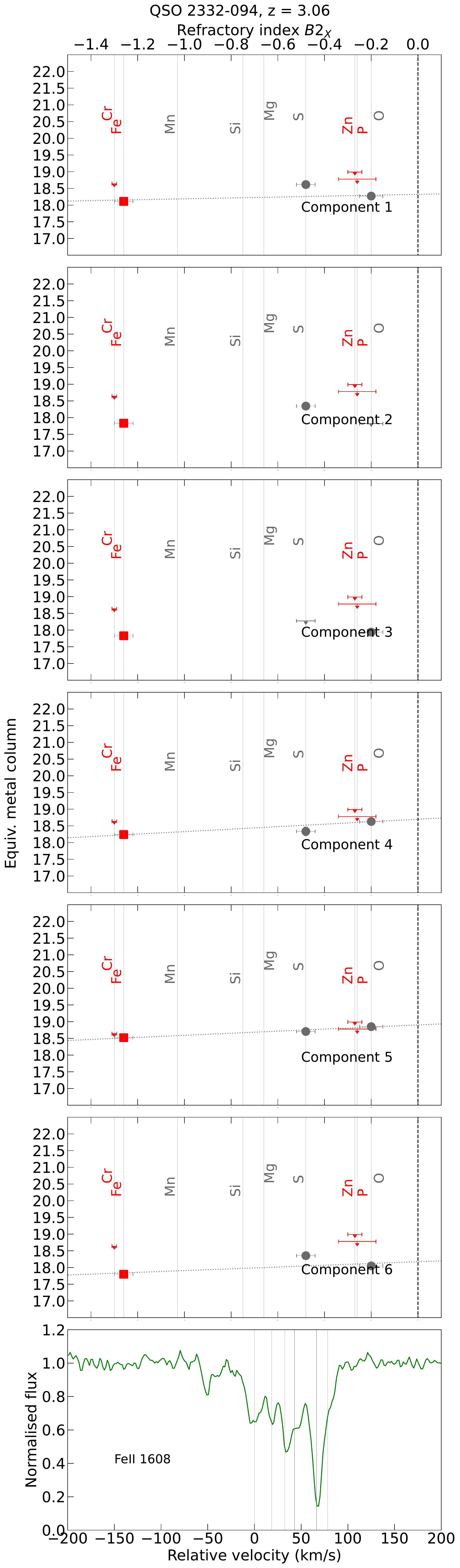}
    \caption{Depletion patterns and respective spectrum for QSO~2332-094} \end{figure}

\clearpage

\begin{figure}[H]
    \centering
    \includegraphics[width=0.425\textwidth]{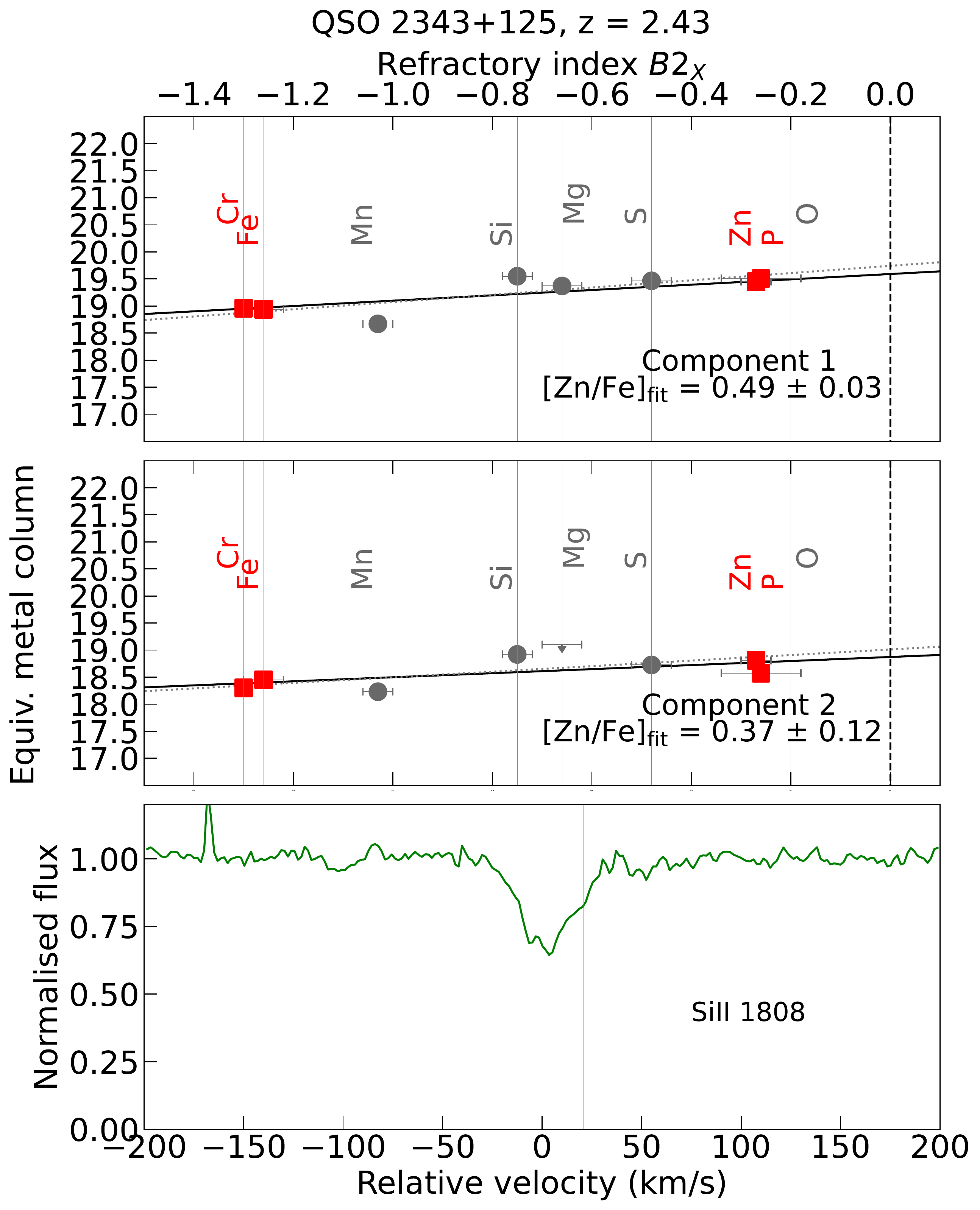}
\caption{Depletion patterns and respective spectrum for QSO~2343+125} \end{figure}

\begin{figure}[H]
    \centering
    \includegraphics[width=0.4\textwidth]{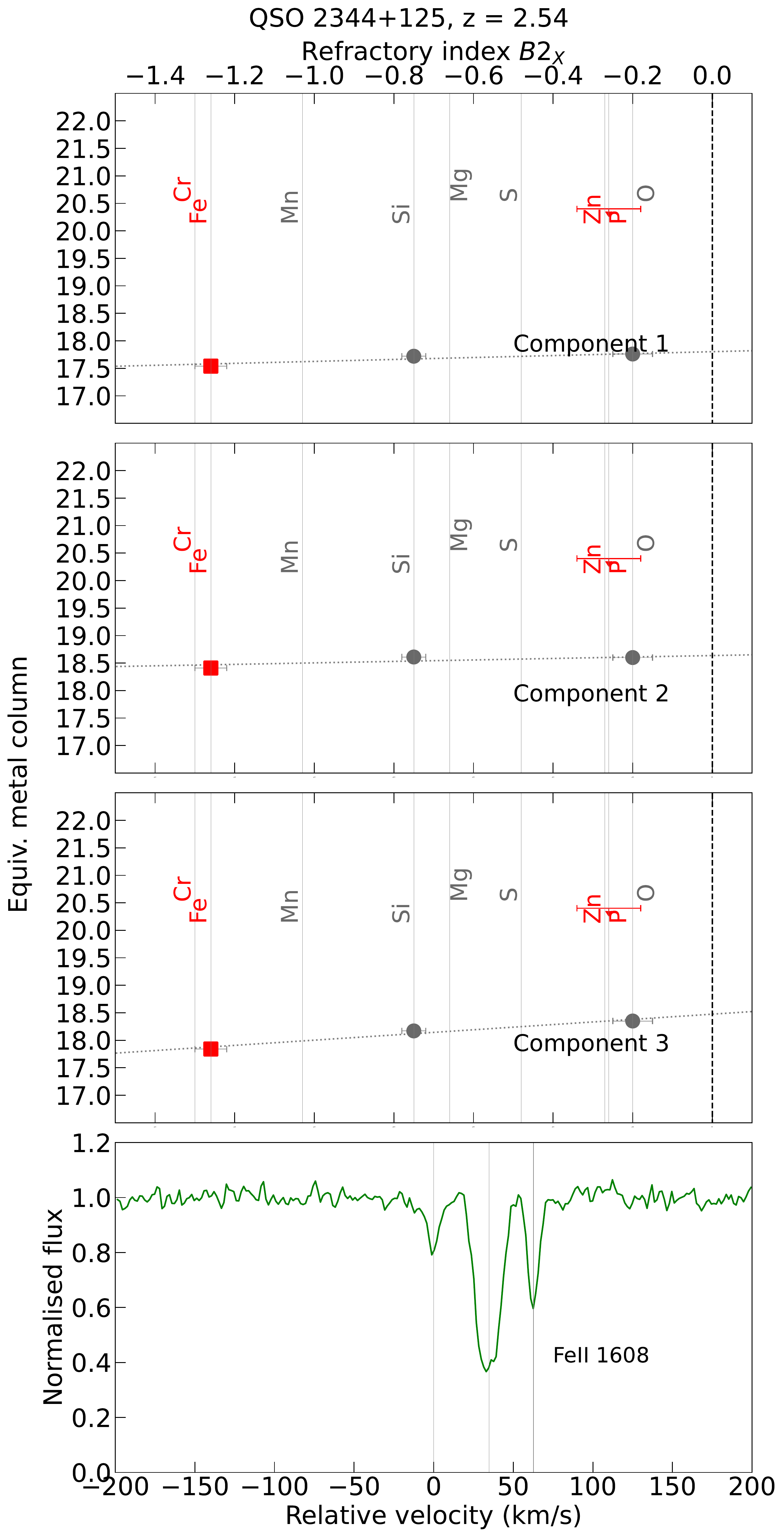}
\caption{Depletion patterns and respective spectrum for QSO~2344+125} \end{figure}

\begin{figure*}
    \centering
    \includegraphics[width=0.8\textwidth]{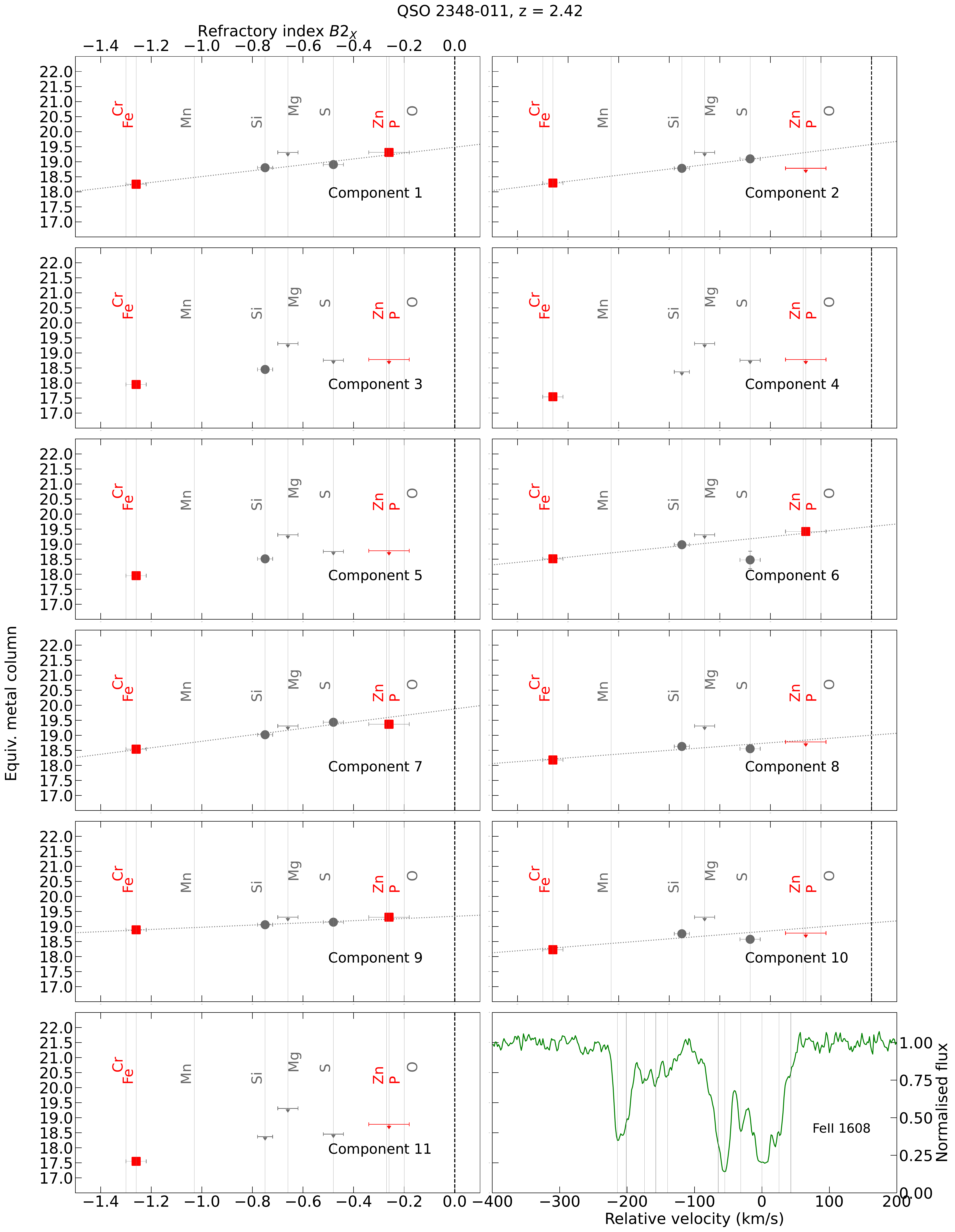}
    \caption{Depletion patterns and respective spectrum for QSO~2348-011} 
\end{figure*}

\clearpage

\begin{figure}[H]
    \centering
    \includegraphics[width=0.4\textwidth]{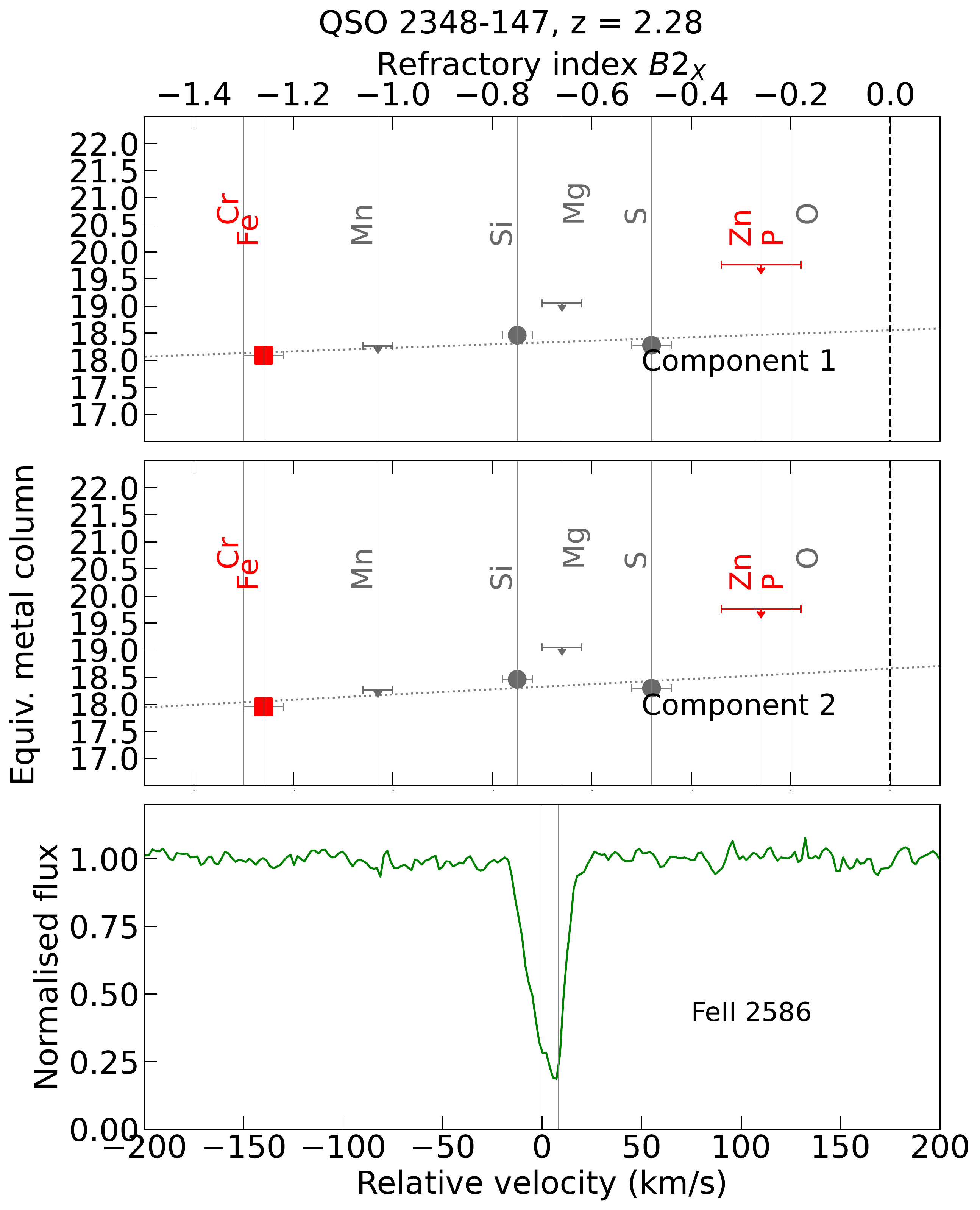}
\caption{Depletion patterns and respective spectrum for QSO~2348-147} \end{figure}

\begin{figure}[H]
    \centering
    \includegraphics[width=0.4\textwidth]{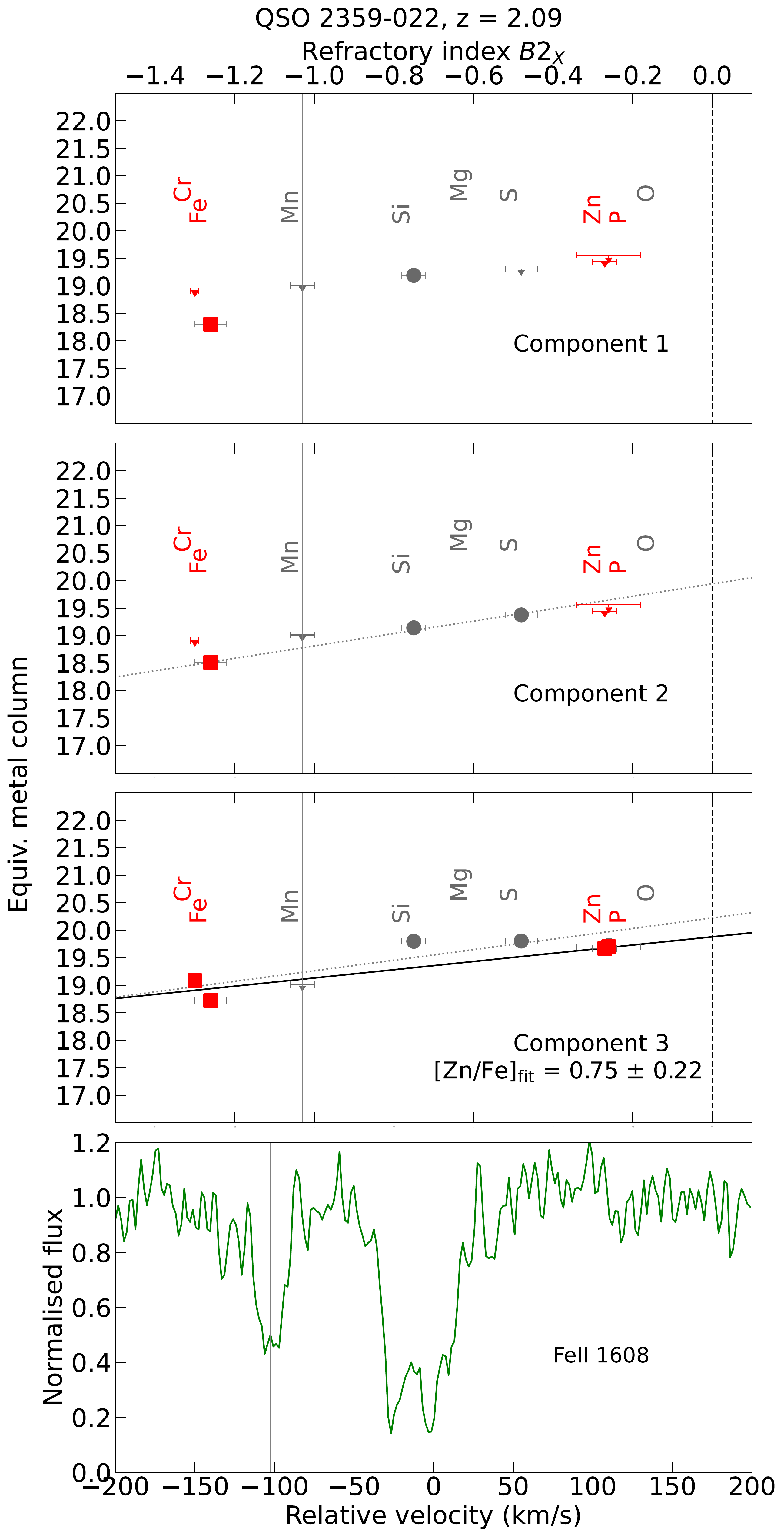}
\caption{Depletion patterns and respective spectrum for QSO~2359-022} 
\label{fig: depl_patt_last}
\end{figure}

\clearpage

% =====================================================

\section{Velocity-depletion plots}
\begin{figure}[H]
    \centering
    \includegraphics[width=0.45\textwidth]{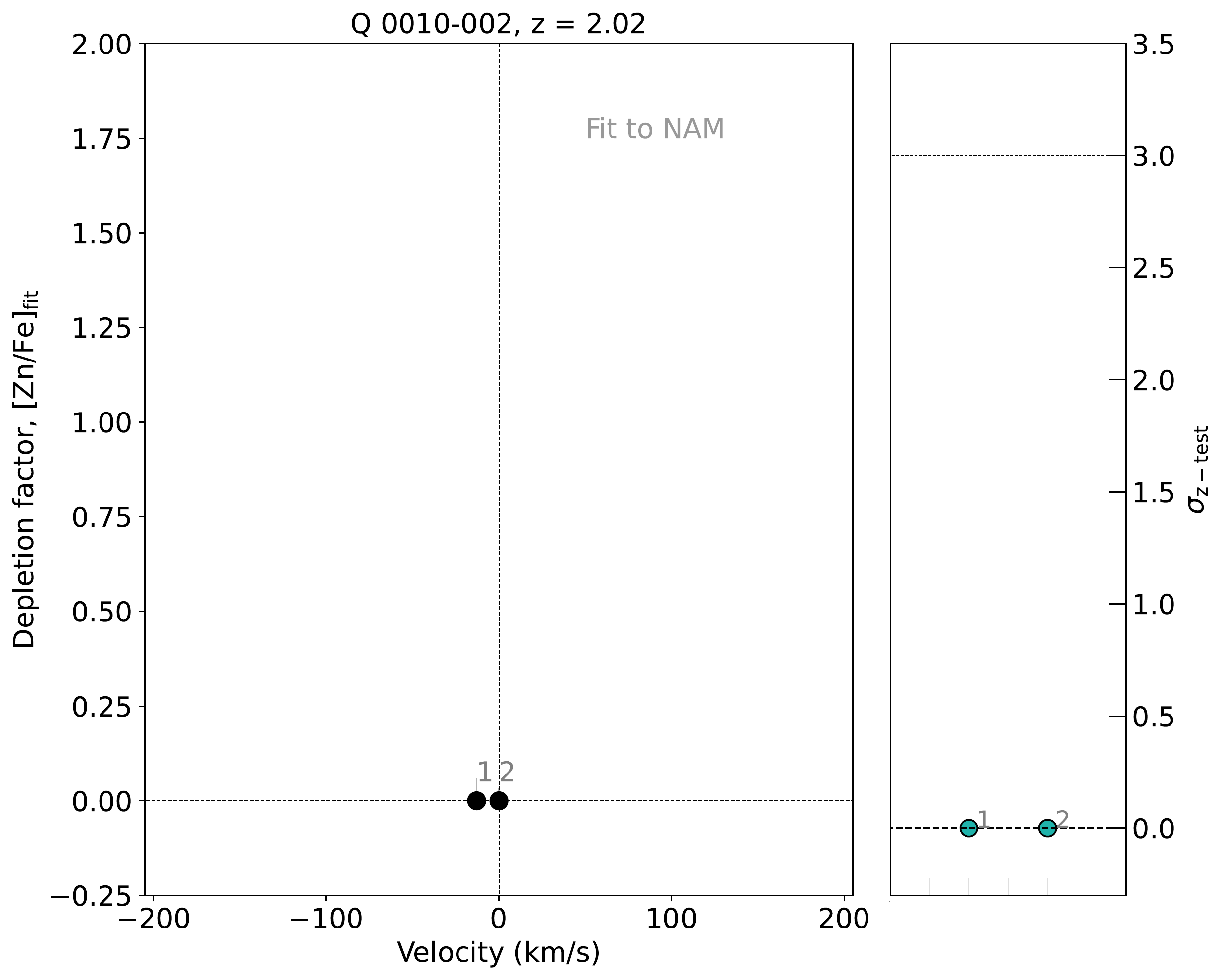}
    \caption{Velocity-depletion plot for DLA system towards QSO~0010-002}
    \label{fig:vel-depl_app1}
    \end{figure}

Figures \ref{fig:vel-depl_app1} to \ref{fig:vel-depl_app-last} show the distribution of the depletion strength [Zn/Fe]$_{\mathrm{fit}}$ with the velocity of the individual components for each system. Only systems with two or more constrained components are shown here, in total 18. 

% =====================================================

%\input{figures/vel-depl_appendix.tex}

\begin{figure}[H]
    \centering
    \includegraphics[width=0.45\textwidth]{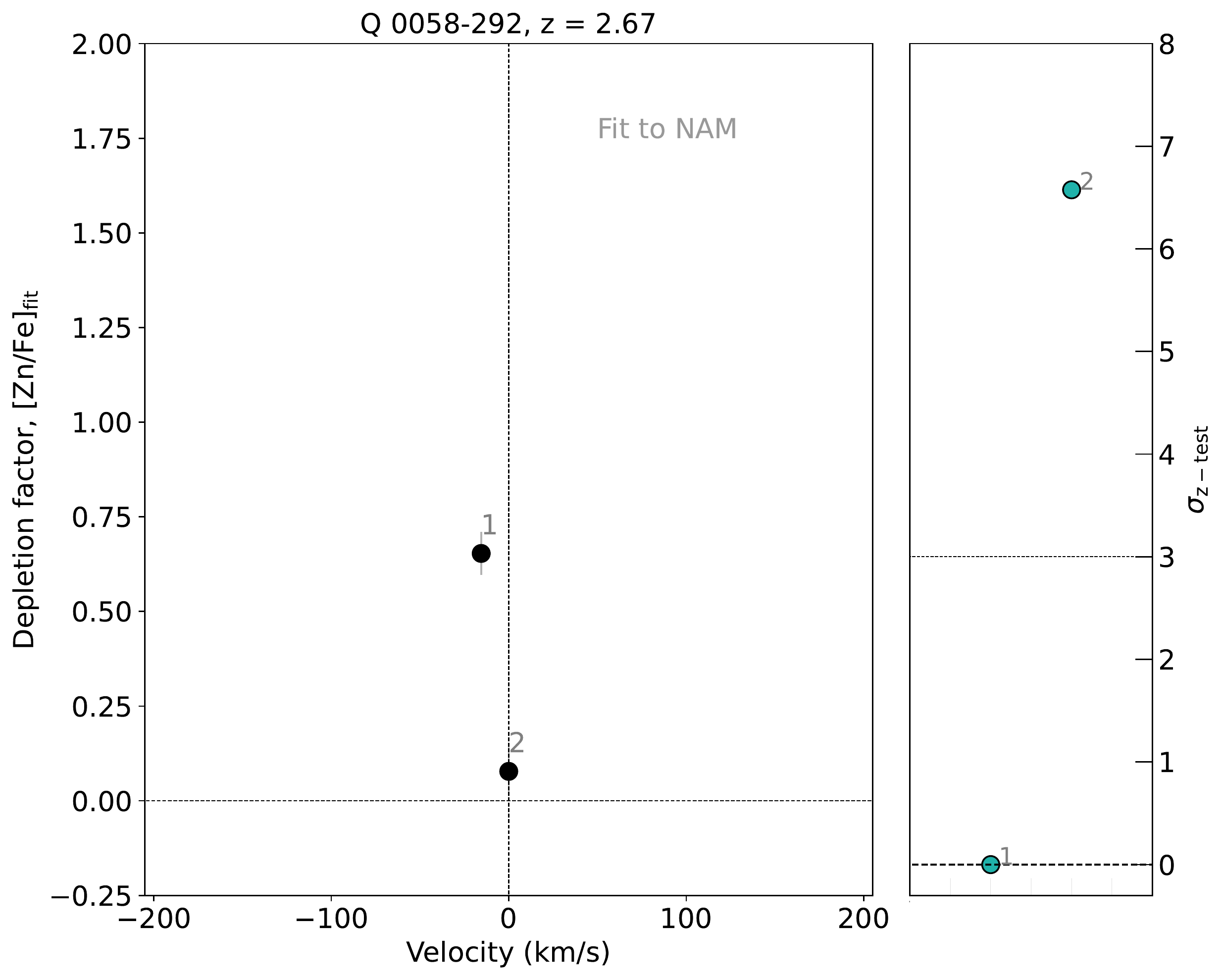}
    \caption{Velocity-depletion plot for DLA system towards QSO~0058-292}
    \end{figure}

\begin{figure}[H]
    \centering
    \includegraphics[width=0.45\textwidth]{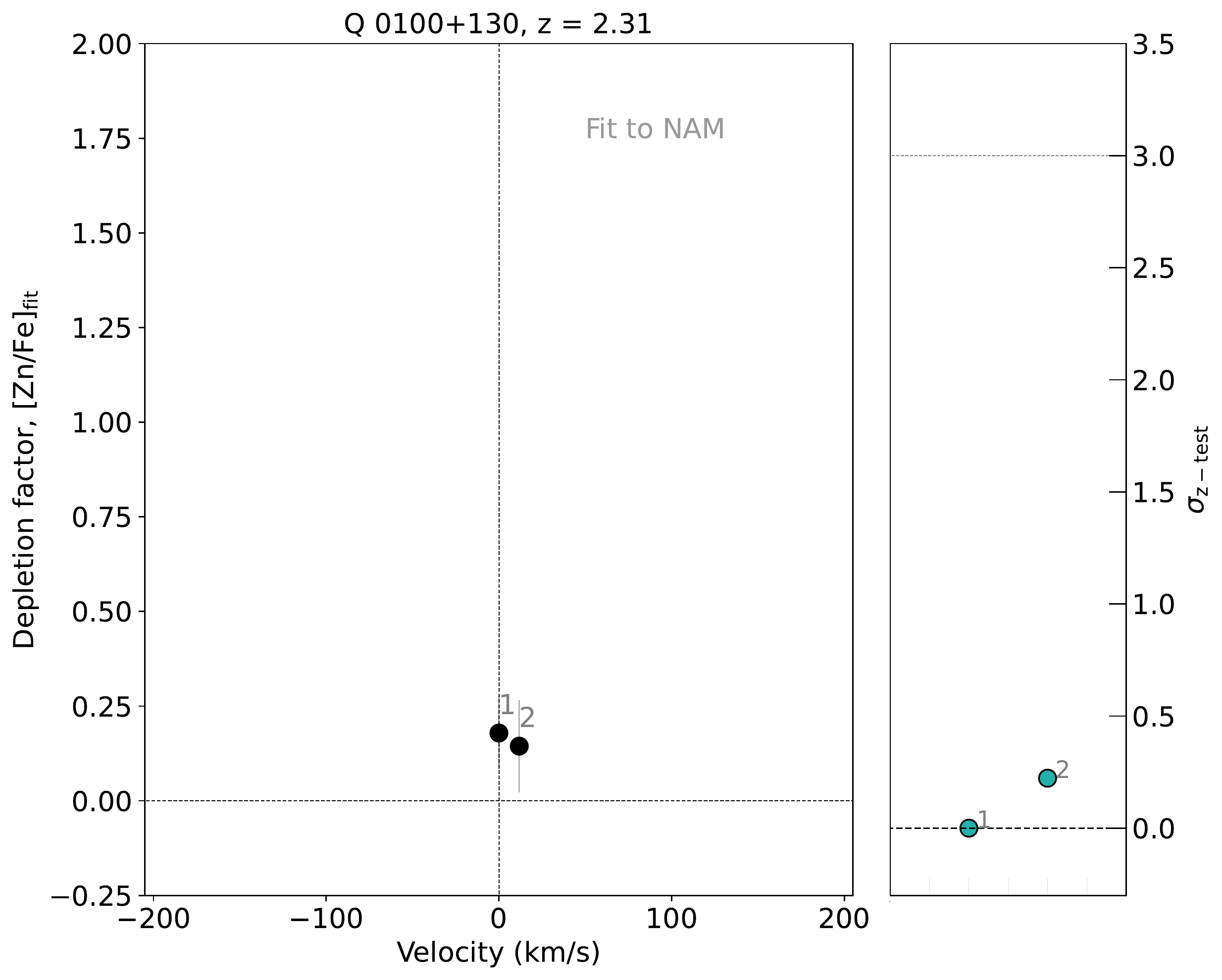}
    \caption{Velocity-depletion plot for DLA system towards QSO~0100+130}
    \end{figure}

\begin{figure}[H]
    \centering
    \includegraphics[width=0.45\textwidth]{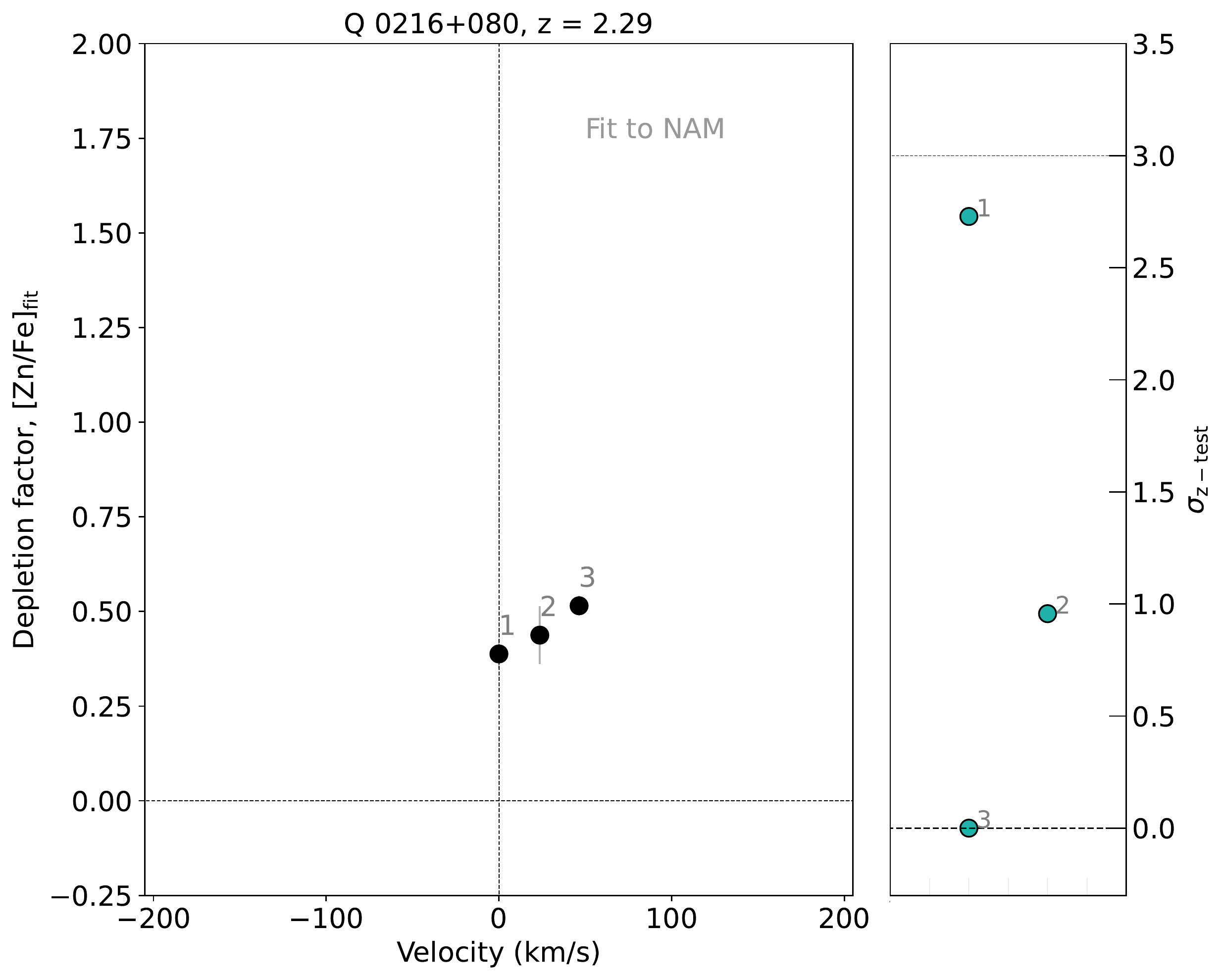}
    \caption{Velocity-depletion plot for DLA system towards QSO~0216+080}
    \label{fig:vel-depl-0216+080}
    \end{figure}

\begin{figure}[H]
    \centering
        \includegraphics[width=0.45\textwidth]{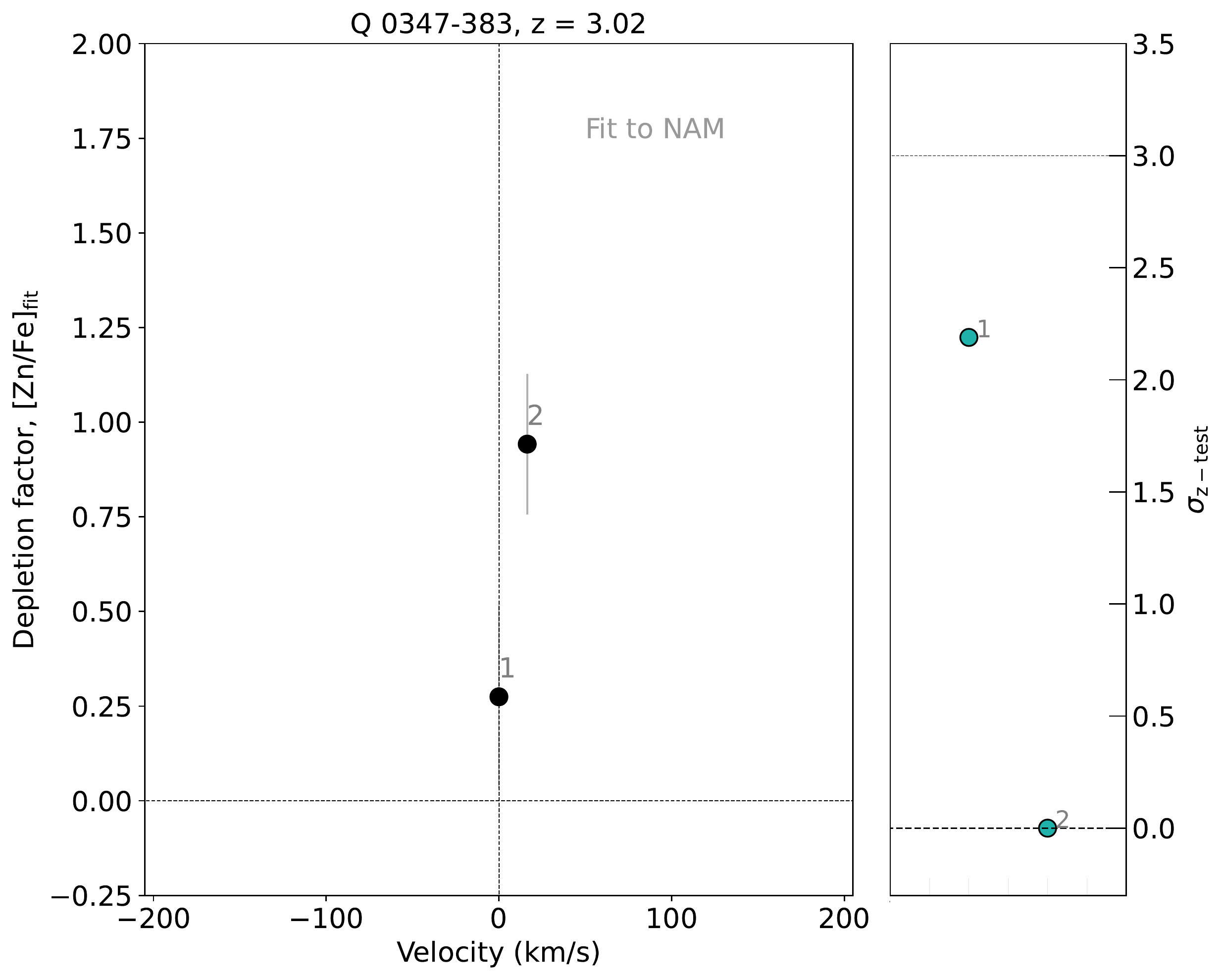}
    \caption{Velocity-depletion plot for DLA system towards QSO~0347-383}
    \label{fig:vel-depl-0347-383}
    \end{figure}

\begin{figure}[H]
    \centering
    \includegraphics[width=0.45\textwidth]{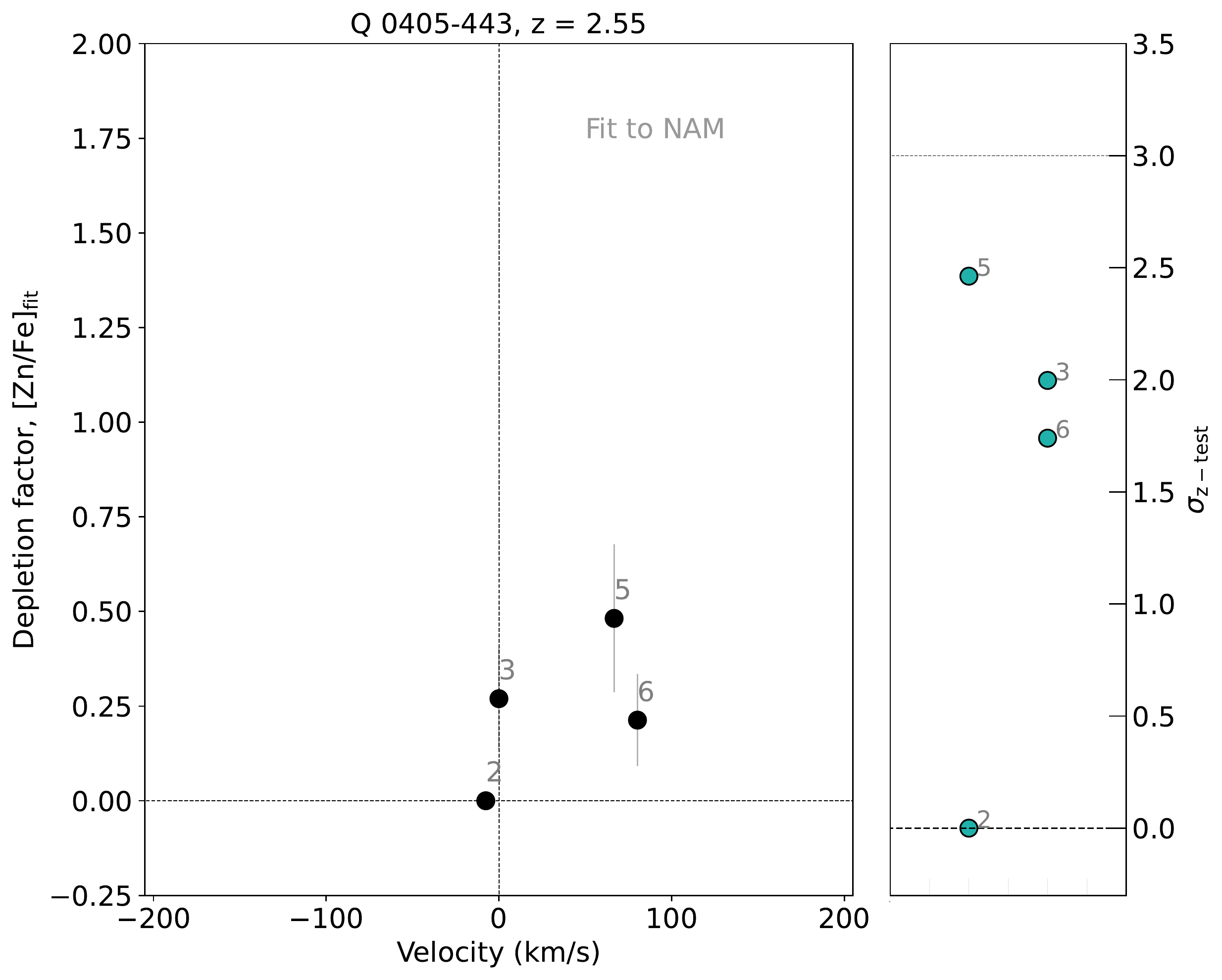}
    \caption{Velocity-depletion plot for DLA system towards QSO~0405-443}
    \end{figure}

\begin{figure}[H]
    \centering
    \includegraphics[width=0.45\textwidth]{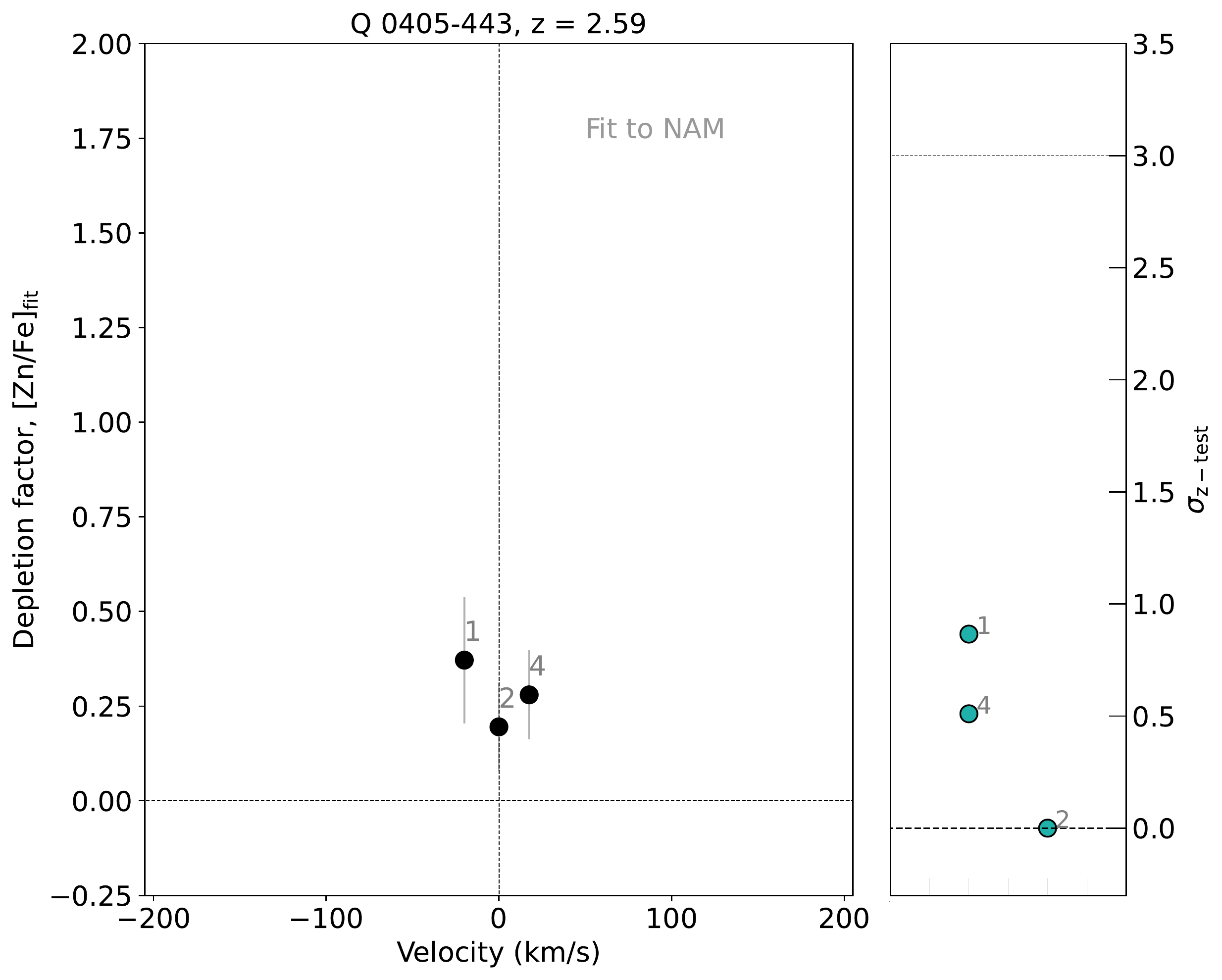}
    \caption{Velocity-depletion plot for DLA system towards QSO~0405-443}
\end{figure}

\begin{figure}[H]
    \centering
    \includegraphics[width=0.45\textwidth]{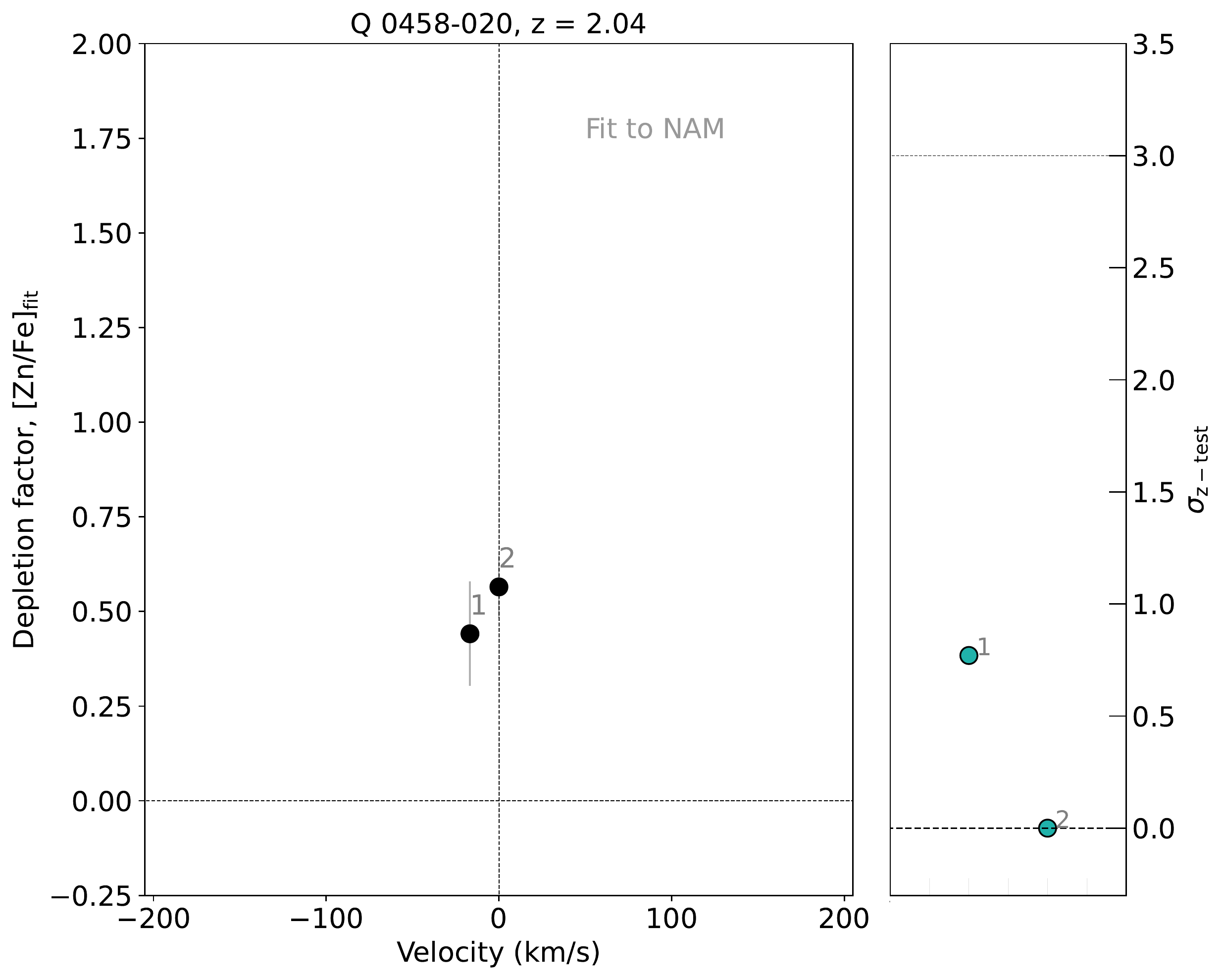}
    \caption{Velocity-depletion plot for DLA system towards QSO~0458-020}
\end{figure}

\begin{figure}[H]
    \centering
    \includegraphics[width=0.45\textwidth]{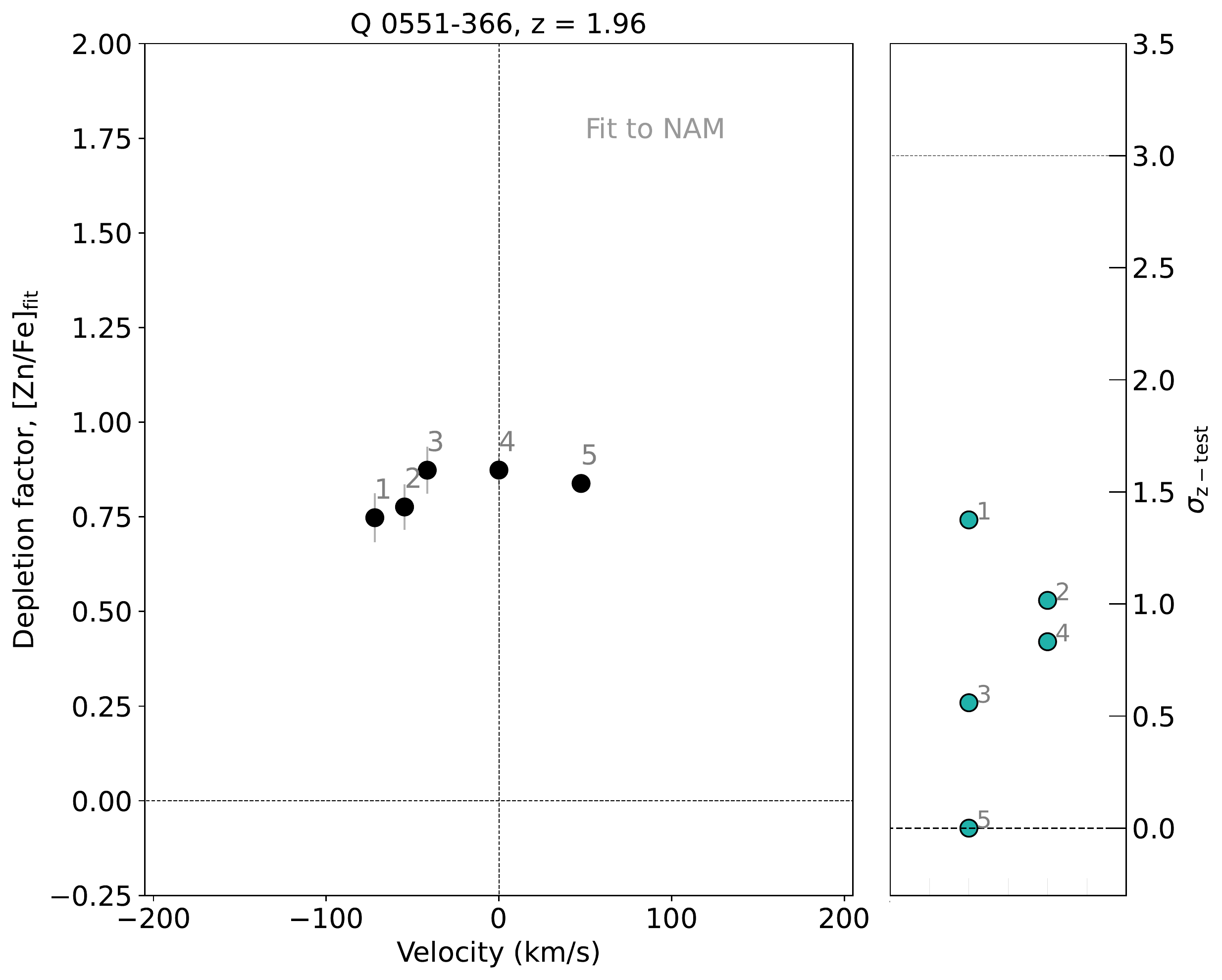}
    \caption{Velocity-depletion plot for DLA system towards QSO~0551-366}
\end{figure}

\begin{figure}[H]
    \centering
    \includegraphics[width=0.45\textwidth]{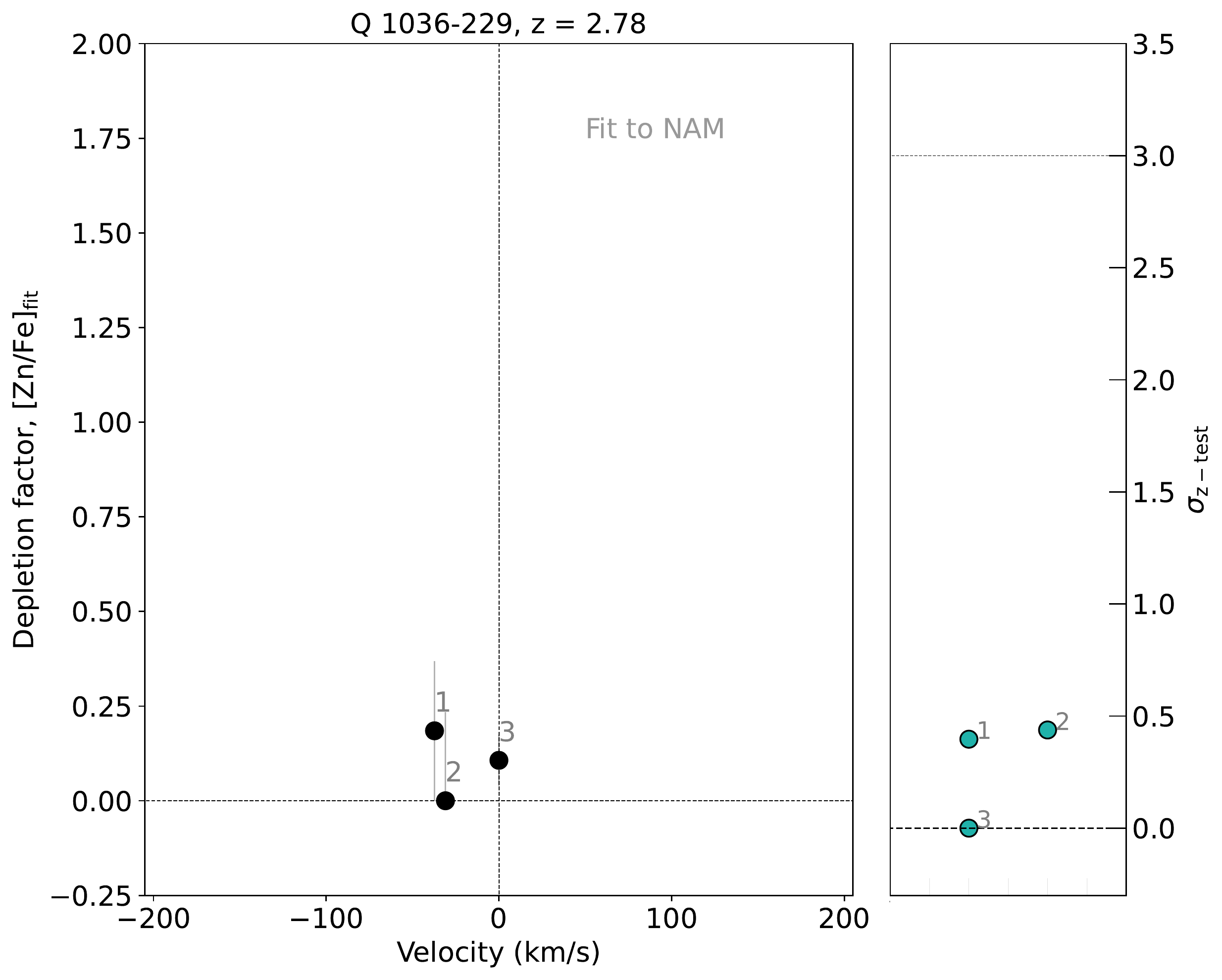}
    \caption{Velocity-depletion plot for DLA system towards QSO~1036-229}
\end{figure}

\begin{figure}[H]
    \centering
    \includegraphics[width=0.45\textwidth]{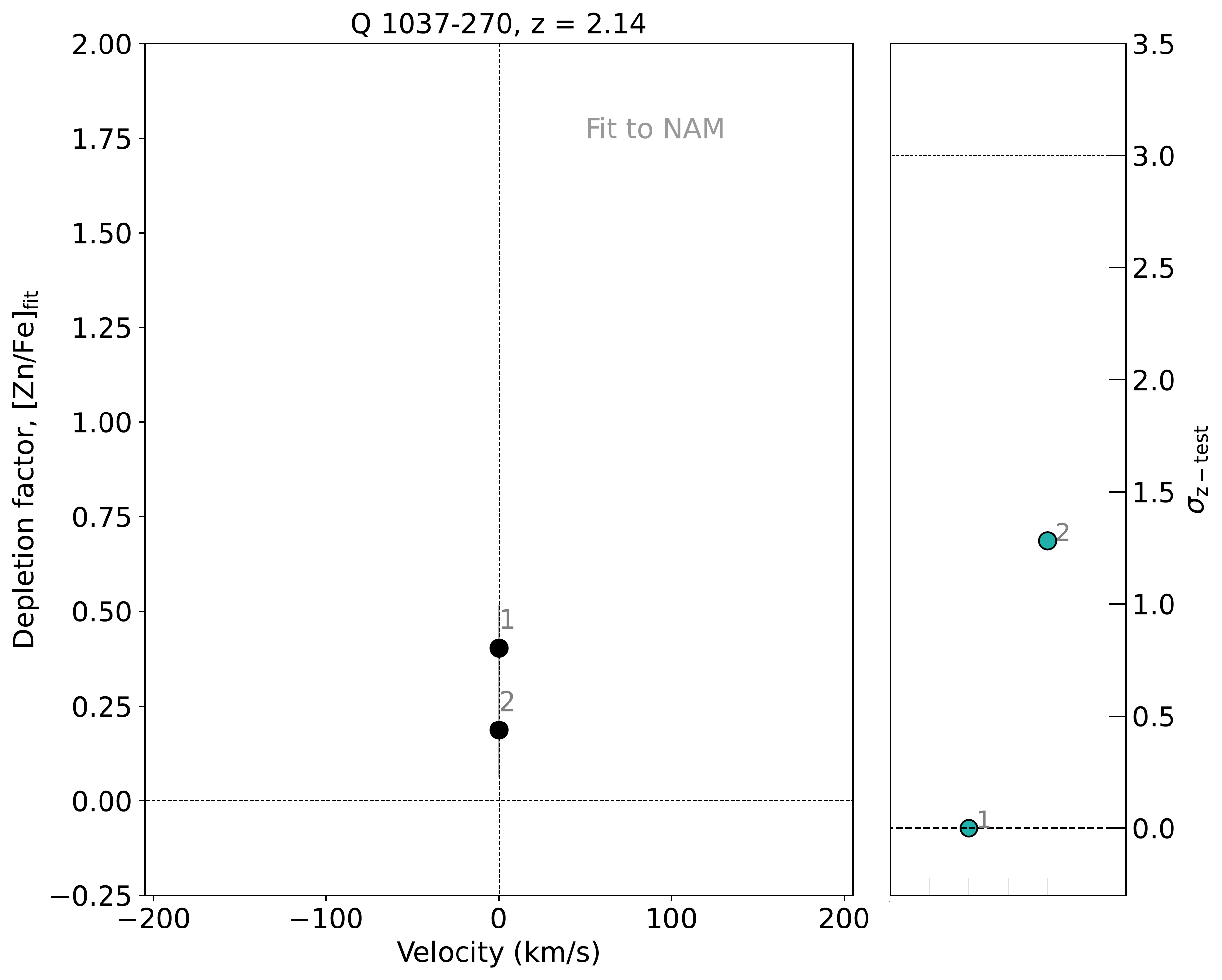}
    \caption{Velocity-depletion plot for DLA system towards QSO~1037-270}
\end{figure}

\begin{figure}[H]
    \centering
    \includegraphics[width=0.45\textwidth]{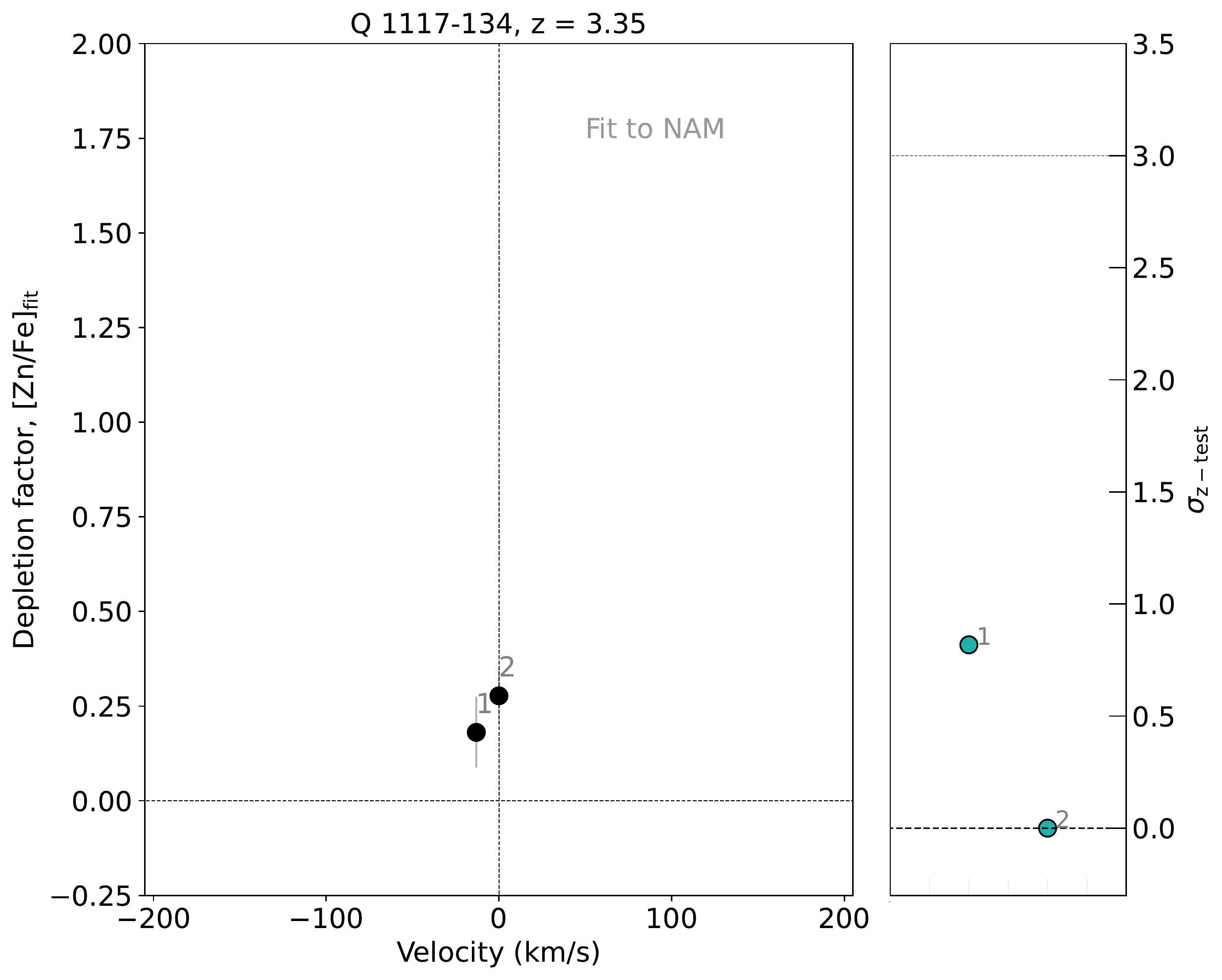}
    \caption{Velocity-depletion plot for DLA system towards QSO~1117-134}
\end{figure}

\begin{figure}[H]
    \centering
    \includegraphics[width=0.45\textwidth]{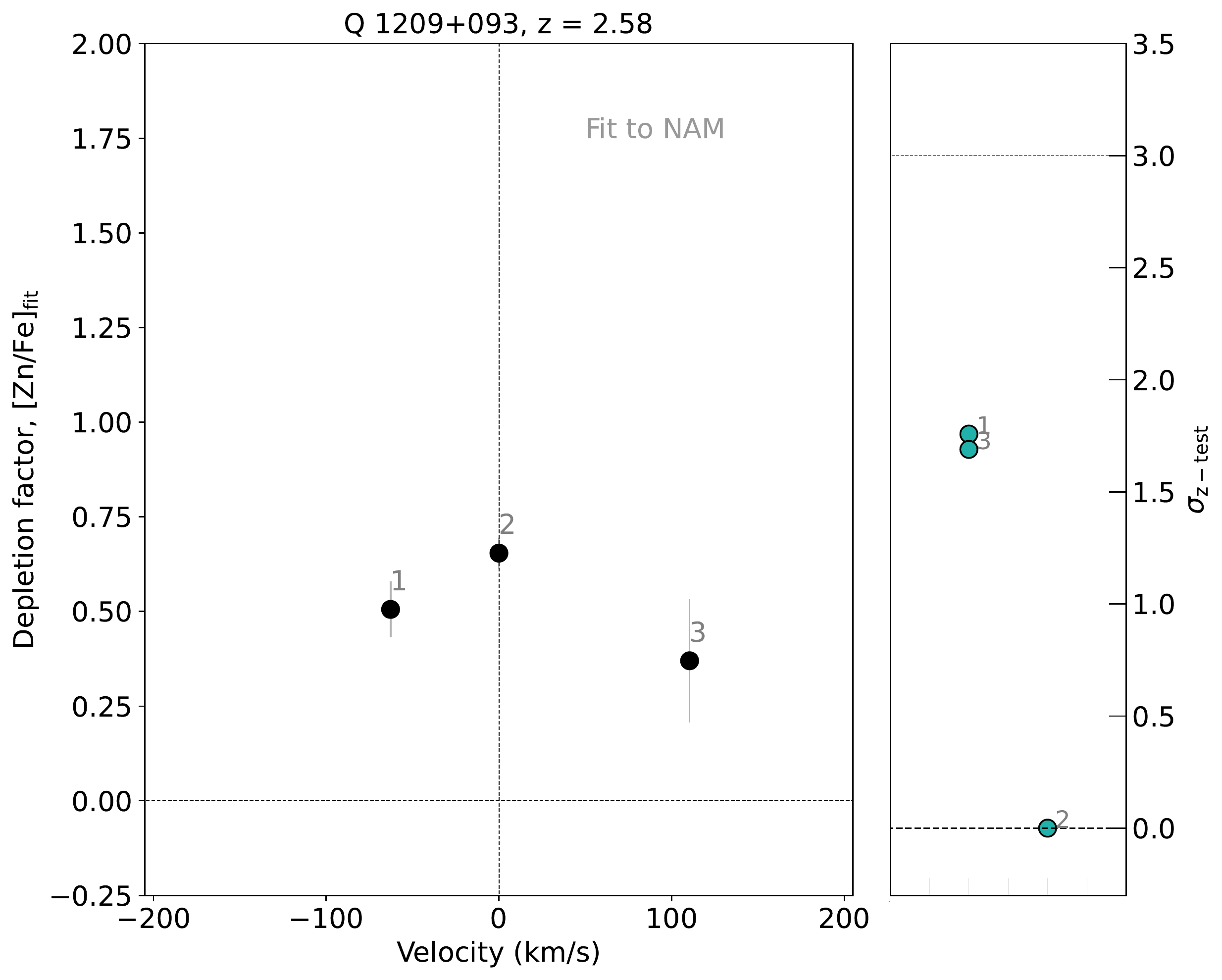}
    \caption{Velocity-depletion plot for DLA system towards QSO1209+093}
\end{figure}

\begin{figure}[H]
    \centering
    \includegraphics[width=0.45\textwidth]{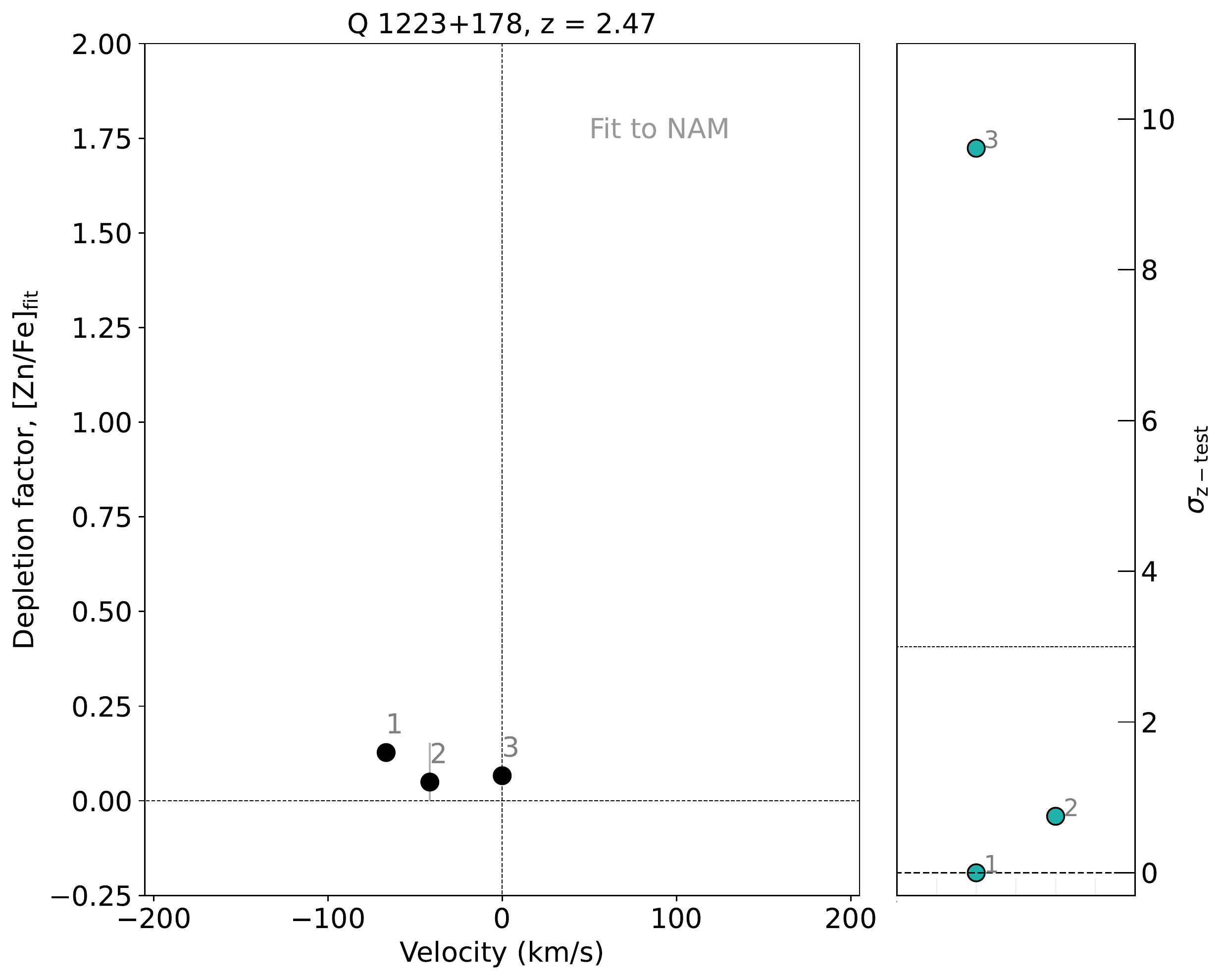}
    \caption{Velocity-depletion plot for DLA system towards QSO~1223+178}
\end{figure}

\begin{figure}[H]
    \centering
    \includegraphics[width=0.45\textwidth]{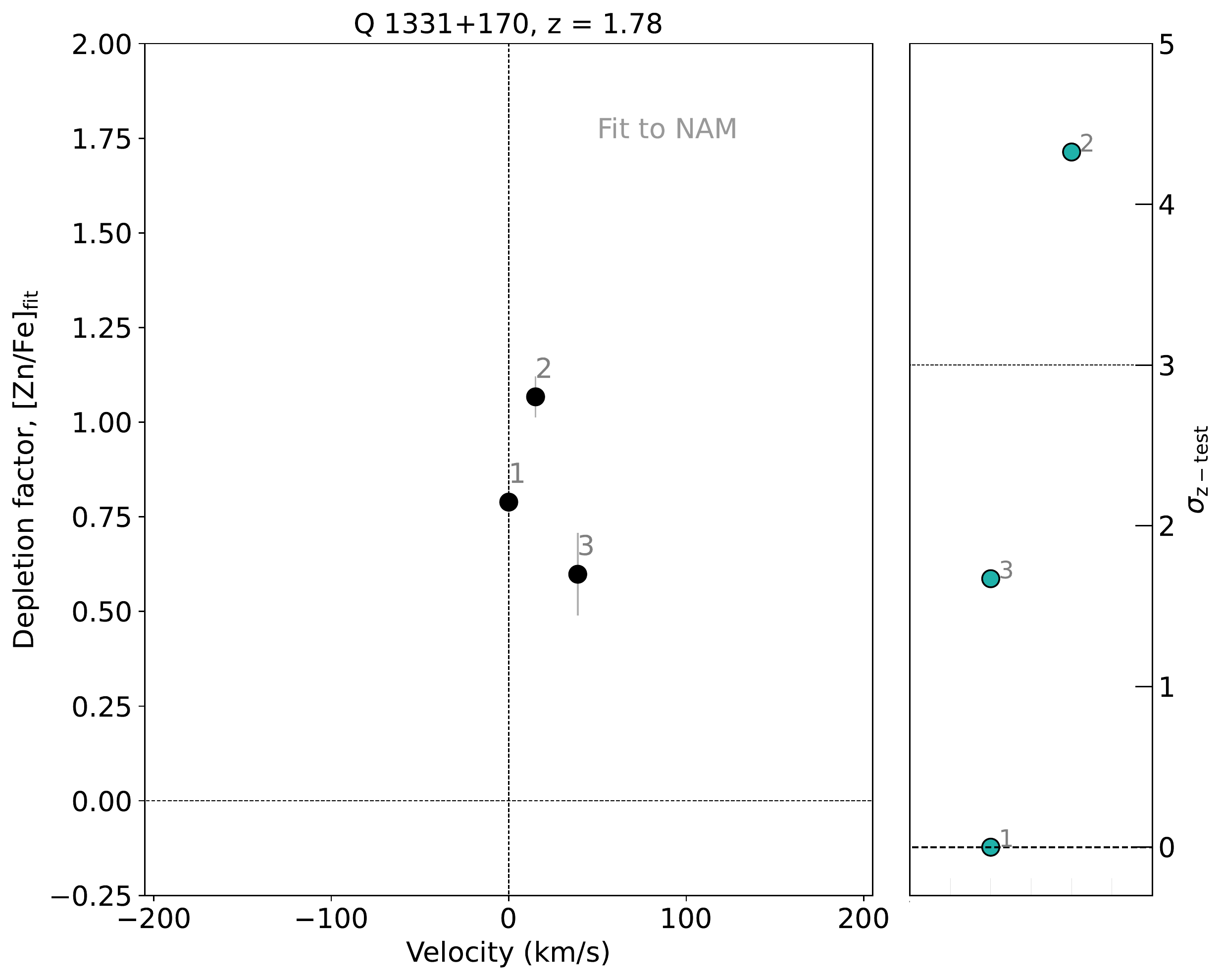}
    \caption{Velocity-depletion plot for DLA system towards QSO~1331+170}
\end{figure}

\begin{figure}[H]
    \centering
    \includegraphics[width=0.45\textwidth]{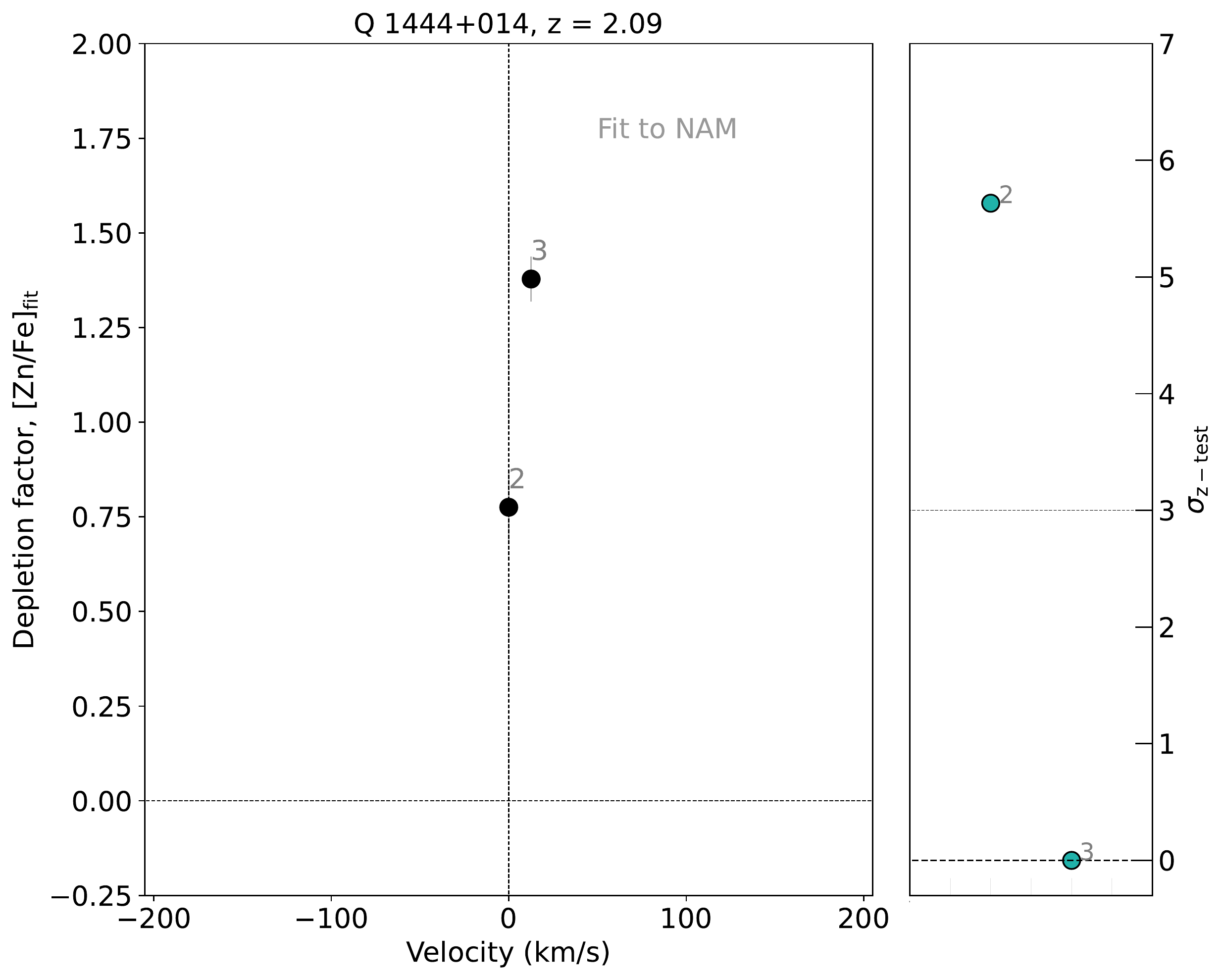}
    \caption{Velocity-depletion plot for DLA system towards QSO~1444+014}
\end{figure}

\begin{figure}[H]
    \centering
    \includegraphics[width=0.45\textwidth]{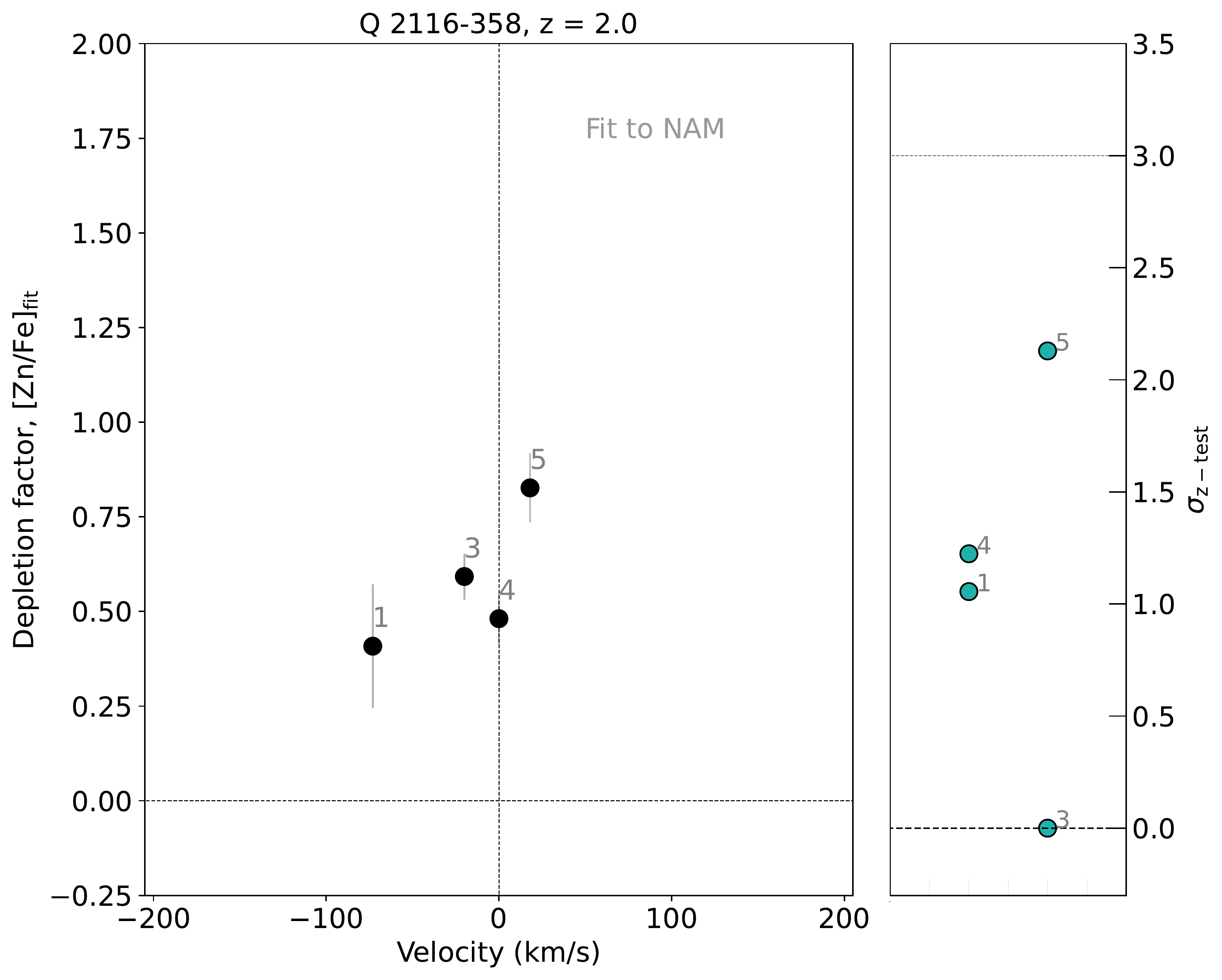}
    \caption{Velocity-depletion plot for DLA system towards QSO~2116-358}
    \label{fig:vel-depl-2116-258}
\end{figure}

\begin{figure}[H]
    \centering
    \includegraphics[width=0.45\textwidth]{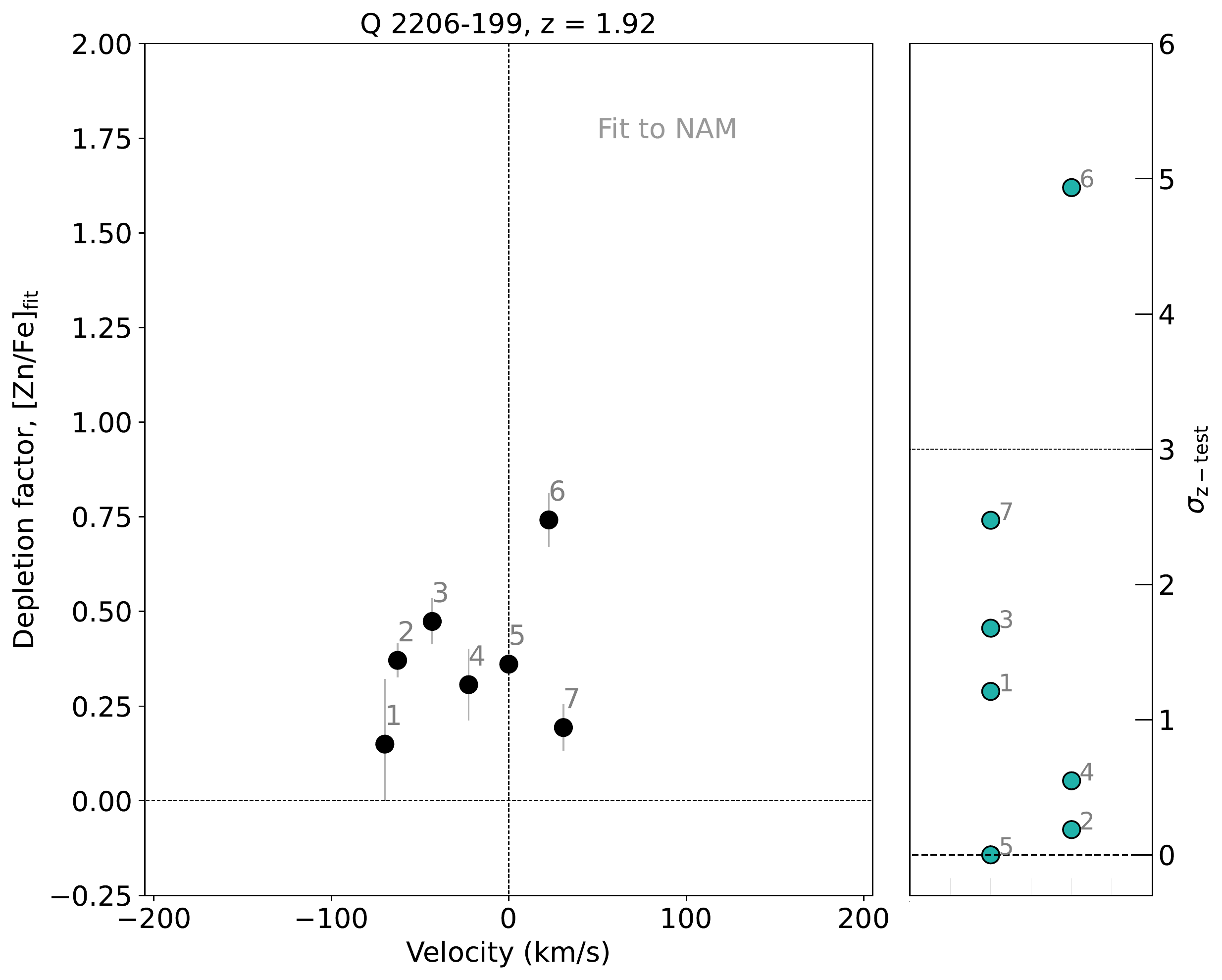}
    \caption{Velocity-depletion plot for DLA system towards QSO~2206-199}
\end{figure}

\begin{figure}[H]
    \centering
    \includegraphics[width=0.45\textwidth]{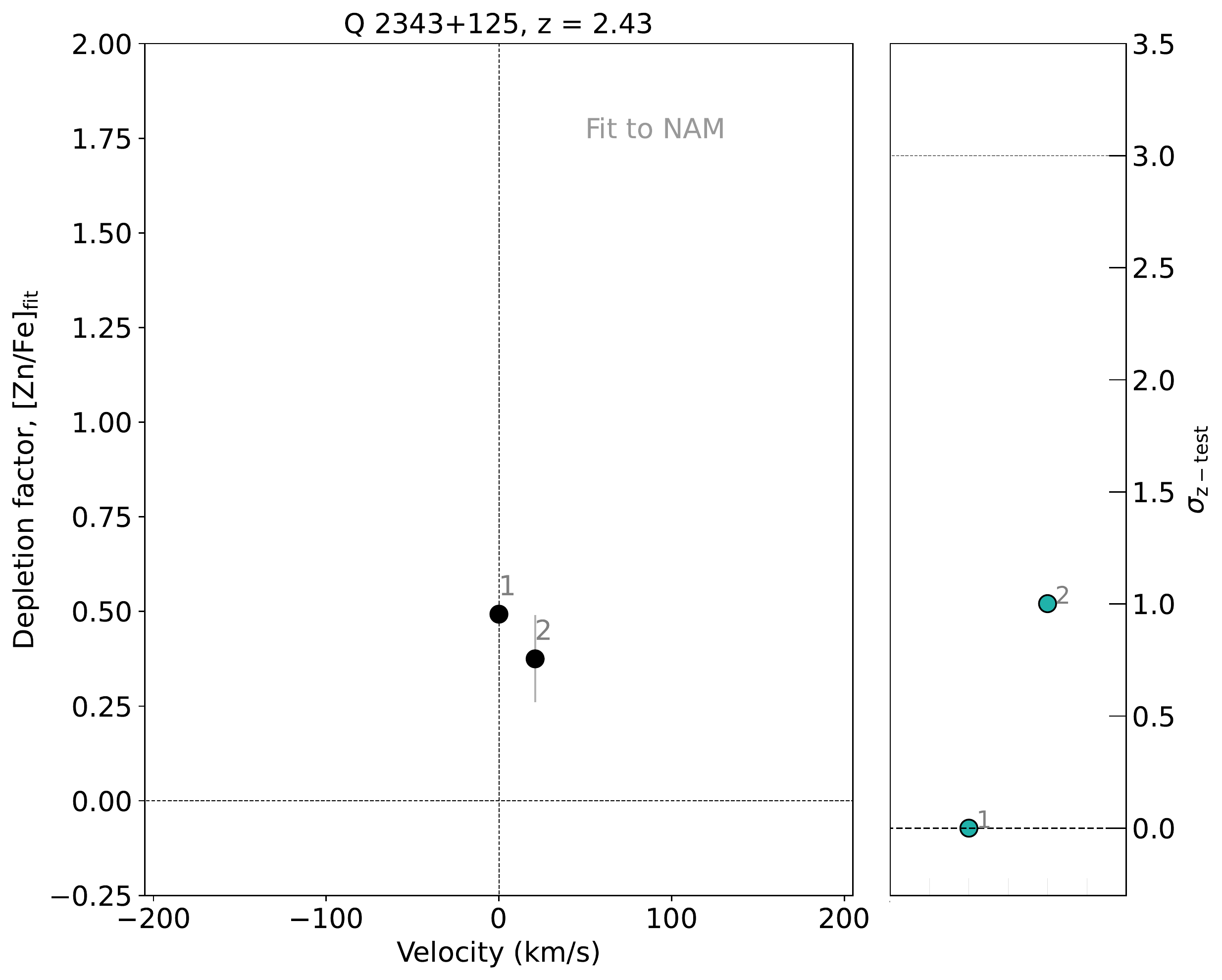}
    \caption{Velocity-depletion plot for DLA system towards QSO~2343+125}
    \label{fig:vel-depl_app-last}
\end{figure}

% =====================================================

\section{Tables}
Table \ref{tab:data} lists the results from the linear fits of the depletion patterns, namely the depletion factors and equivalent metal column densities.

\clearpage
% =====================================================
%\input{tables/final_tbl_appendix.tex}

\onecolumn
\begin{longtable}{ccccccc}

\caption{Depletion factor [Zn/Fe]$_{\mathrm{fit}}$ and equivalent metal column density {[}M/H{]}$_{\mathrm{tot}}$ + $\log{N(\mathrm{H})}$ for each individual component. Values in columns 4 and 5 are calculated from a straight line fit to all the metals in the depletion patterns (all metals). Columns 6 and 7 have values calculated from fits which excluded the $\alpha$-elements and Mn (NAM).  }

\label{tab:data}\\

\hline \hline
\textbf{QSO}       & \textbf{z}       & \textbf{Comp.} & \textbf{{[}Zn/Fe{]}$_{\rm fit}$} & \textbf{{[}M/H{]}$_{\mathrm{tot}}$ + $\log{N(\mathrm{H})}$} & \textbf{{[}Zn/Fe{]}$_{\rm fit}$} & \textbf{{[}M/H{]}$_{\mathrm{tot}}$ + $\log{N(\mathrm{H})}$} \\
 & & & \textbf{All metals}   & \textbf{All metals}   & \textbf{NAM}  & \textbf{NAM} \\ \hline
 
 \hline
\endfirsthead
\caption{continued depletion factors and equivalent metal column densities.}\\
\hline\hline
\textbf{QSO}       & \textbf{z}       & \textbf{Comp.} & \textbf{{[}Zn/Fe{]}$_{\rm fit}$} & \textbf{{[}M/H{]}$_{\mathrm{tot}}$ + $\log{N(\mathrm{H})}$} & \textbf{{[}Zn/Fe{]}$_{\rm fit}$} & \textbf{{[}M/H{]}$_{\mathrm{tot}}$ + $\log{N(\mathrm{H})}$} \\
& & & \textbf{All metals}   & \textbf{All metals}   & \textbf{NAM}  & \textbf{NAM} \\
\hline
\endhead

Q0000-263 & 3.39013 & 1  & 0.11 $\pm$ 0.14 & 19.55 $\pm$ 0.14 & 0 $\pm$ 0.11    & 19.3 $\pm$ 0.11  \\ \hline
Q0010-002 & 2.02471 & 1  & 0.0 $\pm$ 0.12  & 19.33 $\pm$ 0.12 & 0 $\pm$ 0.06    & 19.19 $\pm$ 0.06 \\
          & 2.02484 & 2  & 0.07 $\pm$ 0.15 & 19.57 $\pm$ 0.15 & 0 $\pm$ 0.03    & 19.3 $\pm$ 0.03  \\ \hline
Q0013-004 & 1.96648 & 1  & 0.23 $\pm$ 0.02 & 18.88 $\pm$ 0.02 &                 &                  \\
          & 1.9668  & 2  & 0.26 $\pm$ 0.39 & 18.89 $\pm$ 0.39 & 1.07 $\pm$ 0.29 & 19.28 $\pm$ 0.29 \\
          & 1.96734 & 3  & 0.3 $\pm$ 0.33  & 19.3 $\pm$ 0.33  & 1.29 $\pm$ 0.08 & 19.18 $\pm$ 0.08 \\
          & 1.96743 & 5  & 0.51 $\pm$ 0.51 & 19.29 $\pm$ 0.51 & 0.8 $\pm$ 0.13  & 19.52 $\pm$ 0.13 \\
          & 1.96823 & 6  & 1.45 $\pm$ 0.95 & 19.74 $\pm$ 0.95 &                 &                  \\
          & 1.96999 & 7  & 0.62 $\pm$ 0.01 & 19.31 $\pm$ 0.01 & 1.46 $\pm$ 0.12 & 19.49 $\pm$ 0.12 \\
          & 1.97027 & 8  & 1.4 $\pm$ 0.26  & 19.03 $\pm$ 0.26 & 1.07 $\pm$ 0.25 & 19.2 $\pm$ 0.25  \\
          & 1.97102 & 9  & 1.14 $\pm$ 0.16 & 19.38 $\pm$ 0.16 &                 &                  \\
          & 1.97138 & 10 & 1.39 $\pm$ 0.15 & 19.33 $\pm$ 0.15 & 0.92 $\pm$ 0.07 & 19.86 $\pm$ 0.07 \\
          & 1.97184 & 11 & 0.97 $\pm$ 0.16 & 19.76 $\pm$ 0.16 &                 &                  \\
          & 1.97289 & 14 & 2.32 $\pm$ 0.16 & 20.02 $\pm$ 0.16 & 1.06 $\pm$ 0.01 & 20.18 $\pm$ 0.01 \\
          & 1.97321 & 15 & 1.48 $\pm$ 0.27 & 19.4 $\pm$ 0.27  &                 &                  \\
          & 1.97344 & 16 & 1.02 $\pm$ 0.2  & 19.14 $\pm$ 0.2  &                 &                  \\
          & 1.97367 & 17 & 0.88 $\pm$ 0.16 & 19.34 $\pm$ 0.16 & 0.35 $\pm$ 0.02 & 18.7 $\pm$ 0.02  \\
          & 1.97381 & 18 & 1.03 $\pm$ 0.13 & 19.99 $\pm$ 0.13 &                 &                  \\
          & 1.97398 & 19 & 0.93 $\pm$ 0.16 & 19.16 $\pm$ 0.16 &                 &                  \\ \hline
Q0058-292 & 2.67123 & 1  & 1.13 $\pm$ 0.1  & 20.27 $\pm$ 0.1  & 0.65 $\pm$ 0.06 & 19.5 $\pm$ 0.06  \\
          & 2.67142 & 2  & 1.12 $\pm$ 0.01 & 19.05 $\pm$ 0.01 & 0.08 $\pm$ 0.07 & 19.36 $\pm$ 0.07 \\ \hline
Q0100+130 & 2.30903 & 1  & 1.02 $\pm$ 0.08 & 19.08 $\pm$ 0.08 & 0.18 $\pm$ 0.1  & 19.66 $\pm$ 0.1  \\
          & 2.30916 & 2  & 0.51 $\pm$ 0.14 & 18.95 $\pm$ 0.14 & 0.14 $\pm$ 0.12 & 19.37 $\pm$ 0.12 \\ \hline
Q0102-190 & 2.36958 & 1  & 0.85 $\pm$ 0.17 & 19.38 $\pm$ 0.17 &                 &                  \\
          & 2.36966 & 2  & 0.99 $\pm$ 0.12 & 19.27 $\pm$ 0.12 &                 &                  \\ \hline
Q0102-190 & 2.92625 & 1  & 0.24 $\pm$ 0.41 & 19.29 $\pm$ 0.41 &                 &                  \\
          & 2.92648 & 2  & 0.49 $\pm$ 0.06 & 19.89 $\pm$ 0.06 &                 &                  \\
          & 2.92663 & 3  & 0.43 $\pm$ 0.14 & 19.41 $\pm$ 0.14 &                 &                  \\
          & 2.92727 & 4  & 0.42 $\pm$ 0.15 & 19.76 $\pm$ 0.15 &                 &                  \\
          & 2.92772 & 5  & 0.47 $\pm$ 0.18 & 19.33 $\pm$ 0.18 &                 &                  \\ \hline
Q0112-306 & 2.41844 & 1  & 0.55 $\pm$ 0.17 & 19.53 $\pm$ 0.17 &                 &                  \\
          & 2.41861 & 2  & 0.39 $\pm$ 0.37 & 19.94 $\pm$ 0.37 &                 &                  \\ \hline
Q0112+030 & 2.42235 & 2  & 0.53 $\pm$ 0.13 & 19.49 $\pm$ 0.13 &                 &                  \\
          & 2.42277 & 5  & 0.94 $\pm$ 0.12 & 19.65 $\pm$ 0.12 &                 &                  \\
          & 2.42299 & 6  & 0.36 $\pm$ 0.15 & 18.99 $\pm$ 0.15 &                 &                  \\
          & 2.42322 & 7  & 0.3 $\pm$ 0.17  & 19.66 $\pm$ 0.17 &                 &                  \\
          & 2.42353 & 8  & 0.25 $\pm$ 0.08 & 18.97 $\pm$ 0.08 &                 &                  \\
          & 2.42332 & 9  & 0.46 $\pm$ 0.22 & 19.78 $\pm$ 0.22 &                 &                  \\ \hline
Q0135-273 & 2.10735 & 1  & 0.26 $\pm$ 0.19 & 18.71 $\pm$ 0.19 &                 &                  \\ \hline
Q0135-273 & 2.80004 & 1  & 0.09 $\pm$ 0.26 & 18.78 $\pm$ 0.26 &                 &                  \\ \hline
Q0216+080 & 1.76873 & 1  & 0.2 $\pm$ 0.27  & 19.11 $\pm$ 0.27 & 0.44 $\pm$ 0.05 & 19.4 $\pm$ 0.05  \\ \hline
Q0216+080 & 2.29307 & 1  & 0.22 $\pm$ 0.25 & 19.67 $\pm$ 0.25 & 0.39 $\pm$ 0.04 & 19.67 $\pm$ 0.04 \\
          & 2.29333 & 2  & 0.29 $\pm$ 0.34 & 19.04 $\pm$ 0.34 & 0.44 $\pm$ 0.08 & 19.22 $\pm$ 0.08 \\
          & 2.29358 & 3  & 0.34 $\pm$ 0.16 & 18.51 $\pm$ 0.16 & 0.52 $\pm$ 0.03 & 19.43 $\pm$ 0.03 \\ \hline
Q0336-017 & 3.06209 & 1  & 0.02 $\pm$ 0.34 & 18.43 $\pm$ 0.34 & 0 $\pm$ 0.18    & 19.02 $\pm$ 0.18 \\ \hline
Q0347-383 & 3.02463 & 1  & 0.13 $\pm$ 0.17 & 19.15 $\pm$ 0.17 & 0.27 $\pm$ 0.24 & 19.18 $\pm$ 0.24 \\
          & 3.02485 & 2  & 0.27 $\pm$ 0.13 & 19.45 $\pm$ 0.13 & 0.94 $\pm$ 0.19 & 19.62 $\pm$ 0.19 \\ \hline
Q0405-443 & 1.91193 & 1  & 0.06 $\pm$ 0.2  & 18.67 $\pm$ 0.2  & 0.19 $\pm$ 0.08 & 18.67 $\pm$ 0.08 \\
          & 1.91208 & 2  & 0.39 $\pm$ 0.26 & 19.04 $\pm$ 0.26 & 0.0 $\pm$ 0.03  & 18.7 $\pm$ 0.03  \\
          & 1.91235 & 3  & 0.43 $\pm$ 0.25 & 19.35 $\pm$ 0.25 & 0.07 $\pm$ 0.09 & 18.98 $\pm$ 0.09 \\
          & 1.91267 & 4  & 0.23 $\pm$ 0.01 & 18.79 $\pm$ 0.01 & 0.12 $\pm$ 0.02 & 19.63 $\pm$ 0.02 \\
          & 1.91279 & 5  & 0.29 $\pm$ 0.36 & 18.91 $\pm$ 0.36 & 0.14 $\pm$ 0.09 & 18.92 $\pm$ 0.09 \\
          & 1.91302 & 6  & 0.22 $\pm$ 0.11 & 18.45 $\pm$ 0.11 &                 &                  \\ \hline
Q0405-443 & 2.54971 & 1  & 0.48 $\pm$ 0.21 & 20.17 $\pm$ 0.21 &                 &                  \\
          & 2.54991 & 2  & 0.62 $\pm$ 0.16 & 20.5 $\pm$ 0.16  & 0.0 $\pm$ 0.01  & 18.7 $\pm$ 0.01  \\
          & 2.55    & 3  & 0.61 $\pm$ 0.26 & 20.47 $\pm$ 0.26 & 0.27 $\pm$ 0.13 & 19.43 $\pm$ 0.13 \\
          & 2.55059 & 4  & 0.59 $\pm$ 0.19 & 20.1 $\pm$ 0.19  &                 &                  \\
          & 2.55079 & 5  & 0.93 $\pm$ 0.18 & 20.36 $\pm$ 0.18 & 0.48 $\pm$ 0.2  & 19.06 $\pm$ 0.2  \\
          & 2.55095 & 6  & 0.96 $\pm$ 0.78 & 19.64 $\pm$ 0.78 & 0.21 $\pm$ 0.12 & 19.15 $\pm$ 0.12 \\
          & 2.55123 & 7  & 0.19 $\pm$ 0.15 & 18.75 $\pm$ 0.15 &                 &                  \\ \hline
Q0405-443 & 2.59442 & 1  & 0.35 $\pm$ 0.24 & 19.59 $\pm$ 0.24 & 0.37 $\pm$ 0.17 & 18.97 $\pm$ 0.17 \\
          & 2.59466 & 2  & 0.83 $\pm$ 0.1  & 19.8 $\pm$ 0.1   & 0.2 $\pm$ 0.12  & 19.6 $\pm$ 0.12  \\
          & 2.59476 & 3  & 0.84 $\pm$ 0.15 & 19.35 $\pm$ 0.15 &                 &                  \\
          & 2.59487 & 4  & 0.94 $\pm$ 0.11 & 19.54 $\pm$ 0.11 & 0.28 $\pm$ 0.12 & 19.64 $\pm$ 0.12 \\ \hline
Q0450-131 & 2.06658 & 4  & 0.96 $\pm$ 0.14 & 20.42 $\pm$ 0.14 &                 &                  \\
          & 2.06683 & 5  & 0.87 $\pm$ 0.1  & 20.02 $\pm$ 0.1  &                 &                  \\ \hline
Q0458-020 & 2.03939 & 1  & 0.74 $\pm$ 0.2  & 19.65 $\pm$ 0.2  & 0.44 $\pm$ 0.14 & 20.18 $\pm$ 0.14 \\
          & 2.03956 & 2  & 0.14 $\pm$ 0.1  & 19.45 $\pm$ 0.1  & 0.56 $\pm$ 0.08 & 20.47 $\pm$ 0.08 \\ \hline
Q0528-250 & 2.14062 & 1  & 0.07 $\pm$ 0.81 & 19.11 $\pm$ 0.81 &                 &                  \\
          & 2.14085 & 2  & 0.67 $\pm$ 0.51 & 19.56 $\pm$ 0.51 &                 &                  \\
          & 2.14105 & 3  & 0.34 $\pm$ 0.13 & 19.43 $\pm$ 0.13 & 0.19 $\pm$ 0.03 & 19.48 $\pm$ 0.03 \\ \hline
Q0528-250 & 2.81111 & 1  & 0.32 $\pm$ 0.17 & 19.64 $\pm$ 0.17 & 0.41 $\pm$ 0.0  & 20.28 $\pm$ 0.0  \\
          & 2.81204 & 2  & 0.2 $\pm$ 0.04  & 17.43 $\pm$ 0.04 & 0.46 $\pm$ 0.05 & 19.98 $\pm$ 0.05 \\
          & 2.81378 & 3  & 0.18 $\pm$ 0.1  & 17.77 $\pm$ 0.1  & 0.85 $\pm$ 0.15 & 20.21 $\pm$ 0.15 \\ \hline
Q0551-366 & 1.9615  & 1  & 0.28 $\pm$ 0.18 & 19.75 $\pm$ 0.18 & 0.75 $\pm$ 0.07 & 19.7 $\pm$ 0.07  \\
          & 1.96167 & 2  & 0.23 $\pm$ 0.17 & 19.48 $\pm$ 0.17 & 0.78 $\pm$ 0.06 & 19.27 $\pm$ 0.06 \\
          & 1.9618  & 3  & 0.25 $\pm$ 0.36 & 18.83 $\pm$ 0.36 & 0.87 $\pm$ 0.06 & 19.53 $\pm$ 0.06 \\
          & 1.96221 & 4  & 0.04 $\pm$ 0.46 & 18.83 $\pm$ 0.46 & 0.87 $\pm$ 0.04 & 20.37 $\pm$ 0.04 \\
          & 1.96268 & 5  & 0.0 $\pm$ 0.55  & 16.91 $\pm$ 0.55 & 0.84 $\pm$ 0.01 & 20.01 $\pm$ 0.01 \\ \hline
Q0841+129 & 1.86384 & 1  & 0.03 $\pm$ 0.1  & 18.07 $\pm$ 0.1  &                 &                  \\
          & 1.86392 & 2  & 0.18 $\pm$ 0.24 & 17.99 $\pm$ 0.24 &                 &                  \\ \hline
Q0841+129 & 2.37452 & 1  & 0.0 $\pm$ 0.2   & 17.51 $\pm$ 0.2  & 0.18 $\pm$ 0.06 & 19.55 $\pm$ 0.06 \\ \hline
Q0841+129 & 2.47622 & 1  & 0.0 $\pm$ 0.26  & 17.71 $\pm$ 0.26 & 0.13 $\pm$ 0.12 & 19.24 $\pm$ 0.12 \\ \hline
Q0913+072 & 2.61829 & 1  & 0.35 $\pm$ 0.2  & 19.32 $\pm$ 0.2  &                 &                  \\
          & 2.61843 & 2  & 0.28 $\pm$ 0.3  & 19.1 $\pm$ 0.3   &                 &                  \\ \hline
Q1036-229 & 2.77732 & 1  & 0.33 $\pm$ 0.29 & 19.32 $\pm$ 0.29 & 0.18 $\pm$ 0.18 & 19.14 $\pm$ 0.18 \\
          & 2.7774  & 2  & 0.65 $\pm$ 0.55 & 19.17 $\pm$ 0.55 & 0.0 $\pm$ 0.23  & 18.36 $\pm$ 0.23 \\
          & 2.77779 & 3  & 0.65 $\pm$ 0.22 & 19.74 $\pm$ 0.22 & 0.11 $\pm$ 0.07 & 19.1 $\pm$ 0.07  \\
          & 2.77808 & 4  & 0.38 $\pm$ 0.21 & 19.25 $\pm$ 0.21 &                 &                  \\ \hline
Q1037-270 & 2.139   & 1  & 0.04 $\pm$ 0.35 & 18.73 $\pm$ 0.35 & 0.4 $\pm$ 0.11  & 19.45 $\pm$ 0.11 \\
          & 2.139   & 2  & 0.2 $\pm$ 0.02  & 19.07 $\pm$ 0.02 & 0.19 $\pm$ 0.12 & 19.08 $\pm$ 0.12 \\ \hline
Q1108-077 & 3.60767 & 1  & 0.17 $\pm$ 0.08 & 19.14 $\pm$ 0.08 &                 &                  \\ \hline
Q1111-152 & 3.26536 & 1  & 0.33 $\pm$ 0.04 & 19.09 $\pm$ 0.04 &                 &                  \\
          & 3.26552 & 2  & 0.2 $\pm$ 0.34  & 18.78 $\pm$ 0.34 & 0.2 $\pm$ 0.15  & 19.17 $\pm$ 0.15 \\
          & 3.2657  & 3  & 0.1 $\pm$ 0.79  & 18.25 $\pm$ 0.79 &                 &                  \\
          & 3.26603 & 4  & 0.43 $\pm$ 0.37 & 19.0 $\pm$ 0.37  &                 &                  \\
          & 3.26678 & 6  & 0.21 $\pm$ 0.15 & 19.34 $\pm$ 0.15 &                 &                  \\
          & 3.26704 & 7  & 0.29 $\pm$ 0.07 & 19.37 $\pm$ 0.07 &                 &                  \\ \hline
Q1117-134 & 3.35027 & 1  & 0.4 $\pm$ 0.03  & 18.14 $\pm$ 0.03 & 0.18 $\pm$ 0.09 & 19.27 $\pm$ 0.09 \\
          & 3.35046 & 2  & 0.05 $\pm$ 0.32 & 17.84 $\pm$ 0.32 & 0.28 $\pm$ 0.07 & 19.35 $\pm$ 0.07 \\ \hline
Q1157+014 & 1.94317 & 1  & 0.33 $\pm$ 0.34 & 19.27 $\pm$ 0.34 & 0.05 $\pm$ 0.03 & 18.93 $\pm$ 0.03 \\
          & 1.94349 & 2  & 0.39 $\pm$ 0.19 & 20.04 $\pm$ 0.19 & 0.26 $\pm$ 0.05 & 19.94 $\pm$ 0.05 \\
          & 1.94376 & 3  & 0.69 $\pm$ 0.25 & 20.37 $\pm$ 0.25 & 0.48 $\pm$ 0.07 & 20.2 $\pm$ 0.07  \\
          & 1.94403 & 4  & 0.49 $\pm$ 0.2  & 19.94 $\pm$ 0.2  & 0.36 $\pm$ 0.08 & 19.83 $\pm$ 0.08 \\ \hline
Q1209+093 & 2.58362 & 1  & 0.67 $\pm$ 0.29 & 20.21 $\pm$ 0.29 & 0.51 $\pm$ 0.07 & 20.09 $\pm$ 0.07 \\
          & 2.58437 & 2  & 0.9 $\pm$ 0.28  & 20.61 $\pm$ 0.28 & 0.65 $\pm$ 0.04 & 20.42 $\pm$ 0.04 \\
          & 2.58569 & 3  & 0.57 $\pm$ 0.21 & 19.58 $\pm$ 0.21 & 0.37 $\pm$ 0.16 & 19.33 $\pm$ 0.16 \\ \hline
Q1210+175 & 1.89158 & 1  & 0.77 $\pm$ 0.42 & 19.31 $\pm$ 0.42 &                 &                  \\
          & 1.89177 & 2  & 1.28 $\pm$ 0.68 & 20.54 $\pm$ 0.68 &                 &                  \\
          & 1.89195 & 3  & 0.93 $\pm$ 0.72 & 19.34 $\pm$ 0.72 &                 &                  \\ \hline
Q1223+178 & 2.4653  & 1  & 0.28 $\pm$ 0.22 & 19.48 $\pm$ 0.22 & 0.13 $\pm$ 0.0  & 19.33 $\pm$ 0.0  \\
          & 2.46559 & 2  & 0.4 $\pm$ 0.27  & 19.47 $\pm$ 0.27 & 0.05 $\pm$ 0.1  & 19.08 $\pm$ 0.1  \\
          & 2.46607 & 3  & 0.26 $\pm$ 0.23 & 19.71 $\pm$ 0.23 & 0.07 $\pm$ 0.0  & 19.55 $\pm$ 0.0  \\ \hline
Q1232+082 & 2.33707 & 1  & 0.35 $\pm$ 0.28 & 18.63 $\pm$ 0.28 &                 &                  \\
          & 2.33771 & 2  & 1.23 $\pm$ 0.72 & 20.23 $\pm$ 0.72 &                 &                  \\
          & 2.33776 & 3  & 0.94 $\pm$ 0.09 & 19.6 $\pm$ 0.09  &                 &                  \\ \hline
Q1331+170 & 1.77635 & 1  & 1.01 $\pm$ 0.2  & 20.02 $\pm$ 0.2  & 0.79 $\pm$ 0.03 & 19.81 $\pm$ 0.03 \\
          & 1.77649 & 2  & 1.42 $\pm$ 0.25 & 20.12 $\pm$ 0.25 & 1.07 $\pm$ 0.05 & 19.76 $\pm$ 0.05 \\
          & 1.77671 & 3  & 0.87 $\pm$ 0.32 & 19.51 $\pm$ 0.32 & 0.6 $\pm$ 0.11  & 19.25 $\pm$ 0.11 \\
          & 1.77685 & 4  & 0.89 $\pm$ 0.37 & 19.38 $\pm$ 0.37 &                 &                  \\ \hline
Q1337+113 & 2.50792 & 2  & 0.4 $\pm$ 0.03  & 18.67 $\pm$ 0.03 &                 &                  \\ \hline
Q1337+113 & 2.79557 & 1  & 0.32 $\pm$ 0.23 & 19.25 $\pm$ 0.23 &                 &                  \\
          & 2.79584 & 2  & 0.41 $\pm$ 0.32 & 18.49 $\pm$ 0.32 &                 &                  \\ \hline
Q1340-136 & 3.11835 & 1  & 0.22 $\pm$ 0.2  & 18.88 $\pm$ 0.2  &                 &                  \\ \hline
Q1409+095 & 2.01881 & 1  & 0.09 $\pm$ 0.13 & 18.52 $\pm$ 0.13 & 0.15 $\pm$ 0.08 & 18.95 $\pm$ 0.08 \\ \hline
Q1409+095 & 2.45593 & 1  & 0.34 $\pm$ 0.19 & 18.37 $\pm$ 0.19 &                 &                  \\
          & 2.456   & 2  & 0.1 $\pm$ 0.06  & 17.53 $\pm$ 0.06 &                 &                  \\
          & 2.4564  & 3  & 0.29 $\pm$ 0.08 & 18.21 $\pm$ 0.08 &                 &                  \\
          & 2.45648 & 4  & 0.38 $\pm$ 0.07 & 18.06 $\pm$ 0.07 &                 &                  \\ \hline
Q1409+095 & 2.668   & 1  & 0.23 $\pm$ 0.16 & 19.09 $\pm$ 0.16 &                 &                  \\ \hline
Q1441+276 & 4.22348 & 1  & 0.0 $\pm$ 1.12  & 18.94 $\pm$ 1.12 &                 &                  \\
          & 4.22374 & 2  & 0.65 $\pm$ 0.27 & 20.1 $\pm$ 0.27  &                 &                  \\
          & 4.22401 & 3  & 0.66 $\pm$ 0.18 & 20.4 $\pm$ 0.18  &                 &                  \\
          & 4.22438 & 4  & 0.64 $\pm$ 0.24 & 20.03 $\pm$ 0.24 &                 &                  \\ \hline
Q1444+014 & 2.08667 & 1  & 1.13 $\pm$ 0.24 & 19.07 $\pm$ 0.24 &                 &                  \\
          & 2.08679 & 2  & 0.87 $\pm$ 0.13 & 19.46 $\pm$ 0.13 & 0.78 $\pm$ 0.09 & 19.37 $\pm$ 0.09 \\
          & 2.08692 & 3  & 1.5 $\pm$ 0.19  & 19.59 $\pm$ 0.19 & 1.38 $\pm$ 0.06 & 19.52 $\pm$ 0.06 \\ \hline
Q1451+123 & 2.25466 & 2  & 0.33 $\pm$ 0.19 & 19.25 $\pm$ 0.19 & 0.28 $\pm$ 0.2  & 19.17 $\pm$ 0.2  \\ \hline
Q2059-360 & 2.50734 & 1  & 0.37 $\pm$ 0.1  & 19.03 $\pm$ 0.1  &                 &                  \\ \hline
Q2059-360 & 3.08261 & 1  & 0.36 $\pm$ 0.12 & 19.25 $\pm$ 0.12 &                 &                  \\
          & 3.08291 & 2  & 0.74 $\pm$ 0.15 & 18.92 $\pm$ 0.15 &                 &                  \\ \hline
Q2116-358 & 1.99542 & 1  & 0.52 $\pm$ 0.16 & 19.15 $\pm$ 0.16 & 0.41 $\pm$ 0.16 & 18.98 $\pm$ 0.16 \\
          & 1.99578 & 2  & 0.68 $\pm$ 0.1  & 19.37 $\pm$ 0.1  &                 &                  \\
          & 1.99595 & 3  & 0.59 $\pm$ 0.07 & 19.35 $\pm$ 0.07 & 0.59 $\pm$ 0.06 & 19.35 $\pm$ 0.06 \\
          & 1.99615 & 4  & 0.63 $\pm$ 0.2  & 19.7 $\pm$ 0.2   & 0.48 $\pm$ 0.07 & 19.43 $\pm$ 0.07 \\
          & 1.99633 & 5  & 0.83 $\pm$ 0.07 & 19.52 $\pm$ 0.07 & 0.83 $\pm$ 0.09 & 19.53 $\pm$ 0.09 \\ \hline
Q2138-444 & 2.38279 & 5  & 0.17 $\pm$ 0.23 & 18.78 $\pm$ 0.23 & 0.48 $\pm$ 0.07 & 19.43 $\pm$ 0.07 \\
          & 2.38346 & 8  & 0.42 $\pm$ 0.36 & 19.01 $\pm$ 0.36 &                 &                  \\
          & 2.38376 & 9  & 0.71 $\pm$ 0.3  & 19.27 $\pm$ 0.3  &                 &                  \\ \hline
Q2138-444 & 2.85234 & 1  & 0.17 $\pm$ 0.17 & 19.36 $\pm$ 0.17 & 0.08 $\pm$ 0.11 & 19.29 $\pm$ 0.11 \\ \hline
Q2152+137 & 3.31558 & 1  & 0 $\pm$ 0.38    & 17.51 $\pm$ 0.38 &                 &                  \\ \hline
Q2206-199 & 1.91993 & 1  & 0.43 $\pm$ 0.36 & 19.28 $\pm$ 0.36 & 0.15 $\pm$ 0.17 & 18.98 $\pm$ 0.17 \\
          & 1.92    & 2  & 0.52 $\pm$ 0.16 & 19.69 $\pm$ 0.16 & 0.37 $\pm$ 0.05 & 19.57 $\pm$ 0.05 \\
          & 1.92019 & 3  & 0.66 $\pm$ 0.23 & 19.59 $\pm$ 0.23 & 0.47 $\pm$ 0.06 & 19.4 $\pm$ 0.06  \\
          & 1.92039 & 4  & 0.51 $\pm$ 0.24 & 19.23 $\pm$ 0.24 & 0.31 $\pm$ 0.09 & 19.02 $\pm$ 0.09 \\
          & 1.92061 & 5  & 0.46 $\pm$ 0.15 & 19.93 $\pm$ 0.15 & 0.36 $\pm$ 0.03 & 19.85 $\pm$ 0.03 \\
          & 1.92083 & 6  & 0.79 $\pm$ 0.07 & 19.84 $\pm$ 0.07 & 0.74 $\pm$ 0.07 & 19.8 $\pm$ 0.07  \\
          & 1.92091 & 7  & 0.72 $\pm$ 0.35 & 20.12 $\pm$ 0.35 & 0.19 $\pm$ 0.06 & 19.39 $\pm$ 0.06 \\ \hline
Q2230+025 & 1.86379 & 2  & 0.07 $\pm$ 0.01 & 19.49 $\pm$ 0.01 &                 &                  \\
          & 1.86427 & 4  & 0.3 $\pm$ 0.29  & 19.8 $\pm$ 0.29  &                 &                  \\ \hline
Q2231-002 & 2.06552 & 1  & 0.77 $\pm$ 0.41 & 19.99 $\pm$ 0.41 &                 &                  \\
          & 2.06615 & 2  & 1.13 $\pm$ 0.44 & 20.1 $\pm$ 0.44  &                 &                  \\ \hline
Q2243-605 & 2.32884 & 1  & 0.42 $\pm$ 0.17 & 19.13 $\pm$ 0.17 & 0.28 $\pm$ 0.05 & 18.97 $\pm$ 0.05 \\
          & 2.32917 & 2  & 0.15 $\pm$ 0.22 & 18.59 $\pm$ 0.22 &                 &                  \\
          & 2.32938 & 3  & 0.53 $\pm$ 0.18 & 19.38 $\pm$ 0.18 & 0.42 $\pm$ 0.1  & 19.25 $\pm$ 0.1  \\
          & 2.32959 & 4  & 0.3 $\pm$ 0.08  & 18.9 $\pm$ 0.08  & 0.31 $\pm$ 0.11 & 18.91 $\pm$ 0.11 \\
          & 2.32979 & 5  & 0.47 $\pm$ 0.12 & 18.75 $\pm$ 0.12 & 0.51 $\pm$ 0.04 & 18.76 $\pm$ 0.04 \\
          & 2.32999 & 6  & 0.54 $\pm$ 0.14 & 19.13 $\pm$ 0.14 & 0.5 $\pm$ 0.05  & 19.07 $\pm$ 0.05 \\
          & 2.33022 & 7  & 0.63 $\pm$ 0.25 & 18.69 $\pm$ 0.25 &                 &                  \\
          & 2.33042 & 8  & 0.47 $\pm$ 0.21 & 18.86 $\pm$ 0.21 & 0.52 $\pm$ 0.3  & 18.88 $\pm$ 0.3  \\
          & 2.33062 & 9  & 0.64 $\pm$ 0.22 & 19.61 $\pm$ 0.22 & 0.47 $\pm$ 0.02 & 19.5 $\pm$ 0.02  \\
          & 2.33093 & 11 & 0.8 $\pm$ 0.17  & 19.01 $\pm$ 0.17 &                 &                  \\ \hline
Q2332-094 & 2.28737 & 1  & 0.13 $\pm$ 0.28 & 18.32 $\pm$ 0.28 &                 &                  \\
          & 2.28749 & 2  & 0.37 $\pm$ 0.08 & 18.7 $\pm$ 0.08  & 0.98 $\pm$ 0.01 & 19.96 $\pm$ 0.01 \\ \hline
Q2332-094 & 3.05632 & 1  & 0.31 $\pm$ 0.05 & 18.9 $\pm$ 0.05  &                 &                  \\
          & 3.0569  & 4  & 0.27 $\pm$ 0.29 & 18.18 $\pm$ 0.29 &                 &                  \\
          & 3.05722 & 5  & 1.39 $\pm$ 0.05 & 19.5 $\pm$ 0.05  &                 &                  \\
          & 3.05738 & 6  & 1.13 $\pm$ 0.25 & 20.07 $\pm$ 0.25 &                 &                  \\ \hline
Q2343+125 & 2.43123 & 1  & 0.67 $\pm$ 0.19 & 19.74 $\pm$ 0.19 & 0.49 $\pm$ 0.03 & 19.59 $\pm$ 0.03 \\
          & 2.43147 & 2  & 0.51 $\pm$ 0.21 & 19.01 $\pm$ 0.21 & 0.37 $\pm$ 0.12 & 18.87 $\pm$ 0.12 \\ \hline
Q2344+125 & 2.53746 & 1  & 0.18 $\pm$ 0.07 & 17.8 $\pm$ 0.07  &                 &                  \\
          & 2.53787 & 2  & 0.13 $\pm$ 0.1  & 18.64 $\pm$ 0.1  &                 &                  \\
          & 2.5382  & 3  & 0.47 $\pm$ 0.1  & 18.47 $\pm$ 0.1  &                 &                  \\ \hline
Q2348-011 & 2.4245  & 1  & 0.98 $\pm$ 0.11 & 19.48 $\pm$ 0.11 &                 &                  \\
          & 2.42465 & 2  & 1.02 $\pm$ 0.04 & 19.57 $\pm$ 0.04 &                 &                  \\
          & 2.42621 & 6  & 0.85 $\pm$ 0.25 & 19.59 $\pm$ 0.25 &                 &                  \\
          & 2.42632 & 7  & 1.08 $\pm$ 0.19 & 19.88 $\pm$ 0.19 &                 &                  \\
          & 2.42659 & 8  & 0.63 $\pm$ 0.3  & 19.0 $\pm$ 0.3   &                 &                  \\
          & 2.42695 & 9  & 0.36 $\pm$ 0.05 & 19.34 $\pm$ 0.05 &                 &                  \\
          & 2.42724 & 10 & 0.66 $\pm$ 0.47 & 19.12 $\pm$ 0.47 &                 &                  \\ \hline
Q2348-147 & 2.27932 & 1  & 0.33 $\pm$ 0.36 & 18.55 $\pm$ 0.36 &                 &                  \\
          & 2.27941 & 2  & 0.48 $\pm$ 0.49 & 18.66 $\pm$ 0.49 &                 &                  \\ \hline
Q2359-022 & 2.09485 & 2  & 1.13 $\pm$ 0.08 & 19.94 $\pm$ 0.08 &                 &                  \\
          & 2.0951  & 3  & 0.96 $\pm$ 0.28 & 20.22 $\pm$ 0.28 & 0.75 $\pm$ 0.22 & 19.88 $\pm$ 0.22

\end{longtable}

% =====================================================

\end{appendix}

\end{document}